\documentclass[11pt]{article}
\pdfoutput=1

\usepackage{jheppub} 

\usepackage{amsmath,amssymb,epsfig,amsfonts}
\usepackage{graphicx}
\usepackage{epsf}
\usepackage{makeidx}
\usepackage{multirow}
\usepackage[all]{xy}
\usepackage{hyperref}
\usepackage{verbatim} %for comments
\usepackage{rotating}
\usepackage{xcolor}
\usepackage{multirow}
\usepackage{bm}
\usepackage{cleveref}
\usepackage{slashed}
\usepackage[normalem]{ulem}

\usepackage{graphicx}
\usepackage{latexsym,amsmath,amsfonts,amssymb}
\usepackage{mathrsfs}
\usepackage{bbm}
\usepackage{bm}
\usepackage{subfigure}
\usepackage{paralist}
\usepackage{youngtab}
\usepackage{bclogo}

\usepackage{expl3}
\ExplSyntaxOn
\int_zero_new:N \g__prg_map_int 
\ExplSyntaxOff

\usepackage{tikz}
\usepackage{diagbox}
\usetikzlibrary{positioning}
\usetikzlibrary{calc}
\usetikzlibrary{decorations.pathreplacing,calligraphy}

\usepackage{xstring}
\usetikzlibrary{decorations.pathmorphing} 
\usetikzlibrary{decorations.markings} 
\usetikzlibrary{snakes}
\usetikzlibrary{arrows} 
\usetikzlibrary{shapes} 
\usetikzlibrary{matrix} 
\usetikzlibrary{positioning} 
\usepackage[english]{babel} 
\usepackage[autostyle]{csquotes}

\tikzstyle{brane}=[draw]
\tikzset{D7/.style={circle, draw=black, inner sep=0pt, fill=white, minimum size=3mm}}
\tikzset{hasse/.style={circle, fill,inner sep=2pt}}
\tikzset{flavor/.style={regular polygon,fill=white,regular polygon sides=4,inner sep=2.5pt, draw}}
\tikzset{gauge/.style={circle, draw,inner sep=2.5pt}}
\tikzset{gaugeb/.style={circle, draw,fill=black,inner sep=2.5pt}}
\tikzset{gauger/.style={circle, draw,fill=cyan,inner sep=2.5pt}}
\tikzset{gaugeg/.style={circle, draw,fill=red,inner sep=2.5pt}}
\tikzset{bd/.style={circle, draw=black, inner sep=0pt, fill=black, minimum size=2mm}}
\tikzset{wd/.style={circle, draw=black, inner sep=0pt, fill=white, minimum size=2mm}}
\tikzset{SUd/.style={circle, draw=black, inner sep=0pt, fill=yellow, minimum size=2mm}}
\tikzset{Dynkin/.style={circle, draw=black, inner sep=0pt, fill=white, minimum size=2mm}}
\tikzstyle{ligne}=[draw, thick] 
\tikzset{doublearrow/.style={ draw=black!75, color=black!75, thick, double distance=3pt, }}

\numberwithin{equation}{section}  % make eq labels (sec.num)
\graphicspath{ {figs/} }

\newcommand{\be}{\begin{equation}}
\newcommand{\ee}{\end{equation}}
\newcommand{\ba}{\begin{aligned}}
\newcommand{\ea}{\end{aligned}}

\newcommand{\Spin}{\text{Spin}}

\def\half{{\frac{1}{2}}}

\def\unit{{1\kern-.65ex {\rm l}}}
\def\1{{1\kern-.65ex {\rm l}}}

\def\CA{{\cal A}}

\def\CD{{\cal D}}

\def\CK{{\cal K}}

\def\CN{{\cal N}}

\def\CS{{\cal S}}
\def\CT{{\cal T}}

\def\CW{{\cal W}}

\newcount\hour \newcount\minute
\hour=\time \divide \hour by 60
\minute=\time
\def\now{%
\ifnum \hour<13
  \ifnum \hour=0 \advance \hour by 12 \number\hour:\else \number\hour:\fi%
     \ifnum \minute<10 0\fi%
     \number\minute%
\ A.M.%
\else \advance \hour by -12 \number\hour:%
  \ifnum \minute<10 0\fi%
  \number\minute%
  \ P.M.%
\fi%
}

\makeatother

\def\mb{\mathbb}

\def\mc{\mathcal}
\def\bp{\begin{pmatrix}}
\def\ep{\end{pmatrix}}

\newcommand{\matarray}[1]{\begin{matrix} #1 \end{matrix}}

\newcommand{\bea}{\begin{equation} \begin{aligned}}
 \newcommand{\eea}{\end{aligned} \end{equation}}
\newcommand{\bit}{\begin{itemize}} 
\newcommand{\eit}{\end{itemize}} 

\newcommand{\Z}{\mathbb{Z}}
\newcommand{\C}{\mathbb{C}}
\newcommand{\R}{\mathbb{R}}

\renewcommand{\t}{\widetilde }

\renewcommand{\d}{\partial }

\newcommand{\m}{\mathfrak{m}}
\newcommand{\n}{\mathfrak{n}}

\newcommand{\ov}{\over}

\newcommand{\g}{{\rm g}}

\newcommand{\h}{\widehat}

\newcommand{\MG}{{\mathbf X}} %M-theory singualrity

\newcommand{\FT}{{\mathcal{T}_\MG^{\rm 5d}}} %5d THEORY
\newcommand{\KK}{D_{S^1}}

\newcommand{\FTfour}{{\mathscr{T}_\MG^{\rm 4d}}} %4d THEORY
\newcommand{\FTXfour}[1]{{\mathscr{T}_{#1}^{\rm 4d}}} %4d THEORY X specified

\newcommand{\MQ}{{\text{MQ}}} % electric quiver

\newcommand{\EQfour}{{\text{EQ}^{(4)}}} % electric quiver 4d
\newcommand{\MQfour}{{\text{MQ}^{(4)}}} % magnetic quiver 4d
\newcommand{\EQfive}{{\text{EQ}^{(5)}}} % electric quiver 5d
\newcommand{\MQfive}{{\text{MQ}^{(5)}}} % magnetic quiver 5d

\begin{document}

% format
\baselineskip=18pt  % a la harvmac
\numberwithin{equation}{section}  % make eq labels (sec.num)
\allowdisplaybreaks  % allow page breaks in displayed eqs

%%%%%%%%%%%%%%%%%%%%%%%%%%%%%%%%%%%%%%%%%%%
%%%        TITLE BEGINS HERE
%%%%%%%%%%%%%%%%%%%%%%%%%%%%%%%%%%%%%%%%%%%

%% ========== title (note version) begins here ==========
%
%\vspace*{-1cm}
%\begin{center}
% {\Large\bf Title of the Document}
%\end{center}
%\vspace*{-.5cm}
%
%% ========== title (note version) ends here ==========

%% ========== title (paper version, a la harvmac) begins here ==========

\thispagestyle{empty}

\vspace*{0.8cm} 
\begin{center}
{\huge 5d and 4d SCFTs: \\ 
\bigskip
Canonical Singularities, 
Trinions and S-Dualities}
%3d Quiverines}

 \vspace*{1.5cm}
Cyril Closset$^{\scalebox{0.7}{\bcneige} \scalebox{0.7}{\bcetoile}}$,  Simone Giacomelli$^{\scalebox{0.7}{\bcetoile}}$, Sakura Sch\"afer-Nameki$^{\scalebox{0.7}{\bcetoile}}$, Yi-Nan Wang$^{\scalebox{0.7}{\bcetoile}}$ \\

 \vspace*{0.7cm} 

{\it$^{\scalebox{0.7}{\bcetoile}}$ Mathematical Institute, University of Oxford, \\
Andrew-Wiles Building,  Woodstock Road, Oxford, OX2 6GG, UK}\\
 {\it$^{\scalebox{0.7}{\bcneige}}$  School of Mathematics, University of Birmingham,\\ 
Watson Building, Edgbaston, Birmingham B15 2TT, UK}\\
\end{center}

\vspace*{0.5cm}

\noindent
Canonical threefold singularities in M-theory and Type IIB string theory give rise to superconformal field theories (SCFTs) in 5d and 4d, respectively. In this paper, we study canonical hypersurface singularities whose resolutions contain residual terminal singularities and/or 3-cycles. We focus on a certain class of `trinion' singularities which exhibit these properties.
In Type IIB, they give rise to 4d $\mathcal{N}=2$ SCFTs that we call $D_p^b(G)$-trinions, which are marginal gaugings of three SCFTs with $G$ flavor symmetry. In order to understand the 5d  physics of these trinion singularities in M-theory, we reduce these 4d and 5d SCFTs to 3d $\mathcal{N}=4$ theories, thus determining the electric and magnetic quivers (or, more generally, quiverines). In M-theory, residual terminal singularities give rise to free sectors of massless hypermultiplets, which often are discretely gauged.  These free sectors appear as `ugly' components of the magnetic quiver of the 5d SCFT. The 3-cycles in the crepant resolution also give rise to free hypermultiplets, but their physics is more subtle, and their presence renders the magnetic quiver `bad'. We propose a way to redeem the badness of these quivers using a class $\mathcal{S}$ realization. We also discover new S-dualities between different $D_p^b(G)$-trinions. For instance, a certain $E_8$ gauging of the $E_8$ Minahan-Nemeschansky theory is S-dual to an $E_8$-shaped Lagrangian quiver SCFT. 

\newpage
%%%%%%%%%%%%%%%%%%%%%%%%%%%%%%%%%%%%%%%%%%%
%%%           TITLE ENDS HERE
%%%%%%%%%%%%%%%%%%%%%%%%%%%%%%%%%%%%%%%%%%%

\tableofcontents
%\printindex

%%%%%%%%%%%%%%%%%%%%%%%%%%%%%%%%%%%%%%%%%%%
%%%        MAIN TEXT BEGINS HERE
%%%%%%%%%%%%%%%%%%%%%%%%%%%%%%%%%%%%%%%%%%%

\section{Introduction and Overview}

Five-dimensional superconformal field theories (5d SCFTs) are intrinsically strongly-coupled. Upon mass deformations, they are often related to 5d $\CN=1$ gauge theories, that arise as IR-free low-energy effective descriptions. In this work, we consider 5d SCFTs $\FT$ constructed as the low-energy description of M-theory on $\R^{1,4}\times \MG$, where $\MG$ is a canonical threefold singularity. 
This approach was initiated in \cite{Witten:1996qb, Morrison:1996xf, Douglas:1996xp, Intriligator:1997pq} and further studied {\it e.g.} in \cite{Hayashi:2013lra,Hayashi:2014kca,DelZotto:2017pti,Xie:2017pfl,Jefferson:2017ahm,Jefferson:2018irk,Bhardwaj:2018yhy,Bhardwaj:2018vuu,Apruzzi:2018nre,Closset:2018bjz,Apruzzi:2019vpe,Apruzzi:2019opn,Apruzzi:2019enx,Bhardwaj:2019jtr,Bhardwaj:2019fzv,Bhardwaj:2019ngx,Saxena:2019wuy,Apruzzi:2019kgb,Closset:2019juk,Bhardwaj:2019xeg,Bhardwaj:2020gyu,Eckhard:2020jyr,Morrison:2020ool, Albertini:2020mdx, Bhardwaj:2020kim, vanBeest:2020kou, Hubner:2020uvb,Bhardwaj:2020phs, Bhardwaj:2020ruf,Bhardwaj:2020avz, vanBeest:2020civ}, revealing a rich landscape of 5d SCFTs. 
One of the challenges, in this approach, is to identify the precise map between the canonical threefold singularities and the 5d physics. This is best understood, when the geometry allows a complete Calabi-Yau resolution ({\it i.e.} a crepant resolution, which leaves the canonical class invariant). However, this vanilla situation is far from generic. In many instances, the canonical singularity after resolution still has remnant singularities, so-called terminal singularities, which cannot be  crepantly resolved.  It is the physics of such canonical singularities, and the study of their Higgs and extended Coulomb branches, that are the subject of this paper. 

The 5d $\CN=1$ gauge-theory description, when it exists, corresponds to a (partial) crepant resolution of the singularity \cite{Morrison:1996xf,Intriligator:1997pq}
\be\label{crepant res intro}
\pi\; : \; \t \MG \rightarrow \MG~.
\ee
By definition, a crepant resolution satisfies $K_{\t \MG} = \pi^* K_{\MG}$. Following \cite{Closset:2020scj,CSNWII}, we choose $\MG$ to be a quasi-homogeneous hypersurface singularity in $\C^4$
\be
\MG \cong \big\{  (x_1, x_2, x_3, x_4)\subset \C^4 \; \big| \; F(x_1, x_2, x_3, x_4)=0 \big\}~,
\ee
with an isolated singularity at the origin. A crepant resolution \eqref{crepant res intro} corresponds to probing the extended Coulomb branch (ECB) of $\FT$, while  a generic deformation of the singularity, denoted by $\h\MG$, corresponds to probing the quantum Higgs branch (HB).

The classical geometry of the crepant resolution \eqref{crepant res intro} has a rich and interesting structure. 
It is partially characterized by three integers $r$, $f$ and $b_3$, which correspond to the number of compact cycles within the (partially) resolved threefold
\be
{\rm dim} \, H_2(\t \MG,\R) = r +f~, \qquad  
 {\rm dim} \, H_3(\t \MG,\R) = b_3~,\qquad
 {\rm dim} \, H_4(\t \MG,\R) = r~.
\ee
In particular, $r$ is the number of exceptional divisors ({\it i.e.} compact 4-cycles), which encode the {\it rank} of the 5d SCFT $\FT$ -- by definition, the real dimension of its Coulomb branch. In addition, $f$ is the number of curves (2-cycles calibrated by the K\"ahler form) dual to non-compact divisors, giving rise to a distinguished abelian symmetry  $U(1)^f$ of $\FT$ that survives on its CB. The dimension of the ECB is $r+f$, including $f$ real mass parameters. Furthermore, in the presence of non-trivial 3-cycles (that is, for $b_3>0$) in the resolution, we have additional free hypermultiplets (in their Higgsed phase) at the point of the CB corresponding to $\t \MG$. The integers $r$, $f$ and  $b_3$ are the same for any crepant resolution of $\t\MG$ that results in a smooth geometry (see {\it e.g.} the general CB analysis in \cite{Hayashi:2014kca} and the mathematical analysis of \cite{Caibarb3}), and are therefore intrinsic to the canonical singularity.

The literature on the M-theory engineering of 5d SCFTs has mostly focussed on canonical singularities $\MG$ for which the resolved singularity $\t \MG$ is completely smooth, with its exceptional locus $\pi^{-1}(0)$ consisting of a union of smooth surfaces, and  with $b_3=0$. In this case, $f$ gives the rank of the flavor symmetry $G_H^{\rm 5d}$ of $\FT$. 
Interestingly, this is not the generic situation, since a maximal crepant resolution is generally not entirely smooth.%
\footnote{In 6d SCFTs engineered from elliptic threefolds in F-theory, these effects have been studied in \protect\cite{Arras:2016evy, Grassi:2018rva}, while in 5d and 4d SCFTs this was discussed in \protect\cite{Closset:2020scj}.} 
Various interesting phenomena can occur, including the following two:
\bit
\item The threefold $\t \MG$ has $b_3>0$.  This happens in the presence of exceptional divisors that are ruled surfaces over curves of genus greater than $1$ \cite{Caibarb3, Closset:2020scj}.
\item The threefold $\t\MG$ is not fully resolvable, {\it i.e.} it contains residual  isolated terminal singularities. 
\eit
These two phenomena can co-exist and are logically independent. In this work, we explore them in a number of examples, which are under particularly good control, following the general logic advocated in \cite{Closset:2020scj,CSNWII}. This allows us to bypass the difficulties in interpreting terminal  singularities directly in M-theory. We find that these terminal singularities are associated to free sectors in $\FT$, modulo some discrete gauging. Along the way, we also obtain various new results about certain 4d $\CN=2$ SCFTs and their 3d reductions, as we will explain momentarily.

\subsection{Terminal Singularities, Hypermultiplets and Discrete Gauging}

By definition, a terminal singularity is a canonical singularity that admits a resolution with exceptional divisors $S_i$ such that 
\be\label{Kchange}
K_{\t \MG} = \pi^* K_{\MG} + \sum a_i S_i\,,\quad \text{with}\quad  a_i>0\,, \qquad \text{for all $i$}\,,
\ee
as opposed to $a_i=0$ for all $i$ in the case of a crepant resolution. 
 It thus correspond to  $r=0$, giving us a `rank-zero' 5d theory $\FT$. If, in addition, we have $f=0$, the singularity does not admit any small resolution. More generally, terminal singularities sometimes admit (partial) small resolutions, in which case the exceptional locus of $\t\MG$ is a union of $f$ curves. 
For $r>0$, the singularity allows for $r$ many crepant blowups, however there can be a remnant terminal singularity, {\it i.e.} (\ref{Kchange}) holds true for $a_i=0$ for $i=1, \cdots, r$, and the remaining $a_i >0$.   
In either case, we would like to understand these 5d theories more systematically.
 
The simplest example of a terminal singularity with $f>0$ is the conifold, 
$\MG \cong \{ x_1^2+x_2^2+x_3^2+x_4^2=0\}$,  which has $f=1$. The corresponding 5d theory is the free massless hypermultiplet. Since the resolved conifold is smooth, the usual tools are applicable. The theory is gapped, corresponding to a massive hypermultiplet, with the massive BPS particle realized by the M2-brane wrapped on the exceptional curve. 
 
 A simple example of a terminal singularity with $f=0$ is the so-called $A_2$ singularity $\MG\cong \{ x_1^2+x_2^2+x_3^2+x_4^3=0\}$. It was recently argued that this singularity engineers a single hypermultiplet as well \cite{Closset:2020scj}. In this case, however, we do not know how to see this directly in M-theory. Instead, the conclusion follows from a chain of string theory and field theory dualities, from which we deduce that the Higgs branch of $\FT$ is $\mathbb{H}$, the same as for a single hypermultiplet.

It is tempting to conjecture that terminal singularities, however complicated, always correspond to 5d hypermultiplets,  perhaps up to a decoupled topological sector, or discrete gauging. One of our aims, in this paper, is to substantiate that expectation. For the theories and geometries studied in this paper, we make the following

\paragraph{\bf Observation 1:} 
For $\MG$ a terminal singularity, the Higgs branch of the rank-zero theory $\FT$ is always given by $\mathbb{H}^{d_H}$ or a discrete gauging thereof, where $\mathbb{H} \cong \C^2$ denotes the quaternionic plane
\be
  {\rm HB}[\FT] \cong \mathbb{H}^{d_H} \qquad \text{or} \qquad   \mathbb{H}^{d_H} /\frak{f}~.
\ee
 For hypersurface singularities, the number of hypermultiplets is given by
\be
d_H ={\mu+ f\ov 2}~,
\ee
where $\mu$  is the Milnor number of $\MG$. The discrete gauge group $\frak{f}$ is determined by the singularity itself~\cite{Closset:2020scj}. Depending on a choice of global structure, we can either have the free theory with a distinguished 0-form symmetry $\frak{f}$, or we can gauge that symmetry to obtain the quotient $\mathbb{H}^{d_H} /\frak{f}$. We will determine this quotient in a number of examples.  

{We would naturally like to conjecture that the same is true for any terminal singularity, but we leave a more detailed analysis for future work.}

\medskip
\noindent
More generally, we would like to understand the situation wherein we  have residual terminal singularities upon resolving a larger canonical singularity. We will see that, in a number of examples, the terminal singularities simply contribute  additional free sectors, up to some discrete gauge group that now acts on the interacting sector as well. 
This leads us to the second observation, which we expect to be true more generally:

\paragraph{\bf Observation 2:} 
For $\MG$ a canonical singularity with remnant terminal singularities once it is resolved, the associated rank $0$ sector of the theory is a free sector, modulo discrete gauging, {\it i.e.} the 5d theory is of the form $(\mathcal{T} \otimes \mathbb{H}^k)/\Gamma$, with $\Gamma$ a discrete group (which depends on the singularity and on a choice of global structure).

\medskip
\noindent
Before discussing the specific class of examples that we will study, let us first explain our general strategy \cite{Closset:2020scj}. In particular, let us explain how we determine the quantum Higgs branch of $\FT$.

%%%%%%%%%%%%%%%%%%%%%%%%%%%%%
\begin{figure}
\begin{center}
\begin{tikzpicture}[auto,
    %decision/.style={diamond, draw=black, thick, fill=white,
    %text width=8em, text badly centered,
    %inner sep=1pt, font=\sffamily\small},
    block_1/.style ={rectangle, draw=black, thick, fill=white,
      text width=3.5em, text centered,
      minimum height=2em},
     block_top/.style ={rectangle,  thick, fill=white,
      text width=8em, text centered,
      minimum height=2em}, 
      arrow/.style={           ->,  thick,
           shorten <=2pt, shorten >=2pt},
      line/.style ={draw, thick, -latex', shorten >=0pt}]
 \node [block_top] (top1) at (-4,0) {\underline{dim \& SUSY}};
 \node [block_top] (5d) at (-4,-1) {5d $\CN{=}1$};
 \node [block_top] (4d) at (-4,-3.5)  {4d $\CN{=}2$};
  \node [block_top] (3d) at (-4,-6)  {3d $\CN{=}4$};
 \node [block_top] (top2) at (0,0) {\underline{M-theory on $\MG$}};
 \node [block_1] (FT) at (0,-1) {$\FT$};
 \node [block_1] (KKFT) at (0,-3.5) {$\KK\FT$};
  \node [block_1] (MQfour) at (-1.5,-6) {$\MQfour$}; 
  \node [block_1] (EQfive) at (1.5,-6) {$\EQfive$};
\draw [arrow] (FT) --node[] {\small $S^1$} (KKFT);
\draw [arrow] (KKFT) --node[yshift=-2mm]  {\small $S^1_\beta$} (EQfive);    
   \node [block_top] (top3) at (7,0) {\underline{IIB on $\MG$}};
    \node [block_1] (FTfour) at (7,-3.5) {$\FTfour$};
   \node [block_1] (MQfive) at (5.5,-6) {$\MQfive$}; 
  \node [block_1] (EQfour) at (8.5,-6) {$\EQfour$};  
  \draw [arrow] (FTfour) --node[yshift=-3mm]  {{\small $S^1_{{1\ov \beta}}$}} (EQfour);
\draw [arrow] (MQfour) --node[below] {{\tiny $\cdot/U(1)_T^f$}} (EQfive);
    \draw [arrow] (EQfive.north west) to [out=165,in=15] node[above,sloped]  {{\tiny $\cdot/U(1)^f$}}  (MQfour.north east); 
\draw [arrow] (MQfive) --node[below] {{\tiny $\cdot/U(1)_T^f$}} (EQfour);
  \draw [arrow] (EQfour.north west) to [out=165,in=15] node[above,sloped]  {{\tiny $\cdot/U(1)^f$}}  (MQfive.north east); 
 \draw [arrow,<->] (EQfive) --node[above] {\small 3d mirror} (MQfive);
  \draw [arrow,<->] (EQfour.south) to [out=-165,in=-15] node[above,sloped] {\small 3d mirror} (MQfour.south);   
   \end{tikzpicture}
\caption{Summary of the relations between the supersymmetric QFTs in 5d, 4d and 3d associated to a conical threefold singularity $\MG$ in M-theory and type II string theory. The downward arrows denote dimensional reduction on circles. The 3d $\CN=4$ theories are the `quiverines', which are all related by 3d mirror symmetry and by S-type gauging of $U(1)^f$ flavor symmetries.}\label{summary table 1 Intro}
\end{center}
\end{figure}
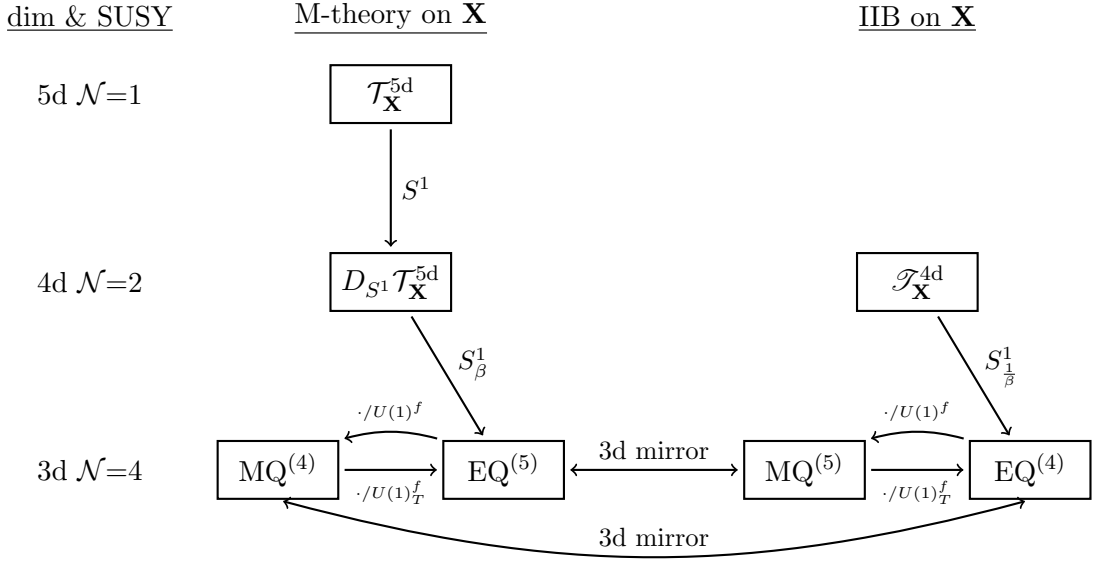
%%%%%%%%%%%%%%%%%%%%%%%%%%%%%

\subsection{Higgs Branches for 5d SCFTs from Quiver(ine)s} 

As alluded to above, deforming the singularity $\MG$ into a smooth threefold $\h\MG$ corresponds to probing the  quantum HB of the 5d SCFT $\FT$.
The precise algebraic structure of the 5d Higgs branch, let alone its hyperk\"ahler metric, is hard to determine in the M-theory engineering, however, due in part to the presence of M2-brane instantons wrapping the compact 3-cycles of the smooth threefold $\h\MG$ (the deformed singularity).  As emphasized in \cite{Ferlito:2017xdq, Cabrera:2018jxt}, a fruitful avenue to study this system is to compactify the 5d theory on a torus, in which case the 5d HB can be recovered as the 3d $\CN=4$ Coulomb branch of a so-called {\it magnetic quiver(ine)}, denoted by $\MQfive$, so that 
\be\label{HB CB rel}
{\rm HB}[\FT] \cong {\rm CB}[\MQfive]~. 
\ee
These theories $\MQfive$ can sometimes be described as quiver gauge theories in 3d,  but they are generally non-Lagrangian in our geometric setup, in which case we call them `quiverines' \cite{Closset:2020scj}.  The 3d $\CN=4$ theory $\MQfive$ is closely related to the circle reduction of a 4d $\CN=2$ SCFT, denoted by $\FTfour$, which arises in Type IIB string theory compactified on the {\it same} canonical singularity. The electric quiverine $\EQfour$ is the 3d $\CN=4$ SCFT defined as the infrared (IR) limit of $\FTfour$ compactified on a circle, and the magnetic quiverine of $\FT$ is obtained by gauging a distinguished abelian flavor symmetry of $\EQfour$ \cite{Closset:2020scj}
\be
\EQfour \stackrel{\rm 3d}{\cong}   \KK \FTfour~, \qquad \qquad 
\MQfive  \cong \EQfour \big/U(1)^f~.
\ee
This 3d gauging is an `S-operation' in the sense of \cite{Witten:2003ya}.
Essentially by construction,  $\MQfive$ is the 3d $\CN=4$ mirror of the `electric quiverine' $\EQfive$ defined as the IR limit of $\FT$ compactified on $T^2 \cong S^1 \times S^1_\beta$, where the first circle is the `M-theory circle' that controls the Type-IIA string coupling. The general setup is summarized in figure~\ref{summary table 1 Intro}. The relations between the 3d $\CN=4$ quiverines essentially follows from T-duality of Type II string theory~\cite{Hori:1997zj}.

We use the neologism {\it quiverine} for an abstract 3d $\CN=4$ (superconformal) field theory obtained from some higher-dimensional field theory upon circle compactification to 3d and after flowing to the IR. In general, the quiverines have no known Lagrangian description, which limits the usefulness of the above framework. In the following, we study a small but rich class of examples, where we understand very well both the 4d SCFT $\FTfour$ in IIB and its 3d reduction, therefore allowing us to determine the Higgs branch of $\FT$. Moreover, the 3d theories $\EQfour$ and $\MQfive$, in that case, can be described by 3d quiver gauge theories with (special) unitary gauge groups. We can then fruitfully compare our results to complementary approaches that use {\it magnetic quivers} derived from brane webs~\cite{Ferlito:2017xdq, Cabrera:2018jxt, Akhond:2020vhc, Bourget:2019aer, Bourget:2019rtl, Bourget:2020gzi, Bourget:2020asf, Bourget:2020xdz, Bourget:2020mez, Bergman:2020myx} or from the dual generalized toric polygons \cite{vanBeest:2020kou, vanBeest:2020civ}. Note that our setup implicitly assumes that we obtain a well-defined 4d SCFT in the IR of $\KK \FT$, at the origin of the 4d CB; in particular, we assume the 5d Higgs branch consists of a single cone.~\footnote{This need not be the case for generic canonical singularities. If the 5d Higgs branch consists of several cones, they can intersect the 4d CB at distinct points due to instanton corrections (see {\it e.g.} \protect\cite{Ganor:1996pc}), but we should still be able to work locally near each point from which a quantum HB emanates. We thank the JHEP referee for pointing it out.}

The relation \eqref{HB CB rel} is realized in this setup because the deformed singularity in IIB now probes the Coulomb branch of $\FTfour$. Conversely, the crepant resolution \eqref{crepant res intro} probes the Higgs branch of the 4d SCFT. In that context, the additional features of the resolution mentioned above -- $b_3>0$ and residual terminal singularities -- are completely understood \cite{Closset:2020scj}. They correspond to additional  low-energy degrees of freedom (free vector multiplets and residual SCFTs, respectively) on the quantum Higgs branch of $\FTfour$. By reducing to 3d, this gives us some additional evidence for their 5d interpretation. 

\subsection{Trinion Singularities and $D_p^b(SU(N))$ SCFTs}

We will study in detail the following families of canonical singularities, which we dub `trinion singularities'
\be\label{hypereq intro}
 \MG \; : \; \; \; F(x, z_1, z_2, z_3)=x^N+x^{N-b_1} z_1^{p_1}+x^{N-b_2} z_2^{p_2}+x^{N-b_3} z_3^{p_3}+ \cdots=0~,
 \ee
 where $p_\alpha$ ($\alpha=1,2,3)$ are positive integers and $b_\alpha \in \{ N, N-1\}$,  
satisfying the condition
\be\label{beta function intro}
\frac{b_1}{p_1}+\frac{b_2}{p_2}+\frac{b_3}{p_3}=N~.
\ee 
The ellipsis in \eqref{hypereq intro} stands for additional terms that are necessary in order to fully specify the isolated singularity whenever at least two of the $b_\alpha$'s are equal to $N-1$ \cite{Kreuzer:1992np, Davenport:2016ggc, CSNWII}. In the case when $b_1=b_2=b_3=N$, the singularity
\be
x^N+ z_1^{p_1}+ z_2^{p_2}+ z_3^{p_3}=0
\ee
 was studied in \cite{DelZotto:2015rca} from the IIB engineering point of view. More generally, for any allowed choice of the $b_\alpha$'s, the 4d SCFT $\FTfour$ associated to \eqref{hypereq intro} takes the form
\be\label{FTfour intro}
 \begin{tikzpicture}[x=.6cm,y=.6cm]
 \node at (-8.5,0.5) {$\FTfour \; \cong$};
\node at (-4,0) {$D_{p_1}^{b_1}(SU(N))$};
\node at (0,0) {$SU(N)$};
\node at (4,0) {$D_{p_3}^{b_3}(SU(N))$};
\node at (0,2) {$D_{p_2}^{b_2}(SU(N))$};
\draw[ligne, black](-2,0)--(-1.3,0);
\draw[ligne, black](1.3,0)--(2,0);
\draw[ligne, black](0,0.7)--(0,1.4);
\end{tikzpicture}
\ee
This theory consists of three $D_p^b(SU(N))$ theories coupled together by gauging their common $SU(N)$ global symmetry.  The 4d SCFTs $D_p^b(G)$, for $G$ an $ADE$ gauge group, were introduced in~\cite{Giacomelli:2017ckh}, generalizing the $D_p(G)$ theories studied in \cite{Cecotti:2012jx, Cecotti:2013lda}. By construction, their flavor symmetry group is (at least) $G$.  For $G=SU(N)$, the 4d SCFT $D_p^b(SU(N))$ corresponds to a Seiberg-Witten (SW) curve\footnote{More precisely, we may formulate the SW geometry in terms of a 2d $\CN=(2,2)$ Landau-Ginzburg (LG) superpotential $W= x^N+ x^{N-b} z^p$, with $(x, z)\in \C\times \C^\ast$.} 
\be
x^N+ x^{N-b} z^p=0 \,.
\ee
One can then gauge the $SU(N)$ flavor symmetries of three distinct $D_p^b(SU(N))$ theories, with the condition \eqref{beta function intro} ensuring the vanishing of the beta function, thus preserving conformal invariance and introducing a marginal gauge coupling.

We will argue that the 3d reduction of the $D_p^b(SU(N))$ theory always admits a Lagrangian description as a linear quiver with unitary and special unitary gauge groups. Upon gauging the central $SU(N)$ as in \eqref{FTfour intro}, this gives us a useful description of the electric quiverine $\EQfour$ as a star-shaped quiver with three legs, with $SU(N)$ gauge groups at the central node. We can then obtain the magnetic quiver $\MQfive$ by gauging a $U(1)^f$ flavor symmetry. In general, one needs to carefully pick the global structure of the gauge group of the quiver $\EQfour$ (and $\MQfive$) to match the non-trivial one-form symmetries of $\FTfour$ \cite{Closset:2020scj,DelZotto:2020esg}. Correspondingly, in 5d, this is related to a choice of discrete 0-form or 3-form symmetry of $\FT$.

In the process of deriving the 3d Lagrangians for $D_p^b(SU(N))$ theories, we also introduce a new family of 4d $\mathcal{N}=2$ SCFTs with $SU(N)\times SU(K)\times U(1)$ global symmetry which we call $\mathcal{D}^b_s(N,K)$ with $s=p/\text{gcd}(p,b)$. These emerge at cusps on the conformal manifold of  $D_p^b(SU(N))$ theories and play a crucial role in deriving the linear quivers in 3d.

\subsection{Trinions, Class $\CS$ Theories and Generalized Toric Polygons}

The trinions we introduce to describe the dimensional reduction of $\FTfour$ result in Lagrangian 3d $\CN=4$ theories which are not always `good theories' in the Gaiotto-Witten sense \cite{Gaiotto:2008ak}. The 3d quivers have $(S)U(N)$ gauge nodes. In particular, for `ugly' quivers or for `bad' quivers (whose nodes are underbalanced), the resulting CB will not be conical, and thus will not correspond immediately to a 3d SCFT. For ugly quivers, where the underbalance is $-1$, we can apply an infrared duality to rebalance the nodes, thus `factoring out' a free sector \cite{Gaiotto:2008ak}.
If the quiver is bad, on the other hand, the theory needs some additional interpretation.

Given the $\EQfour$ so obtained, we find the magnetic quiver $\MQfive$ for the 5d theory by gauging the $U(1)^f$ symmetry. This essentially turns the $SU(n)$ factors in the legs to $U(n)$ factors, which can at most introduce additional ugly nodes.

 For our class of trinion singularities,  we can connect the properties of the magnetic quivers $\MQfive$ to the features of the crepant resolution $\t\MG$.  We make the following: 

\paragraph{\bf Observation 3:} 
If $b_3>0$,  then the resulting  $\MQfive$  is bad. If $b_3=0$ and the geometry is fully resolvable, then the quiver is good. Finally, in the case with $b_3=0$ and with residual terminal singularities, the quiver is ugly.

\medskip
\noindent
We will argue that, given a bad magnetic quivers (corresponding to a singularity with $b_3>0$), the 5d physics can be extracted by considering the most singular locus on the Coulomb branch, using the CB moduli space analysis of \cite{Assel:2017jgo}. 
We will also give a complementary perspective on these 3d quivers by interpreting them as  3d mirror duals of 4d class $\CS$ theories that are trinions (spheres with three regular punctures). The class $\CS$ theories themselves are the 4d reduction of 5d SCFTs engineered from  trinion-shaped $(p,q)$-brane webs, which can be efficiently described by generalized toric polygons (GTP). The specific GTPs are obtained by taking the toric diagram for $T_N$ and including a suitable number of white dots \cite{Benini:2009gi,Eckhard:2020jyr}.
These class-$\CS$/GTP trinions are fully characterized by three partition of $N$: in the GTP, the partitions encode the separation between black dots along the edges of the  polygon, which equivalently encodes the distribution of 5-/7-branes in the brane-web realization. The partitions also determine the class $\CS$ punctures and  the ranks of the $U(n_i)$ gauge nodes in the 3d quiver \cite{Gaiotto:2009we, Chacaltana:2010ks}.

The bad quivers in 3d can be `redeemed' in the following way: they have an alternative interpretation as the higgsing of a theory that is a good, by a process discussed in~\cite{Gaiotto:2012uq}. 
The `unhiggsing' of the bad quivers will be described both in the 4d class $\CS$ setting and in the 5d GTP formalism. Starting from the unhiggsed theory, there are alternative ways  to higgs back to the theory of interest, and this produces several good daughter (higgsed) theories. The badness of the original quiver reflects this fact, and in this sense a bad theory can be thought of as a collection of good theories. We give a prescription  to identify the good theory in the family with the correct 3d magnetic quiver whose CB corresponds to the HB of the 5d theory~$\FT$.

\subsection{Trinions of Type $G=D, E$ and New S-dualities} 

In section \ref{sec7:DandE} and \ref{sec:dualities}, we discuss  trinions in 4d built by gauging together three $D_p^b(G)$ theories as in as in \eqref{FTfour intro}, but with the central node $G$ a special orthogonal or exceptional gauge group. We consider SCFTs for which the $G$ beta-function vanishes. In general, these theories can only be described by a Landau-Ginzburg (LG) model with five fields, which does not correspond to a threefold singularity \cite{Cecotti:2010fi, DelZotto:2015rca}. If however the superpotential is such that at least one of the five fields is massive, we can integrate it out and obtain a threefold hypersurface singularity in $\mathbb{C}^4$, in which case the corresponding 4d SCFT can be engineered by compactifying Type IIB on that singularity.

In this work, we also observe that the Type IIB geometric engineering of 4d SCFTs can be used to discover new S-dualities.  This can be understood as follows. From our database, we find that, in several cases, two trinions with different $G$, which {\it a priori} are distinct SCFTs, are actually described by the same hypersurface singularity (or LG model). In such a situation, the two theories are exactly equivalent -- they correspond to two different-looking descriptions of the same SCFT. Since the gauge groups are different in the two descriptions, we interpret these as nontrivial examples of S-dualities, with the gauge groups in each description becoming weakly coupled at different corners of the conformal manifold as in \cite{Argyres:2007cn, Gaiotto:2009we}. 

We should stress that the dualities we find are extremely nontrivial from a field-theoretic perspective. The most popular tool we have to understand S-dualities is the pair-of-pants decomposition of a Riemann surface in the class $\mathcal{S}$ description  \cite{Gaiotto:2009we}. This however cannot be used to study the examples we will discuss in this paper since the relevant trinion SCFTs, which include special unitary quivers with exceptional shape, are not class $\mathcal{S}$ theories in general. The only other known method to probe S-dualities is the study of the degeneration limits of the Seiberg-Witten (SW) curve as in \cite{Argyres:2007cn} (see also \cite{Buican:2014hfa,  Buican:2016arp, Buican:2017fiq}). This route is not viable here since the curve for quivers with exceptional shape is extremely involved~\cite{Nekrasov:2012xe}. Incidentally, we will also find dualities for orthosymplectic quivers with trivalent nodes, for which the curve is not known at all. 
Nevertheless, the Type IIB (or, equivalently LG) framework renders the S-dualities obvious. We leave a more detailed study of these $D_p^b(G)$ trinions and of their dualities for future work.

\medskip
\medskip
\noindent
The plan of this paper is as follows: in section \ref{sec:Dpbs} we discuss the 4d SCFTs of type $D_p^b(G)$ and their description in terms of trinion singularities as well.  We discuss in detail their 3d reduction. This general analysis is then applied to a large class of examples: in section \ref{sec:Rank0}, we first discuss a class of rank 0 theories. In section \ref{sec:Higher-rank-En}, we study their generalization to higher-rank theories which are close relatives to the higher-rank E-strings. We discuss in detail the interpretation of the terminal singularities in this class of models. 
% and we discuss the role of the terminal singularities and their effect on the Higgs branch and 5d/4d/3d trinion theories.
In section \ref{subsec:classS}, we provide an alternative point of view on the ugly/bad quivers that arise from these trinions in 3d, and propose a way to redeem these theories. A full classification of $D_p^b(SU(N))$ trinions is provided in section \ref{sec:SUN}. For other groups $G= D, E$,  we discuss numerous examples of trinions in section \ref{sec7:DandE}. Finally, in section \ref{sec:dualities}, we discuss non-trivial S-dualities for these $D_p^b(G)$ trinions. 
The appendices contain details on the geometric resolutions of canonical singularities, a proof of the $D_p^b(SU(N))$-classification, as well as a list of $D_p^b(G)$-trinions for $G=D, E$.

%%%%%%%%%%%
\section{4d $\CN=2$ SCFTs $D_p^b(G)$, Trinion Singularities, and 3d Reduction}
\label{sec:Dpbs}

The main protagonists of this  section are the 4d $\CN=2$ SCFTs $D_p^b(G)$ \cite{Giacomelli:2017ckh}, their gauging with $\CN=2$ vector multiplets for the ADE group $G$,  the associated canonical singularities in IIB, and the reduction of these theories to 3d.  It is this class of theories that will play a key role in the following sections,  where we will determine the 5d SCFTs obtained from M-theory on the same singularities.

We will use the following abbreviation in writing the quivers, in 4d and 3d, where the round nodes are gauge nodes while the square node corresponds to flavor nodes:
\bea\label{QuiverDots}
 \begin{tikzpicture}[x=.7cm,y=.7cm]
\node[bd] at (0,0) [label=below:{{\scriptsize$n$}}] {};
\node[] at (3.5,0) {$U(n)$};
\node[SUd] at (0,-1) [label=below:{{\scriptsize$n$}}] {};
\node[] at (3.5,-1) {$SU(n)$};
\end{tikzpicture} \qquad\qquad \qquad
 \begin{tikzpicture}[x=.7cm,y=.7cm]
\node[SUd] at (0,-2) [label=below:{{\scriptsize$n/\mathbb{Z}_k$}}] {};
\node[] at (4,-2) {$SU(n)/\mathbb{Z}_k$};
\node[flavor] at (0,-3.3) [label=below:{{\scriptsize$n$}}] {};
\node[] at (4,-3.3) {Flavor $SU(n)$};
\end{tikzpicture} 
\eea

\subsection{Canonical Singularities and 4d SCFTs}

Consider an isolated canonical singularity $\MG$ defined by a quasi-homogeneous polynomial equation
\be\label{F gen 4d}
F(x, z_1, z_2, z_3)=0~.
\ee
The family of smooth local Calabi-Yau threefolds $\{ \h F =0\}$ obtained by deforming the singularity gives us the Seiberg-Witten geometry of the 4d  $\CN=2$ SCFT $\FTfour$, as engineered in Type IIB string theory (see {\it e.g.}  \cite{Shapere:1999xr, Cecotti:2010fi, DelZotto:2015rca, Xie:2015rpa}). We have
\be\label{F def 4d}
\h F= F +\sum_{l=1}^\mu t_l \, x^{\m_l}=0~, \quad \qquad \Omega_3 = {d x \wedge d z_1 \wedge dz_2 \wedge dz_3\ov d\h F}~,
\ee
where the deformation monomials $x^{\m_l}$ are generators of the Milnor ring of the singularity, i.e. 
$\mathcal{M} = \mathbb{C} [x, z_1, z_2, z_3]/\langle d \hat{F} \rangle$,
 and $\Omega_3$ is the holomorphic 3-form. The condition that $\Omega_3$ have scaling dimension $1$ fixes the scaling dimensions of the variables $x, z_i$, and therefore of the parameters $t_l$ in \eqref{F def 4d}. In particular, the deformation parameters with scaling dimensions $\Delta[t_l]>1$ correspond to the Coulomb branch parameters of the 4d $\CN=2$ SCFT, while the ones with $\Delta[t_l]=1$ correspond to mass terms (coupling to the moment map operators in conserved-current multiplets). We denote by $\h r$ the rank of $\FTfour$ and by $f$ the rank of its flavor symmetry group. Note that $\mu= 2 \h r+f$.  Much of what is summarized here was described in more detail in \cite{Closset:2020scj} and will receive an in depth treatment in \cite{CSNWII}.

For the most parts of this paper, we are chiefly interested in singularities \eqref{F gen 4d} with scaling dimensions
\be
\Delta[x]= 1~, \qquad \Delta[z_\alpha]= {b_\alpha \ov p_\alpha}~,\; \; \alpha=1,2,3~, \qquad\quad \Delta[F]=N~,
\ee
with $b_\alpha\in \{N, N-1\}$, $\forall \alpha$, and for some positive integers $p_\alpha$ and $N$. This spectrum is realized by the singularities \eqref{hypereq intro}, namely
\be
x^N+x^{N-b_1} z_1^{p_1}+x^{N-b_2} z_2^{p_2}+x^{N-b_3} z_3^{p_3}+ \cdots=0~,
\ee
with the condition \eqref{beta function intro} being equivalent to $\Delta[\Omega_3]=1$. Note that the ``$\cdots$'' terms contain necessary marginal deformations to make the singularity isolated. The Milnor ring of this singularity always includes the terms $x^{N-2}, x^{N-3}, \cdots, 1$, and the CB spectrum $\{\Delta\}$ of $\FTfour$ thus contains the dimensions
\be
\{2, 3, \cdots, N\}  \subset \{\Delta\}~.
\ee
We then have at least one marginal coupling, and the spectrum is consistent with $\FTfour$ having a weakly-coupled $SU(N)$ on its conformal manifold. Indeed, we can give such a description as a marginal gauging of three $D_p^b(SU(N))$ theories by this $SU(N)$ \cite{Cecotti:2012jx, Cecotti:2013lda, Giacomelli:2017ckh}. In the following, we review these $D_p^b(SU(N))$ theories and we study their circle reduction to 3d.

 \subsection{$D^b_p(G)$: The Executive Summary}
 \label{subsec:DpSummary}
 
One can engineer an infinite family of 4d $\mathcal{N}=2$ SCFTs with an ADE global symmetry by compactifying Type IIB string theory on threefold hypersurface singularities in $\mathbb{C}^3\times \mathbb{C}^*$
\be 
F(x_1,x_2,x_3,z)=0~,
\ee 
with $(x_1, x_2, x_3)\in \C^3$ and $z \in \C^*$.   The relevant singularities, describing an ADE singularity fibered over $\mathbb{C}^*$, are shown in table~\ref{sing}.  We will be mainly interested in the case $A_{N-1}= SU(N)$ in the following, but we shall also use the other cases in section~ \ref{sec7:DandE}.

\begin{table}
$$\begin{array}{|c|c|c|}
\hline
G &  F & b \\
\hline \hline 
SU(N) & x_1^2+x_2^2+x_3^N+z^p& N \\
\hline
& x_1^2+x_2^2+x_3^N+x_3z^p& N-1 \\
\hline \hline 
SO(2N) & x_1^2+x_2^{N-1}+x_2x_3^2+z^p& 2N-2 \\
\hline 
 & x_1^2+x_2^{N-1}+x_2x_3^2+x_3z^p& N \\
\hline
\end{array}
\qquad\qquad
\begin{array}{|c|c|c|}
\hline
G &  F & b \\
\hline \hline 
E_6 & x_1^2+x_2^3+x_3^4+z^p& 12 \\
\hline
 & x_1^2+x_2^3+x_3^4+x_3z^p& 9 \\
\hline
 & x_1^2+x_2^3+x_3^4+x_2z^p& 8 \\ 
\hline \hline 
E_7 & x_1^2+x_2^3+x_2x_3^3+z^p& 18 \\
\hline
 & x_1^2+x_2^3+x_2x_3^3+x_3z^p& 14 \\ 
\hline \hline 
E_8 & x_1^2+x_2^3+x_3^5+z^p& 30 \\
\hline 
 & x_1^2+x_2^3+x_3^5+x_3z^p& 24 \\
\hline 
 & x_1^2+x_2^3+x_3^5+x_2z^p& 20 \\
\hline
\end{array}
$$
\caption{Hypersurface singularities ($F=0$) realizing the $D_p^b(G)$ theories in Type IIB string theory, with all the allowed values of $b$ for each $G$.}\label{sing}
\end{table}

These models are indexed by the $ADE$ group $G$ (which is, in general, a subgroup of the full flavor symmetry group), a positive integer $p$, and the parameter $b$, which must take one of the values shown in table~\ref{sing}.

Upon deforming the singularity, similarly to \eqref{F def 4d}, we have a smooth threefold with the holomorphic 3-form (note that this is different from \ref{F def 4d})
\be\label{holo}
\Omega_3=\frac{dx_1 \wedge dx_2  \wedge dx_3 \wedge dz}{z d\h F}~.
\ee 
By construction, the deformation ring of the singularity $F=0$ contains a $G=$ADE spectrum, which is now interpreted as mass parameters for the global symmetry $G$ (in particular, they are not paired to deformation parameters of dimension $2-\Delta$, unlike the actual CB parameters). The other deformation parameters with $\Delta >1$ give us the Coulomb branch spectrum of the $D_p^b(G)$ theory, and the corresponding coupling constants arise from deformations of dimensions $\Delta'=2-\Delta <1$. In particular, a coupling constant of dimension $\Delta'=0$ is a marginal coupling. Let us also denote by $f_0$ the number of parameters with $\Delta=1$, corresponding to additional conserved currents (in addition to $G$). Therefore, the rank of the full flavor symmetry, $G_F$, is given by
 \be\label{rankglobal} 
 \text{rk}(G_F)=\text{rk}(G)+f_0~.
\ee 
These SCFTs are denoted by $D_p^b(G)$ \cite{Giacomelli:2017ckh}. The special case with $b= h^\vee(G)$ (the dual Coxeter number of $G$) is often denoted by $D_p(G)$ \cite{Cecotti:2012jx, Cecotti:2013lda}.
The rank $\h r$ of $D_p^b(G)$ can be deduced from
\be
\mu=2\h r+\text{rk}(G_F) =p\frac{\text{rk}(G)  h^{\vee}(G)}{b}~,
\ee
where $\mu$ is the dimension of the ring ${\mb{C}[x_1,x_2,x_3,z]}/{\langle\d_{x_1} F,\d_{x_2} F,\d_{x_3} F, z \d_z F \rangle}$. From the spectrum of CB operators, we may also compute the central charges $a$ and $c$, using the relations~\cite{Shapere:2008zf}
\be
2a - c={1\ov 4}\sum_i (2\Delta_i -1)~,
\ee
and  \cite{Giacomelli:2017ckh}
\be\label{centralc2} 
c=\frac{\text{rk}(G)}{12b}\left(p\, h^{\vee}(G)-b)(h^{\vee}(G)+1\right)-\frac{f_0}{12}~.
\ee 
Another important quantity is the flavor central charge for $G$, which reads:
\be\label{flavorcc} 
k_G=2\left( h^{\vee}(G)-\frac{b}{p}\right)~.
\ee 
Upon gauging the flavor symmetry $G$, this coefficient enters the gauge-coupling beta function.%
\footnote{Recall that, upon gauging a common $G$ symmetry of several SCFTs $\CT_k$, conformal invariance requires $2h^\vee(G) -\half \sum_k k_G^{\CT_k}=0$.}

\subsection{Reduction to Lagrangian Theories for $G=SU(N)$} 
Let us now consider $D_p^b(SU(N))$ on a finite-size circle, and flow to the infrared, where we expect to find a 3d $\CN=4$ fixed point. We will argue that this 3d theory can be described as the IR fixed point of a 3d $\CN=4$  quiver  gauge theory with (special) unitary gauge groups. In this sense, the 3d reduction of $D_p^b(SU(N))$ is a `Lagrangian theory'.%
\footnote{In some instances, the 3d Lagrangian can be also obtained from a 4d $\mathcal{N}=1$ Lagrangian  \protect\cite{Benvenuti:2017bpg, Benvenuti:2017kud}.
 % Do not remove this.
 } 
These 3d theories are also studied in  \cite{Giacomelli:2020ryy}.%

Let us start by stating the main result. The linear quiver describing the 3d reduction of $D_p^b(SU(N))$  takes the  form
\be\label{quiver3d}
\KK D_p^b(SU(N))=
\begin{cases}\;\; \boxed{N}-(n_1)-(n_2)-\dots-(n_\ell) \qquad &\text{if} \; b=N\; \text{and}\; {p\ov N}\notin \Z~,\\
\;\;\boxed{N}-(n_1)-(n_2)-\dots-(n_\ell)-\boxed{1} \qquad\quad &\text{otherwise.} 
\end{cases}
\ee 
The  horizontal lines denote hypermultiplets in bifundamental representations, the boxes denote flavor nodes, and the gauge groups $(n_k)$ are either unitary or special unitary, with
\be\label{nk def}
n_k= \left\lfloor N-{k b \ov p}\right\rfloor~, \qquad
 (n_k)= \begin{cases} U(n_k) \qquad &\text{if} \;  {kb\ov p}\notin \Z~,\\
SU(n_k) \qquad &\text{if} \;  {k b \ov p}\in \Z~,
\end{cases}
\ee
where $\lfloor\;\rfloor$ is the floor function. The length of the quiver is given by
\be\label{ell def}
\ell = \begin{cases}
p - 1 - {p\ov N}  \qquad &\text{if} \;  b=N \; \text{and}\;  {p\ov N}   \in  \Z~,\\
   \left\lfloor p  - {p\ov N} \right\rfloor    \qquad &\text{if} \;  b=N \; \text{and}\;  {p\ov N}   \notin  \Z~,\\
   p - 1  \qquad &\text{if} \;  b=N-1~.
\end{cases}
\ee
The $SU(N)$ symmetry of $D_p^b(SU(N))$ is obviously realized as the flavor group on the left-hand tail of \eqref{quiver3d}. When the quiver only contains $U(n)$ factors (without the extra fundamental on the right-hand tail in \eqref{quiver3d}, which can be viewed as an $SU(1)$ factor), we have $f_0=0$ in \eqref{rankglobal}.  
The $f_0$ additional conserved currents in $D_p^b(SU(N))$ correspond to the $SU(n)$ factors in the 3d quiver, including the $SU(1)$ tail.

To derive this result, we will embed the 4d SCFT in an asymptotically-free Lagrangian model, viewing $D_p^b(SU(N))$ as an Argyres-Douglas (AD) point -- {\it i.e.} as the effective low-energy theory at a special point on the Coulomb branch of the Lagrangian theory. We then dimensionally reduce the Lagrangian theory while taking the scaling limit which brings us to the AD point.  
We first carry out this procedure for the simpler case of  the $D_2(SU(N))$ theories, arguing that the dimensional reduction gives us 3d $\mathcal{N}=4$ SQCD -- that is, we have:
\be\label{D2N red}
\KK D_2(SU(N))=
\begin{cases}\;\; \boxed{2n}-SU(n) \qquad &\text{if} \; N=2n~,\\
\;\;  \boxed{2n-1}- U(n-1)\qquad &\text{if} \; N=2n-1~,
\end{cases}
\ee
where $\KK$ denotes the circle-reduction, which is obviously a special case of \eqref{quiver3d}. The case $N=2n$, $SU(n)$ with $2n$ flavors, is already a Lagrangian superconformal theory in 4d, while the case $N=2n-1$ is a non-trivial result.
 We then generalize the construction to all other $D_p^b(SU(N))$ models.  Note that one can also check our result against other constructions, in a number of special cases. For instance, the result for $D_2(SU(2n-1))$  also follows from the known 3d mirror dual \cite{Xie:2016uqq}. In the case $D_p(SU(2))$, which is the ordinary $D_p$ AD theory, the 3d reduction similarly follow from the known 3d mirror when $p$ is even \cite{Xie:2012hs}. We will discuss the $D_p(SU(2))$ example in more detail momentarily.

\subsubsection{Dimensionally-reduced $D_2(SU(N))$ Theories}\label{subsec: D2N deriv}
The superconformal theories $D^b_p(SU(N))$ can be realized as low-energy effective theories at singular points on the Coulomb branch of  4d Lagrangian theories. The relevant models are linear quivers of special unitary gauge groups. The simplest non-trivial case is that of the $D_2(SU(2n-1))$ theories, which can be embedded in $SU(n)$ SQCD with $N_f= 2n-1$ flavors. The SW curve of the Lagrangian theory can be written as
\be 
y^2=P(x)^2-4\Lambda\prod_{i=1}^{N_f}\left(x-m_i+\frac{\Lambda}{n}\right)~,
\ee  
where $P(x)=x^n+\sum_{i=2}^{n}u_ix^{n-i}$. If we set all the masses $m_i$ to be equal to $\frac{\Lambda}{n}-m$, and tune $P(x)$ to $P(x)=(x-(n-1)m)(x+m)^{n-1}$, the curve degenerates and the low-energy theory is $U(n-1)$ with $2n-1$ flavors. This is true for all values of $m$ except for $m=0$, where the curve degenerates further and the low-energy theory is $D_2(SU(2n-1))$. 

What we have just described is a singular locus of the parameter space of SQCD on which the effective low energy theory  is $U(n-1)$ with $N_f=2n-1$. Furthermore, there exists a special point on that locus, where more degrees of freedom become massless, reconstructing the spectrum of $D_2(SU(2n-1))$. From the perspective of the UV theory, this special point corresponds to the origin of the Coulomb branch (meaning $P(x)=x^n$) and requires setting 
\be\label{condm} 
m_i=\frac{\Lambda}{n}~, \quad \forall i \,.
\ee 
Such tuning of the flavor masses is familiar in AD theories; moreover, in general, a non-trivial quantum Higgs branch arises at the AD point \cite{Argyres:2012fu}.

Next, let us compactify the theory on a circle of radius $\t\beta$ and reduce the theory to 3d.
The discussion of the previous paragraph shows that in 4d the Lagrangian theory is well approximated by $D_2(SU(2n-1))$ at the origin of the Coulomb branch and at energies $E\ll \Lambda$, provided that \eqref{condm} is satisfied. We therefore want to take the 3d limit in such a way that the 4d RG scale remains finite. Since the 3d and 4d gauge coupling constants are related by 
$g_{4}^2=2\pi \t\beta\,  g_{3}^2$, we should keep  the following combination finite
\be
\Lambda = e^{-{4\pi\ov \t\beta g_3^2}}~
\ee 
in the 3d limit, that is, the gauge coupling in 3d should be sent to infinity as $g_3^2 \sim 1/\t\beta$ in the limit $\t\beta \rightarrow 0$. In particular, we should study the behaviour of the 3d $SU(n)$ theory with $N_f=2n-1$ in the infrared, at the origin of the Coulomb branch,  but with non-zero flavor mass terms.

\begin{figure}[t]
\begin{center}
\begin{tikzpicture}[x=.9cm,y=.9cm]
\node [] at (8,0) {\large $\sim$};
\draw [ligne] (0.1, 0) -- (0.9,0) ;
\draw [ligne] (1.1, 0) -- (1.75,0) ;
\draw [ligne] (2.25, 0) -- (2.9,0) ;
\draw [ligne] (3.1, 0) -- (3.9,0) ;
\draw [ligne] (4.1, 0) -- (4.75,0) ;
\draw [ligne] (5.25, 0) -- (5.9,0) ;
\draw [ligne] (3, 0.1) -- (3,0.9) ;
\draw [ligne] (6.1, 0) -- (6.9,0) ;
\draw [ligne] (4,0.1) -- (4,0.9) ;
\draw [ligne] (3.1, 1) -- (3.9,1) ;
\draw [ligne] (9.1, 0) -- (9.9,0) ;
\draw [ligne] (10.1, 0) -- (10.75,0) ;
\draw [ligne] (11.25, 0) -- (11.9,0) ;
\draw [ligne] (12.1, 0) -- (12.9,0) ;
\draw [ligne] (13.1, 0) -- (13.75,0) ;
\draw [ligne] (14.25, 0) -- (14.9,0) ;
\draw [ligne] (12, 0.1) -- (12,0.9) ;
\draw [ligne] (15.1, 0) -- (15.9,0) ;
\draw [ligne] (13,0.1) -- (13,0.9) ;
\draw [ligne] (11.1, 1) -- (11.9,1) ;
\node[bd] at (0,0) [label=below:{{\scriptsize 1}}]{};
\node[bd] at (1,0) [label=below:{{\scriptsize 2}}]{};
\node[] at (2,0) {\dots};
\node[bd] at (3,0) [label=below:{{\scriptsize $n-1$}}]{};
\node[bd] at (4,0) [label=below:{{\scriptsize $n-1$}}]{};
\node [] at (5,0) {\dots};
\node[bd] at (6,0) [label=below:{{\scriptsize 2}}]{};
\node[bd] at (7,0) [label=below:{{\scriptsize 1}}]{};
\node[bd] at (3,1) [label=above:{{\scriptsize 1}}]{};
\node[bd] at (4,1) [label=above:{{\scriptsize 1}}]{};

\node[bd] at (9,0) [label=below:{{\scriptsize 1}}]{};
\node[bd] at (10,0) [label=below:{{\scriptsize 2}}]{};
\node[] at (11,0) {\dots};
\node[bd] at (12,0) [label=below:{{\scriptsize $n-1$}}]{};
\node[bd] at (13,0) [label=below:{{\scriptsize $n-1$}}]{};
\node [] at (14,0) {\dots};
\node[bd] at (15,0) [label=below:{{\scriptsize 2}}]{};
\node[bd] at (16,0) [label=below:{{\scriptsize 1}}]{};
\node[flavor] at (13,1) [label=above:{{\scriptsize 1}}]{};
\node[flavor] at (11,1) [label=above:{{\scriptsize 1}}]{};
\node[bd] at (12,1) [label=above:{{\scriptsize 1}}]{};

\end{tikzpicture} 
\end{center}
\caption{The 3d mirror of $SU(n)$ with $N_f=2n-1$. All nodes are unitary and therefore we should ungauge an overall $U(1)$. We choose to ungauge one of the abelian nodes above.  
 This mirror quiver was obtained in \cite{Bourget:2019rtl} and can be derived {\it e.g.} by starting from the 3d mirror of $SU(n)$ with $2n$ flavors and turning on a $SU(2n-1)$-preserving Fayet-Iliopoulos (FI) deformation.}\label{massless}
\end{figure}
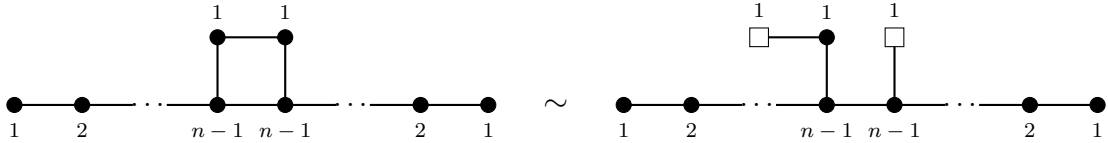

\begin{figure}[t]
\begin{center}
\begin{tikzpicture}[x=.9cm,y=.9cm]
\draw [ligne] (0.1, 0) -- (0.9,0) ;
\draw [ligne] (1.1, 0) -- (1.75,0) ;
\draw [ligne] (2.25, 0) -- (2.9,0) ;
\draw [ligne] (3.1, 0) -- (3.9,0) ;
\draw [ligne] (4.1, 0) -- (4.75,0) ;
\draw [ligne] (5.25, 0) -- (5.9,0) ;
\draw [ligne] (3, 0.1) -- (3,0.9) ;
\draw [ligne] (6.1, 0) -- (6.9,0) ;
\draw [ligne] (4,0.1) -- (4,0.9) ;
\draw [ligne] (2.1, 1) -- (2.9,1) ;
\draw[->, thick] (7.5,0)--(8.5,0);

\draw [ligne] (9.1, 0) -- (9.9,0) ;
\draw [ligne] (10.1, 0) -- (10.75,0) ;
\draw [ligne] (11.25, 0) -- (11.9,0) ;
\draw [ligne] (12.1, 0) -- (12.9,0) ;
\draw [ligne] (13.1, 0) -- (13.75,0) ;
\draw [ligne] (14.25, 0) -- (14.9,0) ;
\draw [ligne] (12, 0.1) -- (12,0.9) ;
\draw [ligne] (15.1, 0) -- (15.9,0) ;
\draw [ligne] (13,0.1) -- (13,0.9) ;
\node[bd] at (0,0) [label=below:{{\scriptsize 1}}]{};
\node[bd] at (1,0) [label=below:{{\scriptsize 2}}]{};
\node[] at (2,0) {\dots};
\node[bd] at (3,0) [label=below:{{\scriptsize $n-1$}}]{};
\node[bd] at (4,0) [label=below:{{\scriptsize $n-1$}}]{};
\node [] at (5,0) {\dots};
\node[bd] at (6,0) [label=below:{{\scriptsize 2}}]{};
\node[bd] at (7,0) [label=below:{{\scriptsize 1}}]{};
\node[flavor] at (4,1) [label=above:{{\scriptsize 1}}]{};
\node[flavor] at (2,1) [label=above:{{\scriptsize 1}}]{};
\filldraw[fill= red] (3,1) circle [radius=0.1] node[above] {\scriptsize 1};
\node[bd] at (9,0) [label=below:{{\scriptsize 1}}]{};
\node[bd] at (10,0) [label=below:{{\scriptsize 2}}]{};
\node[] at (11,0) {\dots};
\node[bd] at (12,0) [label=below:{{\scriptsize $n-1$}}]{};
\node[bd] at (13,0) [label=below:{{\scriptsize $n-1$}}]{};
\node [] at (14,0) {\dots};
\node[bd] at (15,0) [label=below:{{\scriptsize 2}}]{};
\node[bd] at (16,0) [label=below:{{\scriptsize 1}}]{};
\node[flavor] at (13,1) [label=above:{{\scriptsize 1}}]{};
\node[flavor] at (12,1) [label=above:{{\scriptsize 1}}]{};
\end{tikzpicture}
\end{center}
\caption{The FI deformation on the red node leads to the quiver on the right-hand-side, which is the 3d mirror of $U(n-1)$ with $N_f=2n-1$.}\label{mirror quiver massive}
\end{figure}
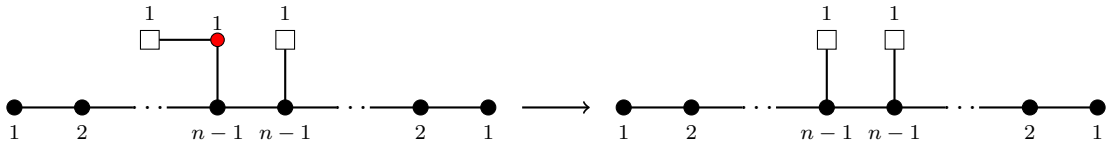

While the 4d Coulomb branch geometry is not scale-invariant due to asymptotic freedom, the 3d gauge theory flows to a 3d $\CN=4$ superconformal fixed point, whose  Coulomb branch is a hyper-K\"ahler cone. For the massless gauge theory, the 3d Coulomb branch can be equivalently described as the Higgs branch of mirror quiver, shown in figure~\ref{massless}.
%%%%%%%%%%%%
%%%%%%%%
Since we are interested in the massive theory, however, we should turn on a $SU(2n-1)$-preserving FI deformation in the mirror quiver, which gives us the quiver shown in figure~\ref{mirror quiver massive}.
This is precisely the 3d mirror of $U(n-1)$ SQCD with $N_f=2n-1$. We therefore conclude that, in contrast to the 4d case, where we had a special (nonzero) value of the mass for which additional massless degrees of freedom appeared at the origin of the Coulomb branch, the low-energy theory in 3d is always $U(n-1)$ SQCD with $N_f=2n-1$, for any non-zero value of the mass. We then conclude that the 3d reduction of the  $D_2(SU(2n-1))$ SCFT can be described as the IR fixed point of 3d $\CN=4$ $U(n-1)$ SQCD with $N_f= 2n-1$ flavors.

%%%%%%%%%%%%

\subsubsection{$D_p^b(SU(N))$ Theories and 3d Quivers}

In order to discuss the general case, let us first discuss the structure of the deformation ring of the Type IIB geometry that engineers the 4d SCFT. We have\footnote{For the case of $SU(N)$, we ignore the terms $x_1$ and $x_2$ in $F(x_1,x_2,x_3,z)$ since they do not appear in deformation terms, and we take $x\equiv x_3$.}
\be\label{def1} 
D_p^b(SU(N))\; : \; \qquad \begin{cases}
x^N+z^p+u_{kl}z^k x^l=0\qquad & \text{if} \; \; b=N~, \\
x^N+xz^p+u_{kl}z^k x^l=0\qquad & \text{if} \; \; b=N-1~.
\end{cases}
\ee 
The deformation coefficients have scaling dimensions
\be
\Delta[u_{kl}] = N- {kb\ov p}-l~.
\ee
We can check that $k$ runs from $1$ to $\ell$ or $\ell+1$, corresponding to the two cases in \eqref{quiver3d}, respectively, with $\ell$ defined in \eqref{ell def}. In the second case, the monomial $z^{\ell+1}$ has dimension $\ell$, and corresponds to the mass term associated to the fundamental flavor on the right-hand tail of the 3d quiver \eqref{quiver3d}. 

Let us collect all the terms with the same power of $z$.  As we already mentioned, the $k=0$ deformations terms are associated with the $SU(N)$ flavor current. On the other hand, for each $k>1$, the monomials
\be
z^k x^l~, \qquad l=0, \cdots, n_k-1~,
\ee
have dimension $\Delta \geq 1$, for $n_k$ as defined in \eqref{nk def}. If, furthermore, ${k b\ov p}\in \Z$, the coefficients $u_{kl}$ have the integer-valued dimensions $\Delta[u_{kl}]=n_k, n_k-1, \cdots, 2, 1$. The last coefficient is then again a mass term, while the other ones are the CB parameters for a conformally-gauged $SU(n_k)$ factor of the 4d SCFT. On the other hand, if ${k b\ov p}\notin \Z$, we have the non-integer dimensions $\Delta= N-{kb\ov p} - l$. We claim that these  $n_k$ CB operators will correspond to a $U(n_k)$ factor in the 3d quiver description. We have already shown this in the case $p=2$. In the following, we argue for our 3d quiver description in the general case.

In order to do this, it is convenient to `break down'  the $D_p^b(SU(N))$ theory into smaller building blocks. Indeed, whenever $p$ and $b$ are not mutually prime, there exists a `partially Lagrangian' description of $D_p^b(N)$ that includes the $SU(n_k)$ gauge groups for $k$ such that ${k b\ov p}\in \Z$.  Indeed, consider
\be
m \equiv {\rm gcd}(p, b) >1~, \qquad \qquad s\equiv {p\ov m}~.
\ee
We then have the first $SU(n_k)\equiv SU(K)$ node at $k=s$, and the SCFT can be broken down as
\be\label{decomposition NK}
D_p^b(SU(N))\; \cong \; \CD^b_{s}(N,K) - SU(K) - D_{\h p}^b(SU(K))~,
\ee
with
\be
K=N- {b\ov m}~, \qquad\qquad  \h p = p -s~.
\ee
The theory $\CD_s(N,K)$ is an `isolated' SCFT without a conformal manifold, and with an $SU(N)\times SU(K) \times U(1)$ flavor symmetry.  Its CB spectrum is entirely non-integer.  It is formally given as a `quotient' of two $D_p^b(SU(N))$ theories
\be
\CD_s^b(N,K) \;  \cong \; {D^b_p(SU(N))\ov D_{\h p}^b(SU(K))}~,
\ee
where, in \eqref{decomposition NK}, we indicated that, to recover $D_p^b(SU(N))$, we must glue $\CD^b_s(N,K)$ to $D^b_{\h p}(SU(K))$ by gauging the common $SU(K)$ flavor symmetry. 
To derive our 3d quiver description, we then need to derive the 3d quiver for $\CD^b_s(N,K)$, which takes the form
\bea
\begin{tikzpicture}[x=.9cm,y=.9cm]
\draw[ligne,black] (0.15,0) -- (2,0);
\draw[ligne,black] (3,0) -- (4.85,0);
\node[flavor] at (0,0) [label=above:{{\scriptsize $N$}}]{};
\node[bd] at (1,0) [label=above:{{\scriptsize $n_1$}}]{};
\node[] at (2.5,0) { $\cdots$};
\node[bd] at (4,0) [label=above:{{\scriptsize $n_{s-1}$}}]{};
\node[flavor] at (5,0) [label=above:{{\scriptsize $K$}}]{};
\end{tikzpicture} \,.
\eea
Finally, we will just need to consider the case $D_p^b(N)$ with ${\rm gcd}(p, b)=1$. 

\medskip
\noindent Let us now assume $b=N$, for simplicity. (The case $b=N-1$ is similar.) Formally, we can view the theory $\CD_s(N,K)$ as engineered from the non-isolated singularity
\be
x^N+ x^K z^s~.
\ee
We then have $\CD_s(N,0)= D_s^N(SU(N))$ and  $\CD_s(N,1)=D_s^{N-1}(SU(N))$ as special cases. 
We would like to realize the SCFTs $\CD_s(N,K)$ at special points on the CB of a Lagrangian theory.

Let us start with the case $K=N-1$, which is simpler to analyze. This model occurs as in \eqref{decomposition NK} for $D_p(SU(N))$ with $p=s N$, for $s$ integer. Then, the 4d Lagrangian theory of interest is the linear quiver
\bea\label{quiver1}
\begin{tikzpicture}[x=.9cm,y=.9cm]
\draw[ligne,black] (0,0) -- (2,0);
\draw[ligne,black] (3,0) -- (5,0);
\node[flavor] at (0,0) [label=above:{{\scriptsize $N$}}]{};
\node[SUd] at (1,0) [label=above:{{\scriptsize $N$}}]{};
\node[] at (2.5,0) { $\cdots$};
\node[SUd] at (4,0) [label=above:{{\scriptsize $N$}}]{};
\node[flavor] at (5,0) [label=above:{{\scriptsize $N-1$}}]{};
\end{tikzpicture}\,,
\eea
where the number of $SU(N)$ gauge groups is $s-1$. Notice that all the $SU(N)$ gauge groups have a vanishing beta function except for the rightmost one, which has $N_f= 2N_c-1$ and is asymptotically free. 
We can therefore apply the results of section~\ref{subsec: D2N deriv}  and argue that, when we go down to 3d, the gauge group is effectively replaced by $U(N-1)$. At this stage, the second rightmost node is no longer conformal, and we can reiterate the procedure. We therefore end up in 3d with the linear quiver
\bea\label{quiver2}
\KK \CD_p(N, N-1) \; \; = \; \; \; \begin{tikzpicture}[x=.9cm,y=.9cm]
\draw[ligne,black] (0,0) -- (2,0);
\draw[ligne,black] (3,0) -- (5,0);
\node[flavor] at (0,0) [label=above:{{\scriptsize $N$}}]{};
\node[bd] at (1,0) [label=above:{{\scriptsize $N-1$}}]{};
\node[] at (2.5,0) { $\cdots$};
\node[bd] at (4,0) [label=above:{{\scriptsize $N-1$}}]{};
\node[flavor] at (5,0) [label=above:{{\scriptsize $N-1$}}]{};
\end{tikzpicture}~.
\eea 
This theory is {\it good} in the  Gaiotto-Witten (GW) sense, and it therefore has a conventional IR fixed point at the origin of its Coulomb branch, with a Higgs-branch global symmetry $SU(N)\times SU(N-1)\times U(1)$, as expected.  Note the enhancement to $SU(2N-1)$ for $p=2$; in that case, the theory is actually equivalent to $D_2(SU(2N-1))$, and we recover the result \eqref{D2N red}. Let us also note that, in this special case with $K=N-1$, we can provide an alternative definition of the theory using class-$\mathcal{S}$ technology, which confirms our result  \eqref{quiver2} for the 3d quiver.%
 \footnote{Consider the $\mathcal{N}=(2,0)$ theory of type $A_{N-1}$ on a sphere with a regular full puncture and an irregular puncture of type III labelled by $s+1$ partitions of $N$.
 In the notation of \protect\cite{Xie:2012hs}, the partitions are $Y_{s+1}=\dots=Y_2=(N-1,1)$ and $Y_1=(1^N)$. It is easy to check that the CB spectra of this class-$\mathcal{S}$ theory and of the $\CD_s(N, N-1)$ model are identical. We then find that the 3d mirror constructed following the algorithm of  \protect\cite{Xie:2012hs} is precisely the mirror dual of  the quiver \protect\eqref{quiver2}.}

For $K<N-1$, we can find a similar Lagrangian UV completion from which we can argue for the 3d quiver description.  The 4d quiver is of the form
\bea\label{quiver3}
\begin{tikzpicture}[x=.9cm,y=.9cm]
\draw[ligne,black] (0,0) -- (3,0);
\draw[ligne,black] (4,0) -- (7,0);
\node[flavor] at (0,0) [label=above:{{\scriptsize $N$}}]{};
\node[SUd] at (1,0) [label=below:{{\scriptsize $N-n$}}]{};
\node[SUd] at (2,0) [label=above:{{\scriptsize $N-2n$}}]{};
\node[] at (3.5,0) { $\cdots$};
\node[SUd] at (5,0) [label=above:{{\scriptsize $K+2n+2$}}]{};
\node[SUd] at (6,0) [label=below:{{\scriptsize $K+n+1$}}]{};
\node[flavor] at (7,0) [label=above:{{\scriptsize $K$}}]{};
\end{tikzpicture}
~,
\eea
where $n= \left\lfloor {N-K \ov s}\right\rfloor$. On the one hand, starting from the left with $N$ fundamentals of $SU(N-n)$, we decrease the number of colors by $n$ at each step. On the other hand, starting from the right-hand tail of the quiver with the  $K$ fundamentals of $SU(K+n+1)$, we increase the number of colors by $n+1$ as we move to the left. These two tails meet at the center at the $(N-K-ns)$-th gauge group from the right, and all the gauge groups are conformal except for the `central' one. 
One can check that the SW curve for this Lagrangian theory includes all possible deformation terms for  the $\CD_s(N,K)$ theory. Locally, the non-conformal `central' node is $SU(N_c)$ with $N_f=2N_c-1$ flavors,  and we can therefore replace that gauge group with an $U(N_c-1)$ node when we dimensionally reduce to 3d. We can then again sequentially replace all $SU$ groups with unitary groups of the same rank.

Once all the gauge groups in the quiver (\ref{quiver3}) have been replaced by unitary groups, some nodes might become   {\it ugly} in the sense of Gaiotto-Witten \cite{Gaiotto:2008ak}\footnote{Recall that by defining $e\equiv N_f-2N_c$ the balance of a node, then quivers with all nodes with $e=0$ are good, if there is $e=-1$ they are ugly and $e<-1$ bad.}, in which case we can apply the local duality
\be\label{uglydual}
U(N_c), \ N_f=2N_c-1  \qquad  \longleftrightarrow\qquad  U(N_c-1), \ N_f= 2N_c-1  \; \oplus \; \mathbb{H}\,,
\ee 
where $\mathbb{H}$ indicates one twisted hypermultiplet. 
This procedure generates a collection of twisted hypermultiplets besides the interacting sector. The interacting sector turns out to be described by the unitary quiver
\bea\label{quiver31}
\begin{tikzpicture}[x=.9cm,y=.9cm]
\draw[ligne,black] (0,0) -- (3,0);
\draw[ligne,black] (4,0) -- (7,0);
\node[flavor] at (0,0) [label=above:{{\scriptsize $N$}}]{};
\node[bd] at (1,0) [label=below:{{\scriptsize $N-n-1$}}]{};
\node[bd] at (2,0) [label=above:{{\scriptsize $N-2n-2$}}]{};
\node[] at (3.5,0) { $\cdots$};
\node[bd] at (5,0) [label=above:{{\scriptsize $K+2n$}}]{};
\node[bd] at (6,0) [label=below:{{\scriptsize $K+n$}}]{};
\node[flavor] at (7,0) [label=above:{{\scriptsize $K$}}]{};
\end{tikzpicture}
~,
\eea
where again $n= \left\lfloor {N-K \ov s}\right\rfloor$ and all nodes are clearly good. We find that (\ref{quiver31}) is equivalent to the interacting sector of the following quiver
\bea\label{quiver4}
\begin{tikzpicture}[x=.9cm,y=.9cm]
\draw[ligne,black] (0,0) -- (3,0);
\draw[ligne,black] (4,0) -- (6,0);
\node[flavor] at (0,0) [label=above:{{\small $N$}}]{};
\node[bd] at (1,0) [label=above:{{\small $n_1$}}]{};
\node[bd] at (2,0) [label=above:{{\small $n_2$}}]{};
\node[] at (3.5,0) { $\cdots$};
\node[bd] at (5,0) [label=above:{{\small $n_{s-1}$}}]{};
\node[flavor] at (6,0) [label=above:{{\small $K$}}]{};
\end{tikzpicture}
~,
\eea 
where $n_1=N-\left \lceil{ \frac{N-K}{s}}\right \rceil$, $n_2=N-\left \lceil{ 2\frac{N-K}{s}}\right \rceil$, and so on. Notice that, although the interacting sectors in the IR are the same, the number of twisted hypermultiplets we find starting from (\ref{quiver3}) and reducing to 3d does not necessarily match the number of twisted hypermultiplets we get from (\ref{quiver4}) by applying the duality  (\ref{uglydual}). The reason is that, in general, the rank of the Lagrangian theory (\ref{quiver3}) is larger than that of the corresponding $\CD_s(N,K)$ model. Therefore, from (\ref{quiver3}), we in general obtain additional  twisted hypermultiplets. This is due to the fact that, besides the $\CD_s(N,K)$ sector, the low-energy theory on the 4d CB of (\ref{quiver3}) includes some abelian vector multiplets, which become twisted hypermultiplets upon dimensional reduction. We therefore claim that (\ref{quiver4}) is the correct dimensional reduction of $\CD_s(N,K)$, since the ranks are the same. The number of colors in \eqref{quiver4}, $n_k= N-\left \lceil{ \frac{(N-K)k}{s}}\right \rceil$, are precisely the ones specified by \eqref{nk def} in our general algorithm.

Next, let us discuss the dimensional reduction of $D_p(SU(N))$ with $N$ and $p$ coprime, which does not have any conformal manifold.  For $p<N$, the analog of (\ref{quiver3}) is a quiver with $p-1$ gauge groups of the form:
\bea
\begin{tikzpicture}[x=.9cm,y=.9cm]
\draw[ligne,black] (0,0) -- (3,0);
\draw[ligne,black] (4,0) -- (6,0);
\node[flavor] at (0,0) [label=above:{{\scriptsize $N$}}]{};
\node[SUd] at (1,0) [label=below:{{\scriptsize $N-n$}}]{};
\node[SUd] at (2,0) [label=above:{{\scriptsize $N-2n$}}]{};
\node[] at (3.5,0) { $\cdots$};
\node[SUd] at (5,0) [label=below:{{\scriptsize $2n+2$}}]{};
\node[SUd] at (6,0) [label=above:{{\scriptsize $n+1$}}]{};
\end{tikzpicture}
~,
\eea 
where $n= \left\lfloor {N\ov p}\right\rfloor$. This can be viewed as the $K=0$ limit of \eqref{quiver3}. There is again only one asymptotically free node, and by the same argument as above we obtain the linear quiver with unitary nodes:
\bea
\begin{tikzpicture}[x=.9cm,y=.9cm]
\draw[ligne,black] (0,0) -- (3,0);
\draw[ligne,black] (4,0) -- (6,0);
\node[flavor] at (0,0) [label=above:{{\scriptsize $N$}}]{};
\node[bd] at (1,0) [label=below:{{\scriptsize $N-n-1$}}]{};
\node[bd] at (2,0) [label=above:{{\scriptsize $N-2n-1$}}]{};
\node[] at (3.5,0) { $\cdots$};
\node[bd] at (5,0) [label=above:{{\scriptsize $2n+1$}}]{};
\node[bd] at (6,0) [label=above:{{\scriptsize $n$}}]{};
\end{tikzpicture}
\eea
 For $p>N$, we take the following 4d linear quiver:
\bea\label{4d quiver SU2 tail}
 \begin{tikzpicture}[x=.9cm,y=.9cm]
\draw[ligne,black] (0,0) -- (3,0);
\draw[ligne,black] (4,0) -- (6,0);
\node[flavor] at (0,0) [label=above:{{\small $N$}}]{};
\node[SUd] at (1,0) [label=above:{{\small $N$}}]{};
\node[SUd] at (2,0) [label=above:{{\small $N$}}]{};
\node[] at (3.5,0) { $\cdots$};
\node[SUd] at (5,0) [label=above:{{\small $3$}}]{};
\node[SUd] at (6,0) [label=above:{{\small $2$}}]{};
\end{tikzpicture}
~,\eea
containing $\ell= \left\lfloor p - {p\ov N} \right\rfloor$ gauge groups. Here, the only new ingredient is the fact that, while turning $SU(n)$ groups into $U(n-1)$ (starting from the last $SU(N)$ factor), we have to deal with an $SU(2)$ gauge group with 2 flavors,  at the right-hand tail of the quiver, which is not of the form $SU(N_c)$ with $2N_c-1$ flavors. In this case, to flow to the $D_p(SU(N))$ theory, we should move onto the Coulomb branch of $SU(2)$ with 2 flavors to its AD point, where the effective theory is $(A_1,A_3)$. Upon dimensional reduction to 3d, the $SU(2)$ group is then replaced by SQED with 2 flavors; in conclusion, the naive substitution $SU(n_k)$ to $U(n_k-1)$ throughout the quiver, from 4d to 3d, is the correct one. The limiting case is $N=2$, the $D_p(SU(2))$ theory, which we discuss separately below, as an example.

By successive application of the decomposition \eqref{decomposition NK}, one can derive the 3d quiver description for arbitrary $N$ and $p$. This then completes our derivation of the 3d `electric quiver' \eqref{quiver3d} for the 4d SCFT $D_p(SU(N))$. The derivation of the 3d quiver $D_p^b(SU(N))$ for $b=N-1$ is similar; in particular, for $m=\text{gcd}(p,N-1)=1$, the simply have the model $\CD_p(N,K)$ discussed above with $K=1$.

\subsubsection{Examples}

Let us illustrate the general discussion from the last section by providing examples for reductions that lead to 3d linear quivers. 

\paragraph{Example 1: $N=2$, $D_p$ AD theory.} As our first example, consider the case $D_p(SU(2))$, which is the ordinary $(A_1, D_p)$ Argyres-Douglas theory. For $p=2n$, we have $D_{2n}(SU(2))= \CD_{2n}(2, 1)$, and the 3d quiver is given by \eqref{quiver2} with $N=2$, namely
\bea
 \begin{tikzpicture}[x=.7cm,y=.7cm]
\draw[ligne,black] (0,0) -- (2,0);
\draw[ligne,black] (3,0) -- (5,0);
\node[flavor] at (0,0) [label=above:{{\scriptsize $2$}}]{};
\node[bd] at (1,0) [label=above:{{\scriptsize $1$}}]{};
\node[] at (2.5,0) { $\cdots$};
\node[bd] at (4,0) [label=above:{{\scriptsize $1$}}]{};
\node[flavor] at (5,0) [label=above:{{\scriptsize $1$}}]{};
\end{tikzpicture}
~,
\eea
with $n-1$ $U(1)$ factors. This agrees with the known result; see {\it e.g.} \cite{Xie:2012hs, Buican:2015hsa, Buican:2015ina} and also \cite{Dedushenko:2019mnd} for a recent discussion. 

For $p=2n+1$, instead, we embed the AD theory within an 4d linear quiver with $SU(2)^n$ gauge group, which is the $N=2$ specialization of \eqref{4d quiver SU2 tail}. The 3d quiver then takes the form
\bea
  \begin{tikzpicture}[x=.7cm,y=.7cm]
\draw[ligne,black] (0,0) -- (2,0);
\draw[ligne,black] (3,0) -- (4,0);
\node[flavor] at (0,0) [label=above:{{\scriptsize $2$}}]{};
\node[bd] at (1,0) [label=above:{{\scriptsize $1$}}]{};
\node[] at (2.5,0) { $\cdots$};
\node[bd] at (4,0) [label=above:{{\scriptsize $1$}}]{};
\end{tikzpicture}~.
\eea
In this case, the rightmost gauge node is `ugly'. Upon repeated use of the GW duality \eqref{uglydual}, we are left with SQED $U(1), N_f=2$ plus $n-1$ free twisted hypermultiplets. This agrees with the 3d reduction of $D_{2n+1}$ first derived in \cite{Dedushenko:2019mnd}.

\paragraph{Example 2: the $D_7(SU(5))$ model.} For $N=5$, $p=7$, the 4d Lagrangian theory is given by \eqref{4d quiver SU2 tail}, with $\ell= 5$. The 3d reduction then gives
\bea
 \begin{tikzpicture}[x=.7cm,y=.7cm]
\draw[ligne,black] (0,0) -- (5,0);
\node[flavor] at (0,0) [label=above:{{\scriptsize $5$}}]{};
\node[bd] at (1,0) [label=above:{{\scriptsize $4$}}]{};
\node[bd] at (2,0) [label=above:{{\scriptsize $3$}}]{};
\node[bd] at (3,0) [label=above:{{\scriptsize $2$}}]{};
\node[bd] at (4,0) [label=above:{{\scriptsize $2$}}]{};
\node[bd] at (5,0) [label=above:{{\scriptsize $1$}}]{};
\end{tikzpicture}~. 
\eea
This quiver is ugly; upon application of the GW duality, we obtain the good quiver 
\bea
T[SU(5)] \cong  \begin{tikzpicture}[x=.7cm,y=.7cm]
\draw[ligne,black] (0,0) -- (4,0);
\node[flavor] at (0,0) [label=above:{{\scriptsize $5$}}]{};
\node[bd] at (1,0) [label=above:{{\scriptsize $4$}}]{};
\node[bd] at (2,0) [label=above:{{\scriptsize $3$}}]{};
\node[bd] at (3,0) [label=above:{{\scriptsize $2$}}]{};
\node[bd] at (4,0) [label=above:{{\scriptsize $1$}}]{};
\end{tikzpicture}
\eea
 plus two twisted hypermultiplets.

\paragraph{Example 3: the $D_6(SU(9))$ model.} In this case, our general algorithm \eqref{quiver3d} gives the 3d quiver
\bea\label{3d quiver example D69}
 \begin{tikzpicture}[x=.7cm,y=.7cm]
\draw[ligne,black] (0,0) -- (5,0);
\node[flavor] at (0,0) [label=above:{{\scriptsize $9$}}]{};
\node[bd] at (1,0) [label=above:{{\scriptsize $7$}}]{};
\node[SUd] at (2,0) [label=above:{{\scriptsize $6$}}]{};
\node[bd] at (3,0) [label=above:{{\scriptsize $4$}}]{};
\node[SUd] at (4,0) [label=above:{{\scriptsize $3$}}]{};
\node[bd] at (5,0) [label=above:{{\scriptsize $1$}}]{};
\end{tikzpicture}
~. 
\eea
Note that the 4d theory is decomposed as:
\be
\CD_2(9,6)-SU(6)-\CD_2(6,3)-SU(3)-D_2(SU(3))~,
\ee
in terms of the `prime factors' $\CD_p(N,K)$.

\paragraph{Example 4: the $D_{14}(SU(8))$ model.}   This example is a bit more elaborate. We have $N=8$, $p=14$, so that $m=\text{gcd}(14,8)=2$ and $s= {p\ov m}=7$; we then have a one-dimensional conformal manifold. According to \eqref{quiver3d}, the 3d quiver reads
 \bea\label{d14su8}
 \begin{tikzpicture}[x=.7cm,y=.7cm]
\draw[ligne,black] (0,0) -- (12,0);
\node[flavor] at (0,0) [label=above:{{\scriptsize $8$}}]{};
\node[bd] at (1,0) [label=above:{{\scriptsize $7$}}]{};
\node[bd] at (2,0) [label=above:{{\scriptsize $6$}}]{};
\node[bd] at (3,0) [label=above:{{\scriptsize $6$}}]{};
\node[bd] at (4,0) [label=above:{{\scriptsize $5$}}]{};
\node[bd] at (5,0) [label=above:{{\scriptsize $5$}}]{};
\node[bd] at (6,0) [label=above:{{\scriptsize $4$}}]{};
\node[SUd] at (7,0) [label=above:{{\scriptsize $4$}}]{};
\node[bd] at (8,0) [label=above:{{\scriptsize $3$}}]{};
\node[bd] at (9,0) [label=above:{{\scriptsize $2$}}]{};
\node[bd] at (10,0) [label=above:{{\scriptsize $2$}}]{};
\node[bd] at (11,0) [label=above:{{\scriptsize $1$}}]{};
\node[bd] at (12,0) [label=above:{{\scriptsize $1$}}]{};
\end{tikzpicture}
 ~.
  \eea
This quiver is ugly, and it is equivalent to the good quiver
\bea\label{d14su8v2}
 \begin{tikzpicture}[x=.7cm,y=.7cm]
\draw[ligne,black] (0,0) -- (10,0);
\node[flavor] at (0,0) [label=above:{{\scriptsize $8$}}]{};
\node[bd] at (1,0) [label=above:{{\scriptsize $7$}}]{};
\node[bd] at (2,0) [label=above:{{\scriptsize $6$}}]{};
\node[bd] at (3,0) [label=above:{{\scriptsize $5$}}]{};
\node[bd] at (4,0) [label=above:{{\scriptsize $4$}}]{};
\node[bd] at (5,0) [label=above:{{\scriptsize $4$}}]{};
\node[bd] at (6,0) [label=above:{{\scriptsize $4$}}]{};
\node[SUd] at (7,0) [label=above:{{\scriptsize $4$}}]{};
\node[bd] at (8,0) [label=above:{{\scriptsize $3$}}]{};
\node[bd] at (9,0) [label=above:{{\scriptsize $2$}}]{};
\node[bd] at (10,0) [label=above:{{\scriptsize $1$}}]{};
\end{tikzpicture}
~,
\eea
plus 6 twisted hypermultiplets. The 4d theory decomposes as
\be
D_{14}(SU(8)) \; \; \cong \; \; \CD_{7}(8,4)-SU(4)-D_7(SU(4))~.
\ee
The $D_7(SU(4))$ tail has the equivalent 3d description as
\bea
\KK D_7(SU(4))\; \cong \; \begin{tikzpicture}[x=.7cm,y=.7cm]
\draw[ligne,black] (0,0) -- (3,0);
\node[flavor] at (0,0) [label=above:{{\scriptsize $4$}}]{};
\node[bd] at (1,0) [label=above:{{\scriptsize $3$}}]{};
\node[bd] at (2,0) [label=above:{{\scriptsize $2$}}]{};
\node[bd] at (3,0) [label=above:{{\scriptsize $1$}}]{};
\end{tikzpicture}\, \oplus\,  3\mb{H}
\eea
The $\CD_{7}(8,4)$ sector has the 3d description
\bea\label{good quiver D84expl}
\begin{tikzpicture}[x=.7cm,y=.7cm]
\draw[ligne,black] (0,0) -- (7,0);
\node[flavor] at (0,0) [label=above:{{\scriptsize $8$}}]{};
\node[bd] at (1,0) [label=above:{{\scriptsize $7$}}]{};
\node[bd] at (2,0) [label=above:{{\scriptsize $6$}}]{};
\node[bd] at (3,0) [label=above:{{\scriptsize $5$}}]{};
\node[bd] at (4,0) [label=above:{{\scriptsize $4$}}]{};
\node[bd] at (5,0) [label=above:{{\scriptsize $4$}}]{};
\node[bd] at (6,0) [label=above:{{\scriptsize $4$}}]{};
\node[flavor] at (7,0) [label=above:{{\scriptsize $4$}}]{};
\end{tikzpicture}
  \, \oplus\, 3\mb{H}
\eea
This gives us a non-trivial example of the discussion below \eqref{quiver4}. According to \eqref{quiver3}, the $\CD_{7}(8,4)$ SCFT can be embedded into the 4d Lagrangian theory
\bea\label{4d Lag example 84}
\begin{tikzpicture}[x=.7cm,y=.7cm]
\draw[ligne,black] (0,0) -- (7,0);
\node[flavor] at (0,0) [label=above:{{\scriptsize $8$}}]{};
\node[SUd] at (1,0) [label=above:{{\scriptsize $8$}}]{};
\node[SUd] at (2,0) [label=above:{{\scriptsize $8$}}]{};
\node[SUd] at (3,0) [label=above:{{\scriptsize $8$}}]{};
\node[SUd] at (4,0) [label=above:{{\scriptsize $7$}}]{};
\node[SUd] at (5,0) [label=above:{{\scriptsize $6$}}]{};
\node[SUd] at (6,0) [label=above:{{\scriptsize $5$}}]{};
\node[flavor] at (7,0) [label=above:{{\scriptsize $4$}}]{};
\end{tikzpicture}
~.\eea
Upon going to 3d, we then find the quiver:
\bea\label{quiver5}
\begin{tikzpicture}[x=.7cm,y=.7cm]
\draw[ligne,black] (0,0) -- (7,0);
\node[flavor] at (0,0) [label=above:{{\scriptsize $8$}}]{};
\node[bd] at (1,0) [label=above:{{\scriptsize $7$}}]{};
\node[bd] at (2,0) [label=above:{{\scriptsize $7$}}]{};
\node[bd] at (3,0) [label=above:{{\scriptsize $7$}}]{};
\node[bd] at (4,0) [label=above:{{\scriptsize $6$}}]{};
\node[bd] at (5,0) [label=above:{{\scriptsize $5$}}]{};
\node[bd] at (6,0) [label=above:{{\scriptsize $4$}}]{};
\node[flavor] at (7,0) [label=above:{{\scriptsize $4$}}]{};
\end{tikzpicture}
~,
\eea
whose rank is equal to the rank of $\CD_{7}(8,4)$ plus 3. But the quiver \eqref{quiver5} is ugly and equivalent to the good quiver shown in \eqref{good quiver D84expl} plus  six twisted hypermultiplets.
Therefore, the 3d infrared description that we obtain starting from \eqref{4d Lag example 84} is equivalent to the theory \eqref{good quiver D84expl} plus three twisted hypermultiplets. This additional free sector accounts for the rank mismatch between the 4d Lagrangian theory  \eqref{4d Lag example 84}  and the $\CD_{7}(8,4)$ model.

%%%%%%%
\subsubsection{Higgs Branch and Anomaly Matching}

The Higgs branch of $D_p^b(SU(N))$ can be directly constructed from the 3d quiver \eqref{quiver3d}, which means that the quantum Higgs branch of the 4d SCFT can be obtained as a hyperk\"ahler quotient. The Higgs branch quaternionic dimension is then given by
\be\label{dH 3d DpN}
\h d_H(D_p^b(SU(N)))=\sum_{k=1}^{\ell +1}(n_k n_{k-1}- n_k^2)+ f_0 \,,
\ee
where we have $n_0\equiv N$ and $n_{k+1}=0$ if $b=N$ and ${p\ov N}\in \Z$ and $n_{k+1}=1$ otherwise. 
This integer should be compared to the quantity
\be
24(c-a) = {(N-1)(2N+5)(Np-b)\ov 2 b}- {5\ov 2} f_0  -6 \sum_{i=1}^{\h r} \Delta_i~,
\ee
which is not always an integer. Whenever the low-energy theory on the 4d Higgs branch consists of only $\h d_H$ hypermultiplets, we must have
\be\label{equality HB DpbN}
24(c-a) = \h d_H(D_p^b(SU(N)))~,
\ee
by anomaly matching on the HB. If this equality fails, this means that the Higgs branch theory is non-trivial: it must contain free vector multiplets and/or a 4d SCFT without  a Higgs branch. 

The equality \eqref{equality HB DpbN} is highly non-trivial, but it can be checked in many example. Moreover, we find that $24(c-a) \neq  \h d_H$  whenever the 3d quiver is ugly in the GW sense -- that is, the IR fixed point contains a free sector of twisted hypermultiplets. This is perhaps expected from the point of view of the 4d Higgs-branch theory, since free vector multiplets will reduce to a free sector and, less trivially, any 4d  $\CN=2$ SCFT with an empty Higgs branch is expected to flows to twisted hypermultiplets as well.%
\footnote{This is assuming that non-trivial 3d $\CN=4$ fixed points with either an empty HB or an empty CB do not exist. (Here, a `trivial theory' is a theory of free (twisted) hypermultiplets or a discrete gauging thereof.)}

On the other hand, the 3d quiver is never bad (in the Gaiotto-Witten sense). This can be seen as follows. The $j$-th gauge group of the quiver is $U(n_j)$ where $n_j=\left\lfloor N-j\frac{N}{p}\right\rfloor$. The number of flavors is given by the rank of the $(j-1)$-th and $(j+1)$-th gauge groups, which are respectively  $\left\lfloor N-(j-1)\frac{N}{p}\right\rfloor$ and $\left\lfloor N-(j+1)\frac{N}{p}\right\rfloor$. Using the inequalities 
\be \left\lfloor N-(j-1)\frac{N}{p}\right\rfloor\geq n_j+\left\lfloor\frac{N}{p}\right\rfloor~, \quad\quad  \left\lfloor N-(j+1)\frac{N}{p}\right\rfloor\geq n_j-1-\left\lfloor\frac{N}{p}\right\rfloor,\ee
it is easy to see that every node in the quiver is either good or ugly.

\subsection{Quiverines from Marginally-gauged $D^b_p(SU(N))$-Trinions}\label{secquiverinesSU}

\subsubsection{Trinion Singularities}

The `trinion singularity' \eqref{hypereq intro} engineers the $SU(N)$ gauging of three $D_p^b(SU(N))$ theories, as displayed in \eqref{FTfour intro}. The condition for the vanishing of the $SU(N)$ beta-function is given by \eqref{beta function intro}
\be\label{beta zero eq}
\ba
\beta_{SU(N)}&= 2 N -\half \sum_{\alpha=1}^3 k_{SU(N)}(D_{p_\alpha}^{b_\alpha}(SU(N)))\cr
&= - N+ \sum_{\alpha=1}^3  {b_\alpha\ov p_\alpha}=0~.
\ea
\ee
The trinion singularity takes the following form, depending on whether we have zero, one, two or three distinct $D_p^b(SU(N))$ factors with $b=N-1$, respectively:
\bea\label{all trinion sings}
& \text{I}(p_1, p_2, p_3, N)\; &&: \; \;  F=x^N+z_1^{p_1}+z_2^{p_2}+z_3^{p_3}  \cr
&  \text{II}(p_1, p_2, N, p_3)\; &&: \; \;  F= x^N+z_1^{p_1}+z_2^{p_2}+xz_3^{p_3}  \cr
 &  \text{VIII} (p_1, N,  p_2, p_3)\; &&: \; \;  F= x^N+z_1^{p_1}+xz_2^{p_2}+xz_3^{p_3}+\cdots  \cr
 &  \text{XIV} (N, p_1, p_2, p_3)\; &&: \; \;   F= x^N+x z_1^{p_1}+xz_2^{p_2}+x z_3^{p_3}+\cdots~.
\eea
The notation in terms of `types' I, II, VIII and XIV follows the Yau-Yu classification \cite{yau2005classification, Kreuzer:1992np}; see \cite{CSNWII} for a detailed discussion.

\medskip
\noindent For the type I singularities, the only solutions to the condition:
\be
{1\ov p_1}+{1\ov p_2}+{1\ov p_3}=1~,
\ee
are well-known to be $(p_\alpha)=(3,3,3)$, $(2,4,4)$ and $(2,3,6)$. This gives three infinite families of trinions, indexed by $N$. These models will be denoted by
\be\label{Ensing}
\ba
& E_6^{(k)}\; : \; \; && x_1^3+x_2^3+x_3^3+x_4^{3+k}~,\\
& E_7^{(k)}\; : \; \; && x_1^2+x_2^4+x_3^4+x_4^{4+k}~,\\
& E_8^{(k)}\; : \; \; && x_1^2+x_2^3+x_3^6+x_4^{6+k}~,
\ea
\ee
with $N=3+k$, $4+k$ and $6+k$, respectively, and for any $N>1$. The corresponding 4d SCFTs $\FT$ were called $(E^{(1,1)}_n, SU(N))$ in \cite{DelZotto:2015rca}.
 The solutions to \eqref{beta zero eq} for the trinions of type II, VIII and XIV will be classified in section~\ref{sec:SUN} below.

The singularities \eqref{all trinion sings}, with holomorphic 3-form given by \eqref{F def 4d}, are such that $\Delta[x]=1$ and $\Delta[z_\alpha]={b_\alpha\ov p_\alpha}$.  Each $D_p^b(SU(N))$ matter sector is described by the terms $x^N+x^{N-b}z^{p}$, and the CB spectrum of $\FT$ is precisely as expected -- it is the direct sum of the CB spectrum of the three $D_p^b(SU(N))$ factors, plus the spectrum $\{2, \cdots, N\}$ from the $SU(N)$ vector multiplet. Note that we could describe marginally-gauged theories involving other $D_p^b(G)$ factors similarly, as we will discuss in section~\ref{sec7:DandE}.

\subsubsection{The 3d Reduction of $\FTfour$}
The trinion singularities \eqref{all trinion sings} engineers the SCFTs $\FT$ in IIB with an $SU(N)$ gauge-theory description: 
\bea\label{general Dp trinion TX}
 \begin{tikzpicture}[x=.6cm,y=.6cm]
 \node at (-7.5,0.5) {$\FTfour \; \cong$};
\node at (-4,0) {$D_{p_1}^{b_1}(SU(N))$};
\node at (0,0) {$SU(N)$};
\node at (4,0) {$D_{p_3}^{b_3}(SU(N))$};
\node at (0,2) {$D_{p_2}^{b_2}(SU(N))$};
\draw[ligne, black](-2,0)--(-1.3,0);
\draw[ligne, black](1.3,0)--(2,0);
\draw[ligne, black](0,0.5)--(0,1.3);
\end{tikzpicture}
\eea
Since we have just given an explicit description of the 3d reduction of $D_p^b(SU(N))$ as a 3d $\CN=4$ linear (special) unitary quiver, it is tempting to simply gauge together three such linear quivers along their common $SU(N)$ flavor node, thus obtaining a star-shaped 3d quiver, which should give us an explicit description of the electric quiverine $\EQfour$ for $\FTfour$ on a circle. This is indeed the prescription we will follow in this work. There are a number of important caveats, however, which we will address in the following. 

%Let us introduce some convenient notation for our (special) unitary quivers. A quiver is a set of nodes connected by edges, corresponding to gauge groups and bifundamental matter, respectively. We denote the $U(n)$ gauge groups by black nodes and the $SU(n)$ gauge groups by yellow nodes, with the number of colors indicated next to the node; flavor nodes are denoted by squares.  For instance, the quiver \eqref{3d quiver example D69} will be denoted by (using the abbreviation in (\ref{QuiverDots}))

If the 4d quiver is already a Lagrangian SCFT,  then there is no issue with reducing the Lagrangian to 3d, and we obtain a star-shaped quiver with $SU(n)$ gauge groups. The only fully Lagrangian $D_p^b(SU(N))$ theories are \cite{Cecotti:2013lda}
\bea\label{DpN Lag 1}
\begin{tikzpicture}[x=.9cm,y=.9cm]
\draw[ligne,black] (0,0) -- (3,0);
\draw[ligne,black] (4,0) -- (5,0);
\node[] at (-3.5,0) {$D_p^{mp}(SU(pm))\;\; \cong \;\; $};
\node[flavor] at (0,0) [label=above:{{\scriptsize $pm$}}]{};
\node[SUd] at (1,0) [label=below:{{\scriptsize $(p-1)m$}}]{};
\node[SUd] at (2,0) [label=above:{{\scriptsize $(p-2)m$}}]{};
\node[] at (3.5,0) {$\cdots$};
\node[SUd] at (5,0) [label=above:{{\scriptsize $m$}}]{};
\end{tikzpicture}~,
\eea
and \cite{Giacomelli:2017ckh}
\bea\label{DpN Lag 2}
\begin{tikzpicture}[x=.9cm,y=.9cm]
\node[] at (-3.5,0) {$ D_p^{mp}(SU(pm+1))\;\; \cong \;\; $};
\draw[ligne,black] (0,0) -- (2,0);
\draw[ligne,black] (3,0) -- (5,0);
\node[flavor] at (0,0) [label=above:{{\scriptsize $pm+1$}}]{};
\node[SUd] at (1,0) [label=below:{{\scriptsize $(p-1)m+1$}}]{};
\node[] at (2.5,0) {$\cdots$};
\node[SUd] at (4,0) [label=above:{{\scriptsize $m+1$}}]{};
\node[flavor] at (5,0) [label=above:{{\scriptsize $1$}}]{};
\end{tikzpicture}
~.
\eea
A well-known class of examples consists of the 4d SCFT associated with the singularities $I(3,3,3,3m)$, $I(2,4,4,4m)$ and $I(2,3,6,6m)$ in \eqref{all trinion sings}, which are $\h E$-shaped 4d superconformal quivers with $\prod_i SU(d_i m)$ gauge group, with $d_i$ the Dynkin labels \cite{Katz:1997eq}:
\be
\label{EnkYellow}
\ba
&
  \begin{tikzpicture}[x=.7cm,y=.7cm]
  \node[] at (-5,1) {$E_6^{(3m-3)}:$};
\draw[ligne, black](-2,0)--(2,0);
\draw[ligne, black](0,0)--(0,2);
\node[SUd] at (-2,0) [label=below:{{\scriptsize$m$}}] {};
\node[SUd] at (-1,0) [label=below:{{\scriptsize$2m$}}] {};
\node[SUd] at (0,0) [label=below:{{\scriptsize$3m$}}] {};
\node[SUd] at (1,0) [label=below:{{\scriptsize$2m$}}] {};
\node[SUd] at (2,0) [label=below:{{\scriptsize$m$}}] {};
\node[SUd] at (0,1) [label=left:{{\scriptsize$2m$}}] {};
\node[SUd] at (0,2) [label=left:{{\scriptsize$m$}}] {};
\end{tikzpicture}~ 
\cr 
&
   \begin{tikzpicture}[x=.7cm,y=.7cm]
     \node[] at (-5,1) {$E_7^{(4m-4)}:$};
\draw[ligne, black](-3,0)--(3,0);
\draw[ligne, black](0,0)--(0,1);
\node[SUd] at (-3,0) [label=below:{{\scriptsize$m$}}] {};
\node[SUd] at (-2,0) [label=below:{{\scriptsize$2m$}}] {};
\node[SUd] at (-1,0) [label=below:{{\scriptsize$3m$}}] {};
\node[SUd] at (0,0) [label=below:{{\scriptsize$4m$}}] {};
\node[SUd] at (1,0) [label=below:{{\scriptsize$3m$}}] {};
\node[SUd] at (2,0) [label=below:{{\scriptsize$2m$}}] {};
\node[SUd] at (3,0) [label=below:{{\scriptsize$m$}}] {};
\node[SUd] at (0,1) [label=above:{{\scriptsize$2m$}}] {};
\end{tikzpicture} \cr 
&
\begin{tikzpicture}[x=.7cm,y=.7cm]
  \node[] at (-5,1) {$E_8^{(6m-6)}:$};
\draw[ligne, black](-2,0)--(5,0);
\draw[ligne, black](0,0)--(0,1);
\node[SUd] at (-2,0) [label=below:{{\scriptsize$2m$}}] {};
\node[SUd] at (-1,0) [label=below:{{\scriptsize$4m$}}] {};
\node[SUd] at (0,0) [label=below:{{\scriptsize$6m$}}] {};
\node[SUd] at (1,0) [label=below:{{\scriptsize$5m$}}] {};
\node[SUd] at (2,0) [label=below:{{\scriptsize$4m$}}] {};
\node[SUd] at (3,0) [label=below:{{\scriptsize$3m$}}] {};
\node[SUd] at (4,0) [label=below:{{\scriptsize$2m$}}] {};
\node[SUd] at (5,0) [label=below:{{\scriptsize$m$}}] {};
\node[SUd] at (0,1) [label=above:{{\scriptsize$3n$}}] {};
\end{tikzpicture}
\ea
\ee
All the legs are of the form \eqref{DpN Lag 1}. E.g. the $\h E_6$-shaped quiver has three legs corresponding to $D_3(SU(3n))$.

In general, however, the $D_p^b(SU(N))$ legs in \eqref{general Dp trinion TX} are not Lagrangian SCFTs, and we are not guaranteed that the 3d star-shaped quiver with unitary groups gives the most useful description of the electric quiverine $\EQfour$. As we will see in examples, sometimes the 3d $\CN=4$ quiver so-obtained is `bad' in the GW sense, and therefore  in need of some additional interpretation. This is expected to occur when the Higgs branch of the 4d $\CN=2$ SCFT $\FTfour$ contains massless degrees of freedom besides the hypermultiplets. Geometrically, this corresponds to having a crepant resolution $\t\MG$ with residual terminal singularities and/or with 3-cycles, as we discussed in the introduction.  In those instances, we will show in the following sections that the interpretation of the magnetic quiver $\MQfive$ requires further refinement -- e.g. we will need to `redeem' the bad quivers, by `unhiggsing' them first and then applying an alternative higgsing as in \cite{Gaiotto:2012uq}.  Let us also note that,  in all the examples below, the `badness' of the magnetic quiver for $\FT$ is related to the 3-cycles rather than to the terminal singularities.

%%%%%%%%%%%%%%%%%%%%%%%%%%
%%%%%%%%%%%%%%%%%%%%%%%%%%

\section{Terminal Singularities and Rank 0 5d SCFTs}
\label{sec:Rank0}

We now turn to 5d SCFTs obtained from M-theory on canonical singularities. The first class of models we study correspond to 
 5d rank-zero SCFTs $\FT$ engineered by terminal singularities in M-theory. 
 We will discuss the singularities shown in table \ref{table:Enk rank 0}, following the logic reviewed in the introduction (see figure~\ref{summary table 1 Intro}). 
 These terminal singularities, denoted by $E_n^{(k)}$ with $k<0$, will appear in later sections as residual singularities on the crepant resolution of  other singularities, including the $E_n^{(k)}$ models with $k>0$.  For specific values of $k$ these were discussed in (\ref{EnkYellow}). 

Using the results of the previous section, we directly find that the 4d theories $\FTfour$ in Type IIB 
can be described as a conformal gauging of three $D_p(SU(n))$ 4d SCFTs.
We then give the electric quiver $\EQfour$ and the magnetic quiver $\MQfive$ as star-shaped quivers. By analysing the Coulomb branch of $\MQfive$, we find that the Higgs branch of $\FT$ takes the form
\be\label{Hd mod Zk}
{\rm HB}[\FT] \cong  \mathbb{H}^{d_H}/\Z_k~.
\ee
The integer $k$ (which could be $k=1$) and the action of $\Z_k$ is determined by the spectrum of line operators in $\FTfour$, as we will see. Therefore, the electric quiverine of $\FT$, $\EQfive$, consists of (in general) discretely gauged 3d $\CN=4$ hypermultiplets. We then conclude that the 5d theory $\FT$ also consists of a discrete gauge theory of hypermultiplets, modulo any `hidden' 5d sector without any moduli space, which, if it exists, cannot be captured by our analysis.

%%%%%%%%%%%%%%%%%%%%%%
\begin{table}[t]
\centering % used for centering table
\begin{tabular}{|c |c| c ||c |c | c|| c|c| c|c|c|| c| c|c|} % centered columns (4 columns)
\hline\hline %inserts double horizontal lines
 $E_{n, 0}^{(k)}$&4d AD&$F$& $r$ &$f$   &$d_H$ &$\h r$& $\h d_H$&$a$&$c$  & $\Delta \CA_r$  & $b_3$ & $\frak{f}$  \\ [0.5ex] % inserts table
%heading
\hline % inserts single horizontal line

$E_{6}^{(-1)}$&$[A_2, D_4]$ &$x_1^3+x_2^3 +x_3^3+ x_4^2$ & $0$ & $0$ & $4$ & $4$ & $0$&$2$&$2$ &$0$&  $0$&{$\Z_2$} \\
\hline
\hline
$E_{7}^{(-2)}$&$[A_3, A_3]$ &$x_1^2+x_2^4 +x_3^4+ x_4^2$ & $0$ & $3$ & $6$ & $3$ & $3$&${15\ov 8}$&$2$ &$0$&  $0$&{$0$} \\
$E_{7}^{(-1)}$&$[A_3, E_6]$ &$x_1^2+x_2^4 +x_3^4+ x_4^3$ & $0$ & $0$ & $9$ & $9$ & $0$&$6$&$6$ &$0$&  $0$&{$\Z_3$} \\
\hline 
\hline
$E_{8}^{(-4)}$&$[A_2, A_5]$ &$x_1^2+x_2^3 +x_3^6+ x_4^2$ & $0$ & $2$ & $6$ & $4$ & $2$&${9\ov4}$&${7\ov3}$ &$0$&  $0$&{$0$} \\
$E_{8}^{(-3)}$&$[A_5, D_4]$ &$x_1^2+x_2^3 +x_3^6+ x_4^3$ & $0$ & $4$ & $12$ & $8$ & $4$&${37\ov6}$&${19\ov3}$ &$0$&  $0$&{$0$} \\
$E_{8}^{(-2)}$&$[A_5, E_6]$ &$x_1^2+x_2^3 +x_3^6+ x_4^{4}$&$0$ & $2$ & $16$ & $14$ & $2$&${49\ov4}$&${37\ov3}$ &$0$&  $0$&{$\Z_2$} \\
$E_{8}^{(-1)}$&$[A_5, E_8]$ &$x_1^2+x_2^3 +x_3^6+ x_4^{5}$&$0$ & $0$ & $20$ & $20$ & $0$&$20$&$20$ &$0$&  $0$&{$\Z_{5}$} \\
\hline
\end{tabular}
\caption{Terminal singularities ($r{=}0$ models) of type $E_n^{(k)}$, with $k<0$, corresponding to $[G, G']$ AD theories as indicated. Note that $\Delta \CA_r \equiv 24(c-a)- \h d_H$.} 
\label{table:Enk rank 0}
\end{table}
%%%%%%%%%

\subsection{$\FTfour$ and Trinion-SCFTs for  Rank 0 $E_n^{(k)}$ Theories}

The 4d theories $\FTfour$ we obtain from these singularities in IIB are all generalized Argyres-Douglas theories of type AD$[G,G']$ (also known as CNV theories \cite{Cecotti:2010fi}), as indicated in table~\ref{table:Enk rank 0}. We can also identify them as $D_p(SU(N))$ trinions, following the discussion in the last section. 

The 4d SCFT $[A_2, D_4]$ obtained from the singularity $E_{6}^{(-1)}$ is equivalent to a marginal gauging of three $D_3(SU(2)) \cong [A_1, A_3]$ AD theories with an $SU(2)$ vector multiplet  \cite{Buican:2016arp, Closset:2020scj}
\bea\label{E6 4d trinion}
 \begin{tikzpicture}[x=.7cm,y=.7cm]
  \node at (-4,0.5) {$\FTXfour{\MG_{E_6^{(-1)}}}\,  = $};
\draw[ligne, black](-1,0)--(1,0);
\draw[ligne, black](0,0)--(0,1);
\node at (-1.6,0) {{$D_3$}};
\node[SUd] at (0,0) [label=below:{{\scriptsize$2$}}] {};
\node at (1.6,0) {{$D_3$}};
\node at (0,1.4) {{$D_3$}};
\end{tikzpicture}
\eea
 By the same token, the $E_7^{(k<0)}$ theories have the 4d trinion description
\bea\label{E7 4d trinions}
 \begin{tikzpicture}[x=.7cm,y=.7cm]
  \node at (-4,0.5) {$\FTXfour{\MG_{E_7^{(-2)}}}\,  = $};
\draw[ligne, black](-1,0)--(1,0);
\draw[ligne, black](0,0)--(0,1);
\node at (-1.6,0) {{$D_4$}};
\node[SUd] at (0,0) [label=below:{{\scriptsize$2$}}] {};
\node at (1.6,0) {{$D_4$}};
\node at (0,1.4) {{$D_2$}};
\end{tikzpicture}~, \qquad \quad
 \begin{tikzpicture}[x=.7cm,y=.7cm]
  \node at (-4,0.5) {$\FTXfour{\MG_{E_7^{(-1)}}}\,  = $};
\draw[ligne, black](-1,0)--(1,0);
\draw[ligne, black](0,0)--(0,1);
\node at (-1.6,0) {{$D_4$}};
\node[SUd] at (0,0) [label=below:{{\scriptsize$3$}}] {};
\node at (1.6,0) {{$D_4$}};
\node at (0,1.4) {{$D_2$}};
\end{tikzpicture}~.
\eea
The $E_8^{(k<0)}$ models are similarly given by
\bea\label{E8 4d trinions}
& \begin{tikzpicture}[x=.7cm,y=.7cm]
  \node at (-4,0.5) {$\FTXfour{\MG_{E_8^{(-4)}}}\,  = $};
\draw[ligne, black](-1,0)--(1,0);
\draw[ligne, black](0,0)--(0,1);
\node at (-1.6,0) {{$D_3$}};
\node[SUd] at (0,0) [label=below:{{\scriptsize$2$}}] {};
\node at (1.6,0) {{$D_6$}};
\node at (0,1.4) {{$D_2$}};
\end{tikzpicture}~, \qquad \quad
&& \begin{tikzpicture}[x=.7cm,y=.7cm]
  \node at (-4,0.5) {$\FTXfour{\MG_{E_8^{(-3)}}}\,  = $};
\draw[ligne, black](-1,0)--(1,0);
\draw[ligne, black](0,0)--(0,1);
\node at (-1.6,0) {{$D_3$}};
\node[SUd] at (0,0) [label=below:{{\scriptsize$3$}}] {};
\node at (1.6,0) {{$D_6$}};
\node at (0,1.4) {{$D_2$}};
\end{tikzpicture}~, \\
& \begin{tikzpicture}[x=.7cm,y=.7cm]
  \node at (-4,0.5) {$\FTXfour{\MG_{E_8^{(-2)}}}\,  = $};
\draw[ligne, black](-1,0)--(1,0);
\draw[ligne, black](0,0)--(0,1);
\node at (-1.6,0) {{$D_3$}};
\node[SUd] at (0,0) [label=below:{{\scriptsize$4$}}] {};
\node at (1.6,0) {{$D_6$}};
\node at (0,1.4) {{$D_2$}};
\end{tikzpicture}~, \qquad \quad
&& \begin{tikzpicture}[x=.7cm,y=.7cm]
  \node at (-4,0.5) {$\FTXfour{\MG_{E_8^{(-1)}}}\,  = $};
\draw[ligne, black](-1,0)--(1,0);
\draw[ligne, black](0,0)--(0,1);
\node at (-1.6,0) {{$D_3$}};
\node[SUd] at (0,0) [label=below:{{\scriptsize$5$}}] {};
\node at (1.6,0) {{$D_6$}};
\node at (0,1.4) {{$D_2$}};
\end{tikzpicture}~.
\eea
Some of these models have a non-trivial one-form symmetry \cite{Closset:2020scj, DelZotto:2020esg}, indicated by $\frak{f}$ in table~\ref{table:Enk rank 0}, which will play an important role below. Given the finite group $\frak{f}\oplus \frak{f}$, which can be computed from the geometry, one must choose a polarization in order to have a well-defined theory, which corresponds to a choice of allowed line operators~\cite{Aharony:2013hda, Gaiotto:2014kfa}. For simplicity, we will only consider purely `electric' or `magnetic' polarizations. For the  `electric' one-form symmetry, we can essentially think of the one-form symmetry of the trinions as (a subgroup of) the center symmetry of the central $SU(N)$ node, in which case the charged objects are Wilson lines. Conversely, having a purely `magnetic' one-form symmetry corresponds to choosing the central gauge group to be $PSU(N)=SU(N)/\Z_N$ (assuming $\frak{f}=\Z_N$, for simplicity), in which case the charged objects are 't Hooft lines \cite{Kapustin:2005py} that carry non-trivial $\Z_N$-valued magnetic charge, i.e. the 't Hooft lines specified by $PSU(N)$ bundles on $S^2$ with a non-trivial Stiefel-Whitney class. In the following, we will denote $SU(n)/\Z_k$ gauge groups by yellow nodes with the label $n/\Z_k$, as in (\ref{QuiverDots}).  See also \cite{Bourget:2020xdz} a for a related recent  discussion.

\subsection{Magnetic Quivers and Higher Form Symmetries}

We are particularly interested in the magnetic quiver $\MQfive= \EQfour/U(1)^f$ for the 5d SCFT $\FT$. 
Let us note that, if we start from a 4d theory $\FTfour$ with an electric one-form symmetry $\Z_N$, we obtain a 3d theory with a one-form symmetry, realized as a center symmetry of the 3d quiver with central gauge group $SU(N)$. On the other hand, if we start from $\FTfour$ with a magnetic one-form symmetry $\Z_N$, the 3d quiver will have a $\Z_N$ zero-form magnetic symmetry. The charged operators are 3d monopole operators with non-trivial $\Z_N$ magnetic charge, which descend from the 't Hooft lines of $\FTfour$ that wrap the $S^1$. By T-duality, the 3d `$SU(N)$ quiver' with an electric symmetry corresponds to the magnetic quiver of a 5d SCFT $\FT$ with a 3-form symmmetry. The `electrically charged' objects in M-theory are M5-branes wrapping $S^3\times T^2$, with $T^2= S^1_M{\times}S^1_\beta$. These map to D3-branes wrapping $S^3$ in IIB.\footnote{It so happens that, for consistency with the 4d gauge-theory language, we have to call these wrapped M5-branes `electric' and the M2-brane `magnetic'. This is only a matter of terminology and it should not cause any confusion.} Conversely, the `$SU(N)/\Z_N$' theory with a magnetic symmetry corresponds to $\FT$ with a $\Z_N$ 0-form symmetry. The `magnetically charged' objects are the M2-branes wrapping $S^3$, which map to D3-branes wrapping $S^3 \times S^1_{1\ov \beta}$ in IIB; the latter objects are the 3d monopole operators on the Coulomb branch of $\MQfive$, which are dual to the M2-brane instantons on the Higgs branch of $\FT$. 

As we will see in examples, whenever the $SU(N)/\Z_N$ `magnetic' theory is allowed by the geometry of the terminal singularity, the magnetic quiver is really equivalent to a unitary quiver, since $PSU(N) \cong U(N)/U(1)$ and the overall $U(1)$ of a unitary quiver is decoupled. It then turns out that these unitary quivers are ugly, and that repeated application of the GW duality shows that the 3d theory consists only of $d_H$ twisted hypermultiplets. Given this result, we can then obtain an explicit description of the magnetic quiver $\MQfive$ for the `electric'  5d theory, which is obtained by gauging the $\Z_N$ 0-form symmetry of the `magnetic'  theory.  In the 3d $\CN=4$ quiver description, the free twisted hypermultiplets are simply monopole operators of conformal dimension $\Delta =\half$. The $\Z_N$ magnetic charge of a monopole operator can be computed as the total $U(N)$ magnetic flux mod $N$, according to:
\be
w_2^{PSU(N)}  \cong c_1^{U(N)} \; \mod N~.
\ee
Therefore, the magnetic charge of the $\Delta=\half$ monopole operators in the $U(N)$ quiver determines the $\Z_N$ action of the 0-form symmetry on the free twisted hypermultiplets. It follows that the electric theory is equivalent to a discrete gauge theory of the form \eqref{Hd mod Zk}, with the $\Z_N$ charges determined as explained.

Let us now apply the above discussion to all the terminal singularities of table~\ref{table:Enk rank 0}.

\subsubsection{The $E_6$ Model}
Consider first the $E_6^{(-1)}$ model, with the trinion description given by \eqref{E6 4d trinion}. Since $f=0$, the electric quiver of $\FTfour$ equals the magnetic quiver of $\FT$, and we have:
\bea\label{E6m1 MQ}
 \begin{tikzpicture}[x=.7cm,y=.7cm]
\node at (-4.7,0.5) {$\EQfour= \MQfive \cong$};
\draw[ligne, black](-1,0)--(1,0);
\draw[ligne, black](0,0)--(0,1);
\node[bd] at (-1,0) [label=below:{{\scriptsize$1$}}] {};
\node[SUd] at (0,0) [label=below:{{\scriptsize$2$}}] {};
\node[bd] at (1,0) [label=below:{{\scriptsize$1$}}] {};
\node[bd] at (0,1) [label=above:{{\scriptsize$1$}}] {};
\node at (2.6,0.5) {$\text{or}$};
\draw[ligne, black](4,0)--(6,0);
\draw[ligne, black](5,0)--(5,1);
\node[bd] at (4,0) [label=below:{{\scriptsize$1$}}] {};
\node[bd] at (5,0) [label=below:{{\scriptsize$2$}}] {};
\node[bd] at (6,0) [label=below:{{\scriptsize$1$}}] {};
\node[bd] at (5,1) [label=above:{{\scriptsize$1$}}] {};
\node at (6.7,0.5) {$\equiv$};
\draw[ligne, black](7.4,0)--(9.4,0);
\draw[ligne, black](8.4,0)--(8.4,1);
\node[bd] at (7.4,0) [label=below:{{\scriptsize$1$}}] {};
\node[SUd] at (8.4,0) [label=below:{{\scriptsize$2/\Z_2$}}] {};
\node[bd] at (9.4,0) [label=below:{{\scriptsize$1$}}] {};
\node[bd] at (8.4,1) [label=above:{{\scriptsize$1$}}] {};
\end{tikzpicture}
\eea
for the `electric' or `magnetic' version of $\FTfour$, respectively.\footnote{As mentioned above, in a quiver with only black nodes, the actual gauge group is $\prod_i U(n_i)/U(1)$, where the $U(1)$ is the overall $U(1)$. It is customary to write the `all-black-dots' quiver as shown here. One can then `ungauge' the $U(1)$ factor at any gauge node and get an equivalent gauge theory.}  As explained above, the quiver on the left in \eqref{E6m1 MQ} is a $\Z_2$ gauging of the unitary quiver on the right. Conversely, we can obtain the quiver on the right by gauging the $\Z_2$ one-form symmetry of the `$SU(2)$ quiver' on the left. Note that the one-form symmetry is a center symmetry that exists as the diagonal $\Z_2$ between the $\Z_2$ center of $SU(2)$ and $\Z_2 \subset U(1)$, which is preserved by the bifundamental matter.

The `$U(2)$ quiver' shown in \eqref{E6m1 MQ} is ugly, since the central node is $U(2)$ with $N_f=3$.
By successive applications of the GW duality \eqref{uglydual}, we find that this theory is simply equivalent to four twisted hypermultiplets:
\bea\label{SU2 quiver GW to H4}
 \begin{tikzpicture}[x=.7cm,y=.7cm]
\draw[ligne, black](-1,0)--(1,0);
\draw[ligne, black](0,0)--(0,1);
\node[bd] at (-1,0) [label=below:{{\scriptsize$1$}}] {};
\node[bd] at (0,0) [label=below:{{\scriptsize$2$}}] {};
\node[bd] at (1,0) [label=below:{{\scriptsize$1$}}] {};
\node[flavor] at (0,1) [label=above:{{\scriptsize$1$}}] {};
\node at (2,0.5) {$\cong$};
\draw[ligne, black](3,0)--(5,0);
\draw[ligne, black](4,0)--(4,1);
\node[bd] at (3,0) [label=below:{{\scriptsize$1$}}] {};
\node[bd] at (4,0) [label=below:{{\scriptsize$1$}}] {};
\node[bd] at (5,0) [label=below:{{\scriptsize$1$}}] {};
\node[flavor] at (4,1) [label=above:{{\scriptsize$1$}}] {};
\node at (5.8,0.5) {$\oplus \, \mathbb{H}$};
\node at (7,0.5) {$\cong$};
\draw[ligne, black](8,0)--(8,1);
\node[bd] at (8,0) [label=below:{{\scriptsize$1$}}] {};
\node[flavor] at (8,1) [label=above:{{\scriptsize$1$}}] {};
\node at (9,0.5) {$\oplus \, \mathbb{H}^3$};
\node at (10.5,0.5) {$\cong$};
\node at (11.5,0.5) {$ \mathbb{H}^4$};
\end{tikzpicture}
\eea
Let us now consider the $SU(2)$ quiver, obtained as a $\Z_2$ gauging of the $U(2)$ quiver. To keep track of the $\Z_2$ charge, consider the four monopole operators of the $U(2)$ quiver that flow to free twisted hypermultiplets (with 3d conformal dimension $\Delta=\half$) in the IR.  It is easily checked that their magnetic charges under $U(2) \times U(1) \times U(1)$, in the quiver notation of the left-hand-side of \eqref{SU2 quiver GW to H4}, are\footnote{Note that we present the fluxes $\m$ for the monopole operators in hypermultiplets, which are each comprised of two 3d $\CN=2$ chiral-multiplet monopole operators with fluxes $\pm \m$.}
\be
(\m_1, \m_2; \n^{(1)};  \n^{(2)})\; =\; (1,0; 0; 0)~, \;  (1,0; 1; 0)~, \; (1,0; 0; 1)~, \; (1,0; 1; 1)~.
\ee
The four operators have $U(2)$ flux $(1,0)$, which means that they have the non-trivial charge under the $\Z_2$ magnetic symmetry. Therefore, we find that the `SU(2) quiver' is a $\Z_2$ gauging of four hypermultiplets
\bea\label{SU2 quiver E60}
 \begin{tikzpicture}[x=.7cm,y=.7cm]
\draw[ligne, black](-1,0)--(1,0);
\draw[ligne, black](0,0)--(0,1);
\node[bd] at (-1,0) [label=below:{{\scriptsize$1$}}] {};
\node[SUd] at (0,0) [label=below:{{\scriptsize$2$}}] {};
\node[bd] at (1,0) [label=below:{{\scriptsize$1$}}] {};
\node[bd] at (0,1) [label=above:{{\scriptsize$1$}}] {};
\node at (2,0.5) {$\cong$};
\node at (4,0.5) {$ \mathbb{H}^4/\Z_2$};
\end{tikzpicture}
\eea
with the $\Z_2$ action
\be
\Z_2 \; : \; \mathbb{H}_i \rightarrow - \mathbb{H}_i~, \qquad i=1,\cdots,4~.
\ee
In particular, the CB of the quiver \eqref{SU2 quiver E60} and, therefore, the HB of the rank-0 5d SCFT, is given by the quotient $\mathbb{H}^4/\Z_2$. This quotient preserves the $Sp(4)$ symmetry of $\mathbb{H}^4$, and therefore the Higgs branch of either the `electric' or the `magnetic' 5d SCFT has an $Sp(4)$ symmetry. 
The Hilbert series of the CB of \eqref{SU2 quiver E60} was computed in \cite{Closset:2020scj} using the monopole formula \cite{Cremonesi:2013lqa}
\be
\ba
{\rm HS}_{E_{6}^{(-1)}}&=\frac{1+28t^2+70t^4+28t^6+t^8}{(1-t^2)^8}\\
&=\sum_{i=0}^\infty \text{dim}([2i,0,0,0])t^{2i} \;=\; 1 +36t^2+ 330 t^4 + \mathcal{O}(t^6) \,.
\ea
\ee
It is exactly the Hilbert Series of the minimal nilpotent orbit of $\mathfrak{sp}(4)$ \cite{Hanany:2016gbz}. This confirms that the flavor symmetry algebra is $\mathfrak{sp}(4)$,-the coefficient of $t^2$ being the number of conserved currents, ${\rm dim} \, \mathfrak{sp}(4)=36$. The appearance only of even spin representation of $\mathfrak{sp}(4)$ indicates in fact that the global form of the flavor symmetry is $Sp(4)/\mathbb{Z}_2$.

Finally, let us mention that this model, at least in its `magnetic' version, is equivalent to the 5d $T_2$ model \cite{Benini:2009gi} -- in particular, the magnetic quiver on the right-hand-side of \eqref{E6m1 MQ} makes the $SU(2)^3$ symmetry of $T_2$ manifest. It is also the $\MQfive$ one obtains for the $T_2$ toric geometry using the methods of \cite{vanBeest:2020kou}. We will generalize this observation to many more models below.

\subsubsection{The $E_7$ Models}
The model $E_7^{(-2)}$ has $f=3$ and $\frak{f}=0$. Using our general results for the 3d reduction of $D_p(N)$, we find
\bea\label{AlphaRomeo}
 \begin{tikzpicture}[x=.7cm,y=.7cm]
\node at (-4.2,0.5) {$\EQfour\cong$};
\draw[ligne, black](-2,0)--(2,0);
\draw[ligne, black](0,0)--(0,1);
\node[flavor] at (-2,0) [label=above:{{\scriptsize$1$}}] {};
\node[bd] at (-1,0) [label=below:{{\scriptsize$1$}}] {};
\node[SUd] at (0,0) [label=below:{{\scriptsize$2$}}] {};
\node[bd] at (1,0) [label=below:{{\scriptsize$1$}}] {};
\node[flavor] at (2,0) [label=above:{{\scriptsize$1$}}] {};
\node[flavor] at (0,1) [label=above:{{\scriptsize$1$}}] {};
\end{tikzpicture}~, \qquad\quad
 \begin{tikzpicture}[x=.7cm,y=.7cm]
\node at (-4.2,0.5) {$\MQfive\cong$};
\draw[ligne, black](-2,0)--(2,0);
\draw[ligne, black](0,0)--(0,1);
\node[bd] at (-2,0) [label=below:{{\scriptsize$1$}}] {};
\node[bd] at (-1,0) [label=below:{{\scriptsize$1$}}] {};
\node[bd] at (0,0) [label=below:{{\scriptsize$2$}}] {};
\node[bd] at (1,0) [label=below:{{\scriptsize$1$}}] {};
\node[bd] at (2,0) [label=below:{{\scriptsize$1$}}] {};
\node[bd] at (0,1) [label=above:{{\scriptsize$1$}}] {};
\end{tikzpicture}~.
\eea
The absence of one-form symmetry is apparent both in $\EQfour$, where the flavor at the center node breaks the $\Z_2$ center symmetry explicitly, and in $\MQfive$. Here, $\MQfive$ is again an ugly quiver, and one can check that it is equivalent to 6 twisted hypermultiplets. Thus, we find that the HB of the 5d $E_7^{(-2)}$ model  is simply $\mathbb{H}^6$.

\medskip
\noindent For the $E_7^{(-1)}$-type model, on the other hand, we have $f=0$ and $\frak{f}=\Z_3$, and we find
\bea\label{E7m1 MQ}
\begin{tikzpicture}[x=.7cm,y=.7cm]
\node at (-5.7,0.5) {$\EQfour = \MQfive\cong$};
\draw[ligne, black](-2,0)--(2,0);
\draw[ligne, black](0,0)--(0,1);
\node[bd] at (-2,0) [label=below:{{\scriptsize$1$}}] {};
\node[bd] at (-1,0) [label=below:{{\scriptsize$2$}}] {};
\node[SUd] at (0,0) [label=below:{{\scriptsize$3$}}] {};
\node[bd] at (1,0) [label=below:{{\scriptsize$2$}}] {};
\node[bd] at (2,0) [label=below:{{\scriptsize$1$}}] {};
\node[bd] at (0,1) [label=above:{{\scriptsize$1$}}] {};
\end{tikzpicture} 
\eea
for the `electric' theory $\FTfour$ reduced on a circle. On the other hand the `$SU(3)/\Z_3$ quiver' is an ugly quiver, corresponding to replacing the central node in \eqref{E7m1 MQ} by a black dot. That unitary quiver is equivalent to 9 twisted hypermultiplets. They arise from 9 monopole operators of dimensions $\Delta=\half$, which carry the $U(3)$ magnetic fluxes
\be
(\m_1, \m_2, \m_3)= (1,0,0)^9 \,,
\ee
where the exponent denote the number of times a given $U(3)$ flux appears. Therefore, by gauging the $\Z_3$ magnetic symmetry, we see that the CB of \eqref{E7m1 MQ} takes the form:
\be
{\rm CB}[\MQfive]\; \cong\; \mathbb{H}^9/\Z_3~, \qquad \qquad \Z_3\, : \, \mathbb{H}_i \rightarrow \omega_3\, \mathbb{H}_i~,\; \; i=1, \cdots, 9~,
\ee
where $\omega_n$ denotes the $n$-th root of unity.  This $\Z_3$ action preserves a subgroup $U(9)$ of the flavor symmetry $Sp(9)$, which is therefore the global symmetry of the 5d `electric' theory, consisting of a $\Z_3$ gauging of 9 hypermultiplets. The dimension of the global symmetry group can be checked by computing the first few terms of the Hilbert series of the CB of \eqref{E7m1 MQ}, which reads
\be
{\rm HS}_{E_{7}^{(-1)}}=1+81 t^2+330 t^3+2025 t^4+\mc{O}(t^5).
\ee
These coefficients can be decomposed in terms of the representations of $\mathfrak{su}(9)$
\be
\ba
\mathbf{81}&=\mathrm{dim}([1,0,0,0,0,0,0,1])+1\cr
\mathbf{330}&=\mathrm{dim}([3,0,0,0,0,0,0,0])+\mathrm{dim}([0,0,0,0,0,0,0,3])\cr
\mathbf{2025}&=\mathrm{dim}([2,0,0,0,0,0,0,2])+\mathrm{dim}([1,0,0,0,0,0,0,1])+1\,.
\ea
\ee
None of these are charged under the center $\mb{Z}_3\subset U(9)$, and hence we postulate that the global form of flavor symmetry in this case is $G_F=U(9)/\mb{Z}_3$.

\subsubsection{The $E_8$ Models}
The $E_8^{(-4)}$ model has $f=2$ and trivial higher-form symmetry. We then find
\bea
 \begin{tikzpicture}[x=.7cm,y=.7cm]
\node at (-3.5,0.5) {$\EQfour\cong$};
\draw[ligne, black](-1,0)--(3,0);
\draw[ligne, black](0,0)--(0,1);
\node[bd] at (-1,0) [label=below:{{\scriptsize$1$}}] {};
\node[SUd] at (0,0) [label=below:{{\scriptsize$2$}}] {};
\node[bd] at (1,0) [label=below:{{\scriptsize$1$}}] {};
\node[bd] at (2,0) [label=below:{{\scriptsize$1$}}] {};
\node[flavor] at (3,0) [label=below:{{\scriptsize$1$}}] {};
\node[flavor] at (0,1) [label=above:{{\scriptsize$1$}}] {};
\end{tikzpicture}~, \qquad\quad
 \begin{tikzpicture}[x=.7cm,y=.7cm]
\node at (-3.5,0.5) {$\MQfive\cong$};
\draw[ligne, black](-1,0)--(3,0);
\draw[ligne, black](0,0)--(0,1);
\node[bd] at (-1,0) [label=below:{{\scriptsize$1$}}] {};
\node[bd] at (0,0) [label=below:{{\scriptsize$2$}}] {};
\node[bd] at (1,0) [label=below:{{\scriptsize$1$}}] {};
\node[bd] at (2,0) [label=below:{{\scriptsize$1$}}] {};
\node[bd] at (3,0) [label=below:{{\scriptsize$1$}}] {};
\node[bd] at (0,1) [label=above:{{\scriptsize$1$}}] {};
\end{tikzpicture}~.
\eea
The quiver $\MQfive$ is ugly and flows to six twisted hypermultiplets. Therefore, we find the 5d Higgs branch HB$[E_8^{(-4)}]\cong \mathbb{H}^6$.

\medskip
\noindent
For the $E_8^{(-3)}$ model, which has $f=4$ and $\frak{f}=0$, we similarly find
\bea
 \begin{tikzpicture}[x=.7cm,y=.7cm]
\node at (-3.5,0.5) {$\EQfour\cong$};
\draw[ligne, black](-2,0)--(4,0);
\draw[ligne, black](0,0)--(0,1);
\node[flavor] at (-2,0) [label=below:{{\scriptsize$1$}}] {};
\node[SUd] at (-1,0) [label=below:{{\scriptsize$2$}}] {};
\node[SUd] at (0,0) [label=below:{{\scriptsize$3$}}] {};
\node[bd] at (1,0) [label=below:{{\scriptsize$2$}}] {};
\node[SUd] at (2,0) [label=below:{{\scriptsize$2$}}] {};
\node[bd] at (3,0) [label=below:{{\scriptsize$1$}}] {};
\node[flavor] at (4,0) [label=below:{{\scriptsize$1$}}] {};
\node[bd] at (0,1) [label=above:{{\scriptsize$1$}}] {};
\end{tikzpicture}~, 
\qquad \begin{tikzpicture}[x=.7cm,y=.7cm]
\node at (-3.5,0.5) {$\MQfive\cong$};
\draw[ligne, black](-2,0)--(4,0);
\draw[ligne, black](0,0)--(0,1);
\node[bd] at (-2,0) [label=below:{{\scriptsize$1$}}] {};
\node[bd] at (-1,0) [label=below:{{\scriptsize$2$}}] {};
\node[bd] at (0,0) [label=below:{{\scriptsize$3$}}] {};
\node[bd] at (1,0) [label=below:{{\scriptsize$2$}}] {};
\node[bd] at (2,0) [label=below:{{\scriptsize$2$}}] {};
\node[bd] at (3,0) [label=below:{{\scriptsize$1$}}] {};
\node[bd] at (4,0) [label=below:{{\scriptsize$1$}}] {};
\node[bd] at (0,1) [label=above:{{\scriptsize$1$}}] {};\end{tikzpicture}~.
\eea
The magnetic quiver is ugly and equivalent to 12 twisted hypermultiplets, and therefore HB$[E_8^{(-3)}]\cong \mathbb{H}^{12}$.

\medskip
\noindent
The $E_8^{(-2)}$ model has $f=2$ and a non-trivial higher-form symmetry, $\frak{f}=\Z_2$. This leads to
\bea\label{E8m2 MQ}
 \begin{tikzpicture}[x=.7cm,y=.7cm]
\node at (-3.5,0.5) {$\EQfour\cong$};
\draw[ligne, black](-2,0)--(4,0);
\draw[ligne, black](0,0)--(0,1);
\node[bd] at (-2,0) [label=below:{{\scriptsize$1$}}] {};
\node[bd] at (-1,0) [label=below:{{\scriptsize$2$}}] {};
\node[SUd] at (0,0) [label=below:{{\scriptsize$4$}}] {};
\node[bd] at (1,0) [label=below:{{\scriptsize$3$}}] {};
\node[bd] at (2,0) [label=below:{{\scriptsize$2$}}] {};
\node[SUd] at (3,0) [label=below:{{\scriptsize$2$}}] {};
\node[bd] at (4,0) [label=below:{{\scriptsize$1$}}] {};
\node[SUd] at (0,1) [label=above:{{\scriptsize$2$}}] {};
\end{tikzpicture}~, 
\qquad \begin{tikzpicture}[x=.7cm,y=.7cm]
\node at (-3.5,0.5) {$\MQfive\cong$};
\draw[ligne, black](-2,0)--(4,0);
\draw[ligne, black](0,0)--(0,1);
\node[bd] at (-2,0) [label=below:{{\scriptsize$1$}}] {};
\node[bd] at (-1,0) [label=below:{{\scriptsize$2$}}] {};
\node[SUd] at (0,0) [label=below:{{\scriptsize$4/\Z_2$}}] {};
\node[bd] at (1,0) [label=below:{{\scriptsize$3$}}] {};
\node[bd] at (2,0) [label=below:{{\scriptsize$2$}}] {};
\node[bd] at (3,0) [label=below:{{\scriptsize$2$}}] {};
\node[bd] at (4,0) [label=below:{{\scriptsize$1$}}] {};
\node[bd] at (0,1) [label=above:{{\scriptsize$2$}}] {};\end{tikzpicture}~.
\eea
Here, the $\EQfour$ is the 3d reduction of the `electric' 4d SCFT $\FTfour$, which has a $\Z_2$ electric one-form symmetry; in the 3d quiver, it corresponds to the diagonal $\Z_2$ center (common to the $\Z_4$ and $\Z_2$ center of the $SU(4)$ and $SU(2)$ gauge nodes, respectively).  The magnetic quiver $\MQfive$ similarly possesses a $\Z_2$ one-form symmetry, corresponding to the center of $SU(4)/\Z_2$, which is inherited from the $\Z_2$ 3-form symmetry of the `electric' version of $\FT$. 

\noindent The Higgs branch of the `magnetic theory' in 5d will instead be described by the quiver
\bea\label{MQ5 E8m2 bis}
\begin{tikzpicture}[x=.7cm,y=.7cm]
%\node at (-3.5,0.5) {$\MQfive\cong$};
\draw[ligne, black](-2,0)--(4,0);
\draw[ligne, black](0,0)--(0,1);
\node[bd] at (-2,0) [label=below:{{\scriptsize$1$}}] {};
\node[bd] at (-1,0) [label=below:{{\scriptsize$2$}}] {};
\node[bd] at (0,0) [label=below:{{\scriptsize$4$}}] {};
\node[bd] at (1,0) [label=below:{{\scriptsize$3$}}] {};
\node[bd] at (2,0) [label=below:{{\scriptsize$2$}}] {};
\node[bd] at (3,0) [label=below:{{\scriptsize$2$}}] {};
\node[bd] at (4,0) [label=below:{{\scriptsize$1$}}] {};
\node[bd] at (0,1) [label=above:{{\scriptsize$2$}}] {};
\end{tikzpicture}%~.
\eea
This is an ugly quiver which is equivalent to 16 twisted hypermultiplets. The $\Delta=\half$ monopole operators of \eqref{MQ5 E8m2 bis} have $U(4)$ flux
\be
(\m_1, \cdots, \m_4) \; = \;  (1,0,0,0)^{8}~, \; (1,1,0,0)^{6}~, \; (0,0,0,0)^{2}~.
\ee
The relevant $\Z_2$ charge is given by the total $U(1)$ flux mod $2$, and therefore we find
\be
{\rm CB}[\MQfive]\cong (\mathbb{H}^8/\Z_2) \times \mathbb{H}^8~,
\ee
where the $\Z_2$ acts by a sign on the first 8 hypermultiplets. 
We then conclude that the `electric' version of $\FT$ has a global symmetry $Sp(8)/\mathbb{Z}_2 \times Sp(8)$.
The Coulomb branch Hilbert series of $[\MQfive]$ from \eqref{E8m2 MQ} reads
\be
\label{E8-3:HS}
{\rm HS}_{E_{8}^{(-2)}}=1+16 t +272 t^2+\mc{O}(t^3)~,
\ee
in agreement with the above discussion. In particular, eight twisted hypermultiplets remain free, which give rise to the fundamental representation $\mathbf{16}$ of $\mathfrak{sp}(8)$, corresponding to the term $16t$ in (\ref{E8-3:HS}). 

\medskip
\noindent
Finally, we consider the $E_8^{(-1)}$, which has $f=0$ and a $\Z_5$ higher-form symmetry. We thus find:
\bea\label{MQ5 E8m1}
 \begin{tikzpicture}[x=.7cm,y=.7cm]
\node at (-5,0.5) {$\EQfour= \MQfive\cong$};
\draw[ligne, black](-2,0)--(4,0);
\draw[ligne, black](0,0)--(0,1);
\node[bd] at (-2,0) [label=below:{{\scriptsize$1$}}] {};
\node[bd] at (-1,0) [label=below:{{\scriptsize$3$}}] {};
\node[SUd] at (0,0) [label=below:{{\scriptsize$5$}}] {};
\node[bd] at (1,0) [label=below:{{\scriptsize$4$}}] {};
\node[bd] at (2,0) [label=below:{{\scriptsize$3$}}] {};
\node[bd] at (3,0) [label=below:{{\scriptsize$2$}}] {};
\node[bd] at (4,0) [label=below:{{\scriptsize$1$}}] {};
\node[bd] at (0,1) [label=above:{{\scriptsize$2$}}] {};
\end{tikzpicture}
\eea
for the `electric' 4d SCFT. Correspondingly, the $\EQfour$ for the `magnetic' theory is given by the same quiver with the $SU(5)$ node replaced  with a $U(5)$ node:
\bea\label{MQ5 magn}
 \begin{tikzpicture}[x=.7cm,y=.7cm]
\node at (-3.7,0.5) {$\MQfive \cong$};
\draw[ligne, black](-2,0)--(4,0);
\draw[ligne, black](0,0)--(0,1);
\node[bd] at (-2,0) [label=below:{{\scriptsize$1$}}] {};
\node[bd] at (-1,0) [label=below:{{\scriptsize$3$}}] {};
\node[bd] at (0,0) [label=below:{{\scriptsize$5$}}] {};
\node[bd] at (1,0) [label=below:{{\scriptsize$4$}}] {};
\node[bd] at (2,0) [label=below:{{\scriptsize$3$}}] {};
\node[bd] at (3,0) [label=below:{{\scriptsize$2$}}] {};
\node[bd] at (4,0) [label=below:{{\scriptsize$1$}}] {};
\node[bd] at (0,1) [label=above:{{\scriptsize$2$}}] {};
\end{tikzpicture}
\eea
This latter quiver is ugly and equivalent to $20$ twisted hypermultiplets. One can indeed find the $20$ monopole operators of \eqref{MQ5 magn} with $\Delta=\half$. They have $U(5)$ flux:
\be
(\m_1, \cdots, \m_5)\; = \;  (1,0,0,0,0)^{10}~, \; (1,1,0,0,0)^{10}~.
\ee
Therefore, the CB of the $\MQfive$ \eqref{MQ5 E8m1}, and the HB of $\FT$, take the form:
\be
{\rm CB}[\MQfive]\; \cong\; \mathbb{H}^{20}/\Z_5~, 
\ee
with the action:
\be
\Z_5\, : \, (\mathbb{H}_i, \mathbb{H}_j) \rightarrow (\omega_5\, \mathbb{H}_i, \omega_5^2\, \mathbb{H}_j)~,\; \; i=1, \cdots, 10~, \; \; j=1, \cdots, 10~.
\ee
This action preserves a $(U(10)\times U(10))/\mb{Z}_5$ subgroup of $Sp(20)$. The Hilbert series is 
\be
{\rm HS}_{E_{8}^{(-1)}}=1+200 t^2+\mc{O}(t^3)\,,
\ee
and confirms the dimension of the symmetry group of the `magnetic' version of $\FT$.

In summary, for this small family of terminal singularities, the Higgs branch of the rank-zero $\FT$ is given by
\be
{\rm HB}[\FT] \; \cong \;  \mathbb{H}^{d_H}/\frak{f} \qquad \text{or} \qquad \mathbb{H}^{d_H}~,
\ee
for the `electric' or `magnetic' version of the theory, respectively,  with the $\frak{f}=\Z_k$ actions specified above. Note that the `electric' theory with a non-trivial $\frak{f}$ has a 3-form symmetry $\Gamma^{(3)}=\frak{f}$, as expected for a discrete gauging of free hypermultiplets in 5d \cite{Gaiotto:2014kfa, Closset:2020scj}.

%%%%%%%%%%%%%%%%%%%%%%%%%%%%%%%%%%%%%%%%%%%%%%%%%%%%%%%%%%%%%%%%%%%%%%%%%%%%%%%%%%%%%%%%%%%%%%%%%%%%%%%%%%%%%%%%%%%%%%%%%%%%%%%%%%%%%%%%%%%%%%%%%%%%%%%%%%%%%%%%%%%%%%%%%%%%%%%%%%%%

\section{Generalized $E_n$ Theories and Discrete Gauging}
\label{sec:Higher-rank-En}

In this section, we study the $E_n^{(k)}$ singularities introduced in \eqref{Ensing}, 
%namely:
%\bea
%& E_6^{(k)}\; : \; \; && x_1^3+x_2^3+x_3^3+x_4^{3+k}~,\\
%& E_7^{(k)}\; : \; \; && x_1^2+x_2^4+x_3^4+x_4^{4+k}~,\\
%& E_8^{(k)}\; : \; \; && x_1^2+x_2^3+x_3^6+x_4^{6+k}~,
%\eea
for $k \geq 0$. They give rise to three infinite families of 5d SCFTs, $\FT= E_n^{(k)}$, $n=6,7,8$, of arbitrary rank $r$, with
\be
r=  \left\lfloor {k\ov \Delta_n} \right\rfloor+1~,\qquad \text{with}\; \; \;
  \Delta_6=3~, \; \;  \Delta_7=4~, \; \;  \Delta_8=6~. 
\ee
Here, the dimension $\Delta_n$ is the conformal dimension of the CB operator of the rank-one 4d $\CN=2$ Minahan-Nemeschanski (MN) theory with $E_n$ flavor symmetry \cite{Minahan:1996fg, Minahan:1996cj}. For $k= 0$, the $E_n^{(k)}$ canonical singularity precisely gives rise to the rank-one $E_n$ 5d SCFT, which flows to the 4d $E_n$ MN theory upon compactification on a circle. For ${k\ov \Delta_n}= \ell\in \Z$, one engineers the higher-rank $E_n$ theories (with $r= \ell +1$). Finally, for ${k\ov \Delta_n}$ not an integer, we will argue that we simply obtain the $E_n$ theories tensored with free hypermultiplets, modulo some discrete gauge group. 

As we argued in section~\ref{secquiverinesSU}, the 4d SCFT $\FTfour$ is a trinion with central node $SU(N)$, with $N= \Delta_n+k$. We therefore find  \cite{DelZotto:2015rca}
\bea\label{T4d Enk general k}
& \begin{tikzpicture}[x=.7cm,y=.7cm]
  \node at (-4,0.5) {$\FTXfour{\MG_{E_6^{(k)}}}\,  = $};
\draw[ligne, black](-1,0)--(1,0);
\draw[ligne, black](0,0)--(0,1);
\node at (-1.6,0) {{$D_3$}};
\node[SUd] at (0,0) [label=below:{{\scriptsize$N=3{+}k$}}] {};
\node at (1.6,0) {{$D_3$}};
\node at (0,1.4) {{$D_3$}};
\end{tikzpicture}~,
\quad
&& \begin{tikzpicture}[x=.7cm,y=.7cm]
  \node at (-4,0.5) {$\FTXfour{\MG_{E_7^{(k)}}}\,  = $};
\draw[ligne, black](-1,0)--(1,0);
\draw[ligne, black](0,0)--(0,1);
\node at (-1.6,0) {{$D_4$}};
\node[SUd] at (0,0) [label=below:{{\scriptsize$N=4{+}k$}}] {};
\node at (1.6,0) {{$D_4$}};
\node at (0,1.4) {{$D_2$}};
\end{tikzpicture}~,\\
& \begin{tikzpicture}[x=.7cm,y=.7cm]
  \node at (-4,0.5) {$\FTXfour{\MG_{E_8^{(k)}}}\,  = $};
\draw[ligne, black](-1,0)--(1,0);
\draw[ligne, black](0,0)--(0,1);
\node at (-1.6,0) {{$D_3$}};
\node[SUd] at (0,0) [label=below:{{\scriptsize$N=6{+}k$}}] {};
\node at (1.6,0) {{$D_6$}};
\node at (0,1.4) {{$D_2$}};
\end{tikzpicture}~.\\
\eea

For $N=m\Delta_n$, the 4d theory is Lagrangian, see the quivers in (\ref{EnkYellow}).

 % We first discuss the rank-one theories, which share most of the interesting properties of the general case. We then discuss the higher-rank case.  We also propose an explicit description of the 4d $\CN=2$ SCFT $\CS_{E_n^{(k)}}$, obtained by reducing the 5d SCFT $E_n^{(k)}$ on a circle, as a class-$\CS$ theory. This gives us a useful complementary point of view on the 5d physics of these canonical singularities. 

%%%%%%%%%%%%%%%%%%%%%%%%%%%%%%%%%%%%
\subsection{The Rank-one $E_n^{(k)}$ Theories and their Magnetic Quivers}

Let us first consider the $r=1$ $E_n^{(k)}$ models, with $k=0, \cdots, \Delta_n{-}1$; their basic geometric data is summarized in  table~\ref{table:Enk}. For $k=0$, these models engineer the rank-one $E_n$ 5d SCFTs. 
Indeed, these singularities admit  crepant resolutions with a single exceptional divisor. The resolved singularity $\t\MG$ takes the form:
\bea
&E_6^{(k)}\; : \;  &&  x_1^3+x_2^3+x_3^3+\delta_1^k x_4^{3+k}~, \\
&E_7^{(k)}\; : \;  &&  x_1^2+x_2^4+x_3^4+\delta_1^k x_4^{4+k}~, \\
&E_8^{(k)}\; : \;  &&  x_1^2+x_2^3+x_3^6+\delta_1^k x_4^{6+k}~.
\eea
For $k=0$, the exceptional divisor $\{\delta_1=0\}$ is a smooth del Pezzo surface, $dP_n$.
For $k>0$, the exceptional locus is a singular surface:
\bea
&E_6^{(k)}\; : \;  && \delta_1=0\; : \; \;   \{x_1^3+x_2^3+x_3^3= 0 \} \; && \subset \, \mathbb{P}^3~, \\
&E_7^{(k)}\; : \;  && \delta_1=0\; : \; \;    \{x_1^2+x_2^4+x_3^4= 0 \} \; && \subset \, \mathbb{P}_{[2,1,1,1]}^3~, \\
&E_8^{(k)}\; : \;  && \delta_1=0\; : \; \;     \{x_1^2+x_2^3+x_3^6= 0 \} \; && \subset \, \mathbb{P}_{[3,2,1,1]}^3~.
\eea
In addition, for $k>1$, we have a residual terminal singularity at $\delta_1= x_1=x_2=x_3=0$ (on the patch $x_4 =1$). These terminal singularities are of the type studied in the previous section: the $r=1$ model $E_n^{(k)}$ with $k>1$ contains the terminal singularity $E_n^{(k-\Delta_n)}$.

%%%%%%%%%%%%%%%%%%%%%%
\begin{table}[t]
\centering % used for centering table
\begin{tabular}{|c | c ||c |c | c|| c|c| c|c|c|| c| c|c|} % centered columns (4 columns)
\hline\hline %inserts double horizontal lines
 $E_n^{(k)}$&$F$& $r$ &$f$   &$d_H$ &$\h r$& $\h d_H$ &$a$&$c$ & $\Delta \CA_r$  & $b_3$ & $\frak{f}$  \\ [0.5ex] % inserts table
%heading
\hline % inserts single horizontal line
$E_6$ &$x_1^3+x_2^3 +x_3^3+ x_4^3$ & $1$ & $6$ & $11$ & $5$ & $7$&${109 \ov 24}$ &${29\ov 6}$  &$0$&  $0$&{$0$} \\
$E_6^{(1)}$ & $x_1^3+x_2^3 +x_3^3+ x_4^4$ & $1$ & $0$ & $12$ & $12$ & $1$&$10$ &$10$  &$-1$&  $2$&{$\Z_4$} \\
$E_6^{(2)}$ &$x_1^3+x_2^3 +x_3^3+ x_4^5$ & $1$ & $0$ & $16$ & $16$ & $1$&$16$ &$16$  &$-1$&  $2$&{$\Z_5$} \\
\hline
\hline
$E_7$ & $x_1^2+x_2^4 +x_3^4+ x_4^4$ & $1$ & $7$ & $17$ & $10$ & $8$&${31\ov3}$ &${32\ov3}$  &$0$&  $0$&{$0$} \\
$E_7^{(1)}$ &$x_1^2+x_2^4 +x_3^4+ x_4^5$ & $1$ & $0$ & $18$ & $18$ & $1$&${18}$ &${18}$  &$-1$&  $2$&{$\Z_5$} \\
$E_7^{(2)}$ &$x_1^2+x_2^4 +x_3^4+ x_4^6$ & $1$ & $3$ & $24$ & $21$ & $4$&${207\ov8}$ &$26$  &$-1$&  $2$&{$\Z_3$} \\
$E_7^{(3)}$ &$x_1^2+x_2^4 +x_3^4+ x_4^7$ & $1$ & $0$ & $27$ & $27$ & $1$&$36$ &$36$  &$-1$&  $2$&{$\Z_7$} \\
\hline 
\hline
$E_8$ & $x_1^2+x_2^3 +x_3^6+ x_4^6$ & $1$ & $8$ & $29$ & $21$ & $9$&${225\ov8}$ &${57\ov2}$  &$0$&  $0$&{$0$} \\
$E_8^{(1)}$ &$x_1^2+x_2^3 +x_3^6+ x_4^7$ & $1$ & $0$ & $30$ & $30$ & $1$&$40$ &$40$  &$-1$&  $2$&{$\Z_7$} \\
$E_8^{(2)}$ &$x_1^2+x_2^3 +x_3^6+ x_4^8$ & $1$ & $2$ & $36$ & $34$ & $3$&${209\ov4}$ &${157\ov3}$  &$-1$&  $2$&{$\Z_4$} \\
$E_8^{(3)}$ &$x_1^2+x_2^3 +x_3^6+ x_4^9$ & $1$ & $4$ & $42$ & $38$ & $5$&${397\ov6}$ &${199\ov3}$  &$-1$&  $2$&{$\Z_3$} \\
$E_8^{(4)}$ &$x_1^2+x_2^3 +x_3^6+ x_4^{10}$&$1$ & $2$ & $46$ & $44$ & $3$&${329\ov4}$ &${247\ov3}$  &$-1$&  $2$&{$\Z_5$} \\
$E_8^{(5)}$ &$x_1^2+x_2^3 +x_3^6+ x_4^{11}$&$1$ & $0$ & $50$ & $50$ & $1$&$100$ &$100$  &$-1$&  $2$&{$\Z_{11}$} \\
\hline
\end{tabular}
\caption{The $r=1$  $E_n^{(k)}$ models.} 
\label{table:Enk}
\end{table}
%%%%%%%%%

As we see from table~\ref{table:Enk}, these model display an interesting pattern of higher-form symmetries when $k>0$, since $\frak{f}$ is  non-trivial. As in the previous section, we should distinguish between the `electric'/$SU(N)$-type theory and `magnetic'/$U(N)$-type theories. 

We argued above that the terminal singularities $E_n^{(k<0)}$ give rise to free 5d hypermultiplets in M-theory, in the `magnetic' version of the theory. Therefore, from the above pattern of crepant resolutions, we expect that the $r=1$ $E_n^{(k>1)}$ 5d SCFTs have a Coulomb-branch low-energy theory that contains additional free hypermultiplets. By studying the Higgs branch of these models through their magnetic quivers, we will further argue that these free hypermultiplets actually arise as decoupled free sectors. 

%%%%%%%%%%%%%%%%%%%%%%
\begin{table}
\centering
$
\begin{array}{|c||c|c|c|c|}\hline
\text{Theory} &\EQfour& f& \MQfive & \Gamma^{(1)}_{\text{3d}} \cr \hline \hline
E_6&   \begin{tikzpicture}[x=.7cm,y=.7cm]
\draw[ligne, black](-2,0)--(2,0);
\draw[ligne, black](0,0)--(0,2);
\node[SUd] at (-2,0) [label=below:{{\scriptsize$1$}}] {};
\node[SUd] at (-1,0) [label=below:{{\scriptsize$2$}}] {};
\node[SUd] at (0,0) [label=below:{{\scriptsize$3$}}] {};
\node[SUd] at (1,0) [label=below:{{\scriptsize$2$}}] {};
\node[SUd] at (2,0) [label=below:{{\scriptsize$1$}}] {};
\node[SUd] at (0,1) [label=left:{{\scriptsize$2$}}] {};
\node[SUd] at (0,2) [label=left:{{\scriptsize$1$}}] {};
\end{tikzpicture} 
&
6
&   \begin{tikzpicture}[x=.7cm,y=.7cm]
\draw[ligne, black](-2,0)--(2,0);
\node[bd] at (-2,0) [label=below:{{\scriptsize$1$}}] {};
\node[bd] at (-1,0) [label=below:{{\scriptsize$2$}}] {};
\node[bd] at (0,0) [label=below:{{\scriptsize$3$}}] {};
\node[bd] at (1,0) [label=below:{{\scriptsize$2$}}] {};
\node[bd] at (2,0) [label=below:{{\scriptsize$1$}}] {};
\draw[ligne, black](0,0)--(0,2);
\node[bd] at (0,1) [label=left:{{\scriptsize$2$}}] {};
\node[bd] at (0,2) [label=left:{{\scriptsize$1$}}] {};
\end{tikzpicture} 
& \emptyset 
 \cr\hline 
E_6^{(1)} &   
\begin{tikzpicture}[x=.7cm,y=.7cm]
\draw[ligne, black](-2,0)--(2,0);
\draw[ligne, black](0,0)--(0,2);
\node[bd] at (-2,0) [label=below:{{\scriptsize$1$}}] {};
\node[bd] at (-1,0) [label=below:{{\scriptsize$2$}}] {};
\node[SUd] at (0,0) [label=below:{{\scriptsize$4$}}] {};
\node[bd] at (1,0) [label=below:{{\scriptsize$2$}}] {};
\node[bd] at (2,0) [label=below:{{\scriptsize$1$}}] {};
\node[bd] at (0,1) [label=left:{{\scriptsize$2$}}] {};
\node[bd] at (0,2) [label=left:{{\scriptsize$1$}}] {};
\end{tikzpicture}
&
\, 0\,
&   
\begin{tikzpicture}[x=.7cm,y=.7cm]
\draw[ligne, black](-2,0)--(2,0);
\draw[ligne, black](0,0)--(0,2);
\node[bd] at (-2,0) [label=below:{{\scriptsize$1$}}] {};
\node[bd] at (-1,0) [label=below:{{\scriptsize$2$}}] {};
\node[SUd] at (0,0) [label=below:{{\scriptsize$4$}}] {};
\node[bd] at (1,0) [label=below:{{\scriptsize$2$}}] {};
\node[bd] at (2,0) [label=below:{{\scriptsize$1$}}] {};
\node[bd] at (0,1) [label=left:{{\scriptsize$2$}}] {};
\node[bd] at (0,2) [label=left:{{\scriptsize$1$}}] {};
\end{tikzpicture}
%%
%% \begin{tikzpicture}[x=.7cm,y=.7cm]
%%\draw[ligne, black](-2,0)--(2,0);
%%\draw[ligne, black](0,0)--(0,2);
%%\node[bd] at (-2,0) [label=below:{{\scriptsize$1$}}] {};
%%\node[bd] at (-1,0) [label=below:{{\scriptsize$2$}}] {};
%%\node[bd] at (0,0) [label=below:{{\scriptsize$3$}}] {};
%%\node[bd] at (1,0) [label=below:{{\scriptsize$2$}}] {};
%%\node[bd] at (2,0) [label=below:{{\scriptsize$1$}}] {};
%%\node[bd] at (0,1) [label=left:{{\scriptsize$2$}}] {};
%%\node[bd] at (0,2) [label=left:{{\scriptsize$1$}}] {};
%%\end{tikzpicture}
%%+\mathbb{H} 
& \mathbb{Z}_4 
 \cr\hline 
E_6^{(2)} &  
\begin{tikzpicture}[x=.7cm,y=.7cm]
\draw[ligne, black](-2,0)--(2,0);
\draw[ligne, black](0,0)--(0,2);
\node[bd] at (-2,0) [label=below:{{\scriptsize$1$}}] {};
\node[bd] at (-1,0) [label=below:{{\scriptsize$3$}}] {};
\node[SUd] at (0,0) [label=below:{{\scriptsize$5$}}] {};
\node[bd] at (1,0) [label=below:{{\scriptsize$3$}}] {};
\node[bd] at (2,0) [label=below:{{\scriptsize$1$}}] {};
\node[bd] at (0,1) [label=left:{{\scriptsize$3$}}] {};
\node[bd] at (0,2) [label=left:{{\scriptsize$1$}}] {};
\end{tikzpicture}
&  
0
&
\begin{tikzpicture}[x=.7cm,y=.7cm]
\draw[ligne, black](-2,0)--(2,0);
\draw[ligne, black](0,0)--(0,2);
\node[bd] at (-2,0) [label=below:{{\scriptsize$1$}}] {};
\node[bd] at (-1,0) [label=below:{{\scriptsize$3$}}] {};
\node[SUd] at (0,0) [label=below:{{\scriptsize$5$}}] {};
\node[bd] at (1,0) [label=below:{{\scriptsize$3$}}] {};
\node[bd] at (2,0) [label=below:{{\scriptsize$1$}}] {};
\node[bd] at (0,1) [label=left:{{\scriptsize$3$}}] {};
\node[bd] at (0,2) [label=left:{{\scriptsize$1$}}] {};
\end{tikzpicture}
& 
\mathbb{Z}_5 \cr \hline 
\end{array}
$
\caption{$\EQfour$ and $\MQfive$ for the $r=1$ $E_6$  theories. Here we present the `electric' (`$SU(N)$ quiver') version of each theory, which has a one-form symmetry $\Gamma_{\rm 3d}^{(1)}=\frak{f}$ in 3d. \label{tab:E6MQ5}}
\end{table}
The quiverines $\EQfour$ and $\MQfive$ are given in tables~\ref{tab:E6MQ5}, \ref{tab:E7MQ5} and  \ref{tab:E8MQ5}, in their `electric' versions. With that choice of global structure, the electric 1-form symmetry of $\FTfour$ maps to a one-form symmetry of the 3d quiver,  $\Gamma^{(1)}_{\rm 3d}=\frak{f}$; correspondingly, the 5d theory $\FT$ has a 3-form symmetry $\Gamma^{(3)}_{\rm 5d}=\frak{f}$.

On the other hand, the `magnetic' version of the magnetic quivers $\MQfive$, corresponding to having a five-dimensional one-form symmetry $\Gamma^{(0)}_{\rm 5d}=\frak{f}$, are given by the same quivers with only black nodes (only $U(n)$ gauge groups, modulo an overall $U(1)$). These magnetic quivers are shown in tables~\ref{tab:E6kMQ}, \ref{tab:E7kMQ} and  \ref{tab:E8kMQ} below, for general $k$.

\subsubsection{The $r=1$ $E_6^{(k)}$ theories} 
Let us first consider the $E_6$ series in detail. The $\MQfive$ for $E_6 \equiv E_6^{(0)}$ in table~\ref{tab:E6MQ5} is the well-known magnetic quiver for the $E_6$ theory, in the shape of the affine $E_6$ Dynkin diagram; its 3d $\CN=4$ Coulomb branch is the  one-$E_6$-instanton moduli space, of dimension $d_H=11$. Note that is was crucial to gauge the $U(1)^6$ flavor symmetry of $\EQfour$ to obtain $\MQfive$, as explained in \cite{Closset:2020scj, CSNWII}; the resulting $U(1)_T^6$ topological symmetry becomes the Cartan of the $E_6$ 3d CB flavor symmetry. 

\medskip
\noindent
The $E_6^{(1)}$ model has $r=1$ and $d_H=12$. The $U(N)$ version of the $\MQfive$ is given by
\bea\label{MQ5 E61 0}
\begin{tikzpicture}[x=.6cm,y=.6cm]
\draw[ligne, black](-2,0)--(2,0);
\draw[ligne, black](0,0)--(0,2);
\node[bd] at (-2,0) [label=below:{{\scriptsize$1$}}] {};
\node[bd] at (-1,0) [label=below:{{\scriptsize$2$}}] {};
\node[bd] at (0,0) [label=below:{{\scriptsize$4$}}] {};
\node[bd] at (1,0) [label=below:{{\scriptsize$2$}}] {};
\node[bd] at (2,0) [label=below:{{\scriptsize$1$}}] {};
\node[bd] at (0,1) [label=left:{{\scriptsize$2$}}] {};
\node[bd] at (0,2) [label=left:{{\scriptsize$1$}}] {};
\end{tikzpicture}
\eea
Note that the central node is `bad' in the GW sense, since it is locally $U(4)$ SQCD with $N_f=6$ flavors. Correspondingly, the 3d UV description contains monopole operators of vanishing dimension. Therefore, this quiver description does not give a direct handle on the IR theory in 3d. 

In this case, one can check that the `badness' of the quiver \eqref{MQ5 E61 0} only comes from its central node. Consider $U(N)$ SQCD with $N_f= 2N-2$ flavors. It is known that the most singular locus on the Coulomb branch of that theory is $\mathbb{H} \cong \mathbb{C}^2$, and that along that locus  the low-energy effective theory is $U(N-1)$ SQCD with $N_f= 2N-2$, plus a free twisted hypermultiplet whose vacuum expectation value (VEV) parametrizes the $\mathbb{C}^2$ locus \cite{Assel:2017jgo}.%
\footnote{This Seiberg-like duality was first suggested by \protect\cite{Yaakov:2013fza}, and its meaning was clarified in \protect\cite{Assel:2017jgo}. See also \protect\cite{Dey:2017fqs}.} 
This is the effect of a non-perturbative Higgs mechanism:
\be
U(N) \rightarrow U(N-1) \times U(1)~,
\ee
as we tune the CB parameters to the most singular locus.  At the central node of the quiver, we locally have such a situation, with $N=4$. We then propose that the relevant moduli space of the $E_6^{(1)}$ magnetic quiverine is obtained by zooming in onto that most singular locus. By effectively replacing the $U(4)$ node by an $U(3)$ node in \eqref{MQ5 E61 0}, the relevant 3d CB of $\MQfive$ is then obtained from the 3d theory:
\bea\label{MQ5 E61 1}
 \begin{tikzpicture}[x=.6cm,y=.6cm]
 \node at (-5,0.5) {$\MQfive[E_6^{(1)}] \cong $ };
\draw[ligne, black](-2,0)--(2,0);
\draw[ligne, black](0,0)--(0,2);
\node[bd] at (-2,0) [label=below:{{\scriptsize$1$}}] {};
\node[bd] at (-1,0) [label=below:{{\scriptsize$2$}}] {};
\node[bd] at (0,0) [label=below:{{\scriptsize$3$}}] {};
\node[bd] at (1,0) [label=below:{{\scriptsize$2$}}] {};
\node[bd] at (2,0) [label=below:{{\scriptsize$1$}}] {};
\node[bd] at (0,1) [label=left:{{\scriptsize$2$}}] {};
\node[bd] at (0,2) [label=left:{{\scriptsize$1$}}] {};
\node at (3,0.5) {$\oplus \, \mathbb{H}$};
\end{tikzpicture} 
\eea
which is the $\MQfive$ of the rank-1 $E_6$ theory plus a decoupled (twisted) hypermultiplet.

This interpretation is corroborated by looking at the Higgs branch of the $\MQfive$, which describes the Coulomb branch of $\FT$ compactfied on a torus. Both descriptions \eqref{MQ5 E61 0} and \eqref{MQ5 E61 1} give the same result: the HB of $\MQfive$ is given by $\C^2/\Gamma_{E_6}$, the Kleinian singularity of type $E_6$, plus a free twisted hypermultiplet. Indeed, it is well-known that the HB of the quiver with $U(3)$ central node is $\C^2/\Gamma_{E_6}$, and one can check that the same is true for the quiver \eqref{MQ5 E61 0} (using, for instance, the methods of \cite{Lindstrom:1999pz}). In the latter case, however, there is a residual `unhiggsed' $U(1)$ vector multiplet at low-energy on the HB, which gives us the free twisted hypermultiplet in the IR.
This simple structure of the HB of $\MQfive$ is also consistent with the structure of the smooth resolved threefold $\t \MG_{E_6^{(1)}}$. Importantly, the resolved threefold  contains a pair of three-cycles ($b_3=2$), which, from the M-theory perspective, give rise to a free (uncharged) hypermultiplet on the 5d Coulomb branch, in perfect agreement with the above $\MQfive$.

These considerations lead us to conjecture that the `magnetic' version of the $E_6^{(1)}$ 5d SCFT is simply the $E_6$ SCFT tensored with a free hypermultiplet:
\be\label{E61 magn}
{\mathcal{T}}_{\MG_{E_6^{(1)}}}^{\rm 5d} \Big|_{\text{`magnetic'}} \; \cong\;  {\mathcal{T}}_{\MG_{E_6}}^{\rm 5d}  \otimes \mathbb{H}~.
\ee
We will present further evidence for this claim in section~\ref{subsec:classS}.

The asymptotic geometry of the canonical singularity tells us that the 5d theory \eqref{E61 magn} has a distinguished $\Z_4$ 0-form symmetry. By gauging it, we expect to find the `electric' version of the theory, with a $\Z_4$ 3-form symmetry in 5d:
\be\label{E61 elec}
{\mathcal{T}}_{\MG_{E_6^{(1)}}}^{\rm 5d} \Big|_{\text{`electric'}} \; \cong\;  \left({\mathcal{T}}_{\MG_{E_6}}^{\rm 5d}  \otimes \mathbb{H}\right)/\Z_4~.
\ee
The $\Z_4$ action used in \eqref{E61 elec} can be understood at the level of the quantum Higgs branch. Indeed, the $\MQfive$ for the `electric' $E_6^{(1)}$ is the one with $SU(4)$ central node shown in table~\ref{tab:E6MQ5}, which is a $\Z_4$ gauging of the unitary quiver \eqref{MQ5 E61 0}.

\begin{table}
\centering
$
\begin{array}{|c||c|c|c|c|}\hline
\text{Theory} &\EQfour& f& \MQfive & \Gamma^{(1)}_{\text{3d}} \cr \hline \hline
E_7
&   \begin{tikzpicture}[x=.7cm,y=.7cm]
\draw[ligne, black](-3,0)--(3,0);
\draw[ligne, black](0,0)--(0,1);
\node[SUd] at (-3,0) [label=below:{{\scriptsize$1$}}] {};
\node[SUd] at (-2,0) [label=below:{{\scriptsize$2$}}] {};
\node[SUd] at (-1,0) [label=below:{{\scriptsize$3$}}] {};
\node[SUd] at (0,0) [label=below:{{\scriptsize$4$}}] {};
\node[SUd] at (1,0) [label=below:{{\scriptsize$3$}}] {};
\node[SUd] at (2,0) [label=below:{{\scriptsize$2$}}] {};
\node[SUd] at (3,0) [label=below:{{\scriptsize$1$}}] {};
\node[SUd] at (0,1) [label=above:{{\scriptsize$2$}}] {};
\end{tikzpicture}
&
\,7\,
&   \begin{tikzpicture}[x=.7cm,y=.7cm]
\draw[ligne, black](-3,0)--(3,0);
\draw[ligne, black](0,0)--(0,1);
\node[bd] at (-3,0) [label=below:{{\scriptsize$1$}}] {};
\node[bd] at (-2,0) [label=below:{{\scriptsize$2$}}] {};
\node[bd] at (-1,0) [label=below:{{\scriptsize$3$}}] {};
\node[bd] at (0,0) [label=below:{{\scriptsize$4$}}] {};
\node[bd] at (1,0) [label=below:{{\scriptsize$3$}}] {};
\node[bd] at (2,0) [label=below:{{\scriptsize$2$}}] {};
\node[bd] at (3,0) [label=below:{{\scriptsize$1$}}] {};
\node[bd] at (0,1) [label=above:{{\scriptsize$2$}}] {};
\end{tikzpicture}
&
\emptyset 
 \cr\hline
E_7^{(1)}&  
 \begin{tikzpicture}[x=.7cm,y=.7cm]
\draw[ligne, black](-3,0)--(3,0);
\draw[ligne, black](0,0)--(0,1);
\node[bd] at (-3,0) [label=below:{{\scriptsize$1$}}] {};
\node[bd] at (-2,0) [label=below:{{\scriptsize$2$}}] {};
\node[bd] at (-1,0) [label=below:{{\scriptsize$3$}}] {};
\node[SUd] at (0,0) [label=below:{{\scriptsize$5$}}] {};
\node[bd] at (1,0) [label=below:{{\scriptsize$3$}}] {};
\node[bd] at (2,0) [label=below:{{\scriptsize$2$}}] {};
\node[bd] at (3,0) [label=below:{{\scriptsize$1$}}] {};
\node[bd] at (0,1) [label=above:{{\scriptsize$2$}}] {};
\end{tikzpicture}
& 
0
&  
 \begin{tikzpicture}[x=.7cm,y=.7cm]
\draw[ligne, black](-3,0)--(3,0);
\draw[ligne, black](0,0)--(0,1);
\node[bd] at (-3,0) [label=below:{{\scriptsize$1$}}] {};
\node[bd] at (-2,0) [label=below:{{\scriptsize$2$}}] {};
\node[bd] at (-1,0) [label=below:{{\scriptsize$3$}}] {};
\node[SUd] at (0,0) [label=below:{{\scriptsize$5$}}] {};
\node[bd] at (1,0) [label=below:{{\scriptsize$3$}}] {};
\node[bd] at (2,0) [label=below:{{\scriptsize$2$}}] {};
\node[bd] at (3,0) [label=below:{{\scriptsize$1$}}] {};
\node[bd] at (0,1) [label=above:{{\scriptsize$2$}}] {};
\end{tikzpicture}

%% \begin{tikzpicture}[x=.7cm,y=.7cm]
%%\draw[ligne, black](-3,0)--(3,0);
%%\draw[ligne, black](0,0)--(0,1);
%%\node[bd] at (-3,0) [label=below:{{\scriptsize$1$}}] {};
%%\node[bd] at (-2,0) [label=below:{{\scriptsize$2$}}] {};
%%\node[bd] at (-1,0) [label=below:{{\scriptsize$3$}}] {};
%%\node[bd] at (0,0) [label=below:{{\scriptsize$4$}}] {};
%%\node[bd] at (1,0) [label=below:{{\scriptsize$3$}}] {};
%%\node[bd] at (2,0) [label=below:{{\scriptsize$2$}}] {};
%%\node[bd] at (3,0) [label=below:{{\scriptsize$1$}}] {};
%%\node[bd] at (0,1) [label=above:{{\scriptsize$2$}}] {};
%%\end{tikzpicture}
%%+\mathbb{H} 
& 
\mathbb{Z}_5
 \cr\hline
E_7^{(2)}&   
\begin{tikzpicture}[x=.7cm,y=.7cm]
\draw[ligne, black](-3,0)--(3,0);
\draw[ligne, black](0,0)--(0,1);
\node[bd] at (-3,0) [label=below:{{\scriptsize$1$}}] {};
\node[SUd] at (-2,0) [label=below:{{\scriptsize$3$}}] {};
\node[bd] at (-1,0) [label=below:{{\scriptsize$4$}}] {};
\node[SUd] at (0,0) [label=below:{{\scriptsize$6$}}] {};
\node[bd] at (1,0) [label=below:{{\scriptsize$4$}}] {};
\node[SUd] at (2,0) [label=below:{{\scriptsize$3$}}] {};
\node[bd] at (3,0) [label=below:{{\scriptsize$1$}}] {};
\node[SUd] at (0,1) [label=above:{{\scriptsize$3$}}] {};
\end{tikzpicture}
&
3
&   
\begin{tikzpicture}[x=.7cm,y=.7cm]
\draw[ligne, black](-3,0)--(3,0);
\draw[ligne, black](0,0)--(0,1);
\node[bd] at (-3,0) [label=below:{{\scriptsize$1$}}] {};
\node[bd] at (-2,0) [label=below:{{\scriptsize$3$}}] {};
\node[bd] at (-1,0) [label=below:{{\scriptsize$4$}}] {};
\node[SUd] at (0,0) [label=below:{{\scriptsize$6/\mathbb{Z}_2$}}] {};
\node[bd] at (1,0) [label=below:{{\scriptsize$4$}}] {};
\node[bd] at (2,0) [label=below:{{\scriptsize$3$}}] {};
\node[bd] at (3,0) [label=below:{{\scriptsize$1$}}] {};
\node[bd] at (0,1) [label=above:{{\scriptsize$3$}}] {};
\end{tikzpicture}
%%
%%\begin{tikzpicture}[x=.7cm,y=.7cm]
%%\draw[ligne, black](-3,0)--(3,0);
%%\draw[ligne, black](0,0)--(0,1);
%%\node[bd] at (-3,0) [label=below:{{\scriptsize$1$}}] {};
%%\node[bd] at (-2,0) [label=below:{{\scriptsize$2$}}] {};
%%\node[bd] at (-1,0) [label=below:{{\scriptsize$4$}}] {};
%%\node[SUd] at (0,0) [label=below:{{\scriptsize$6/\mathbb{Z}_2$}}] {};
%%\node[bd] at (1,0) [label=below:{{\scriptsize$4$}}] {};
%%\node[bd] at (2,0) [label=below:{{\scriptsize$2$}}] {};
%%\node[bd] at (3,0) [label=below:{{\scriptsize$1$}}] {};
%%\node[bd] at (0,1) [label=above:{{\scriptsize$3$}}] {};
%%\end{tikzpicture}
%%+2\mathbb{H} 
&
 \mathbb{Z}_3
 \cr\hline
E_7^{(3)} &   \begin{tikzpicture}[x=.7cm,y=.7cm]
\draw[ligne, black](-3,0)--(3,0);
\draw[ligne, black](0,0)--(0,1);
\node[bd] at (-3,0) [label=below:{{\scriptsize$1$}}] {};
\node[bd] at (-2,0) [label=below:{{\scriptsize$3$}}] {};
\node[bd] at (-1,0) [label=below:{{\scriptsize$5$}}] {};
\node[SUd] at (0,0) [label=below:{{\scriptsize$7$}}] {};
\node[bd] at (1,0) [label=below:{{\scriptsize$5$}}] {};
\node[bd] at (2,0) [label=below:{{\scriptsize$3$}}] {};
\node[bd] at (3,0) [label=below:{{\scriptsize$1$}}] {};
\node[bd] at (0,1) [label=above:{{\scriptsize$3$}}] {};
\end{tikzpicture}
& 
0
&   \begin{tikzpicture}[x=.7cm,y=.7cm]
\draw[ligne, black](-3,0)--(3,0);
\draw[ligne, black](0,0)--(0,1);
\node[bd] at (-3,0) [label=below:{{\scriptsize$1$}}] {};
\node[bd] at (-2,0) [label=below:{{\scriptsize$3$}}] {};
\node[bd] at (-1,0) [label=below:{{\scriptsize$5$}}] {};
\node[SUd] at (0,0) [label=below:{{\scriptsize$7$}}] {};
\node[bd] at (1,0) [label=below:{{\scriptsize$5$}}] {};
\node[bd] at (2,0) [label=below:{{\scriptsize$3$}}] {};
\node[bd] at (3,0) [label=below:{{\scriptsize$1$}}] {};
\node[bd] at (0,1) [label=above:{{\scriptsize$3$}}] {};
\end{tikzpicture}
& 
\mathbb{Z}_7 \cr \hline
\end{array}
$
\caption{$\EQfour$ and $\MQfive$ for the $r=1$ $E_7$  theories and their 1-form symmetry. \label{tab:E7MQ5}}
\end{table}
%%%%%%%%%%%%%

\medskip
\noindent
The $E_6^{(2)}$ model has $r=1$ and $d_H=16$. Now, the resolved geometry both contains a 3-cycle and a terminal singularity of type $E_6^{(-1)}$. The physics of this model is again easiest to understand in its `magnetic' version. The unitary $\MQfive$ has a $U(5)$ central node with $N_f=9$ flavors, so that one can dualize it to a $U(4)$ node plus a free hypermultiplet. At this stage, the $U(3)$ nodes on the three legs become ugly, and can each be dualized:
\bea\label{MQ5 E62 0}
\begin{tikzpicture}[x=.6cm,y=.6cm]
\draw[ligne, black](-2,0)--(2,0);
\draw[ligne, black](0,0)--(0,2);
\node[bd] at (-2,0) [label=below:{{\scriptsize$1$}}] {};
\node[bd] at (-1,0) [label=below:{{\scriptsize$3$}}] {};
\node[bd] at (0,0) [label=below:{{\scriptsize$5$}}] {};
\node[bd] at (1,0) [label=below:{{\scriptsize$3$}}] {};
\node[bd] at (2,0) [label=below:{{\scriptsize$1$}}] {};
\node[bd] at (0,1) [label=left:{{\scriptsize$3$}}] {};
\node[bd] at (0,2) [label=left:{{\scriptsize$1$}}] {};
\node at (3,0.5) {$\cong$};
\end{tikzpicture}
\; \;
\begin{tikzpicture}[x=.6cm,y=.6cm]
\draw[ligne, black](-2,0)--(2,0);
\draw[ligne, black](0,0)--(0,2);
\node[bd] at (-2,0) [label=below:{{\scriptsize$1$}}] {};
\node[bd] at (-1,0) [label=below:{{\scriptsize$3$}}] {};
\node[bd] at (0,0) [label=below:{{\scriptsize$4$}}] {};
\node[bd] at (1,0) [label=below:{{\scriptsize$3$}}] {};
\node[bd] at (2,0) [label=below:{{\scriptsize$1$}}] {};
\node[bd] at (0,1) [label=left:{{\scriptsize$3$}}] {};
\node[bd] at (0,2) [label=left:{{\scriptsize$1$}}] {};
\node at (3.5,0.5) {$\oplus\; \mathbb{H}\; \; \; \cong$};
\end{tikzpicture}
\;\;
\begin{tikzpicture}[x=.6cm,y=.6cm]
\draw[ligne, black](-2,0)--(2,0);
\draw[ligne, black](0,0)--(0,2);
\node[bd] at (-2,0) [label=below:{{\scriptsize$1$}}] {};
\node[bd] at (-1,0) [label=below:{{\scriptsize$2$}}] {};
\node[bd] at (0,0) [label=below:{{\scriptsize$4$}}] {};
\node[bd] at (1,0) [label=below:{{\scriptsize$2$}}] {};
\node[bd] at (2,0) [label=below:{{\scriptsize$1$}}] {};
\node[bd] at (0,1) [label=left:{{\scriptsize$2$}}] {};
\node[bd] at (0,2) [label=left:{{\scriptsize$1$}}] {};
\node at (3.5,0.5) {$\oplus\; \mathbb{H}^4~.$};
\end{tikzpicture}
\eea
We then obtain the $\MQfive$ of the $E_6^{(1)}$ theory plus four twisted hypermultiplets. We claim that this free sector is associated to the terminal singularity $E_6^{(-1)}$ that arises on the resolved threefold $\t\MG_{E_6^{(2)}}$. We then treat the $E_6^{(1)}$ factor as above, and conclude that the `magnetic' version of the $E_6^{(2)}$ theory in 5d is equivalent to the $E_6$ theory tensored with five free hypermultiplets. 
We can summarize this decomposition pictorially as
\bea\label{E62 MQ decomp}
\begin{tikzpicture}[x=.6cm,y=.6cm]
\draw[ligne, black](-2,0)--(2,0);
\draw[ligne, black](0,0)--(0,2);
\node[bd] at (-2,0) [label=below:{{\scriptsize$1$}}] {};
\node[bd] at (-1,0) [label=below:{{\scriptsize$3$}}] {};
\node[bd] at (0,0) [label=below:{{\scriptsize$5$}}] {};
\node[bd] at (1,0) [label=below:{{\scriptsize$3$}}] {};
\node[bd] at (2,0) [label=below:{{\scriptsize$1$}}] {};
\node[bd] at (0,1) [label=left:{{\scriptsize$3$}}] {};
\node[bd] at (0,2) [label=left:{{\scriptsize$1$}}] {};
\node at (4,0.5) {$\cong$};
\end{tikzpicture}
\quad
\begin{tikzpicture}[x=.6cm,y=.6cm]
\draw[ligne, black](-2,0)--(2,0);
\draw[ligne, black](0,0)--(0,2);
\node[bd] at (-2,0) [label=below:{{\scriptsize$1$}}] {};
\node[bd] at (-1,0) [label=below:{{\scriptsize$2$}}] {};
\node[bd] at (0,0) [label=below:{{\scriptsize$3$}}] {};
\node[bd] at (1,0) [label=below:{{\scriptsize$2$}}] {};
\node[bd] at (2,0) [label=below:{{\scriptsize$1$}}] {};
\node[bd] at (0,1) [label=left:{{\scriptsize$2$}}] {};
\node[bd] at (0,2) [label=left:{{\scriptsize$1$}}] {};
\node at (3.6,0.5) {$\oplus\;\; \mathbb{H}\;\; \oplus$};
\draw[ligne, black](5,0)--(7,0);
\draw[ligne, black](6,0)--(6,1);
\node[bd] at (5,0) [label=below:{{\scriptsize$1$}}] {};
\node[bd] at (6,0) [label=below:{{\scriptsize$2$}}] {};
\node[bd] at (7,0) [label=below:{{\scriptsize$1$}}] {};
\node[bd] at (6,1) [label=left:{{\scriptsize$1$}}] {};
\end{tikzpicture}
\eea
This pattern is again consistent with the analysis of the Higgs branch for the $\MQfive$ on the left of \eqref{E62 MQ decomp}, which is still $\C^2/\Gamma_{E_6}$ together with a larger unhiggsable sector that is equivalent to the $E_6^{(-1)}\cong T_2$ magnetic quiver (plus a free twisted hypermultiplet) on the right of \eqref{E62 MQ decomp}.
We also note, heuristically, that this result can be obtained by a naive application of the rules of quiver substraction \cite{Cabrera:2018ann, Bourget:2019aer}.

Finally, the `electric version' of the theory is obtained from the `magnetic' version by gauging the $\Z_5$ $0$-form symmetry. In conclusion, we find
\be
{\mathcal{T}}_{\MG_{E_6^{(2)}}}^{\rm 5d} \Big|_{\text{`magnetic'}} \; \cong\;  {\mathcal{T}}_{\MG_{E_6}}^{\rm 5d}  \otimes \mathbb{H}^5~, \qquad \qquad
{\mathcal{T}}_{\MG_{E_6^{(2)}}}^{\rm 5d} \Big|_{\text{`electric'}} \; \cong\;  \left({\mathcal{T}}_{\MG_{E_6}}^{\rm 5d}  \otimes \mathbb{H}^5\right)/\Z_5~.
\ee

%%%%%%%%%%%%
\subsubsection{The $r=1$ $E_7^{(k)}$ and  $E_8^{(k)}$ theories} 
%%%%%%%%
\begin{table}
$
\begin{array}{|c||c|c|c|c|}\hline
\text{Theory} &\EQfour& f& \MQfive & \Gamma^{(1)}_{\text{3d}} \cr \hline \hline
E_8 &   \begin{tikzpicture}[x=.7cm,y=.7cm]
\draw[ligne, black](-2,0)--(5,0);
\draw[ligne, black](0,0)--(0,1);
\node[SUd] at (-2,0) [label=below:{{\scriptsize$2$}}] {};
\node[SUd] at (-1,0) [label=below:{{\scriptsize$4$}}] {};
\node[SUd] at (0,0) [label=below:{{\scriptsize$6$}}] {};
\node[SUd] at (1,0) [label=below:{{\scriptsize$5$}}] {};
\node[SUd] at (2,0) [label=below:{{\scriptsize$4$}}] {};
\node[SUd] at (3,0) [label=below:{{\scriptsize$3$}}] {};
\node[SUd] at (4,0) [label=below:{{\scriptsize$2$}}] {};
\node[SUd] at (5,0) [label=below:{{\scriptsize$1$}}] {};
\node[SUd] at (0,1) [label=above:{{\scriptsize$3$}}] {};
\end{tikzpicture}
& 
\, 8\, 
&   \begin{tikzpicture}[x=.7cm,y=.7cm]
\draw[ligne, black](-2,0)--(5,0);
\node[bd] at (-2,0) [label=below:{{\scriptsize$2$}}] {};
\node[bd] at (-1,0) [label=below:{{\scriptsize$4$}}] {};
\node[bd] at (0,0) [label=below:{{\scriptsize$6$}}] {};
\node[bd] at (1,0) [label=below:{{\scriptsize$5$}}] {};
\node[bd] at (2,0) [label=below:{{\scriptsize$4$}}] {};
\node[bd] at (3,0) [label=below:{{\scriptsize$3$}}] {};
\node[bd] at (4,0) [label=below:{{\scriptsize$2$}}] {};
\node[bd] at (5,0) [label=below:{{\scriptsize$1$}}] {};
\draw[ligne, black](0,0)--(0,1);
\node[bd] at (0,1) [label=above:{{\scriptsize$3$}}] {};
\end{tikzpicture}
&
\emptyset 
\cr\hline 
E_8^{(1)}
& 
 \begin{tikzpicture}[x=.7cm,y=.7cm]
\draw[ligne, black](-2,0)--(5,0);
\draw[ligne, black](0,0)--(0,1);
\node[bd] at (-2,0) [label=below:{{\scriptsize$2$}}] {};
\node[bd] at (-1,0) [label=below:{{\scriptsize$4$}}] {};
\node[SUd] at (0,0) [label=below:{{\scriptsize$7$}}] {};
\node[bd] at (1,0) [label=below:{{\scriptsize$5$}}] {};
\node[bd] at (2,0) [label=below:{{\scriptsize$4$}}] {};
\node[bd] at (3,0) [label=below:{{\scriptsize$3$}}] {};
\node[bd] at (4,0) [label=below:{{\scriptsize$2$}}] {};
\node[bd] at (5,0) [label=below:{{\scriptsize$1$}}] {};
\node[bd] at (0,1) [label=above:{{\scriptsize$3$}}] {};
\end{tikzpicture} 
& 
0
& 
 \begin{tikzpicture}[x=.7cm,y=.7cm]
\draw[ligne, black](-2,0)--(5,0);
\draw[ligne, black](0,0)--(0,1);
\node[bd] at (-2,0) [label=below:{{\scriptsize$2$}}] {};
\node[bd] at (-1,0) [label=below:{{\scriptsize$4$}}] {};
\node[SUd] at (0,0) [label=below:{{\scriptsize$7$}}] {};
\node[bd] at (1,0) [label=below:{{\scriptsize$5$}}] {};
\node[bd] at (2,0) [label=below:{{\scriptsize$4$}}] {};
\node[bd] at (3,0) [label=below:{{\scriptsize$3$}}] {};
\node[bd] at (4,0) [label=below:{{\scriptsize$2$}}] {};
\node[bd] at (5,0) [label=below:{{\scriptsize$1$}}] {};
\node[bd] at (0,1) [label=above:{{\scriptsize$3$}}] {};
\end{tikzpicture}
%
% \begin{tikzpicture}[x=.7cm,y=.7cm]
%\draw[ligne, black](-2,0)--(5,0);
%\draw[ligne, black](0,0)--(0,1);
%\node[bd] at (-2,0) [label=below:{{\scriptsize$2$}}] {};
%\node[bd] at (-1,0) [label=below:{{\scriptsize$4$}}] {};
%\node[bd] at (0,0) [label=below:{{\scriptsize$6$}}] {};
%\node[bd] at (1,0) [label=below:{{\scriptsize$5$}}] {};
%\node[bd] at (2,0) [label=below:{{\scriptsize$4$}}] {};
%\node[bd] at (3,0) [label=below:{{\scriptsize$3$}}] {};
%\node[bd] at (4,0) [label=below:{{\scriptsize$2$}}] {};
%\node[bd] at (5,0) [label=below:{{\scriptsize$1$}}] {};
%\node[bd] at (0,1) [label=above:{{\scriptsize$3$}}] {};
%\end{tikzpicture}  
%+\mathbb{H} 
& 
\mathbb{Z}_7 
\cr \hline 
E_8^{(2)}
&   \begin{tikzpicture}[x=.7cm,y=.7cm]
\draw[ligne, black](-2,0)--(5,0);
\draw[ligne, black](0,0)--(0,1);
\node[bd] at (-2,0) [label=below:{{\scriptsize$2$}}] {};
\node[bd] at (-1,0) [label=below:{{\scriptsize$5$}}] {};
\node[SUd] at (0,0) [label=below:{{\scriptsize$8$}}] {};
\node[bd] at (1,0) [label=below:{{\scriptsize$6$}}] {};
\node[bd] at (2,0) [label=below:{{\scriptsize$5$}}] {};
\node[SUd] at (3,0) [label=below:{{\scriptsize$4$}}] {};
\node[bd] at (4,0) [label=below:{{\scriptsize$2$}}] {};
\node[bd] at (5,0) [label=below:{{\scriptsize$1$}}] {};
\node[SUd] at (0,1) [label=above:{{\scriptsize$4$}}] {};
\end{tikzpicture}
& 
2
&   
\begin{tikzpicture}[x=.7cm,y=.7cm]
\draw[ligne, black](-2,0)--(5,0);
\draw[ligne, black](0,0)--(0,1);
\node[bd] at (-2,0) [label=below:{{\scriptsize$2$}}] {};
\node[bd] at (-1,0) [label=below:{{\scriptsize$5$}}] {};
\node[SUd] at (0,0) [label=below:{{\scriptsize$8/\mathbb{Z}_2$}}] {};
\node[bd] at (1,0) [label=below:{{\scriptsize$6$}}] {};
\node[bd] at (2,0) [label=below:{{\scriptsize$5$}}] {};
\node[bd] at (3,0) [label=below:{{\scriptsize$4$}}] {};
\node[bd] at (4,0) [label=below:{{\scriptsize$2$}}] {};
\node[bd] at (5,0) [label=below:{{\scriptsize$1$}}] {};
\node[bd] at (0,1) [label=above:{{\scriptsize$4$}}] {};
\end{tikzpicture}
%
% \begin{tikzpicture}[x=.7cm,y=.7cm]
%\draw[ligne, black](-2,0)--(5,0);
%\draw[ligne, black](0,0)--(0,1);
%\node[bd] at (-2,0) [label=below:{{\scriptsize$2$}}] {};
%\node[bd] at (-1,0) [label=below:{{\scriptsize$5$}}] {};
%\node[SUd] at (0,0) [label=below:{{\scriptsize$8/\mathbb{Z}_2$}}] {};
%\node[bd] at (1,0) [label=below:{{\scriptsize$6$}}] {};
%\node[bd] at (2,0) [label=below:{{\scriptsize$4$}}] {};
%\node[bd] at (3,0) [label=below:{{\scriptsize$3$}}] {};
%\node[bd] at (4,0) [label=below:{{\scriptsize$2$}}] {};
%\node[bd] at (5,0) [label=below:{{\scriptsize$1$}}] {};
%\node[bd] at (0,1) [label=above:{{\scriptsize$4$}}] {};
%\end{tikzpicture}
%+2\mathbb{H} 
& 
\mathbb{Z}_4
\cr \hline 
E_8^{(3)}
&   \begin{tikzpicture}[x=.7cm,y=.7cm]
\draw[ligne, black](-2,0)--(5,0);
\draw[ligne, black](0,0)--(0,1);
\node[SUd] at (-2,0) [label=below:{{\scriptsize$3$}}] {};
\node[SUd] at (-1,0) [label=below:{{\scriptsize$6$}}] {};
\node[SUd] at (0,0) [label=below:{{\scriptsize$9$}}] {};
\node[bd] at (1,0) [label=below:{{\scriptsize$7$}}] {};
\node[SUd] at (2,0) [label=below:{{\scriptsize$6$}}] {};
\node[bd] at (3,0) [label=below:{{\scriptsize$4$}}] {};
\node[SUd] at (4,0) [label=below:{{\scriptsize$3$}}] {};
\node[bd] at (5,0) [label=below:{{\scriptsize$1$}}] {};
\node[bd] at (0,1) [label=above:{{\scriptsize$4$}}] {};
\end{tikzpicture}
&
4
&  
 \begin{tikzpicture}[x=.7cm,y=.7cm]
\draw[ligne, black](-2,0)--(5,0);
\draw[ligne, black](0,0)--(0,1);
\node[bd] at (-2,0) [label=below:{{\scriptsize$3$}}] {};
\node[SUd] at (-1,0) [label=below:{{\scriptsize$6/\mathbb{Z}_2$}}] {};
\node[SUd] at (0,0) [label=below:{{\scriptsize$9/\mathbb{Z}_3$}}] {};
\node[bd] at (1,0) [label=below:{{\scriptsize$7$}}] {};
\node[SUd] at (2,0) [label=below:{{\scriptsize$6/\mathbb{Z}_2$}}] {};
\node[bd] at (3,0) [label=below:{{\scriptsize$4$}}] {};
\node[bd] at (4,0) [label=below:{{\scriptsize$3$}}] {};
\node[bd] at (5,0) [label=below:{{\scriptsize$1$}}] {};
\node[bd] at (0,1) [label=above:{{\scriptsize$4$}}] {};
\end{tikzpicture}
%
%\begin{tikzpicture}[x=.7cm,y=.7cm]
%\draw[ligne, black](-2,0)--(5,0);
%\draw[ligne, black](0,0)--(0,1);
%\node[bd] at (-2,0) [label=below:{{\scriptsize$3$}}] {};
%\node[SUd] at (-1,0) [label=below:{{\scriptsize$6/\mathbb{Z}_2$}}] {};
%\node[SUd] at (0,0) [label=below:{{\scriptsize$9/\mathbb{Z}_3$}}] {};
%\node[bd] at (1,0) [label=below:{{\scriptsize$7$}}] {};
%\node[SUd] at (2,0) [label=below:{{\scriptsize$6/\mathbb{Z}_2$}}] {};
%\node[bd] at (3,0) [label=below:{{\scriptsize$4$}}] {};
%\node[bd] at (4,0) [label=below:{{\scriptsize$2$}}] {};
%\node[bd] at (5,0) [label=below:{{\scriptsize$1$}}] {};
%\node[bd] at (0,1) [label=above:{{\scriptsize$4$}}] {};
%\end{tikzpicture}
%+\mathbb{H} 
&
\mathbb{Z}_3
\cr \hline 
E_8^{(4)}
&   \begin{tikzpicture}[x=.7cm,y=.7cm]
\draw[ligne, black](-2,0)--(5,0);
\draw[ligne, black](0,0)--(0,1);
\node[bd] at (-2,0) [label=below:{{\scriptsize$3$}}] {};
\node[bd] at (-1,0) [label=below:{{\scriptsize$6$}}] {};
\node[SUd] at (0,0) [label=below:{{\scriptsize$10$}}] {};
\node[bd] at (1,0) [label=below:{{\scriptsize$8$}}] {};
\node[bd] at (2,0) [label=below:{{\scriptsize$6$}}] {};
\node[SUd] at (3,0) [label=below:{{\scriptsize$5$}}] {};
\node[bd] at (4,0) [label=below:{{\scriptsize$3$}}] {};
\node[bd] at (5,0) [label=below:{{\scriptsize$1$}}] {};
\node[SUd] at (0,1) [label=above:{{\scriptsize$5$}}] {};
\end{tikzpicture}
&
2
&  
 \begin{tikzpicture}[x=.7cm,y=.7cm]
\draw[ligne, black](-2,0)--(5,0);
\draw[ligne, black](0,0)--(0,1);
\node[bd] at (-2,0) [label=below:{{\scriptsize$3$}}] {};
\node[bd] at (-1,0) [label=below:{{\scriptsize$6$}}] {};
\node[SUd] at (0,0) [label=below:{{\scriptsize$10/\mathbb{Z}_2$}}] {};
\node[bd] at (1,0) [label=below:{{\scriptsize$8$}}] {};
\node[bd] at (2,0) [label=below:{{\scriptsize$6$}}] {};
\node[bd] at (3,0) [label=below:{{\scriptsize$5$}}] {};
\node[bd] at (4,0) [label=below:{{\scriptsize$3$}}] {};
\node[bd] at (5,0) [label=below:{{\scriptsize$1$}}] {};
\node[bd] at (0,1) [label=above:{{\scriptsize$5$}}] {};
\end{tikzpicture}
%
%\begin{tikzpicture}[x=.7cm,y=.7cm]
%\draw[ligne, black](-2,0)--(5,0);
%\draw[ligne, black](0,0)--(0,1);
%\node[bd] at (-2,0) [label=below:{{\scriptsize$3$}}] {};
%\node[bd] at (-1,0) [label=below:{{\scriptsize$6$}}] {};
%\node[SUd] at (0,0) [label=below:{{\scriptsize$10/\mathbb{Z}_2$}}] {};
%\node[bd] at (1,0) [label=below:{{\scriptsize$8$}}] {};
%\node[bd] at (2,0) [label=below:{{\scriptsize$6$}}] {};
%\node[bd] at (3,0) [label=below:{{\scriptsize$4$}}] {};
%\node[bd] at (4,0) [label=below:{{\scriptsize$2$}}] {};
%\node[bd] at (5,0) [label=below:{{\scriptsize$1$}}] {};
%\node[bd] at (0,1) [label=above:{{\scriptsize$5$}}] {};
%\end{tikzpicture}
%+ 2\mathbb{H} 
&
 \mathbb{Z}_5
\cr \hline 
E_8^{(5)}
&   \begin{tikzpicture}[x=.7cm,y=.7cm]
\draw[ligne, black](-2,0)--(5,0);
\draw[ligne, black](0,0)--(0,1);
\node[bd] at (-2,0) [label=below:{{\scriptsize$3$}}] {};
\node[bd] at (-1,0) [label=below:{{\scriptsize$7$}}] {};
\node[SUd] at (0,0) [label=below:{{\scriptsize$11$}}] {};
\node[bd] at (1,0) [label=below:{{\scriptsize$9$}}] {};
\node[bd] at (2,0) [label=below:{{\scriptsize$7$}}] {};
\node[bd] at (3,0) [label=below:{{\scriptsize$5$}}] {};
\node[bd] at (4,0) [label=below:{{\scriptsize$3$}}] {};
\node[bd] at (5,0) [label=below:{{\scriptsize$1$}}] {};
\node[bd] at (0,1) [label=above:{{\scriptsize$5$}}] {};
\end{tikzpicture}
& 
0
&   \begin{tikzpicture}[x=.7cm,y=.7cm]
\draw[ligne, black](-2,0)--(5,0);
\draw[ligne, black](0,0)--(0,1);
\node[bd] at (-2,0) [label=below:{{\scriptsize$3$}}] {};
\node[bd] at (-1,0) [label=below:{{\scriptsize$7$}}] {};
\node[SUd] at (0,0) [label=below:{{\scriptsize$11$}}] {};
\node[bd] at (1,0) [label=below:{{\scriptsize$9$}}] {};
\node[bd] at (2,0) [label=below:{{\scriptsize$7$}}] {};
\node[bd] at (3,0) [label=below:{{\scriptsize$5$}}] {};
\node[bd] at (4,0) [label=below:{{\scriptsize$3$}}] {};
\node[bd] at (5,0) [label=below:{{\scriptsize$1$}}] {};
\node[bd] at (0,1) [label=above:{{\scriptsize$5$}}] {};
\end{tikzpicture}
& 
\mathbb{Z}_{11}\cr \hline
\end{array}
$
\caption{$\EQfour$ and $\MQfive$ for the $r=1$ $E_8$ series of theories and their 1-form symmetry. \label{tab:E8MQ5}}
\end{table}
%%%%%%%%%%%%%%%%%%%%%%%

The $r=1$ $E_7$ and $E_8$ series are completely similar. The magnetic quivers with 3d 1-form symmetries are  shown in tables~\ref{tab:E7MQ5} and~\ref{tab:E8MQ5}. For $k=1$, the central node is again `bad' with a balance $e\equiv  N_f-2N_c=-2$, and we should go to the most singular locus on the CB, where we obtain the effective description
\bea
& \begin{tikzpicture}[x=.6cm,y=.6cm]
 \node at (-4.5,0.5) {$E_7^{(1)}\; \;:$};
\draw[ligne, black](-3,0)--(3,0);
\draw[ligne, black](0,0)--(0,1);
\node[bd] at (-3,0) [label=below:{{\scriptsize$1$}}] {};
\node[bd] at (-2,0) [label=below:{{\scriptsize$2$}}] {};
\node[bd] at (-1,0) [label=below:{{\scriptsize$3$}}] {};
\node[bd] at (0,0) [label=below:{{\scriptsize$5$}}] {};
\node[bd] at (1,0) [label=below:{{\scriptsize$3$}}] {};
\node[bd] at (2,0) [label=below:{{\scriptsize$2$}}] {};
\node[bd] at (3,0) [label=below:{{\scriptsize$1$}}] {};
\node[bd] at (0,1) [label=above:{{\scriptsize$2$}}] {};
\end{tikzpicture}
&&
 \begin{tikzpicture}[x=.6cm,y=.6cm]
 \node at (-5,0.5) {$\rightarrow$};
\draw[ligne, black](-3,0)--(3,0);
\draw[ligne, black](0,0)--(0,1);
\node[bd] at (-3,0) [label=below:{{\scriptsize$1$}}] {};
\node[bd] at (-2,0) [label=below:{{\scriptsize$2$}}] {};
\node[bd] at (-1,0) [label=below:{{\scriptsize$3$}}] {};
\node[bd] at (0,0) [label=below:{{\scriptsize$4$}}] {};
\node[bd] at (1,0) [label=below:{{\scriptsize$3$}}] {};
\node[bd] at (2,0) [label=below:{{\scriptsize$2$}}] {};
\node[bd] at (3,0) [label=below:{{\scriptsize$1$}}] {};
\node[bd] at (0,1) [label=above:{{\scriptsize$2$}}] {};
\node at (4,0.5) {$\oplus\; \mathbb{H}$};
\end{tikzpicture}\\
&\begin{tikzpicture}[x=.6cm,y=.6cm]
 \node at (-3.5,0.5) {$E_8^{(1)}\; \;:$};
\draw[ligne, black](-2,0)--(5,0);
\draw[ligne, black](0,0)--(0,1);
\node[bd] at (-2,0) [label=below:{{\scriptsize$2$}}] {};
\node[bd] at (-1,0) [label=below:{{\scriptsize$4$}}] {};
\node[bd] at (0,0) [label=below:{{\scriptsize$7$}}] {};
\node[bd] at (1,0) [label=below:{{\scriptsize$5$}}] {};
\node[bd] at (2,0) [label=below:{{\scriptsize$4$}}] {};
\node[bd] at (3,0) [label=below:{{\scriptsize$3$}}] {};
\node[bd] at (4,0) [label=below:{{\scriptsize$2$}}] {};
\node[bd] at (5,0) [label=below:{{\scriptsize$1$}}] {};
\node[bd] at (0,1) [label=above:{{\scriptsize$3$}}] {};
\end{tikzpicture}
&& 
 \begin{tikzpicture}[x=.6cm,y=.6cm]
  \node at (-4,0.5) {$\rightarrow$};
\draw[ligne, black](-2,0)--(5,0);
\draw[ligne, black](0,0)--(0,1);
\node[bd] at (-2,0) [label=below:{{\scriptsize$2$}}] {};
\node[bd] at (-1,0) [label=below:{{\scriptsize$4$}}] {};
\node[bd] at (0,0) [label=below:{{\scriptsize$6$}}] {};
\node[bd] at (1,0) [label=below:{{\scriptsize$5$}}] {};
\node[bd] at (2,0) [label=below:{{\scriptsize$4$}}] {};
\node[bd] at (3,0) [label=below:{{\scriptsize$3$}}] {};
\node[bd] at (4,0) [label=below:{{\scriptsize$2$}}] {};
\node[bd] at (5,0) [label=below:{{\scriptsize$1$}}] {};
\node[bd] at (0,1) [label=above:{{\scriptsize$3$}}] {};
\node at (6,0.5) {$\oplus\; \mathbb{H}$};
\end{tikzpicture}  
\eea
We then argue that the 5d SCFT $E_n^{(1)}$ with a non-trivial $0$-form symmetry is given by
\be
E_n^{(1)} \cong   E_n\, \otimes\, \mathbb{H}~,
\ee
for the same reasons as before.
 Similarly, for $k>1$, the  unitary magnetic quivers can be reduced to the case $E_n^{(1)}$ by repeated application of the GW duality on ugly nodes. We then find
\be
\MQfive[E_n^{(k)}] \cong   \MQfive[E_n]\, \oplus\, \mathbb{H} \, \oplus \, \MQfive[E_n^{(k-\Delta_n)}]~.
\ee
Therefore, the `magnetic' version of the $r=1$ $E_n^{(k)}$ SCFT $\FT$ consists of the $E_n$ SCFT tensored with the appropriate number of  free hypermultiplets.  The electric version is then obtained by gauging $\Gamma_{\rm 5d}^{(0)}=\frak{f}$.

%
%%The naive Hasse diagram is computed by subtracting the affine $e_6$ quiver. Note that, for $E_6^{(1)}$, we are left with a central node ``$1$'' corresponding supposedly to a free hypermultiplet (from the 3-cycle in the resolution). For $E_6^{(2)}$, we are left with the $\MQfive$ of rank-0 theory shown in \eqref{E62 rank0}.
%
%For the $E_7$ models, we find the MQ and 1-form symmetries in table \ref{tab:E7MQ5}.
%Now, substracting the affine $e_7$ quiver gives us the rank-0 theories discussed above. 
%For the $E_8$ series, we have the magnetic quivers in table \ref{tab:E8MQ5}. 
%Here, substracting the affine $e_8$ quiver gives us the rank-0 theories discussed above. 
%Note that we can directly generalize this discussion to $r=N >1$ as well. 
%
%For the general cases we find the following magnetic quivers: tables  \ref{tab:E6kMQ}, \ref{tab:E7kMQ}, and \ref{tab:E8kMQ}. We will discuss how to interpret these bad quivers in general in section \ref{badths}. 

%%%%%%%%%%%%%%%%%%%%%%%%%%%%%
\subsection{The Higher-rank $E_n^{(k)}$ Theories}

Consider the $E_n^{(k)}$ singularities with:
\be
k= \Delta_n \ell + \h k~, \qquad \h k =0, \cdots, \Delta_n -1~,
\ee
which give rise to 5d SCFTs of rank $r=\ell +1$. For the resolution geometry, see appendix~\ref{app:resolution} for more details.

%%%%%%%%%%%%%%%%%%%%%%
\begin{table}
$$
\begin{array}{|c|c|c|}\hline
\, k\,  & \MQfive \; \& \;   \bm{\lambda}  \; \& \; \bm{\lambda}' &\; \Gamma^{(0)}_{\rm 5d}{=}\frak{f} \; \cr \hline\hline
3 \ell+0 & 
  \begin{tikzpicture}[x=.75cm,y=.75cm]
\draw[ligne, black](-2,0)--(2,0);
\draw[ligne, black](0,0)--(0,2);
\node[bd] at (-2,0) [label=below:{{\scriptsize$\ell{+}1$}}] {};
\node[bd] at (-1,0) [label=below:{{\scriptsize$2\ell{+}2$}}] {};
\node[bd] at (0,0) [label=below:{{\scriptsize$3\ell{+}3$}}] {};
\node[bd] at (1,0) [label=below:{{\scriptsize$2\ell{+}2$}}] {};
\node[bd] at (2,0) [label=below:{{\scriptsize$\ell{+}1$}}] {};
\node[bd] at (0,1) [label=left:{{\scriptsize$2\ell{+}2$}}] {};
\node[bd] at (0,2) [label=left:{{\scriptsize$\ell{+}1$}}] {};
\end{tikzpicture}
 &\Z_{\ell+1}   \cr
 &\scriptstyle{\{[\ell{+}1, \ell{+}1, \ell{+}1]^3\} }&\cr \hline 
 3\ell+1 &
 \begin{tikzpicture}[x=.75cm,y=.75cm]
\draw[ligne, black](-2,0)--(2,0);
\draw[ligne, black](0,0)--(0,2);
\node[bd] at (-2,0) [label=below:{{\scriptsize$\ell{+}1$}}] {};
\node[bd] at (-1,0) [label=below:{{\scriptsize$2\ell{+}2$}}] {};
\node[bd] at (0,0) [label=below:{{\scriptsize$3\ell{+}4$}}] {};
\node[bd] at (1,0) [label=below:{{\scriptsize$2\ell{+}2$}}] {};
\node[bd] at (2,0) [label=below:{{\scriptsize$\ell{+}1$}}] {};
\node[bd] at (0,1) [label=left:{{\scriptsize$2\ell{+}2$}}] {};
\node[bd] at (0,2) [label=left:{{\scriptsize$\ell{+}1$}}] {};
\end{tikzpicture}
 & \Z_{3\ell+4}  \cr
 &\scriptstyle{\{   [\ell{+}2, \ell{+}1, \ell{+}1]^3 \}  }&\cr 
  & \scriptstyle{\{ [\ell{+}2, \ell{+}1 ,\ell{+}1]^2 [\ell{+}1, \ell{+}1 ,\ell{+}1, 1] \} } & \cr \hline 
 3\ell+2 &
  \begin{tikzpicture}[x=.75cm,y=.75cm]
\draw[ligne, black](-2,0)--(2,0);
\draw[ligne, black](0,0)--(0,2);
\node[bd] at (-2,0) [label=below:{{\scriptsize$\ell{+}1$}}] {};
\node[bd] at (-1,0) [label=below:{{\scriptsize$2\ell{+}3$}}] {};
\node[bd] at (0,0) [label=below:{{\scriptsize$3\ell{+}5$}}] {};
\node[bd] at (1,0) [label=below:{{\scriptsize$2\ell{+}3$}}] {};
\node[bd] at (2,0) [label=below:{{\scriptsize$\ell{+}1$}}] {};
\node[bd] at (0,1) [label=left:{{\scriptsize$2\ell{+}3$}}] {};
\node[bd] at (0,2) [label=left:{{\scriptsize$\ell{+}1$}}] {};
\end{tikzpicture}
 & \Z_{3\ell+5}    \cr
 &\scriptstyle{ \{  [\ell{+}2, \ell{+}2, \ell{+}1]^3   \}}&\cr 
 &\scriptstyle{ \{[\ell{+}2, \ell{+}2 ,\ell{+}1]^2 [ \ell{+}2 ,\ell{+}1, \ell{+}1, 1] \}} &
\cr  \hline 
\end{array}
%%%
$$
\caption{Magnetic quivers for the $E_{6}^{(k)}$  models. The three partitions $\bm{\lambda}$ are partitions of $N\equiv \Delta_n +k$  associated with the three legs of the quiver. The second partition (if listed) is the one after the box-lifting, later denoted by $\bm{\lambda}'$.  These are introduced here for later reference.   \label{tab:E6kMQ}}
\end{table}
%%%%%%%%

%%%%%%%%%%%%%%%%%%%%%%
\begin{table}
$$
%%%
\begin{array}{|c|c|c|}\hline
\, k\,  & \MQfive \; \& \;   \bm{\lambda}  \; \& \; \bm{\lambda}' &\; \Gamma^{(0)}_{\rm 5d}{=}\frak{f} \; \cr \hline\hline
3 \ell+0 & 
  \begin{tikzpicture}[x=.75cm,y=.75cm]
\draw[ligne, black](-3,0)--(3,0);
\draw[ligne, black](0,0)--(0,1);
\node[bd] at (-3,0) [label=below:{{\scriptsize$\ell{+}1$}}] {};
\node[bd] at (-2,0) [label=below:{{\scriptsize$2\ell{+}2$}}] {};
\node[bd] at (-1,0) [label=below:{{\scriptsize$3\ell{+}3$}}] {};
\node[bd] at (0,0) [label=below:{{\scriptsize$4\ell{+}4$}}] {};
\node[bd] at (1,0) [label=below:{{\scriptsize$3 \ell{+}3$}}] {};
\node[bd] at (2,0) [label=below:{{\scriptsize$2\ell{+}2$}}] {};
\node[bd] at (3,0) [label=below:{{\scriptsize$\ell{+}1$}}] {};
\node[bd] at (0,1) [label=above:{{\scriptsize$2\ell{+}2$}}] {};
\end{tikzpicture} & \Z_{\ell+1} \\
 &\scriptstyle{ \{   [2\ell{+}2, 2\ell{+}2], [\ell{+}1, \ell{+}1,\ell{+}1, \ell{+}1]^2 \}} &
  \cr\hline 
 4\ell+1 &
 \begin{tikzpicture}[x=.75cm,y=.75cm]
\draw[ligne, black](-3,0)--(3,0);
\draw[ligne, black](0,0)--(0,1);
\node[bd] at (-3,0) [label=below:{{\scriptsize$\ell{+}1$}}] {};
\node[bd] at (-2,0) [label=below:{{\scriptsize$2\ell{+}2$}}] {};
\node[bd] at (-1,0) [label=below:{{\scriptsize$3\ell{+}3$}}] {};
\node[bd] at (0,0) [label=below:{{\scriptsize$4\ell{+}5$}}] {};
\node[bd] at (1,0) [label=below:{{\scriptsize$3 \ell{+}3$}}] {};
\node[bd] at (2,0) [label=below:{{\scriptsize$2\ell{+}2$}}] {};
\node[bd] at (3,0) [label=below:{{\scriptsize$\ell{+}1$}}] {};
\node[bd] at (0,1) [label=above:{{\scriptsize$2\ell{+}2$}}] {};
\end{tikzpicture}
  & \Z_{4\ell+5}  \\
  & \scriptstyle{\{[2\ell{+}3, 2\ell{+}2] [\ell{+}2, \ell{+}1,\ell{+}1, \ell{+}1]^2 \}}&\cr
  &\scriptstyle{\{[2\ell{+}3, 2 \ell{+}2] [(\ell{+}2)(\ell{+}1)^3][(\ell{+}1)^4,1]\} }&\cr \hline 
  4\ell+2 &
  \begin{tikzpicture}[x=.75cm,y=.75cm]
\draw[ligne, black](-3,0)--(3,0);
\draw[ligne, black](0,0)--(0,1);
\node[bd] at (-3,0) [label=below:{{\scriptsize$\ell{+}1$}}] {};
\node[bd] at (-2,0) [label=below:{{\scriptsize$2\ell{+}3$}}] {};
\node[bd] at (-1,0) [label=below:{{\scriptsize$3\ell{+}4$}}] {};
\node[bd] at (0,0) [label=below:{{\scriptsize$4\ell{+}6$}}] {};
\node[bd] at (1,0) [label=below:{{\scriptsize$3 \ell{+}4$}}] {};
\node[bd] at (2,0) [label=below:{{\scriptsize$2\ell{+}3$}}] {};
\node[bd] at (3,0) [label=below:{{\scriptsize$\ell{+}1$}}] {};
\node[bd] at (0,1) [label=above:{{\scriptsize$2\ell{+}3$}}] {};
\end{tikzpicture}
& \Z_{2\ell+3} \\
&\scriptstyle{ \{ [2\ell{+}3, 2\ell{+}3] [\ell{+}2, \ell{+}1,\ell{+}2, \ell{+}1]^2 \}}& \cr 
&\scriptstyle{\{ [2\ell{+}3, 2 \ell{+}3] [(\ell{+}2)(\ell{+}1)(\ell{+}2)(\ell{+}1)][(\ell{+}2),(\ell{+}1)^3,1] \}} &\cr\hline 
 4\ell +3&
  \begin{tikzpicture}[x=.75cm,y=.75cm]
\draw[ligne, black](-3,0)--(3,0);
\draw[ligne, black](0,0)--(0,1);
\node[bd] at (-3,0) [label=below:{{\scriptsize$\ell{+}1$}}] {};
\node[bd] at (-2,0) [label=below:{{\scriptsize$2\ell{+}2$}}] {};
\node[bd] at (-1,0) [label=below:{{\scriptsize$3\ell{+}5$}}] {};
\node[bd] at (0,0) [label=below:{{\scriptsize$4\ell{+}7$}}] {};
\node[bd] at (1,0) [label=below:{{\scriptsize$3 \ell{+}5$}}] {};
\node[bd] at (2,0) [label=below:{{\scriptsize$2\ell{+}3$}}] {};
\node[bd] at (3,0) [label=below:{{\scriptsize$\ell{+}1$}}] {};
\node[bd] at (0,1) [label=above:{{\scriptsize$ 2\ell{+}3$}}] {};
\end{tikzpicture}
& \Z_{4\ell+7} \\
&\scriptstyle{\{ [2\ell{+}4, 2\ell{+}3] [\ell{+}2, \ell{+}2,\ell{+}2, \ell{+}1]^2 \} }&\cr 
&\scriptstyle{\{ [2\ell{+}4, 2 \ell{+}3] [(\ell{+}2)^3,(\ell{+}1)][(\ell{+}2)^2,(\ell{+}1)^2,1]\} }&\cr\hline 
\end{array}
$$
\caption{Magnetic quivers for the $E_7^{(k)}$ models. The three partitions $\bm{\lambda}$ are partitions of $N\equiv \Delta_n +k$  associated with the three legs of the quiver. The second partition (if listed) is the one after the box-lifting, later denoted by $\bm{\lambda}'$.  These are introduced here for later reference.   \label{tab:E7kMQ}}
\end{table}
%%%%%%%%

\begin{table}[h!]
$$
\begin{array}{|c|c|c|}\hline
\, k\,  & \MQfive \; \& \;   \bm{\lambda}\; \& \;   \bm{\lambda}'  &\; \Gamma^{(0)}_{\rm 5d}{=}\frak{f} \; \cr \hline\hline
  6\ell+0 & 
    \begin{tikzpicture}[x=.75cm,y=.75cm]
\draw[ligne, black](-2,0)--(5,0);
\node[bd] at (-2,0) [label=below:{{\scriptsize$2\ell{+}2$}}] {};
\node[bd] at (-1,0) [label=below:{{\scriptsize$4\ell{+}4$}}] {};
\node[bd] at (0,0) [label=below:{{\scriptsize$6\ell{+}6$}}] {};
\node[bd] at (1,0) [label=below:{{\scriptsize$5\ell{+}5$}}] {};
\node[bd] at (2,0) [label=below:{{\scriptsize$4\ell{+}4$}}] {};
\node[bd] at (3,0) [label=below:{{\scriptsize$3\ell{+}3$}}] {};
\node[bd] at (4,0) [label=below:{{\scriptsize$2\ell{+}2$}}] {};
\node[bd] at (5,0) [label=below:{{\scriptsize$\ell{+}1$}}] {};
\draw[ligne, black](0,0)--(0,1);
\node[bd] at (0,1) [label=above:{{\scriptsize$3\ell{+}3$}}] {};
\end{tikzpicture}
 &\Z_{\ell+1} \\
 & \scriptstyle{  \{[3\ell{+}3, 3\ell{+}3] [2\ell{+}2, 2\ell{+}2,2\ell{+}2][(\ell{+}1)^6] \}} &\cr\hline 
  6\ell+1 &
     \begin{tikzpicture}[x=.75cm,y=.75cm]
\draw[ligne, black](-2,0)--(5,0);
\node[bd] at (-2,0) [label=below:{{\scriptsize$2\ell{+}2$}}] {};
\node[bd] at (-1,0) [label=below:{{\scriptsize$4\ell{+}4$}}] {};
\node[bd] at (0,0) [label=below:{{\scriptsize$6\ell{+}7$}}] {};
\node[bd] at (1,0) [label=below:{{\scriptsize$5\ell{+}5$}}] {};
\node[bd] at (2,0) [label=below:{{\scriptsize$4\ell{+}4$}}] {};
\node[bd] at (3,0) [label=below:{{\scriptsize$3\ell{+}3$}}] {};
\node[bd] at (4,0) [label=below:{{\scriptsize$2\ell{+}2$}}] {};
\node[bd] at (5,0) [label=below:{{\scriptsize$\ell{+}1$}}] {};
\draw[ligne, black](0,0)--(0,1);
\node[bd] at (0,1) [label=above:{{\scriptsize$3\ell{+}3$}}] {};
\end{tikzpicture}
  &\Z_{6\ell+7} \\
  & \scriptstyle{   \{[3\ell{+}4, 3\ell{+}3] [2\ell{+}3, 2\ell{+}2,2\ell{+}2][(\ell{+}2)(\ell{+}1)^5]\} }& \cr 
  & \scriptstyle{ \{ [3\ell{+}4,3 \ell{+}3] [2\ell{+}3, 2\ell{+}2, 2 \ell{+}2][(\ell{+}1)^6,1]\}}&\cr\hline 
 6\ell+2 &
   \begin{tikzpicture}[x=.75cm,y=.75cm]
\draw[ligne, black](-2,0)--(5,0);
\node[bd] at (-2,0) [label=below:{{\scriptsize$2\ell{+}2$}}] {};
\node[bd] at (-1,0) [label=below:{{\scriptsize$4\ell{+}5$}}] {};
\node[bd] at (0,0) [label=below:{{\scriptsize$6\ell{+}8$}}] {};
\node[bd] at (1,0) [label=below:{{\scriptsize$5\ell{+}6$}}] {};
\node[bd] at (2,0) [label=below:{{\scriptsize$4\ell{+}5$}}] {};
\node[bd] at (3,0) [label=below:{{\scriptsize$3\ell{+}4$}}] {};
\node[bd] at (4,0) [label=below:{{\scriptsize$2\ell{+}2$}}] {};
\node[bd] at (5,0) [label=below:{{\scriptsize$\ell{+}1$}}] {};
\draw[ligne, black](0,0)--(0,1);
\node[bd] at (0,1) [label=above:{{\scriptsize$3\ell{+}4$}}] {};
\end{tikzpicture}
  &\Z_{3\ell+4}   \\
  & \scriptstyle{    \{[3\ell{+}4, 3\ell{+}4] [2\ell{+}3, 2\ell{+}3,2\ell{+}2][(\ell{+}2)^2(\ell{+}1)^4] \} }& \cr 
  & \scriptstyle{  \{[3\ell{+}4,3 \ell{+}4] [2\ell{+}3, 2\ell{+}3, 2 \ell{+}2][(\ell{+}2)(\ell{+}1)^5,1]\}}& \cr\hline 
%%  \end{array}
%%  %%
%%  \quad
%%  %%
%% \begin{array}{|c|c|c|}\hline
%%\, k\,  & \MQfive \; \& \;   \bm{\lambda}  &\; \Gamma^{(0)}_{\rm 5d} \; \cr \hline\hline 
 6\ell+3 &
  \begin{tikzpicture}[x=.75cm,y=.75cm]
\draw[ligne, black](-2,0)--(5,0);
\node[bd] at (-2,0) [label=below:{{\scriptsize$2\ell{+}3$}}] {};
\node[bd] at (-1,0) [label=below:{{\scriptsize$4\ell{+}6$}}] {};
\node[bd] at (0,0) [label=below:{{\scriptsize$6\ell{+}9$}}] {};
\node[bd] at (1,0) [label=below:{{\scriptsize$5\ell{+}7$}}] {};
\node[bd] at (2,0) [label=below:{{\scriptsize$4\ell{+}6$}}] {};
\node[bd] at (3,0) [label=below:{{\scriptsize$3\ell{+}4$}}] {};
\node[bd] at (4,0) [label=below:{{\scriptsize$2\ell{+}3$}}] {};
\node[bd] at (5,0) [label=below:{{\scriptsize$\ell{+}1$}}] {};
\draw[ligne, black](0,0)--(0,1);
\node[bd] at (0,1) [label=above:{{\scriptsize$3\ell{+}4$}}] {};
\end{tikzpicture}
  &\Z_{2\ell+3} \\
  &  \scriptstyle{  \{[3\ell{+}5, 3\ell{+}4] [2\ell{+}3, 2\ell{+}3,2\ell{+}3][(\ell{+}2)^3(\ell{+}1)^3] \}}& \cr 
  & \scriptstyle{  \{[3\ell{+}5,3 \ell{+}4] [(2\ell{+}3)^3][(\ell{+}2)^2(\ell{+}1)^4,1] \}}& \cr\hline 
6\ell +4&
   \begin{tikzpicture}[x=.75cm,y=.75cm]
\draw[ligne, black](-2,0)--(5,0);
\node[bd] at (-2,0) [label=below:{{\scriptsize$2\ell{+}3$}}] {};
\node[bd] at (-1,0) [label=below:{{\scriptsize$4\ell{+}6$}}] {};
\node[bd] at (0,0) [label=below:{{\scriptsize$6\ell{+}10$}}] {};
\node[bd] at (1,0) [label=below:{{\scriptsize$5\ell{+}8$}}] {};
\node[bd] at (2,0) [label=below:{{\scriptsize$4\ell{+}6$}}] {};
\node[bd] at (3,0) [label=below:{{\scriptsize$3\ell{+}5$}}] {};
\node[bd] at (4,0) [label=below:{{\scriptsize$2\ell{+}3$}}] {};
\node[bd] at (5,0) [label=below:{{\scriptsize$\ell{+}1$}}] {};
\draw[ligne, black](0,0)--(0,1);
\node[bd] at (0,1) [label=above:{{\scriptsize$3\ell{+}5$}}] {};
\end{tikzpicture}
  &\Z_{3\ell+5} \\
  &\scriptstyle{    \{[3\ell{+}5, 3\ell{+}5] [2\ell{+}4, 2\ell{+}3,2\ell{+}3][(\ell{+}2)^4(\ell{+}1)^2] \}}& \cr
  & \scriptstyle{  \{[3\ell{+}5,3 \ell{+}5] [2\ell{+}4, 2\ell{+}3, 2 \ell{+}3][(\ell{+}2)^3(\ell{+}1)^3,1] \} }& \cr \hline 
 6\ell +5&
    \begin{tikzpicture}[x=.75cm,y=.75cm]
\draw[ligne, black](-2,0)--(5,0);
\node[bd] at (-2,0) [label=below:{{\scriptsize$2\ell{+}3$}}] {};
\node[bd] at (-1,0) [label=below:{{\scriptsize$4\ell{+}7$}}] {};
\node[bd] at (0,0) [label=below:{{\scriptsize$6\ell{+}11$}}] {};
\node[bd] at (1,0) [label=below:{{\scriptsize$5\ell{+}9$}}] {};
\node[bd] at (2,0) [label=below:{{\scriptsize$4\ell{+}7$}}] {};
\node[bd] at (3,0) [label=below:{{\scriptsize$3\ell{+}5$}}] {};
\node[bd] at (4,0) [label=below:{{\scriptsize$2\ell{+}3$}}] {};
\node[bd] at (5,0) [label=below:{{\scriptsize$\ell{+}1$}}] {};
\draw[ligne, black](0,0)--(0,1);
\node[bd] at (0,1) [label=above:{{\scriptsize$3\ell{+}5$}}] {};
\end{tikzpicture}
  &\Z_{6\ell+11}  \\
  &\scriptstyle{ \{[3\ell{+}6, 3\ell{+}5] [2\ell{+}4, 2\ell{+}4,2\ell{+}3][(\ell{+}2)^5 (\ell{+}1)] \}}& \cr 
  &\scriptstyle{ \{ [3\ell{+}6,3 \ell{+}5] [2\ell{+}4, 2\ell{+}4, 2 \ell{+}3][(\ell{+}2)^4(\ell{+}1)^2,1]\} }&\cr\hline 
\end{array}
$$
\caption{Magnetic quivers for the $E_{8}^{(k)}$ models. We list the partitions $\bm{\lambda}$ that underlies these models, the one after the box-lifting $\bm{\lambda}'$, as well as the 0-form symmetry of the associated 5d SCFT.   \label{tab:E8kMQ}}
\end{table}
%%%%%%%%%%%%%

The analysis of the 4d SCFTs $\FTfour$ and of the electric quiverines $\EQfour$ is similar to the $\ell=0$ case. Here, we will mainly discuss the magnetic quivers $\MQfive$ with unitary gauge groups, for the `magnetic version' of the theory. They are displayed in tables~\ref{tab:E6kMQ}, \ref{tab:E7kMQ} and~\ref{tab:E8kMQ}. As before, the `electric version' is obtained by discretely gauging the 5d $0$-form symmetry.

\medskip
\noindent
Let us first consider the case $\h k=0$. The magnetic quivers are given by:
\bea\label{MQ En rank m}
  \begin{tikzpicture}[x=.6cm,y=.6cm]
\draw[ligne, black](-2,0)--(2,0);
\draw[ligne, black](0,0)--(0,2);
\node[bd] at (-2,0) [label=below:{{\scriptsize$m$}}] {};
\node[bd] at (-1,0) [label=below:{{\scriptsize$2m$}}] {};
\node[bd] at (0,0) [label=below:{{\scriptsize$3m$}}] {};
\node[bd] at (1,0) [label=below:{{\scriptsize$2m$}}] {};
\node[bd] at (2,0) [label=below:{{\scriptsize$m$}}] {};
\node[bd] at (0,1) [label=left:{{\scriptsize$2m$}}] {};
\node[bd] at (0,2) [label=left:{{\scriptsize$m$}}] {};
\end{tikzpicture}~, \quad
   \begin{tikzpicture}[x=.6cm,y=.6cm]
\draw[ligne, black](-3,0)--(3,0);
\draw[ligne, black](0,0)--(0,1);
\node[bd] at (-3,0) [label=below:{{\scriptsize$m$}}] {};
\node[bd] at (-2,0) [label=below:{{\scriptsize$2m$}}] {};
\node[bd] at (-1,0) [label=below:{{\scriptsize$3m$}}] {};
\node[bd] at (0,0) [label=below:{{\scriptsize$4m$}}] {};
\node[bd] at (1,0) [label=below:{{\scriptsize$3m$}}] {};
\node[bd] at (2,0) [label=below:{{\scriptsize$2m$}}] {};
\node[bd] at (3,0) [label=below:{{\scriptsize$m$}}] {};
\node[bd] at (0,1) [label=above:{{\scriptsize$2m$}}] {};
\end{tikzpicture}~, \quad
\begin{tikzpicture}[x=.6cm,y=.6cm]
\draw[ligne, black](-2,0)--(5,0);
\draw[ligne, black](0,0)--(0,1);
\node[bd] at (-2,0) [label=below:{{\scriptsize$2m$}}] {};
\node[bd] at (-1,0) [label=below:{{\scriptsize$4m$}}] {};
\node[bd] at (0,0) [label=below:{{\scriptsize$6m$}}] {};
\node[bd] at (1,0) [label=below:{{\scriptsize$5m$}}] {};
\node[bd] at (2,0) [label=below:{{\scriptsize$4m$}}] {};
\node[bd] at (3,0) [label=below:{{\scriptsize$3m$}}] {};
\node[bd] at (4,0) [label=below:{{\scriptsize$2m$}}] {};
\node[bd] at (5,0) [label=below:{{\scriptsize$m$}}] {};
\node[bd] at (0,1) [label=above:{{\scriptsize$3m$}}] {};
\end{tikzpicture}
\eea
for the $E_6$, $E_7$ and $E_8$ theories, respectively, with  $m \equiv \ell+1$. The Higgs branch of this 3d $\CN=4$ quiver is ${\rm Sym}^m\left(\C^2/\Gamma_{E_n}\right)$, and its low-energy theory also contains $\ell=m-1$ free twisted hypermultiplets. Correspondingly, we have $\ell$ free hypermultiplets on the Coulomb branch of $\FT$, which arise in the resolved geometry from the $\ell$ pairs of 3-cycles associated to the $\ell$ exceptional divisors that are ruled surfaces of genus $1$.

Importantly, although each node is individually balanced, the quivers \eqref{MQ En rank m} are nonetheless `bad', since they have monopole operators of dimension $0$. These $\Delta=0$ monopole operators are precisely related to the unhigssable $U(1)^\ell$ factors.  These quivers have been studied from various point of view in the literature  \cite{Gaiotto:2012uq}.  We will discuss them further in section~\ref{subsec:classS}.

Next, we consider the case $\h k>0$. We can  easily see that the quivers with $\h k\geq 2$ are equivalent to the $\h k=1$ quivers by successive dualization of the `ugly' nodes. For instance, the $E_6^{(3 \ell +2)}$ quiver can be reduced to the $E_6^{(3\ell+1)}$ quiver exactly as in the $\ell=0$ case \eqref{MQ5 E62 0}. The resulting free sector is  the one associated with the residual terminal singularity $E_n^{(\h k-\Delta_n)}$, for any $\ell\geq 0$.

We therefore only need to understand the $\h k=1$ models. In that case and for any $\ell\geq 0$, the central node is bad with a balance $e=-2$. By the same argument as before, we claim that the relevant physics occurs on the most singular locus of the CB, whose effective description is given by
\be
\MQ[E_n^{(\Delta_n \ell +1)}] \;\; \rightarrow \; \; \MQ[E_n^{(\Delta_n \ell)}] \oplus \; \mathbb{H}~.
\ee
The magnetic quiver $\MQ[E_n^{(\Delta_n \ell+1)}]$ is given in \eqref{MQ En rank m}. 

\medskip
\noindent
In conclusion, we see that the `magnetic' 5d theory is always a rank-$(\ell{+}1)$ $E_n$ theory tensored with free hypermultiplets, while the `electric' theory is necessarily a discrete gauging thereof:
\be
{\mathcal{T}}_{\MG_{E_n^{(k)}}}^{\rm 5d} \Big|_{\text{`magnetic'}} \; \cong\;  {\mathcal{T}}_{\MG_{E_n}}^{\rm 5d}  \otimes \mathbb{H}^{d_H(\h k)+1}~, \quad 
{\mathcal{T}}_{\MG_{E_n^{(k)}}}^{\rm 5d} \Big|_{\text{`electric'}} \; \cong\;  \left({\mathcal{T}}_{\MG_{E_n}}^{\rm 5d}  \otimes \mathbb{H}^{d_H(\h k)+1}\right)/\frak{f}~.
\ee
Here we have $k= \Delta_n \ell + \h k$, the integer $d_H(\h k)$ is defined as the dimension of the 5d HB of the rank-zero theory $E_n^{(\h k-\Delta_n)}$ for $\h k>1$, and by $d_H(\h k)=0$ if $\h k=1$; finally, the discrete group $\frak{f}$ is given for any  $E_n^{(k)}$  in the tables~\ref{tab:E6kMQ}, \ref{tab:E7kMQ} and~\ref{tab:E8kMQ}.

It is also interesting to note that, for the  canonical singularities $\MG_{E_n^{(k)} }$, the residual terminal singularities in the crepant resolution $\t\MG$ are related to the `local ugliness' of the magnetic quiver, while the 3-cycles in the crepant resolution are associated with the `badness' of the magnetic quiver for $E_n^{(1)}$ obtained after getting rid of all the ugly nodes ({\it i.e.} with balance $e=-1$) by repeated dualization.

%%%%%%%%%%%%%%%%%%%%%%%%%%%%%%%%%%%%
\section{Class $\CS$ Description, Brane-webs and Partial Higgsing} 
\label{subsec:classS}

 Let $\CS_{N, \bm{\lambda}}^{\rm 4d}$ denote a 4d $\CN=2$ SCFT of class $\CS$  \cite{Gaiotto:2009we}, defined as the $A_{N-1}$ 6d $\CN=(0,2)$ theory with three regular punctures,
  indexed by three partitions of $N$:
\be
\bm{\lambda}  \equiv   \{ \lambda^{(\alpha)}\}_{\alpha=1}^3 =\{ [\lambda_1^{(1)}, \lambda_2^{(1)}, \cdots ]  [\lambda_1^{(2)}, \lambda_2^{(2)}, \cdots ]  [\lambda_1^{(3)}, \lambda_2^{(3)}, \cdots ] \}~.
\ee
We propose that our 5d SCFTs ${\mathcal{T}}_{\MG_{E_n^{(k)}}}^{\rm 5d}$ compactified on a circle give rise to such a class $\CS$ theory in the IR:
\be\label{class S description 5d 4d}
\KK {\mathcal{T}}_{\MG_{E_n^{(k)}}}^{\rm 5d}  \Big|_{\rm IR} \quad \longleftrightarrow\quad  \CS_{N, \bm{\lambda}}^{\rm 4d}~, \qquad\qquad N= \Delta_n+k~.
\ee
The main piece of evidence for this conjecture is that the magnetic quivers for the 5d SCFT $E_n^{(k)}$  are identical to the magnetic quivers (also known as 3d mirrors) of these class $\CS$ theories~\cite{Benini:2010uu, Chacaltana:2010ks}. For $k=0$, the  duality \eqref{class S description 5d 4d} is the well-studied relation between the 5d rank-one $E_n$ SCFTs and the 4d $E_n$ Minahan-Nemeschansky (MN)-theories. In this section, we explore the generalization to any $k$, which sheds some additional light on this family of models. We expect \eqref{class S description 5d 4d} to be true at least for the `magnetic' version of $\FT$ with (in general) a non-trivial $0$-form symmetry $\Gamma_{\rm 5d}^{(0)}=\frak{f}$. Identifying (and gauging) that $0$-form symmetry in the class $\CS$ description is left for future work.

The partitions $\bm{\lambda}$ are obtained from the three unitary linear quivers that form the three tails of $\MQfive$, which correspond to the 3d $T_\lambda[SU(N)]$ theories \cite{Gaiotto:2008ak}. They are written down explicitly in tables~\ref{tab:E6kMQ}, \ref{tab:E7kMQ},   and~\ref{tab:E8kMQ}. Each leg consists of a quiver $[N]-U(n_1)-U(n_2) \cdots$ attached to the central $U(N)$, and it determines the partition $\lambda=[\lambda_a]$ of $N$ according to $\lambda_a= n_{a-1}-n_a$, with $n_0\equiv N$. Such partitions obtained from our $\MQfive$ are not all ordered when $1<\h k< \Delta_n-1$. This is due to the ugliness of some of the tails. A GW dualization at the ugly nodes along the tails gives us properly ordered partitions (and a few twisted hypermultiplets).

Such trinion class $\CS$ theories, also called fixtures \cite{Chacaltana:2010ks}, can be uplifted to 5d theories described by type-IIB $(p,q)$-webs \cite{Aharony:1997bh, DeWolfe:1999hj},  corresponding to a junctions of three stacks of $N$ D5-branes, $N$ NS5-branes and $N$ $(1,1)$ 5-branes, respectively, with the three stacks of five-branes ending on 7-branes with boundary conditions specified by the partitions $\lambda^{(\alpha)}$ \cite{Benini:2009gi}.\footnote{Recall that, locally, each stack of 5-branes can be T-dualized (and S-dualized) to the D3-brane boundary condition described by the 3d $\CN=4$ $T_\lambda[SU(N)]$ theory. This is how the magnetic quiver is most easily understood, in this setup.} 
We then expect  this brane-web construction to be dual to our canonical singularity in M-theory:
\be
 {\mathcal{T}}_{\MG_{E_n^{(k)}}}^{\rm 5d}   \quad \longleftrightarrow\quad   \text{trinion-shaped brane web in IIB} \quad \longleftrightarrow \quad \text{triangle-shaped GTP}\,,
\ee
with the last description to be discussed momentarily. 
In other words, the duality \eqref{class S description 5d 4d} uplifts to 5d. In the special case when all every external 5-brane ends on its own 7-brane, this is the well-known duality between the $T_N \cong \C^3/(\Z_N \times \Z_N)$ toric geometry and 5-brane webs~\cite{Leung:1997tw, Benini:2009gi}. The general case is a conjecture, based on a fair amount of circumstantial evidence.

These branes webs can be efficiently represented as dot diagrams \cite{Benini:2009gi}, also known as a generalized toric polygons (GTP) \cite{vanBeest:2020kou, vanBeest:2020civ}. In particular, given a GTP, there is a systematic algorithm (either based on brane-web technology \cite{Cabrera:2018ann, Bourget:2019aer} or directly in the GTP \cite{vanBeest:2020kou}) for computing the magnetic quiver of the SCFT defined by the brane web/GTP, whenever that magnetic quiver is `good' or `ugly'  \cite{vanBeest:2020kou, vanBeest:2020civ}.  In the present setup, all GTP are triangles related to the $T_N$ theory. The partitions $\bm{\lambda}$ give the successive spacing lengths between black dots along the three edges of length $N$, starting from the lower-left corner and going counter-clockwise.

\subsection{Rank-zero Trinions}
To illustrate the class $\CS$/GTP perspective (and our notation), let us first consider the rank-zero 5d models of section~\ref{sec:Rank0}. As already pointed out, the $E_6^{(-1)}$ model is the $T_2$ theory, whose corresponding GTP reads
 \bea
\begin{tikzpicture}[x=.5cm,y=.5cm]
\node[] at (-5,1) {$E_6^{(-1)}=T_2 \; :$ };
\draw[step=.5cm,gray,very thin] (0,0) grid (2,2);
\draw[ligne] (0,0)--(0,2)--(2,0)--(0,0); 
\node[bd] at (0,0) {}; 
\node[bd] at (0,1) {}; 
\node[bd] at (0,2) {}; 
\node[bd] at (1,1) {};
\node[bd] at (2,0) {}; 
\node[bd] at (1,0) {}; 
\node[] at (7,1) {$\bm{\lambda}= \{[1,1]^3\}$ };
\end{tikzpicture} \,.
\eea
The $E_7^{(-2)}$ magnetic quiver appearing in (\ref{AlphaRomeo}) has two ugly legs, and it is then equivalent to $T_2$ plus two hypermultiplets. The $E_7^{(-1)}$ model, on the other hand, is given by the trinion
 \bea
\begin{tikzpicture}[x=.5cm,y=.5cm]
\node[] at (-5,1.5) {$E_7^{(-1)}\; :$ };
\draw[step=.5cm,gray,very thin] (0,0) grid (3,3);
\draw[ligne] (0,0)--(0,3)--(3,0)--(0,0); 
\node[bd] at (0,0) {}; 
\node[bd] at (0,1) {}; 
\node[bd] at (0,2) {}; 
\node[bd] at (0,3) {}; 
\node[bd] at (1,2) {};
\node[bd] at (2,1) {};
\node[bd] at (3,0) {}; 
\node[bd] at (2,0) {}; 
\node[wd] at (1,0) {}; 
\node[] at (9,1.5) {$\bm{\lambda}= \{[2,1][1,1,1]^2\}$ };
\end{tikzpicture} 
\eea
This describes 9 hypermultiplets, and corresponds to the second $A_2$ free-field fixture in \cite{Chacaltana:2010ks}. 
Similarly, for the $E_8$ models, we have $E_8^{(-4)} \cong T_2\oplus \mathbb{H}^2$ and $E_8^{(-3)} \cong E_7^{(-1)}\oplus \mathbb{H}^3$ after dualizing the ugly legs. We also find
\bea\label{E8m2 GTP}
 \begin{tikzpicture}[x=.5cm,y=.5cm]
\node[] at (-5,2.5) {$E_8^{(-2)}\; :$ };
\draw[step=.5cm,gray,very thin] (0,0) grid (4,4);
\draw[ligne] (0,0)--(0,4)--(4,0)--(0,0); 
\node[bd] at (0,0) {}; 
\node[bd] at (0,1) {}; 
\node[bd] at (0,2) {}; 
\node[bd] at (0,3) {}; 
\node[bd] at (0,4) {}; 
\node[bd] at (1,3) {};
\node[bd] at (2,2) {};
\node[wd] at (3,1) {}; 
\node[bd] at (4,0) {}; 
\node[wd] at (3,0) {}; 
\node[bd] at (2,0) {}; 
\node[wd] at (1,0) {}; 
\node[] at (5,2) {$\; \oplus \; \mathbb{H}^2$ };
\node[] at (12,2) {$\bm{\lambda}= \{[2,2][2,1,1][1,1,1,1]\}$ };
\end{tikzpicture} \,.
\eea
Here, we first dualize the ugly nodes on the legs of $\MQfive$, which gives us the two hypermultiplets. We are then left with the $A_4$ free-field fixture corresponding to 14 hypermultiplets~\cite{Chacaltana:2010ks}.
Finally, we have
\bea\label{fixture 20}
 \begin{tikzpicture}[x=.5cm,y=.5cm]
\node[] at (-5,2.5) {$E_8^{(-1)}\; :$ };
\draw[step=.5cm,gray,very thin] (0,0) grid (5,5);
\draw[ligne] (0,0)--(0,5)--(5,0)--(0,0); 
\node[bd] at (0,0) {}; 
\node[bd] at (0,1) {}; 
\node[bd] at (0,2) {}; 
\node[bd] at (0,3) {}; 
\node[bd] at (0,4) {}; 
\node[bd] at (0,5) {}; 
\node[bd] at (1,4) {};
\node[wd] at (2,3) {};
\node[bd] at (3,2) {}; 
\node[wd] at (4,1) {}; 
\node[bd] at (5,0) {}; 
\node[wd] at (4,0) {}; 
\node[bd] at (3,0) {}; 
\node[wd] at (2,0) {}; 
\node[wd] at (1,0) {}; 
\node[] at (12,2.5) {$\bm{\lambda}= \{[3,2][2,2,1][1,1,1,1,1]\}$ };
\end{tikzpicture} \,.
\eea
This is the $A_5$ free-field fixture corresponding to $20$ hypermultiplets \cite{Chacaltana:2010ks}. 

\subsection{Coulomb Branch Spectrum and `Badness' in the Class $\CS$ Description}

The class $\CS$ description leads to a complementary point of view on the potential `badness' of the magnetic quiver. Recall that the Coulomb branch operators of  dimension $\Delta \in \Z$ of a 4d $\CN=2$ SCFT of  class $\CS$ correspond to $\Delta$-differentials on the Gaiotto curve -- in our case, on the 3-punctured sphere -- with poles at the punctures specified by some integers $p_\Delta^{(\alpha)}$. The virtual number of operators of dimension $\Delta$, denoted by $d_\Delta$, is given by the Riemann-Roch theorem \cite{Gaiotto:2009we, Chacaltana:2010ks}
\be\label{dDelta}
d_\Delta = h^0(L_\Delta) - h^1(L_\Delta) = 1-2 \Delta + \sum_{\alpha=1}^3 p_{\Delta}^{(\alpha)}~,
\ee
with $L_\Delta= \CK^\Delta(p_\Delta^{(\alpha)})$. Here, we specialized to the three-punctured sphere. The integers $p_\Delta$  associated to a given (regular) puncture are obtained as follows. Let us draw the transposed Young tableau associated to the partition $\lambda$, and let us start numbering the boxes from the upper-left corner, with the first box labeled by $p_1 \equiv 0$, the second box to the right labelled by $p_2=1$, and so forth from left to right and downwards; moreover,  each time we go to the next row, we start the new row with the same value of $p$ as in the last box on the previous row. This defines  a vector $p_\Delta$, with $\Delta=2, \cdots, N$.%
\footnote{For instance, for the partitions of $N=5$ appearing in \protect\eqref{fixture 20}, we have:
$[3,2] \rightarrow {\scriptsize\young(01,12,2)} \, : \, p_\Delta= (1,1,2,2)$, $[2,2,1] \rightarrow {\scriptsize\young(012,23)} \, : \, p_\Delta= (1,2,2,3)$, and $[1,1,1,1,1] \rightarrow {\scriptsize\young(01234)}\, : \, p_\Delta= (1,2,3,4)$.
}

In particular, \eqref{dDelta} gives a formula for the `virtual rank' of the class $\CS$ theory:
\be
r\big|_{\rm virtual} = \sum_{\Delta=2}^N d_\Delta~.
\ee
This equals the rank, $r$, if and only if the virtual dimension $d_\Delta$ equal the actual dimension, $d_\Delta = h^0(L_\Delta)$, for all $\Delta$. Using the relation
\be
\sum_{\Delta=2}^N p_\Delta= {N(N-1)\ov 2}- \half \sum_a \lambda_a (\lambda_a-1)~,
\ee
which holds for any partition of $N$, $\lambda= [\lambda_a]$, we find the general expression for the virtual rank of our trinion theories: 
\be
r\big|_{\rm virtual} = {(N-1)(N-2)\ov 2} -  \half \sum_{\alpha=1}^3 \sum_a \lambda_a^{(\alpha)} (\lambda_a^{(\alpha)}-1)~.
\ee
This matches the rank computed from the GTP using the generalised s-rule \cite{vanBeest:2020kou, vanBeest:2020civ}.

In general, however, some $d_\Delta$ could be negative, in which case the virtual rank is smaller than the true rank. For all the $E_n^{(k)}$ theories, we find that $d_\Delta \in \{1, 0, -1\}$, $\forall \Delta$. We then propose the following heuristic formula for the actual rank:
\be\label{r actual}
r=  \sum_{\Delta\, |\, d_\Delta >0} d_\Delta~.
\ee
It turns out that the number of $\Delta$'s such that $d_\Delta=-1$ is precisely equal to $\half b_3$, the number of pairs of 3-cycles in the resolved threefold $\t \MG$. Indeed, $d_\Delta=-1$ really means that $h^1(L_\Delta)=1$, which should be interpreted as an additional degree of freedom on the Coulomb branch of the class $\CS$ theory. Since these 3-cycles give rise to free hypermultiplets on the Coulomb branch of $\FT$, we propose that these $\CK$-valued $\Delta$-differential on the Gaiotto curve should also correspond to free hypermultiplets on the CB of the class $\CS$ theory  $\CS_{N, \bm{\lambda}}^{\rm 4d}$.

\medskip
\noindent 
As an example of the above discussion, consider the $r=1$ $E_6^{(k)}$ models. The corresponding GTPs and partitions are
\bea\label{E6 r1 GTP}
& \begin{tikzpicture}[x=.5cm,y=.5cm]
\node[] at (-2,2) {$E_6$:\qquad };
\draw[step=.5cm,gray,very thin] (0,0) grid (3,3);
\draw[ligne] (0,0)--(0,3)--(3,0)--(0,0); 
\node[bd] at (0,0) {}; 
\node[bd] at (0,1) {}; 
\node[bd] at (0,2) {}; 
\node[bd] at (0,3) {}; 
\node[bd] at (1,2) {};
\node[bd] at (2,1) {};
\node[bd] at (3,0) {}; 
\node[bd] at (2,0) {}; 
\node[bd] at (1,0) {}; 
\end{tikzpicture} \quad
    &&   \begin{tikzpicture}[x=.5cm,y=.5cm]
\node[] at (-2,2) {$E_6^{(1)}$:\qquad };
\draw[step=.5cm,gray,very thin] (0,0) grid (4,4);
\draw[ligne] (0,0)--(0,4)--(4,0)--(0,0); 
\node[bd] at (0,0) {}; 
\node[bd] at (0,1) {}; 
\node[bd] at (0,2) {}; 
\node[wd] at (0,3) {}; 
\node[bd] at (0,4) {}; 
\node[bd] at (1,3) {};
\node[bd] at (2,2) {};
\node[wd] at (3,1) {};
\node[bd] at (4,0) {}; 
\node[bd] at (3,0) {}; 
\node[bd] at (2,0) {}; 
\node[wd] at (1,0) {}; 
\end{tikzpicture}   \quad
 &&
 \begin{tikzpicture}[x=.5cm,y=.5cm]
\node[] at (-2,2) {$E_6^{(2)}$:\qquad };
\draw[step=.5cm,gray,very thin] (0,0) grid (5,5);
\draw[ligne] (0,0)--(0,5)--(5,0)--(0,0); 
\node[bd] at (0,0) {}; 
\node[bd] at (0,1) {}; 
\node[wd] at (0,2) {}; 
\node[bd] at (0,3) {}; 
\node[wd] at (0,4) {}; 
\node[bd] at (0,5) {}; 
\node[bd] at (1,4) {};
\node[wd] at (2,3) {};
\node[bd] at (3,2) {};
\node[wd] at (4,1) {};
\node[bd] at (5,0) {}; 
\node[bd] at (4,0) {}; 
\node[wd] at (3,0) {}; 
\node[bd] at (2,0) {}; 
\node[wd] at (1,0) {}; 
\end{tikzpicture}  \\
&\qquad {\small \bm{\lambda}= \{ [1,1,1]^3 \}}  
&&\qquad\;\;{\small   \bm{\lambda}= \{ [2,1,1]^3\} }    
&&\qquad\quad {\small   \bm{\lambda}= \{ [2,2,1]^3 \} }   
\eea
The vectors of virtual dimensions are then
\be
d_\Delta= (0,1)~, \qquad  \quad d_\Delta=(0,1,-1)~, \qquad\qquad d_\Delta= (0,1,-1,0)~,
\ee
respectively. The first model is the $E_6$ MN-theory, with a single CB operator of dimension $\Delta=3$. The other two models also have a CB operator of dimension $3$, but there is also a single negative entry at $\Delta=4$, so that their virtual rank vanishes.

The higher-rank theories are also interesting. For the $r=2$ $E_6$ theory, we have
\bea\label{E63 bad}
& \begin{tikzpicture}[x=.5cm,y=.5cm]
\node[] at (-2,2) {$E_6^{(3)}$:\qquad };
\draw[step=.5cm,gray,very thin] (0,0) grid (6,6);
\draw[ligne] (0,0)--(0,6)--(6,0)--(0,0); 
\node[bd] at (0,0) {}; 
\node[wd] at (0,1) {}; 
\node[bd] at (0,2) {}; 
\node[wd] at (0,3) {}; 
\node[bd] at (0,4) {}; 
\node[wd] at (0,5) {}; 
\node[bd] at (0,6) {}; 
\node[wd] at (1,5) {};
\node[bd] at (2,4) {};
\node[wd] at (3,3) {}; 
\node[bd] at (4,2) {}; 
\node[wd] at (5,1) {}; 
\node[bd] at (6,0) {}; 
\node[wd] at (5,0) {}; 
\node[bd] at (4,0) {}; 
\node[wd] at (3,0) {}; 
\node[bd] at (2,0) {}; 
\node[wd] at (1,0) {}; 
\end{tikzpicture} \qquad\quad
    &&   \begin{tikzpicture}[x=.5cm,y=.5cm]
\node[] at (-2,2) {$E_6^{(4)}$:\qquad };
\draw[step=.5cm,gray,very thin] (0,0) grid (7,7);
\draw[ligne] (0,0)--(0,7)--(7,0)--(0,0); 
\node[bd] at (0,0) {}; 
\node[wd] at (0,1) {}; 
\node[bd] at (0,2) {}; 
\node[wd] at (0,3) {}; 
\node[bd] at (0,4) {}; 
\node[wd] at (0,5) {}; 
\node[wd] at (0,6) {}; 
\node[bd] at (0,7) {}; 
\node[wd] at (1,6) {};
\node[bd] at (2,5) {};
\node[wd] at (3,4) {}; 
\node[bd] at (4,3) {}; 
\node[wd] at (5,2) {}; 
\node[wd] at (6,1) {}; 
\node[bd] at (7,0) {}; 
\node[wd] at (6,0) {}; 
\node[bd] at (5,0) {}; 
\node[wd] at (4,0) {}; 
\node[bd] at (3,0) {}; 
\node[wd] at (2,0) {}; 
\node[wd] at (1,0) {}; 
\end{tikzpicture}  \\
&\qquad {\small \bm{\lambda}= \{ [2,2,2]^3 \}}~, 
&&\qquad\;\;{\small   \bm{\lambda}= \{ [3,2,2]^3\} }~,  \\
&\qquad d_\Delta= (0, 1, -1, 0, 1)~, 
&&  \qquad d_\Delta=(0, 1, -1, 0, 1, -1)~,
\eea
and also $d_\Delta=(0, 1, -1, 0, 1, -1,0)$ for $E_6^{(5)}$. The rank-2 $E_6$ theory, $E_6^{(3)}$, indeed has rank $2$ according to the formula \eqref{r actual}. We also have one $d_\Delta=-1$ at $\Delta=4$, in agreement with $b_3=2$. Similarly, for $E_6^{(4)}$, we have two $-1$ entries in $d_\Delta$, corresponding to $b_3=4$ in the geometry.

This discussion readily generalizes to all the $E_n^{(k)}$ trinions with $k=\Delta_n \ell+\h k$. We have
\bea
&d_\Delta=1\;\;   &&\text{for} \quad &&  \Delta=(j+1) \Delta_n~, \;\; \quad &&j=0, \cdots, \ell~, \\
&d_\Delta=-1\;\;   &&\text{for} \quad && \Delta=(j+1)\Delta_n+1~, \; \;&&  j=0, \cdots, \ell - \delta_{\h k, 0}~.
\eea
The central node of the $\MQfive$ is bad whenever $d_N=-1$, which corresponds to $\h k=1$. For a fixed $\ell$, the vector $d_\Delta$ of the models with $\h k>1$ is the same as for the model with $\h k=1$, with $\h k-1$ zeros appended.

\subsection{Redeeming Bad Theories by Box-Lifting}
Up to this point, we have seen the badness of our $E_n^{(k)}$ theories for $k>0$ appear in  at least three  complementary ways: from the magnetic quivers (which describes the HB of $\FT$), from the presence of three-cycles in the resolved geometry $\t\MG$ (which engineers the CB of $\FT$), and from the virtual dimensions of $\Delta$-differentials on the Gaiotto curve (which engineers the CB of $\CS_{N, \bm{\lambda}}^{\rm 4d}$).

In section~\ref{sec:Higher-rank-En}, we have also argued that we could redeem part of the badness of the magnetic quiver -- essentially, for $k=1$ -- by considering a special locus of the CB of the bad $\MQfive$. Here, we would like to further discuss how we may cure the bad magnetic quivers more generally, using the class $\CS$ intuition. This discussion closely follows the proposal of \cite{Gaiotto:2012uq} on how to cure the higher-rank $E_n$ theories. See also \cite{Gaiotto:2011xs} for a related discussion from the Higgs-branch point of view. 

The basic idea is to embed the bad theory, $\CS$, into a larger `good' theory, $\CS'$, so that we can go back from $\CS'$ to $\CS$ by partial Higgsing (that is, by taking a particular limit onto the HB of $\CS'$). Indeed, any trinion $\CS_{N, \bm{\lambda}}^{\rm 4d}$ can be obtained by partially closing the punctures (or, equivalently, by lifting boxes in the partition) of the $T_N$ theory, which is the theory corresponding to three full punctures, $\bm{\lambda}=\{ [1^N]^3\}$. The GTP of $T_N$ is a triangle of length $N$ with only black dots, and any reduction of the punctures correspond to adding white dots along the edges.

This embedding $\CS_{N, \bm{\lambda}}^{\rm 4d} \subset \CS'$ implicitly defines the bad theory. In favorables cases, we can then confirm that the original `bad' theory is indeed equivalent to a (higher-rank) $E_n$ theory tensored with a free sector.   In terms of the partitions $\bm{\lambda}$ defining $\CS_{N, \bm{\lambda}}^{\rm 4d}$, the proposal of \cite{Gaiotto:2012uq} was to consider $\CS'$ defined by the partitions obtained by `lifting up one box' of one Young tableaux associated to $\bm{\lambda}$. In terms of the GTP, this corresponds to `filling in' a white dot. We will see concrete examples of this in the following.

\subsubsection{Rank-One $E_n^{(k)}$ Theories and GTPs}

Let us again consider the simplest bad theory in detail. This is the $E_6^{(1)}$ theory, whose GTP is shown in \eqref{E6 r1 GTP}. 
We propose to uplift this theory to the class $\CS$ theory defined by the partitions
\bea\label{boxliftedclassS E61}
  \begin{tikzpicture}[x=.5cm,y=.5cm]
\node[] at (-10,2) {$E_6^{(1)'}:\qquad \bm{\lambda}'= \{ [2,1,1] [2,1,1] [1,1,1,1] \}\, :$};
\draw[step=.5cm,gray,very thin] (0,0) grid (4,4);
\draw[ligne] (0,0)--(0,4)--(4,0)--(0,0); 
\node[bd] at (0,0) {}; 
\node[bd] at (0,1) {}; 
\node[bd] at (0,2) {}; 
\node[bd] at (0,3) {}; 
\node[bd] at (0,4) {}; 
\node[bd] at (1,3) {};
\node[bd] at (2,2) {};
\node[wd] at (3,1) {};
\node[bd] at (4,0) {}; 
\node[bd] at (3,0) {}; 
\node[bd] at (2,0) {}; 
\node[wd] at (1,0) {}; 
\end{tikzpicture} 
\eea
This is a good theory, in particular,  the virtual dimensions $d_\Delta$ in this case are $d_\Delta=(0, 1, 0)$. 
This theory can be higgsed back to  $E_6^{(1)}$  by partially closing the full puncture, $\lambda=[1,1,1,1]$, to $[2,1,1]$. Physically, this corresponds to activating a minimal nilpotent VEV for the moment map of the $SU(4)$ global symmetry associated with the full  puncture. Such a higgsing can lead to different theories in the IR, depending on how it is performed, as we will now explain. 

The trinion  \eqref{boxliftedclassS E61} describes the rank-1 $E_6$ theory plus four free hypermultiplets \cite{Chacaltana:2010ks}. Indeed, to these partitions we associate the magnetic quiver:
\bea\label{E6p MQ5}
 \begin{tikzpicture}[x=.6cm,y=.6cm]
 \node at (-5,0.5) {$\MQfive[E_6^{(1)'}] \cong $ };
\draw[ligne, black](-2,0)--(3,0);
\draw[ligne, black](0,0)--(0,2);
\node[bd] at (-2,0) [label=below:{{\scriptsize$1$}}] {};
\node[bd] at (-1,0) [label=below:{{\scriptsize$2$}}] {};
\node[bd] at (0,0) [label=below:{{\scriptsize$4$}}] {};
\node[bd] at (1,0) [label=below:{{\scriptsize$3$}}] {};
\node[bd] at (2,0) [label=below:{{\scriptsize$2$}}] {};
\node[bd] at (3,0) [label=below:{{\scriptsize$1$}}] {};
\node[bd] at (0,1) [label=left:{{\scriptsize$2$}}] {};
\node[bd] at (0,2) [label=left:{{\scriptsize$1$}}] {};
\end{tikzpicture}~,
\eea
This   3d $\CN=4$ quiver is ugly and equivalent to the $r=1$ $E_6$ quiver plus four twisted hypermultiplets, which transform in the fundamental of the $SU(4)$ flavor symmetry. In the $\MQfive$ \eqref{E6p MQ5}, this flavor $SU(4)$ arises as the CB symmetry associated to the right-hand leg, which is a $T[SU(4)]$ quiver, and the 4d  twisted hypermultiplets correspond to the $\Delta=\half$ monopole operators, with the flavor weights determined by their topological charges. Let us denote these hypermultiplets by $\widetilde{Q}_i$, $Q_i$,  $i=1,\cdots, 4$.

The actual global symmetry of this theory $\CS'$ is  therefore $E_6\times Sp(4)$, with only the subgroup $SU(4) \times SU(2)^2$ manifest in the class $\CS$ description \eqref{boxliftedclassS E61}. Crucially for our purpose,  the $SU(4)$ symmetry associated with the full puncture is also embedded inside $E_6$.%
\footnote{More precisely, $SU(4)$ is embedded in the $SU(6)$ factor of the maximal subgroup $SU(6)\times SU(2)\subset E_6$.} 
 Consequently, its moment map takes the form
 \be\label{mmap}
 \mu_{SU(4)}=\mu^{E_6}\bigr|_{SU(4)}+\widetilde{Q}_iQ_j~,
 \ee 
where $\mu^{E_6}\bigr|_{SU(4)}$ denotes the restriction of the $E_6$ moment map to the $SU(4)$ subgroup.
To recover the bad theory $E_6^{(1)}$, we need to activate a minimal nilpotent VEV for $\mu_{SU(4)}$
\be\label{vevbadS} 
\mu^{E_6}\bigr|_{SU(4)}+\widetilde{Q}_iQ_j=\left(\begin{array}{cccc}0&1&0&0\\ 0&0&0&0\\ 0&0&0&0\\ 0&0&0&0\\\end{array}\right)~.
\ee 
There is an obvious ambiguity in implementing this Higgsing.  Since the matrix \eqref{vevbadS} has rank $1$,  we can set $ \mu^{E_6}\bigr|_{SU(4)}=0$, which corresponds to a VEV for the free sector only. In this case, we flow to the rank-1 $E_6$ theory plus a free hypermultiplet.%
\footnote{The other three are identified with Goldstone modes for the Higgsing and we can discard them.} We then recover the description of $E_6^{(1)}$ argued for in section~\ref{sec:Higher-rank-En}.

On the other hand, if we turn on a nilpotent VEV for $\mu^{E_6}\bigr|_{SU(4)}$ instead, the $E_6$ theory is higgsed to a free theory, resulting in a collection of free hypermultiplets in the infrared. Both options are possible, but only the former option gives rise to an $r=1$ theory.

All the other $r=1$ trinions can be treated in the same way. For instance, the box-lifted version of $E_6^{(2)}$ reads
\bea
&\begin{tikzpicture}[x=.5cm,y=.5cm]
\node[] at (-3,2) {$E_6^{(2)}$:\qquad };
\draw[step=.5cm,gray,very thin] (0,0) grid (5,5);
\draw[ligne] (0,0)--(0,5)--(5,0)--(0,0); 
\node[bd] at (0,0) {}; 
\node[bd] at (0,1) {}; 
\node[wd] at (0,2) {}; 
\node[bd] at (0,3) {}; 
\node[wd] at (0,4) {}; 
\node[bd] at (0,5) {}; 
\node[bd] at (1,4) {};
\node[wd] at (2,3) {};
\node[bd] at (3,2) {};
\node[wd] at (4,1) {};
\node[bd] at (5,0) {}; 
\node[bd] at (4,0) {}; 
\node[wd] at (3,0) {}; 
\node[bd] at (2,0) {}; 
\node[wd] at (1,0) {}; 
\end{tikzpicture} 
\qquad 
&&\begin{tikzpicture}[x=.5cm,y=.5cm]
\node[] at (-5,2) {$\rightarrow \qquad  E_6^{(2)'}$:\quad };
\draw[step=.5cm,gray,very thin] (0,0) grid (5,5);
\draw[ligne] (0,0)--(0,5)--(5,0)--(0,0); 
\node[bd] at (0,0) {}; 
\node[bd] at (0,1) {}; 
\node[bd] at (0,2) {}; 
\node[bd] at (0,3) {}; 
\node[wd] at (0,4) {}; 
\node[bd] at (0,5) {}; 
\node[bd] at (1,4) {};
\node[wd] at (2,3) {};
\node[bd] at (3,2) {};
\node[wd] at (4,1) {};
\node[bd] at (5,0) {}; 
\node[bd] at (4,0) {}; 
\node[wd] at (3,0) {}; 
\node[bd] at (2,0) {}; 
\node[wd] at (1,0) {}; 
\end{tikzpicture} \\
&\qquad\qquad\bm{\lambda}= \{ [2,2,1]^3\} 
&&\qquad \qquad \qquad \qquad \bm{\lambda}'= \{ [2,2,1]^2 [2,1,1,1]\} \\
&\qquad\qquad d_\Delta= (0, 1, -1, 0) 
&&\qquad \qquad \qquad \qquad   d_\Delta= (0, 1, 0, 0)  \\
\eea
The magnetic quiver associated to the $E_6^{(2)'}$ theory is again ugly. It is equivalent to the $E_6$ quiver plus $7$ twisted hypermultiplets.  

For the rank-one $E_7^{(k)}$ theories, we have the GTPs
\bea
\begin{tikzpicture}[x=.5cm,y=.5cm]
\node[] at (2,5) {$E_7^{(0)}$};
\draw[step=.5cm,gray,very thin] (0,0) grid (4,4);
\draw[ligne] (0,0)--(0,4)--(4,0)--(0,0); 
\node[bd] at (0,0) {}; 
\node[bd] at (1,0) {}; 
\node[bd] at (2,0) {}; 
\node[bd] at (3,0) {}; 
\node[bd] at (4,0) {}; 
\node[bd] at (3,1) {}; 
\node[bd] at (2,2) {}; 
\node[bd] at (1,3) {}; 
\node[bd] at (0,4) {}; 
\node[wd] at (0,3) {}; 
\node[bd] at (0,2) {}; 
\node[wd] at (0,1) {}; 
\end{tikzpicture} 
\quad 
\begin{tikzpicture}[x=.5cm,y=.5cm]
\node[] at (2.5,6) {$E_7^{(1)}$};
\draw[step=.5cm,gray,very thin] (0,0) grid (5,5);
\draw[ligne] (0,0)--(0,5)--(5,0)--(0,0); 
\node[bd] at (0,0) {}; 
\node[wd] at (1,0) {}; 
\node[bd] at (2,0) {}; 
\node[bd] at (3,0) {}; 
\node[bd] at (4,0) {}; 
\node[bd] at (5,0) {}; 
\node[wd] at (4,1) {}; 
\node[bd] at (3,2) {}; 
\node[bd] at (2,3) {}; 
\node[bd] at (1,4) {}; 
\node[bd] at (0,5) {}; 
\node[wd] at (0,4) {}; 
\node[wd] at (0,3) {}; 
\node[bd] at (0,2) {}; 
\node[wd] at (0,1) {}; 
\end{tikzpicture} 
\quad 
\begin{tikzpicture}[x=.5cm,y=.5cm]
\node[] at (3,7) {$E_7^{(2)}$ };
\draw[step=.5cm,gray,very thin] (0,0) grid (6,6);
\draw[ligne] (0,0)--(0,6)--(6,0)--(0,0); 
\node[bd] at (0,0) {}; 
\node[wd] at (1,0) {}; 
\node[bd] at (2,0) {}; 
\node[bd] at (3,0) {}; 
\node[wd] at (4,0) {}; 
\node[bd] at (5,0) {}; 
\node[bd] at (6,0) {}; 
\node[wd] at (5,1) {}; 
\node[bd] at (4,2) {}; 
\node[bd] at (3,3) {}; 
\node[wd] at (2,4) {}; 
\node[bd] at (1,5) {}; 
\node[bd] at (0,6) {}; 
\node[wd] at (0,5) {}; 
\node[wd] at (0,4) {}; 
\node[bd] at (0,3) {}; 
\node[wd] at (0,2) {}; 
\node[wd] at (0,1) {}; 
\end{tikzpicture} 
\quad 
\begin{tikzpicture}[x=.5cm,y=.5cm]
\node[] at (3.5,8) {$E_7^{(3)}$ };
\draw[step=.5cm,gray,very thin] (0,0) grid (7,7);
\draw[ligne] (0,0)--(0,7)--(7,0)--(0,0); 
\node[bd] at (0,0) {}; 
\node[wd] at (1,0) {}; 
\node[bd] at (2,0) {}; 
\node[wd] at (3,0) {}; 
\node[bd] at (4,0) {}; 
\node[wd] at (5,0) {}; 
\node[bd] at (6,0) {}; 
\node[bd] at (7,0) {}; 
\node[wd] at (6,1) {}; 
\node[bd] at (5,2) {}; 
\node[wd] at (4,3) {}; 
\node[bd] at (3,4) {}; 
\node[wd] at (2,5) {}; 
\node[bd] at (1,6) {}; 
\node[bd] at (0,7) {}; 
\node[wd] at (0,6) {}; 
\node[wd] at (0,5) {}; 
\node[bd] at (0,4) {}; 
\node[wd] at (0,3) {}; 
\node[wd] at (0,2) {}; 
\node[wd] at (0,1) {}; 
\end{tikzpicture} 
\eea
The first one is simply the rank-one $E_7$ theory \cite{Benini:2009gi}; it has a dimension vector $d_\Delta=(0,0,1)$. The other theories have bad magnetic quivers, and correspondingly $d_\Delta= (0,0,1,-1,0, \cdots)$. Note also that, for the $E_7^{(2)}$ GTP, the partitions along the edges are unordered, and we should reorder them before interpreting them as class $\CS$ models (this introduces free hypermultiplet, as in the discussion around \eqref{E8m2 GTP}). The box-lifting/white-dot-filling prescription gives us the following good $r=1$ models:
\bea
%\begin{tikzpicture}[x=.5cm,y=.5cm]
%\node[] at (2,5) {$E_7^{(0)}$};
%\draw[step=.5cm,gray,very thin] (0,0) grid (4,4);
%\draw[ligne] (0,0)--(0,4)--(4,0)--(0,0); 
%\node[bd] at (0,0) {}; 
%\node[bd] at (1,0) {}; 
%\node[bd] at (2,0) {}; 
%\node[bd] at (3,0) {}; 
%\node[bd] at (4,0) {}; 
%\node[bd] at (3,1) {}; 
%\node[bd] at (2,2) {}; 
%\node[bd] at (1,3) {}; 
%\node[bd] at (0,4) {}; 
%\node[wd] at (0,3) {}; 
%\node[bd] at (0,2) {}; 
%\node[wd] at (0,1) {}; 
%\end{tikzpicture} 
%\quad 
\begin{tikzpicture}[x=.5cm,y=.5cm]
\node[] at (2.5,6) {$E_7^{(1)'}$};
\draw[step=.5cm,gray,very thin] (0,0) grid (5,5);
\draw[ligne] (0,0)--(0,5)--(5,0)--(0,0); 
\node[bd] at (0,0) {}; 
\node[bd] at (1,0) {}; 
\node[bd] at (2,0) {}; 
\node[bd] at (3,0) {}; 
\node[bd] at (4,0) {}; 
\node[bd] at (5,0) {}; 
\node[wd] at (4,1) {}; 
\node[bd] at (3,2) {}; 
\node[bd] at (2,3) {}; 
\node[bd] at (1,4) {}; 
\node[bd] at (0,5) {}; 
\node[wd] at (0,4) {}; 
\node[wd] at (0,3) {}; 
\node[bd] at (0,2) {}; 
\node[wd] at (0,1) {}; 
\end{tikzpicture} 
\qquad 
\begin{tikzpicture}[x=.5cm,y=.5cm]
\node[] at (3,7) {$E_7^{(2)'}$ };
\draw[step=.5cm,gray,very thin] (0,0) grid (6,6);
\draw[ligne] (0,0)--(0,6)--(6,0)--(0,0); 
\node[bd] at (0,0) {}; 
\node[bd] at (1,0) {}; 
\node[bd] at (2,0) {}; 
\node[bd] at (3,0) {}; 
\node[wd] at (4,0) {}; 
\node[bd] at (5,0) {}; 
\node[bd] at (6,0) {}; 
\node[wd] at (5,1) {}; 
\node[bd] at (4,2) {}; 
\node[bd] at (3,3) {}; 
\node[wd] at (2,4) {}; 
\node[bd] at (1,5) {}; 
\node[bd] at (0,6) {}; 
\node[wd] at (0,5) {}; 
\node[wd] at (0,4) {}; 
\node[bd] at (0,3) {}; 
\node[wd] at (0,2) {}; 
\node[wd] at (0,1) {}; 
\end{tikzpicture} 
\qquad 
\begin{tikzpicture}[x=.5cm,y=.5cm]
\node[] at (3.5,8) {$E_7^{(3)'}$ };
\draw[step=.5cm,gray,very thin] (0,0) grid (7,7);
\draw[ligne] (0,0)--(0,7)--(7,0)--(0,0); 
\node[bd] at (0,0) {}; 
\node[bd] at (1,0) {}; 
\node[bd] at (2,0) {}; 
\node[wd] at (3,0) {}; 
\node[bd] at (4,0) {}; 
\node[wd] at (5,0) {}; 
\node[bd] at (6,0) {}; 
\node[bd] at (7,0) {}; 
\node[wd] at (6,1) {}; 
\node[bd] at (5,2) {}; 
\node[wd] at (4,3) {}; 
\node[bd] at (3,4) {}; 
\node[wd] at (2,5) {}; 
\node[bd] at (1,6) {}; 
\node[bd] at (0,7) {}; 
\node[wd] at (0,6) {}; 
\node[wd] at (0,5) {}; 
\node[bd] at (0,4) {}; 
\node[wd] at (0,3) {}; 
\node[wd] at (0,2) {}; 
\node[wd] at (0,1) {}; 
\end{tikzpicture} \,.
\eea
Their magnetic quivers are all IR-equivalent to the $E_7$ magnetic quiver plus free twisted hypermultiplets. As always, we can focus on the $k=1$ case, which has an $SU(5)$ symmetry associated to the full puncture. The partial closure of the puncture is similar to the $E_6^{(1)}$ case above. The $E_8$ series is completely similar. For $k=1$, we have:
\bea
\begin{tikzpicture}[x=.5cm,y=.5cm]
\node[] at (3.5,8) {$E_8^{(1)}$ };
\draw[step=.5cm,gray,very thin] (0,0) grid (7,7);
\draw[ligne] (0,0)--(0,7)--(7,0)--(0,0); 
\node[bd] at (0,0) {}; 
\node[wd] at (1,0) {}; 
\node[bd] at (2,0) {}; 
\node[bd] at (3,0) {}; 
\node[bd] at (4,0) {}; 
\node[bd] at (5,0) {}; 
\node[bd] at (6,0) {}; 
\node[bd] at (7,0) {}; 
\node[wd] at (6,1) {}; 
\node[wd] at (5,2) {}; 
\node[bd] at (4,3) {}; 
\node[wd] at (3,4) {}; 
\node[bd] at (2,5) {}; 
\node[wd] at (1,6) {}; 
\node[bd] at (0,7) {}; 
\node[wd] at (0,6) {}; 
\node[wd] at (0,5) {}; 
\node[wd] at (0,4) {}; 
\node[bd] at (0,3) {}; 
\node[wd] at (0,2) {}; 
\node[wd] at (0,1) {}; 
\end{tikzpicture} 
\qquad 
\begin{tikzpicture}[x=.5cm,y=.5cm]
\node[] at (-3,4) {$\rightarrow\quad $ };
\node[] at (3.5,8) {$E_8^{(1)'}$ };
\draw[step=.5cm,gray,very thin] (0,0) grid (7,7);
\draw[ligne] (0,0)--(0,7)--(7,0)--(0,0); 
\node[bd] at (0,0) {}; 
\node[bd] at (1,0) {}; 
\node[bd] at (2,0) {}; 
\node[bd] at (3,0) {}; 
\node[bd] at (4,0) {}; 
\node[bd] at (5,0) {}; 
\node[bd] at (6,0) {}; 
\node[bd] at (7,0) {}; 
\node[wd] at (6,1) {}; 
\node[wd] at (5,2) {}; 
\node[bd] at (4,3) {}; 
\node[wd] at (3,4) {}; 
\node[bd] at (2,5) {}; 
\node[wd] at (1,6) {}; 
\node[bd] at (0,7) {}; 
\node[wd] at (0,6) {}; 
\node[wd] at (0,5) {}; 
\node[wd] at (0,4) {}; 
\node[bd] at (0,3) {}; 
\node[wd] at (0,2) {}; 
\node[wd] at (0,1) {}; 
\end{tikzpicture} 
\eea
The pattern of partial closure from $E_8^{(1)'}$ to  $E_8^{(1)}$ is the same as before, with a nilpotent VEV given to free hypermultiplets charged under the full-puncture $SU(7)$ symmetry of the $E_8^{(1)'}$ model.

\subsubsection{Higher-rank Theories}

Finally, let us briefly consider the general case with rank $r=\ell+1$. The important new ingredient is the presence of additional sources of badness, as in the example \eqref{E63 bad}. Let us first review the case $k= \ell \Delta_n$ ($\h k=0$) with $\ell>0$.  The partitions associated with the $E_n^{(\Delta_n \ell)}$ models are:
\bea\label{En hr lambda}
& \bm{\lambda}_{E_6^{(3\ell)}}  =\{ [\ell +1, \ell+1, \ell+1]^3\}~, \\
& \bm{\lambda}_{E_7^{(4\ell)}} =\{ [2\ell +2, 2 \ell+2] [(\ell+1)^4]^2\}~,  \\
& \bm{\lambda}_{E_8^{(6\ell)}} =\{ [3\ell +3,3 \ell+3] [2\ell +2, 2\ell+2, 2 \ell+2][(\ell+1)^6]\}~.
\eea
The `box-lifting' prescription of \cite{Gaiotto:2012uq} reads:
\bea\label{En hr lambda prime}
&  \bm{\lambda}_{E_6^{(3\ell)'}} =\{ [\ell +1, \ell+1, \ell+1]^2 [\ell +1, \ell+1, \ell, 1]\}~,\\
 &   \bm{\lambda}_{E_7^{(4\ell)'}}' = \{[2\ell +2, 2 \ell+2] [(\ell+1)^4][(\ell+1)^3,\ell,1]\}~,\\
&  \bm{\lambda}_{E_8^{(6\ell)'}}' =\{ [3\ell +3,3 \ell+3] [2\ell +2, 2\ell+2, 2 \ell+2][(\ell+1)^5,\ell,1]\}~.
\eea
For instance, for the rank-2 $E_6$ theory, we have:
\bea
& \begin{tikzpicture}[x=.5cm,y=.5cm]
\node[] at (-3,2) {$E_6^{(3)}$:\qquad };
\draw[step=.5cm,gray,very thin] (0,0) grid (6,6);
\draw[ligne] (0,0)--(0,6)--(6,0)--(0,0); 
\node[bd] at (0,0) {}; 
\node[wd] at (1,0) {}; 
\node[bd] at (2,0) {}; 
\node[wd] at (3,0) {}; 
\node[bd] at (4,0) {}; 
\node[wd] at (5,0) {}; 
\node[bd] at (6,0) {}; 
\node[wd] at (5,1) {}; 
\node[bd] at (4,2) {}; 
\node[wd] at (3,3) {}; 
\node[bd] at (2,4) {}; 
\node[wd] at (1,5) {}; 
\node[bd] at (0,6) {}; 
\node[wd] at (0,5) {}; 
\node[bd] at (0,4) {}; 
\node[wd] at (0,3) {}; 
\node[bd] at (0,2) {}; 
\node[wd] at (0,1) {}; 
\end{tikzpicture} 
\qquad 
&&\begin{tikzpicture}[x=.5cm,y=.5cm]
\node[] at (-4,2) {$\rightarrow \qquad  E_6^{(3)'}$:\quad };
\draw[step=.5cm,gray,very thin] (0,0) grid (6,6);
\draw[ligne] (0,0)--(0,6)--(6,0)--(0,0); 
\node[bd] at (0,0) {}; 
\node[wd] at (1,0) {}; 
\node[bd] at (2,0) {}; 
\node[wd] at (3,0) {}; 
\node[bd] at (4,0) {}; 
\node[wd] at (5,0) {}; 
\node[bd] at (6,0) {}; 
\node[wd] at (5,1) {}; 
\node[bd] at (4,2) {}; 
\node[wd] at (3,3) {}; 
\node[bd] at (2,4) {}; 
\node[wd] at (1,5) {}; 
\node[bd] at (0,6) {}; 
\node[wd] at (0,5) {}; 
\node[bd] at (0,4) {}; 
\node[wd] at (0,3) {}; 
\node[bd] at (0,2) {}; 
\node[bd] at (0,1) {}; 
\end{tikzpicture} \\
&\qquad\qquad\bm{\lambda}= \{ [2,2,2]^3\} 
&&\qquad \quad \qquad \qquad \bm{\lambda}'= \{ [2,2,2]^2 [2,2,1,1]\} \\
&\qquad\qquad d_\Delta= (0, 1, -1, 0, 1) 
&&\qquad \quad \qquad \qquad   d_\Delta= (0, 1, 0, 0, 1)  \\
\eea
The magnetic quivers associated to \eqref{En hr lambda}  are given by $\ell+1$ times the affine $E_n$ quiver (that is, the ranks are given by $(\ell+1)d_i$, with $d_i$ the Dynkin labels) while the magnetic quivers associated to \eqref{En hr lambda prime} are given by:
\bea\label{MQ En hr prime}
& \begin{tikzpicture}[x=0.8cm,y=0.8cm]
\draw[ligne, black](-2,0)--(3,0);
\draw[ligne, black](0,0)--(0,2);
\node[] at (-6,1) {$ \qquad  E_6^{(3\ell)'}$:\quad };
\node[bd] at (3,0) [label=below:{{\scriptsize$1$}}] {};
\node[bd] at (-2,0) [label=below:{{\scriptsize$\ell{+}1$}}] {};
\node[bd] at (-1,0) [label=below:{{\scriptsize$2\ell{+}2$}}] {};
\node[bd] at (0,0) [label=below:{{\scriptsize$3\ell{+}3$}}] {};
\node[bd] at (1,0) [label=below:{{\scriptsize$2 \ell{+}2$}}] {};
\node[bd] at (2,0) [label=below:{{\scriptsize$\ell{+}1$}}] {};
\node[bd] at (0,1) [label=left:{{\scriptsize$2\ell{+}2$}}] {};
\node[bd] at (0,2) [label=left:{{\scriptsize$\ell{+}1$}}] {};
\end{tikzpicture} \\
& \begin{tikzpicture}[x=.8cm,y=.8cm]
 \node[] at (-7,1) {$ \qquad  E_7^{(4\ell)'}$:\quad };
\draw[ligne, black](-3,0)--(4,0);
\draw[ligne, black](0,0)--(0,1);
\node[bd] at (-3,0) [label=below:{{\scriptsize$\ell{+}1$}}] {};
\node[bd] at (-2,0) [label=below:{{\scriptsize$2\ell{+}2$}}] {};
\node[bd] at (-1,0) [label=below:{{\scriptsize$3\ell{+}3$}}] {};
\node[bd] at (0,0) [label=below:{{\scriptsize$ 4\ell{+}4$}}] {};
\node[bd] at (1,0) [label=below:{{\scriptsize$3 \ell{+}4$}}] {};
\node[bd] at (2,0) [label=below:{{\scriptsize$2\ell{+}2$}}] {};
\node[bd] at (3,0) [label=below:{{\scriptsize$\ell{+}1$}}] {};
\node[bd] at (4,0) [label=below:{{\scriptsize$1$}}] {};
\node[bd] at (0,1) [label=left:{{\scriptsize$2\ell{+}2$}}] {};
\end{tikzpicture}
\\
&   \begin{tikzpicture}[x=.8cm,y=.8cm]
\node[] at (-6,1) {$ \qquad  E_8^{(6\ell)'}$:\quad };
\draw[ligne, black](-2,0)--(6,0);
\draw[ligne, black](0,0)--(0,1);
\node[bd] at (-2,0) [label=below:{{\scriptsize$2\ell{+}2$}}] {};
\node[bd] at (-1,0) [label=below:{{\scriptsize$4\ell{+}4$}}] {};
\node[bd] at (0,0) [label=below:{{\scriptsize$6\ell{+}6$}}] {};
\node[bd] at (1,0) [label=below:{{\scriptsize$5\ell{+}5$}}] {};
\node[bd] at (2,0) [label=below:{{\scriptsize$4\ell{+}4$}}] {};
\node[bd] at (3,0) [label=below:{{\scriptsize$3\ell{+}3$}}] {};
\node[bd] at (4,0) [label=below:{{\scriptsize$2\ell{+}2$}}] {};
\node[bd] at (5,0) [label=below:{{\scriptsize$\ell{+}1$}}] {};
\node[bd] at (6,0) [label=below:{{\scriptsize$1$}}] {};
\node[bd] at (0,1) [label=left:{{\scriptsize$3\ell{+}3$}}] {};
\end{tikzpicture}\,.
\eea
These magnetic quivers are `good', due to the additional $U(1)$ factor at the end of the left-hand tail. The corresponding Higgs-branch flavor symmetry of the class $\CS$ theory is $E_n \times SU(2)$. The CB of the magnetic quivers \eqref{MQ En hr prime} reproduce the correct moduli space of $(\ell+1)$ $E_n$ instantons, as studied {\it e.g.} in \cite{Hori:1997zj, Cremonesi:2014xha}. 

The uplifting procedure for $\h k>0$ is similar. It gives rise to the same magnetic quivers \eqref{MQ En hr prime}  plus free twisted hypermultiplets. For completeness, we list the partitions after the box-lifting in tables \ref{tab:E6kMQ},  \ref{tab:E7kMQ} and \ref{tab:E8kMQ}.
There we listed the unordered partitions, corresponding to GTP, which are equivalent to ordered partitions plus free hypermultiplets.

%%%%%%%%%%%%%%%%%%%%%%%%%%%%%%%%%%%%

\section{Classification of the $SU(N)$ Trinions} 
\label{sec:SUN}
%%%%%%%%%%%
In the previous sections, we focussed on `trinion' singularities of type I, in the notation of~\eqref{all trinion sings}. The corresponding 4d theory $\FTfour$ is  a $D^b_p(SU(N))$-trinions with $b=N$ for the three legs. In this section, we give a complete classification of the $D_p^b(SU(N))$ trinions
\bea
 \begin{tikzpicture}[x=.7cm,y=.7cm]
\draw[ligne, black](-1,0)--(1,0);
\draw[ligne, black](0,0)--(0,1);
\node at (-2.4,0) {{\scriptsize $D_{p_1}^{b_1}(SU(N))$}};
\node[SUd] at (0,0) [label=below:{{\scriptsize$N$}}] {};
\node at (2.5,0) {{\scriptsize $D_{p_3}^{b_3}(SU(N))$}};
\node at (0,1.4) {{\scriptsize $D_{p_2}^{b_2}(SU(N))$}};
\end{tikzpicture}
\eea
As discussed in section~\ref{sec:Dpbs}, the vanishing of the $SU(N)$ $\beta$-function is equivalent to the condition
\be\label{beta zero cond gen}
\sum_{\alpha=1}^3 {b_\alpha \ov p_\alpha} = N~.
\ee
Given a solution, the $\CN=2$ SCFT is realized by the isolated hypersurface singularity (IHS)
\be\label{gen trinion sing}
x^N + x^{N-b_1} z_1^{p_1} + x^{N-b_2} z_2^{p_2} + x^{N-b_3} z_3^{p_3}+ \cdots =0~.
\ee
There are four subfamilies of solutions, depending on the choices for $b_\alpha \in \{N, N-1\}$. In terms of the Yau-Yu classification~\cite{yau2005classification}, we have
\bea
&(b_\alpha)= (N, N, N)\quad && : \qquad &  \text{I} &\; (p_1, p_2, p_3, N)~, \cr
&(b_\alpha)= (N, N, N-1)\quad && : \qquad &  \text{II}&\; (p_1, p_2, N, p_3)~, \cr
&(b_\alpha)= (N, N-1, N-1)\quad && : \qquad &  \text{VIII}&\; (p_1, N,  p_2, p_3)~, \cr
&(b_\alpha)= (N-1, N-1, N-1)\quad && : \qquad & \text{XIV}&\; (N, p_1, p_2, p_3)~.
\eea
Note that, for the cases of type VIII and XIV, the singularity \eqref{gen trinion sing} contains additional monomials so that the singularity is isolated~\cite{Kreuzer:1992np,Davenport:2016ggc}. This introduces some subtypes in the YY classification, which will be discussed in more detail elsewhere~\cite{CSNWII}.

We already discussed at length the type I solutions, which are the $E_n^{(k)}$ models. 
The other solutions to \eqref{beta zero cond gen} are easily classified, and the corresponding theories will be described below. They are summarized in table~\ref{tab:trinion series} and in tables~\ref{tab:trinion series 1},~\ref{tab:trinion series 2},~\ref{tab:trinion series 3} and \ref{tab:trinion series 4}, where we display the numbers $r$, $f$, $d_H$, which characterize the 5d theory $\FT$, as well as the electric quivers $\EQfour$ for the $D_p^b(SU(N))$ trinions. 
We have also indicated in the tables whether the resolved Calabi-Yau threefold is smooth, and whether any of the divisors of the resolution are singular. The details of the crepant resolutions are presented in appendix~\ref{app:resolution}.  From these results, we observe that $\FTfour$ is a Lagrangian 4d $\CN=2$ SCFT (a conformal special unitary quiver) if the resolved geometry is smooth, and the divisors have at most codimension-one singularities on them (no point singularity). 
On the other hand, if there are residual terminal singularities after the resolution, the 4d theory $\FTfour$ is non-Lagrangian and the $\EQfour$ is a mixed $SU(n){-}U(m)$ quiver.

%%%%%%%%%%%%%%%%%%%%%%%%%%%%%%%%%%%%%%%%%%%%%%%%
%%%%%%%%%%%%%%%%%%%%%%%%%%%%%%%%%%%%%%%%%%%%%%%%
\begin{table}
$$
\small
\begin{array}{|c|c|c|c|c|c||c|c|}
\hline
\text{Type} & (a,b,c,d) & \text{AD}[G,G'] &\, r\, &  f&d_H & \widehat{r} & \widehat{d_H} \cr \hline \hline
 \text{II} & \{2,2N,N,2\} & [A_{2N{-}1}, D_{N{+}1}] & 0  & N+1 & N(N{+}1) &N^2-1 & N+1 \\
  \hline
    \hline
\text{VIII}_2 & \{2k-1,2k-1,2,2\} & [D_{2k},D_{2k}] & k-1 & 2k+2 & 2k^2+k+1 & (2k+1)(k-1) & 3k+1\\ \hline
\text{VIII}_2 & \{2k,2k,2,2\} & [D_{2k+1},D_{2k+1}] & k-1 & 2k+1 & (k+1)(2k+1) & k(2k+1) & 3k\\ \hline
 \end{array}
$$
\caption{Basic data for the infinite series of $SU(N)$ trinions. Here $(a,b,c,d)$ indicate the powers in the hypersurface equation.  For the $[D_{N+1}, D_{N+1}]$ series, which are given by type VIII$_2$ $(N,N,2,2)$ singularities, or equivalently by type IV $(N,2,N,2)$ singularities, we look at odd and even $N$ separately. They all have $\Delta \CA_r =0$, $b_3=0$ and trivial one-form symmetry.  \label{tab:trinion series}}
\end{table}
%%%%%%%%%%%%%%%%%%%%%%%%%%%%%%%%%%%%%%%%%%%%%%%%
%%%%%%%%%%%%%%%%%%%%%%%%%%%%%%%%%%%%%%%%%%%%%%%%

\subsection{Case $(b_\alpha)= (N,N, N)$}
In this case, the $N$ dependence drops out from \eqref{beta zero cond gen}, and we therefore obtain infinite families of solutions indexed by $N$: 
\be
\sum_{\alpha=1}^3 {1 \ov p_\alpha} =1 \qquad \Rightarrow \qquad (p_\alpha)= (3,3,3)~, \; (2,4,4)~, \; (2,3,6)~.
\ee
When $(p_\alpha)= (3,3,3)$, the theory corresponds to the $E_6^{(N-3)}$ theory.
When $(p_\alpha)= (2,4,4)$, the theory corresponds to the $E_7^{(N-4)}$. When $(p_\alpha)= (2,3,6)$, the theory corresponds to the $E_8^{(N-6)}$ theory. These models were already discussed at length in sections~\ref{sec:Rank0} and~\ref{sec:Higher-rank-En}.

\subsection{Case $(b_\alpha)= (N,N, N-1)$}
Consider the condition
\be
\label{Diophantine-eq2}
{N\ov p_1} +{N\ov p_2}+{N-1\ov p_3}  =N~.
\ee
This has the following obvious solutions  for any $N$:
\be\label{series su i}
(p_\alpha) = (2,2N,2)~, \; \; \forall N~.
\ee
They are terminal singularities of Type II$(2,2N,N,2)$.  These singularities engineer the generalized Argyres-Douglas theory  AD$[A_{2N-1}, D_{N+1}]$ in Type IIB \cite{Cecotti:2010fi}. After the dimensional reduction of $D_p^b(SU(N))$ to 3d, we therefore obtain an explicit `electric quiver description' for these 4d SCFTs (for $k>1$)
\be\label{EQ4 series A D}
\EQfour = \begin{cases} 
 \begin{tikzpicture}[x=.6cm,y=.6cm]
\draw[ligne, black](-2,0)--(4.4,0);
\draw[ligne, black,dotted](4.6,0)--(5.4,0);
\draw[ligne, black](5.6,0)--(9,0);
\draw[ligne, black](0,0)--(0,1);
%central node:
\node[SUd] at (0,0) [label=below:{{\scriptsize$N$}}] {};
%left:
\node[bd] at (-1,0) [label=above:{{\scriptsize$k$}}] {};
\node[flavor] at (-2,0) [label=below:{{\scriptsize$1$}}] {};
%up:
\node[SUd] at (0,1) [label=above:{{\scriptsize$k$}}] {};
%right:
\node[bd] at (1,0) [label=above:{{\scriptsize$2k{-}1$}}] {};
\node[SUd] at (2,0) [label=below:{{\scriptsize$2k{-}1$}}] {};
\node[bd] at (3,0) [label=above:{{\scriptsize$2k{-}2$}}] {};
\node[SUd] at (4,0) [label=below:{{\scriptsize$2k{-}2$}}] {};
\node[bd] at (6,0) [label=above:{{\scriptsize$2$}}] {};
\node[SUd] at (7,0) [label=below:{{\scriptsize$2$}}] {};
\node[bd] at (8,0) [label=above:{{\scriptsize$1$}}] {};
\node[flavor] at (9,0) [label=below:{{\scriptsize$1$}}] {};
\end{tikzpicture}  & \; \;   \text{if} \;\; N= 2k~,\\
%%%%%%------------%%%%%
 \begin{tikzpicture}[x=.6cm,y=.6cm]
\draw[ligne, black](-2,0)--(4.4,0);
\draw[ligne, black,dotted](4.6,0)--(5.4,0);
\draw[ligne, black](5.6,0)--(9,0);
\draw[ligne, black](0,0)--(0,1);
%central node:
\node[SUd] at (0,0) [label=below:{{\scriptsize$N$}}] {};
%left:
\node[SUd] at (-1,0) [label=above:{{\scriptsize$k$}}] {};
\node[flavor] at (-2,0) [label=below:{{\scriptsize$1$}}] {};
%up:
\node[bd] at (0,1) [label=above:{{\scriptsize$k{-}1$}}] {};
%right:
\node[bd] at (1,0) [label=above:{{\scriptsize$2k{-}2$}}] {};
\node[SUd] at (2,0) [label=below:{{\scriptsize$2k{-}2$}}] {};
\node[bd] at (3,0) [label=above:{{\scriptsize$2k{-}3$}}] {};
\node[SUd] at (4,0) [label=below:{{\scriptsize$2k{-}3$}}] {};
\node[bd] at (6,0) [label=above:{{\scriptsize$2$}}] {};
\node[SUd] at (7,0) [label=below:{{\scriptsize$2$}}] {};
\node[bd] at (8,0) [label=above:{{\scriptsize$1$}}] {};
\node[flavor] at (9,0) [label=below:{{\scriptsize$1$}}] {};
\end{tikzpicture}  & \; \; \text{if}\;\; N= 2k{-}1~,
\end{cases}
\ee
Since these singularities have $r=0$, the 5d theory consists of $d_H=N(N+1)$ free hypermultiplets. To see this explicitly, we can obtain the $\MQfive$ from \eqref{EQ4 series A D}, corresponding to replacing all the $SU(n)$ nodes by $U(n)$ nodes, including the central one, because there is no higher-form symmetry associated to the singularity ($\frak{f}=0$). The resulting unitary quiver is ugly, and repeated applications of the GW duality gives us exactly $d_H$ twisted hypermultiplets.

\medskip
\noindent
In addition, there are three sporadic solutions to \eqref{Diophantine-eq2}
\be
\begin{tabular}{ c| c |c c  c} % centered columns (4 columns)
\text{Sporadic Models} {(\#)}& $N$ & $p_1$ & $p_2$ & $p_3$ \\%heading
 \hline % inserts single horizontal line
\scriptsize{(1)}&6& 3&4&2\\
\scriptsize{(2)}& 10& 2 & 5 &3\\
\scriptsize{(3)}&15& 3 & 5 &2\\
\end{tabular}
\ee
The basic properties and the $\EQfour$ for these models are shown in table~\ref{tab:trinion series 1}. In appendix \ref{app:Diophantine}, we prove that this set of solutions is complete.

For the model $\scriptstyle{(1)}$, from the resolution of the geometry in appendix~\ref{app:SporadicRes}, 
we see that this is a rank-1 $E_6$ theory (trivially) coupled to the rank-0 theory of type $E_{8}^{(-3)}$, which is indicated by the fact that there is a remnant terminal singularity. This again means that the `naive' $\MQfive$ is ugly. 
Since there is no one-form symmetry,  we replace all the $SU(N)$ by $U(N)$ nodes in  $\EQfour$. We can then check that the $\MQfive$ is equivalent to the $E_6$ quiver plus $12$ twisted hypermultiplets
\be
    \begin{tikzpicture}[x=.7cm,y=.7cm]
    \node[] at (-4,1)[] {$\MQfive\cong$};
\draw[ligne, black](-2,0)--(3,0);
\draw[ligne, black](0,0)--(0,2);
\node[bd] at (-2,0) [label=below:{{\scriptsize$2$}}] {};
\node[bd] at (-1,0) [label=below:{{\scriptsize$4$}}] {};
\node[bd] at (0,0) [label=below:{{\scriptsize$6$}}] {};
\node[bd] at (1,0) [label=below:{{\scriptsize$4$}}] {};
\node[bd] at (2,0) [label=below:{{\scriptsize$3$}}] {};
\node[bd] at (3,0) [label=below:{{\scriptsize$1$}}] {};
\node[bd] at (0,1) [label=left:{{\scriptsize$3$}}] {};
\node[bd] at (0,2) [label=left:{{\scriptsize$1$}}] {};
\end{tikzpicture}
\;\;
    \begin{tikzpicture}[x=.7cm,y=.7cm]
    \node[] at (-3.6,1)[] {$\cong$};
\draw[ligne, black](-2,0)--(2,0);
\draw[ligne, black](0,0)--(0,2);
\node[bd] at (-2,0) [label=below:{{\scriptsize$1$}}] {};
\node[bd] at (-1,0) [label=below:{{\scriptsize$2$}}] {};
\node[bd] at (0,0) [label=below:{{\scriptsize$3$}}] {};
\node[bd] at (1,0) [label=below:{{\scriptsize$2$}}] {};
\node[bd] at (2,0) [label=below:{{\scriptsize$1$}}] {};
\node[bd] at (0,1) [label=left:{{\scriptsize$3$}}] {};
\node[bd] at (0,2) [label=left:{{\scriptsize$1$}}] {};
    \node[] at (3,1)[] {$\quad  \oplus \quad \mb{H}^{12}$};
\end{tikzpicture}\,.
\ee

\medskip
\noindent
For the sporadic models $\scriptstyle{(2)}$ and $\scriptstyle{(3)}$, there is no residual terminal singularity, but there are singular divisors.  The singularities are codimension one on the divisors.
 The model $\scriptstyle{(2)}$ has $r=2$ and was studied in \cite{Closset:2020scj}. Its Higgs branch, as determined from the $\MQfive$, is the next-to-minimal nilpotent orbit of $\frak{e}_8$  \cite{Hanany:2017ooe}. The $\MQfive$ of model $\scriptstyle{(3)}$  appeared before in \cite{Cabrera:2018uvz}.

In addition, we also have the solutions $N=2$, $(p_\alpha)=(2,3,3)$, $N_3$, $N=3$, $(p_\alpha)=(3,3,2)$, $N=4$, $(p_\alpha)=(2,4,3)$  and  $N=6$, $(p_\alpha)=(2,3,5)$, but they are equivalent to the type-I singularities $(2,2,3,6)$, $(3,3,3,3)$, $(2,4,4,4)$ and $(2,3,6,6)$, respectively. The first one is the rank-0 theory $E_{8}^{(-4)}$, and the other theories are rank-1 $E_6$, $E_7$ and $E_8$ theories respectively.

%%%%%%%%%%%%%%%%%%%%%%%%%%%%%%%%%%%%%%%%%%%%%%%%
%%%%%%%%%%%%%%%%%%%%%%%%%%%%%%%%%%%%%%%%%%%%%%%%
\begin{table}
$$
\begin{array}{|c|c|c|c|c|c|c|}
\hline
$ \scriptsize{(\#)}$& \matarray{ \text{Type} \;   (a,b,c,d)  \\  F(x_1, x_2, x_3, x_4) } &\, r\, &  f&d_H &  \text{Smooth CY/Divs} & \EQfour \cr \hline \hline
 %%---------------------------------%
%%---------------------------------%
 $\scriptsize{(1)}$&  \matarray{\text{II}_1 \;  (3,4,6,2) \\x_1^3 +x_2^4+ x_3^6+x_4^2 x_3\\ \phantom{x}}  
        & 1 & 4 & 23 & (\text{Singular}, \text{Singular})
&
   \begin{tikzpicture}[x=.7cm,y=.7cm]
\draw[ligne, black](-2,0)--(3,0);
\draw[ligne, black](0,0)--(0,2);
\node[SUd] at (-2,0) [label=below:{{\scriptsize$2$}}] {};
\node[SUd] at (-1,0) [label=below:{{\scriptsize$4$}}] {};
\node[SUd] at (0,0) [label=below:{{\scriptsize$6$}}] {};
\node[bd] at (1,0) [label=below:{{\scriptsize$4$}}] {};
\node[SUd] at (2,0) [label=below:{{\scriptsize$3$}}] {};
\node[bd] at (3,0) [label=below:{{\scriptsize$1$}}] {};
\node[bd] at (0,1) [label=left:{{\scriptsize$3$}}] {};
\node[flavor] at (0,2) [label=left:{{\scriptsize$1$}}] {};
\end{tikzpicture} \\    \hline
%%---------------------------------%
$ \scriptsize{(2)}$& \matarray{\text{II}_1 \;  (2,5,10,3) \\x_1^2+ x_2^5+ x_3^{10}+x_4^3 x_3 \\ \phantom{x}}
  & 2 & 8 & 46 & (\text{Smooth}, \text{Singular})
&    \begin{tikzpicture}[x=.7cm,y=.7cm]
\draw[ligne, black](-4,0)--(3,0);
\draw[ligne, black](0,0)--(0,1);
%central node:
\node[SUd] at (0,0) [label=below:{{\scriptsize$10$}}] {};
%left:
\node[SUd] at (-1,0) [label=below:{{\scriptsize$8$}}] {};
\node[SUd] at (-2,0) [label=below:{{\scriptsize$6$}}] {};
\node[SUd] at (-3,0) [label=below:{{\scriptsize$4$}}] {};
\node[SUd] at (-4,0) [label=below:{{\scriptsize$2$}}] {};
\node[bd] at (3,0) [label=below:{{\scriptsize$1$}}] {};
%up:
\node[SUd] at (0,1) [label=left:{{\scriptsize$5$}}] {};
%right:
\node[SUd] at (1,0) [label=below:{{\scriptsize$7$}}] {};
\node[SUd] at (2,0) [label=below:{{\scriptsize$4$}}] {};
\node[flavor] at (3,0) [label=below:{{\scriptsize$1$}}] {};
\end{tikzpicture}  \\    \hline
%%---------------------------------%
$ \scriptsize{(3)}$& \matarray{\text{II}_1 \;  (3,5,15,2) \\x_1^3+x_2^5+ x_3^{15}+x_4^2 x_3 \\ \phantom{x}}
 & 4 & 8 & 68 & (\text{Smooth}, \text{Singular})
&
%%%%%%%%
%%%%%%%%
 \begin{tikzpicture}[x=.7cm,y=.7cm]
\draw[ligne, black](-4,0)--(2,0);
\draw[ligne, black](0,0)--(0,2);
%central node:
\node[SUd] at (0,0) [label=below:{{\scriptsize$15$}}] {};
%left:
\node[SUd] at (-1,0) [label=below:{{\scriptsize$12$}}] {};
\node[SUd] at (-2,0) [label=below:{{\scriptsize$9$}}] {};
\node[SUd] at (-3,0) [label=below:{{\scriptsize$6$}}] {};
\node[SUd] at (-4,0) [label=below:{{\scriptsize$3$}}] {};
%up:
\node[SUd] at (0,1) [label=left:{{\scriptsize$8$}}] {};
\node[flavor] at (0,2) [label=left:{{\scriptsize$1$}}] {};
%right:
\node[SUd] at (1,0) [label=below:{{\scriptsize$10$}}] {};
\node[SUd] at (2,0) [label=below:{{\scriptsize$5$}}] {};
\end{tikzpicture} 
%%%%%%%%
%%%%%%%%
    \\    \hline  
%%---------------------------------%
 \end{array}
$$
\caption{Basic data and $\EQfour$ for the first three `sporadic' $SU(N)$ trinions. They all have $\Delta \CA_r =0$, $b_3=0$ and trivial one-form symmetry.  The $\EQfour$ that only have $SU(N)$ gauge groups (yellow nodes) correspond to Lagrangian $\CN=2$ SCFTs in 4d. The (smooth/singular, smooth/singular) indicates whether the Calabi-Yau threefold is singular and there are still singular divisors in the resolution, respectively. 
 \label{tab:trinion series 1}}
\end{table}

%%%%%%%%%%%%%%%%%%%%%%%
\begin{table}
$$
\begin{array}{|c|c|c|c|c|c|c|}
\hline
$ \scriptsize{(\#)}$& \matarray{ \text{Type} \;   (a,b,c,d)  \\  F(x_1, x_2, x_3, x_4) } &\, r\, &  f&d_H &  \text{Smooth CY/Divs} & \EQfour \cr \hline \hline
%%---------------------------------%
 $\scriptsize{(4)}$& \matarray{\text{VIII}_2 \;  (3,5,2,3) \\ x_1^3+ x_2^5+x_3^2 x_2+x_4^3 x_2 \\ +x_1 x_3 x_4}
   & 1 & 3 & 18 & (\text{Singular}, \text{Singular})
   & \begin{tikzpicture}[x=.7cm,y=.7cm]
\draw[ligne, black](-2,0)--(3,0);
\draw[ligne, black](0,0)--(0,2);
%central node:
\node[SUd] at (0,0) [label=below:{{\scriptsize$5$}}] {};
%left:
\node[bd] at (-1,0) [label=below:{{\scriptsize$3$}}] {};
\node[bd] at (-2,0) [label=below:{{\scriptsize$1$}}] {};
%up:
\node[SUd] at (0,1) [label=left:{{\scriptsize$3$}}] {};
\node[flavor] at (0,2) [label=left:{{\scriptsize$1$}}] {};
%right:
\node[bd] at (1,0) [label=below:{{\scriptsize$3$}}] {};
\node[bd] at (2,0) [label=below:{{\scriptsize$2$}}] {};
\node[flavor] at (3,0) [label=below:{{\scriptsize$1$}}] {};
\end{tikzpicture}  \\ \hline
   %%---------------------------------%     
 $\scriptsize{(5)}$&\matarray{\text{VIII}_1 \;  (2,7,3,4) \\ x_1^2+x_2^7+x_3^3 x_2+x_4^4 x_2\\+x_3^2 x_4^2}
   & 1 & 5 & 30 & (\text{Singular}, \text{Singular})
   & \begin{tikzpicture}[x=.7cm,y=.7cm]
\draw[ligne, black](-3,0)--(4,0);
\draw[ligne, black](0,0)--(0,1);
%central node:
\node[SUd] at (0,0) [label=below:{{\scriptsize$7$}}] {};
%left:
\node[SUd] at (-1,0) [label=below:{{\scriptsize$5$}}] {};
\node[SUd] at (-2,0) [label=below:{{\scriptsize$3$}}] {};
\node[flavor] at (-3,0) [label=below:{{\scriptsize$1$}}] {};
%up:
\node[bd] at (0,1) [label=left:{{\scriptsize$3$}}] {};
%right:
\node[bd] at (1,0) [label=below:{{\scriptsize$5$}}] {};
\node[SUd] at (2,0) [label=below:{{\scriptsize$4$}}] {};
\node[bd] at (3,0) [label=below:{{\scriptsize$2$}}] {};
\node[flavor] at (4,0) [label=below:{{\scriptsize$1$}}] {};
\end{tikzpicture} \\ \hline
   %%---------------------------------%     
 $\scriptsize{(6)}$&\matarray{\text{VIII}_2 \; (3,9,2,4) \\ x_1^3+x_2^9+x_3^2 x_2+x_4^4 x_2\\ +x_1 x_4^3}
 & 3 & 8 & 39 & (\text{Smooth}, \text{Smooth})
 & \begin{tikzpicture}[x=.7cm,y=.7cm]
\draw[ligne, black](-2,0)--(4,0);
\draw[ligne, black](0,0)--(0,2);
%central node:
\node[SUd] at (0,0) [label=below:{{\scriptsize$9$}}] {};
%left:
\node[SUd] at (-1,0) [label=below:{{\scriptsize$6$}}] {};
\node[SUd] at (-2,0) [label=below:{{\scriptsize$3$}}] {};
%up:
\node[SUd] at (0,1) [label=left:{{\scriptsize$5$}}] {};
\node[flavor] at (0,2) [label=left:{{\scriptsize$1$}}] {};
%right:
\node[SUd] at (1,0) [label=below:{{\scriptsize$7$}}] {};
\node[SUd] at (2,0) [label=below:{{\scriptsize$5$}}] {};
\node[SUd] at (3,0) [label=below:{{\scriptsize$3$}}] {};
\node[flavor] at (4,0) [label=below:{{\scriptsize$1$}}] {};
\end{tikzpicture}   \\ \hline
 %%---------------------------------%     
  $\scriptsize{(7)}$&\matarray{\text{VIII}_2 \;  (4,10,2,3) \\ x_1^4+ x_2^{10}+x_3^2 x_2+x_4^3 x_2 \\+x_1 x_3
   x_4 }
   & 3 & 5 & 41&  (\text{Singular}, \text{Singular})
 & \begin{tikzpicture}[x=.7cm,y=.7cm]
\draw[ligne, black](-3,0)--(3,0);
\draw[ligne, black](0,0)--(0,2);
%central node:
\node[SUd] at (0,0) [label=below:{{\scriptsize$10$}}] {};
%left:
\node[bd] at (-1,0) [label=below:{{\scriptsize$7$}}] {};
\node[SUd] at (-2,0) [label=below:{{\scriptsize$5$}}] {};
\node[bd] at (-3,0) [label=below:{{\scriptsize$2$}}] {};
%up:
\node[bd] at (0,1) [label=left:{{\scriptsize$5$}}] {};
\node[flavor] at (0,2) [label=left:{{\scriptsize$1$}}] {};
%right:
\node[SUd] at (1,0) [label=below:{{\scriptsize$7$}}] {};
\node[SUd] at (2,0) [label=below:{{\scriptsize$4$}}] {};
\node[flavor] at (3,0) [label=below:{{\scriptsize$1$}}] {};
\end{tikzpicture}  \\ \hline
 \end{array}
$$
\caption{Basic data and $\EQfour$ for the `sporadic' $SU(N)$ trinions, continued (2/4). \label{tab:trinion series 2}}
\end{table}

%%%%%%%%%%%%%%%%%%%%%%%
\begin{table}
$$
\begin{array}{|c|c|c|c|c|c|c|}
\hline
$ \scriptsize{(\#)}$& \matarray{ \text{Type} \;   (a,b,c,d)  \\  F(x_1, x_2, x_3, x_4) } &\, r\, &  f&d_H &  \text{Smooth CY/Divs}& \EQfour \cr \hline \hline
      %%---------------------------------%     
 $\scriptsize{(8)}$&\matarray{\text{VIII}_1 \; (2,16,3,5) \\ x_1^2+x_2^{16}+x_3^3 x_2 +x_4^5 x_2\\+x_3^2
   x_4^2 \\ \phantom{x}}
   & 4 & 9 & 76 &  (\text{Smooth}, \text{Singular})
   & \begin{tikzpicture}[x=.7cm,y=.7cm]
\draw[ligne, black](-3,0)--(5,0);
\draw[ligne, black](0,0)--(0,1);
%central node:
\node[SUd] at (0,0) [label=below:{{\scriptsize$16$}}] {};
%left:
\node[SUd] at (-1,0) [label=below:{{\scriptsize$11$}}] {};
\node[SUd] at (-2,0) [label=below:{{\scriptsize$6$}}] {};
\node[flavor] at (-3,0) [label=below:{{\scriptsize$1$}}] {};
%up:
\node[SUd] at (0,1) [label=left:{{\scriptsize$8$}}] {};
%right:
\node[SUd] at (1,0) [label=below:{{\scriptsize$13$}}] {};
\node[SUd] at (2,0) [label=below:{{\scriptsize$10$}}] {};
\node[SUd] at (3,0) [label=below:{{\scriptsize$7$}}] {};
\node[SUd] at (4,0) [label=below:{{\scriptsize$4$}}] {};
\node[flavor] at (5,0) [label=below:{{\scriptsize$1$}}] {};
\end{tikzpicture}  \\ \hline 
   %%---------------------------------%     
 $\scriptsize{(9)}$&\matarray{\text{VIII}_2 \; (3,21,2,5) \\ x_1^3+ x_2^{21}+x_3^2 x_2+x_4^5 x_2\\+x_1 x_3 x_4\\ \phantom{x}}
   & 7 & 9 & 98 & (\text{Smooth}, \text{Singular})
      & \begin{tikzpicture}[x=.7cm,y=.7cm]
\draw[ligne, black](-2,0)--(5,0);
\draw[ligne, black](0,0)--(0,2);
%central node:
\node[SUd] at (0,0) [label=below:{{\scriptsize$21$}}] {};
%left:
\node[SUd] at (-1,0) [label=below:{{\scriptsize$14$}}] {};
\node[SUd] at (-2,0) [label=below:{{\scriptsize$7$}}] {};
%up:
\node[SUd] at (0,1) [label=left:{{\scriptsize$11$}}] {};
\node[flavor] at (0,2) [label=left:{{\scriptsize$1$}}] {};
%right:
\node[SUd] at (1,0) [label=below:{{\scriptsize$17$}}] {};
\node[SUd] at (2,0) [label=below:{{\scriptsize$13$}}] {};
\node[SUd] at (3,0) [label=below:{{\scriptsize$9$}}] {};
\node[SUd] at (4,0) [label=below:{{\scriptsize$5$}}] {};
\node[flavor] at (5,0) [label=below:{{\scriptsize$1$}}] {};
\end{tikzpicture}  \\ \hline
  $\scriptsize{(10)}$&\matarray{\text{VIII}_2 \; (5, 25, 2, 3) \\ x_1^5+x_2^{25}+x_3^2 x_2+x_4^3 x_2 \\+x_1 x_3 x_4\\ \phantom{x} }
   & 10 & 9 & 115 & (\text{Smooth}, \text{Singular})
      & \begin{tikzpicture}[x=.7cm,y=.7cm]
\draw[ligne, black](-4,0)--(3,0);
\draw[ligne, black](0,0)--(0,2);
%central node:
\node[SUd] at (0,0) [label=below:{{\scriptsize$25$}}] {};
%left:
\node[SUd] at (-1,0) [label=below:{{\scriptsize$20$}}] {};
\node[SUd] at (-2,0) [label=below:{{\scriptsize$15$}}] {};
\node[SUd] at (-3,0) [label=below:{{\scriptsize$10$}}] {};
\node[SUd] at (-4,0) [label=below:{{\scriptsize$5$}}] {};
%up:
\node[SUd] at (0,1) [label=left:{{\scriptsize$13$}}] {};
\node[flavor] at (0,2) [label=left:{{\scriptsize$1$}}] {};
%right:
\node[SUd] at (1,0) [label=below:{{\scriptsize$17$}}] {};
\node[SUd] at (2,0) [label=below:{{\scriptsize$9$}}] {};
\node[flavor] at (3,0) [label=below:{{\scriptsize$1$}}] {};
\end{tikzpicture}  \\ \hline
  \end{array}
$$
\caption{Basic data and $\EQfour$ for the `sporadic' $SU(N)$ trinions, continued (3/4). \label{tab:trinion series 3}}
\end{table}

 %%%%%%%%%%%%%%%%%%%%%%%
\begin{table}
$$
\begin{array}{|c|c|c|c|c|c|c|}
\hline
$ \scriptsize{(\#)}$& \matarray{ \text{Type} \;   (a,b,c,d)  \\  F(x_1, x_2, x_3, x_4) } &\, r\, &  f&d_H &  \text{Smooth CY/Divs}& \EQfour \cr \hline \hline
     %%---------------------------------%
 $\scriptsize{(11)}$&\matarray{\text{XIV}_3 \; (7, 2, 3, 3) \\x_1^7+x_2^2 x_1+x_3^3 x_1+x_4^3 x_1\\+x_2 x_3^2+x_2 x_4^2+x_2 x_3 x_4\\ \phantom{x} } & 3 & 8 & 29 & (\text{Smooth}, \text{Smooth}) 
 &\begin{tikzpicture}[x=.7cm,y=.7cm]
\draw[ligne, black](-3,0)--(3,0);
\draw[ligne, black](0,0)--(0,2);
%central node:
\node[SUd] at (0,0) [label=below:{{\scriptsize$7$}}] {};
%left:
\node[SUd] at (-1,0) [label=below:{{\scriptsize$5$}}] {};
\node[SUd] at (-2,0) [label=below:{{\scriptsize$3$}}] {};
\node[flavor] at (-3,0) [label=below:{{\scriptsize$1$}}] {};
%up:
\node[SUd] at (0,1) [label=left:{{\scriptsize$4$}}] {};
\node[flavor] at (0,2) [label=left:{{\scriptsize$1$}}] {};
%right:
\node[SUd] at (1,0) [label=below:{{\scriptsize$5$}}] {};
\node[SUd] at (2,0) [label=below:{{\scriptsize$3$}}] {};
\node[flavor] at (3,0) [label=below:{{\scriptsize$1$}}] {};
\end{tikzpicture} 
 \\ \hline
 $\scriptsize{(12)}$&\matarray{\text{XIV}_4 \; (13,  2,3, 4) \\ x_1^{13}+x_2^2 x_1+x_3^3 x_1+x_4^4 x_1\\+x_3 x_4^3+2 x_2 x_3 x_4\\  \phantom{x} } & 6 & 9 & 57  & (\text{Smooth}, \text{Smooth}) 
&\begin{tikzpicture}[x=.7cm,y=.7cm]
\draw[ligne, black](-3,0)--(4,0);
\draw[ligne, black](0,0)--(0,2);
%central node:
\node[SUd] at (0,0) [label=below:{{\scriptsize$13$}}] {};
%left:
\node[SUd] at (-1,0) [label=below:{{\scriptsize$9$}}] {};
\node[SUd] at (-2,0) [label=below:{{\scriptsize$5$}}] {};
\node[flavor] at (-3,0) [label=below:{{\scriptsize$1$}}] {};
%up:
\node[SUd] at (0,1) [label=left:{{\scriptsize$7$}}] {};
\node[flavor] at (0,2) [label=left:{{\scriptsize$1$}}] {};
%right:
\node[SUd] at (1,0) [label=below:{{\scriptsize$10$}}] {};
\node[SUd] at (2,0) [label=below:{{\scriptsize$7$}}] {};
\node[SUd] at (3,0) [label=below:{{\scriptsize$4$}}] {};
\node[flavor] at (4,0) [label=below:{{\scriptsize$1$}}] {};
\end{tikzpicture}  \\
  \hline
   $\scriptsize{(13)}$&\matarray{\text{XIV}_4 \; (31, 2, 3, 5) \\  x_1^{31}+x_2^2 x_1+x_3^3 x_1+x_4^5 x_1\\+3 x_2 x_3 x_4 \\  \phantom{x} } & 15 & 10 & 145  
   & (\text{Smooth}, \text{Singular}) 
&\begin{tikzpicture}[x=.7cm,y=.7cm]
\draw[ligne, black](-3,0)--(5,0);
\draw[ligne, black](0,0)--(0,2);
%central node:
\node[SUd] at (0,0) [label=below:{{\scriptsize$31$}}] {};
%left:
\node[SUd] at (-1,0) [label=below:{{\scriptsize$21$}}] {};
\node[SUd] at (-2,0) [label=below:{{\scriptsize$11$}}] {};
\node[flavor] at (-3,0) [label=below:{{\scriptsize$1$}}] {};
%up:
\node[SUd] at (0,1) [label=left:{{\scriptsize$16$}}] {};
\node[flavor] at (0,2) [label=left:{{\scriptsize$1$}}] {};
%right:
\node[SUd] at (1,0) [label=below:{{\scriptsize$25$}}] {};
\node[SUd] at (2,0) [label=below:{{\scriptsize$19$}}] {};
\node[SUd] at (3,0) [label=below:{{\scriptsize$13$}}] {};
\node[SUd] at (4,0) [label=below:{{\scriptsize$7$}}] {};
\node[flavor] at (5,0) [label=below:{{\scriptsize$1$}}] {};
\end{tikzpicture}  \\
  \hline
 \end{array}
$$
\caption{Basic data and $\EQfour$ for the `sporadic' $SU(N)$ trinions, continued (4/4). \label{tab:trinion series 4}}
\end{table}
%%%%%%%%%%%%%%%%%%%%%%%%%%%%%%%%%%%%%%%%%%%%%%%%
%%%%%%%%%%%%%%%%%%%%%%%%%%%%%%%%%%%%%%%%%%%%%%%%

\subsection{Case $(b_\alpha)= (N,N-1, N-1)$}
Next,  consider the condition
\be
\label{Diophantine-eq3}
{N\ov p_1} +{N-1\ov p_2}+{N-1\ov p_3}  =N~.
\ee
There are two infinite series of solutions
\be\label{series su ii}
(p_1, p_2, p_3)= (N,2,2)\,,\   (2, 2, 2N-2)~, \; \; \forall N~.
\ee
For the first series, one needs to add marginal deformation terms in \eqref{gen trinion sing} to obtain the following IHS of type VIII$_2${$(N,N,2,2)$}
\be
x_1^N+x_2^N+x_1 x_3^2+x_1 x_4^2+x_2 x_3 x_4=0\,.
\ee
In fact, this isolated singularity is isomorphic to a singularity of type IV$(N,2,N,2)$
\be
x_1^N+x_1 x_2^2+x_3^N+x_3 x_4^2=0\,,
\ee
which gives rise to AD$[D_{N+1},D_{N+1}]$ theories \cite{Cecotti:2010fi, Xie:2015rpa}. We list its geometric data in table \ref{tab:trinion series}. For $N=2k-1$, the 5d SCFT is the UV fixed point of the 5d $\CN=1$ gauge theory
\be
SU(k)_{\pm {3\ov 2}}+(2k+1)\bm{F}\,,
\ee
which has flavor symmetry algebra is $\mathfrak{g}_F=\mathfrak{so}(4k+2)\oplus \mathfrak{u}(1)$. For $N=2k$, the theory is the same SCFT coupled to $2k$ free hypermultiplets, which correspond to $2k$ conifold singularities in the  crepant resolution. The $\EQfour$ for these infinite sequences are
\be
\EQfour  = \begin{cases} 
 \begin{tikzpicture}[x=.6cm,y=.6cm]
\draw[ligne, black](-2,0)--(2.4,0);
\draw[ligne, black,dotted](2.6,0)--(3.4,0);
\draw[ligne, black](3.6,0)--(5,0);
\draw[ligne, black](0,0)--(0,2);
%central node:
\node[SUd] at (0,0) [label=below:{{\scriptsize$N$}}] {};
%left:
\node[SUd] at (-1,0) [label=above:{{\scriptsize$k$}}] {};
\node[flavor] at (-2,0) [label=below:{{\scriptsize$1$}}] {};
%up:
\node[SUd] at (0,1) [label=left:{{\scriptsize$k$}}] {};
\node[flavor] at (0,2) [label=above:{{\scriptsize$1$}}] {};
%right:
\node[SUd] at (1,0) [label=above:{{\scriptsize$2k{-}2$}}] {};
\node[SUd] at (2,0) [label=below:{{\scriptsize$2k{-}3$}}] {};
\node[SUd] at (4,0) [label=above:{{\scriptsize$2$}}] {};
\node[flavor] at (5,0) [label=below:{{\scriptsize$1$}}] {};
\end{tikzpicture}  & \; \; \text{if}\;\; N= 2k{-}1~,\\
%%%%%%------------%%%%%
 \begin{tikzpicture}[x=.6cm,y=.6cm]
\draw[ligne, black](-2,0)--(2.4,0);
\draw[ligne, black,dotted](2.6,0)--(3.4,0);
\draw[ligne, black](3.6,0)--(5,0);
\draw[ligne, black](0,0)--(0,2);
%central node:
\node[SUd] at (0,0) [label=below:{{\scriptsize$N$}}] {};
%left:
\node[bd] at (-1,0) [label=above:{{\scriptsize$k$}}] {};
\node[flavor] at (-2,0) [label=below:{{\scriptsize$1$}}] {};
%up:
\node[bd] at (0,1) [label=left:{{\scriptsize$k$}}] {};
\node[flavor] at (0,2) [label=above:{{\scriptsize$1$}}] {};
%right:
\node[SUd] at (1,0) [label=above:{{\scriptsize$2k{-}1$}}] {};
\node[SUd] at (2,0) [label=below:{{\scriptsize$2k{-}2$}}] {};
\node[SUd] at (4,0) [label=above:{{\scriptsize$2$}}] {};
\node[flavor] at (5,0) [label=below:{{\scriptsize$1$}}] {};
\end{tikzpicture}  & \; \;   \text{if} \;\; N= 2k~,

\end{cases}
\ee.
The $\MQfive$ is obtained by replacing all the $SU(n)$ nodes in $\EQfour$ by $U(n)$ nodes. The quiver for $N=2k$ is ugly, and can be dualized to extract the free sector, so that we obtain
\be\label{R2N5d}
\MQfive  = \begin{cases} 

 \begin{tikzpicture}[x=.6cm,y=.6cm]
\draw[ligne, black](-2,0)--(2.4,0);
\draw[ligne, black,dotted](2.6,0)--(3.4,0);
\draw[ligne, black](3.6,0)--(5,0);
\draw[ligne, black](0,0)--(0,2);
%central node:
\node[bd] at (0,0) [label=below:{{\scriptsize$2k{-}1$}}] {};
%left:
\node[bd] at (-1,0) [label=above:{{\scriptsize$k$}}] {};
\node[bd] at (-2,0) [label=below:{{\scriptsize$1$}}] {};
%up:
\node[bd] at (0,1) [label=left:{{\scriptsize$k$}}] {};
\node[bd] at (0,2) [label=above:{{\scriptsize$1$}}] {};
%right:
\node[bd] at (1,0) [label=above:{{\scriptsize$2k{-}2$}}] {};
\node[bd] at (2,0) [label=below:{{\scriptsize$2k{-}3$}}] {};
\node[bd] at (4,0) [label=above:{{\scriptsize$2$}}] {};
\node[bd] at (5,0) [label=below:{{\scriptsize$1$}}] {};
\end{tikzpicture}  
& \qquad  \text{if}\quad N= 2k{-}1~,\\

%%%%%%------------%%%%%
  \begin{tikzpicture}[x=.6cm,y=.6cm]
\draw[ligne, black](-2,0)--(2.4,0);
\draw[ligne, black,dotted](2.6,0)--(3.4,0);
\draw[ligne, black](3.6,0)--(5,0);
\draw[ligne, black](0,0)--(0,2);
%central node:
\node[bd] at (0,0) [label=below:{{\scriptsize$2k{-}1$}}] {};
%left:
\node[bd] at (-1,0) [label=above:{{\scriptsize$k$}}] {};
\node[bd] at (-2,0) [label=below:{{\scriptsize$1$}}] {};
%up:
\node[bd] at (0,1) [label=left:{{\scriptsize$k$}}] {};
\node[bd] at (0,2) [label=above:{{\scriptsize$1$}}] {};
%right:
\node[bd] at (1,0) [label=above:{{\scriptsize$2k{-}2$}}] {};
\node[bd] at (2,0) [label=below:{{\scriptsize$2k{-}3$}}] {};
\node[bd] at (4,0) [label=above:{{\scriptsize$2$}}] {};
\node[bd] at (5,0) [label=below:{{\scriptsize$1$}}] {};
\node[] at (7,0) [] {$\, \oplus \quad  \mb{H}^{2k}  $} ;
\end{tikzpicture} 
 & \qquad \text{if} \quad  N= 2k~,
\end{cases}
\ee
One can check that, for $N=2k-1$, the magnetic quiver is indeed the ones of $SU(k)_{\pm {3\ov 2}}+(2k+1)\bm{F}$~\cite{Bourget:2019aer}. This gives a nice consistency check on our geometric approach. 
The second series in \eqref{series su ii} is equivalent to \eqref{series su i}, {\it i.e.} we have the equivalences
\be
\text{VIII}\, (2, 2, N, 2N-2) \; \cong \; \text{II}\,(2, 2N, N, 2)~.  
\ee
In addition, we find the sporadic solutions
\be
\begin{tabular}{ c| c |c c  c} % centered columns (4 columns)
\scriptsize{(\#)}& $N$ & $p_1$ & $p_2$ & $p_3$ \\%heading
 \hline % inserts single horizontal line
\scriptsize{(4)}&5 & 3  &  2 &   3  \\
\scriptsize{(5)}&7 &  2 &  3 &    4 \\
\scriptsize{(6)}&9 &  3 &  2 &     4\\
\scriptsize{(7)}&10  & 4  & 2  &    3 \\
\scriptsize{(8)}&16 & 2  & 3  & 5   \\
\scriptsize{(9)}&21 &  3 &  2 &    5 \\
\scriptsize{(10)}&25 &  5 &  2 &    3 
\end{tabular}
\ee
Here, we omitted the solution $N=4$, $(p_\alpha)=(2,3,3)$ because it is equivalent to the model I$(2,4,4,4)$, the rank-1 $E_7$ theory.

\medskip
\noindent
For the model  $\scriptstyle{(4)}$, we deduce from the resolution geometry that it is a rank-1 $E_6$ theory coupled to a rank-0 theory $E_{8}^{(-4)}$ plus another conifold (a free hypermultiplet). 
From $\EQfour$, after gauging and rebalancing ugly nodes, we indeed get
\bea
\begin{tikzpicture}[x=.7cm,y=.7cm]
\draw[ligne, black](-2,0)--(2,0);
\draw[ligne, black](0,0)--(0,2);
%central node:
\node[] at (-4,1) [] {$\MQfive=$};
\node[bd] at (0,0) [label=below:{{\scriptsize$3$}}] {};
%left:
\node[bd] at (-1,0) [label=below:{{\scriptsize$2$}}] {};
\node[bd] at (-2,0) [label=below:{{\scriptsize$1$}}] {};
%up:
\node[bd] at (0,1) [label=left:{{\scriptsize$2$}}] {};
\node[bd] at (0,2) [label=left:{{\scriptsize$1$}}] {};
%right:
\node[bd] at (1,0) [label=below:{{\scriptsize$2$}}] {};
\node[bd] at (2,0) [label=below:{{\scriptsize$1$}}] {};
\node[] at (4,1) [] {$\oplus \quad \mb{H}^7$};
\end{tikzpicture}\,.
\eea

\medskip
\noindent
Similarly, the model $\scriptstyle{(5)}$  is a rank-1 $E_7$ theory coupled to a rank-0 theory $E_{8}^{(-3)}$ and another conifold. From gauging and rebalancing $\EQfour$, we get
\bea
\begin{tikzpicture}[x=.7cm,y=.7cm]
\draw[ligne, black](-3,0)--(3,0);
\draw[ligne, black](0,0)--(0,1);
\node[] at (-5,1) [] {$\MQfive=$};
%central node:
\node[bd] at (0,0) [label=below:{{\scriptsize$4$}}] {};
%left:
\node[bd] at (-1,0) [label=below:{{\scriptsize$3$}}] {};
\node[bd] at (-2,0) [label=below:{{\scriptsize$2$}}] {};
\node[bd] at (-3,0) [label=below:{{\scriptsize$1$}}] {};
%up:
\node[bd] at (0,1) [label=left:{{\scriptsize$2$}}] {};
%right:
\node[bd] at (1,0) [label=below:{{\scriptsize$3$}}] {};
\node[bd] at (2,0) [label=below:{{\scriptsize$2$}}] {};
\node[bd] at (3,0) [label=below:{{\scriptsize$1$}}] {};
\node[] at (5,1) [] {$\oplus \quad \mb{H}^{13}$};
\end{tikzpicture}\,.
\eea

\medskip
\noindent
The model $\scriptstyle{(6)}$ has a smooth resolution, and correspondingly $\FTfour$ is a Lagrangian theory, with the  $SU(n)$ quiver shown in table~\ref{tab:trinion series 3}. The corresponding 5d SCFT $\FT$ has $r=3$ and $d_H=39$. The structure of its Higgs branch can be studied using the magnetic quiver
\bea
\begin{tikzpicture}[x=.7cm,y=.7cm]
\draw[ligne, black](-2,0)--(4,0);
\draw[ligne, black](0,0)--(0,2);
%central node:
\node[bd] at (0,0) [label=below:{{\scriptsize$9$}}] {};
%left:
\node[bd] at (-1,0) [label=below:{{\scriptsize$6$}}] {};
\node[bd] at (-2,0) [label=below:{{\scriptsize$3$}}] {};
%up:
\node[bd] at (0,1) [label=left:{{\scriptsize$5$}}] {};
\node[bd] at (0,2) [label=left:{{\scriptsize$1$}}] {};
%right:
\node[bd] at (1,0) [label=below:{{\scriptsize$7$}}] {};
\node[bd] at (2,0) [label=below:{{\scriptsize$5$}}] {};
\node[bd] at (3,0) [label=below:{{\scriptsize$3$}}] {};
\node[bd] at (4,0) [label=below:{{\scriptsize$1$}}] {};
\end{tikzpicture} 
\eea
In particular, the 5d flavor symmetry is $\mathfrak{g}_F=\mathfrak{e}_7\oplus \mathfrak{u}(1)$.  

\medskip
\noindent
The model $\scriptstyle{(7)}$ has a crepant resolutions with a residual terminal singularity  $E_8^{(-3)}$.  After gauging the $U(1)^f$ of $\EQfour$ and rebalancing, we get
\bea
\begin{tikzpicture}[x=.7cm,y=.7cm]
\draw[ligne, black](-3,0)--(3,0);
\draw[ligne, black](0,0)--(0,2);
\node[] at (-5,1) [] {$\MQfive=$};
%central node:
\node[bd] at (0,0) [label=below:{{\scriptsize$7$}}] {};
%left:
\node[bd] at (-1,0) [label=below:{{\scriptsize$5$}}] {};
\node[bd] at (-2,0) [label=below:{{\scriptsize$3$}}] {};
\node[bd] at (-3,0) [label=below:{{\scriptsize$1$}}] {};
%up:
\node[bd] at (0,1) [label=left:{{\scriptsize$4$}}] {};
\node[bd] at (0,2) [label=left:{{\scriptsize$1$}}] {};
%right:
\node[bd] at (1,0) [label=below:{{\scriptsize$5$}}] {};
\node[bd] at (2,0) [label=below:{{\scriptsize$3$}}] {};
\node[bd] at (3,0) [label=below:{{\scriptsize$1$}}] {};
\node[] at (5,1) [] {$\oplus\quad \mb{H}^{12}$};
\end{tikzpicture} \,.
\eea
Note that the magnetic quiver of the interacting sector is exactly the same as the model  $\scriptstyle{(11)}$ in table~\ref{tab:trinion series 4}, which corresponds to a $r=3$, $f=8$ 5d SCFT. Thus, we conclude that the 5d SCFT for model $\scriptstyle{(7)}$ is equivalent to the  one of model $\scriptstyle{(11)}$ plus 12 hypermultiplets. 

\medskip
\noindent
The model $\scriptstyle{(8)}$, $\scriptstyle{(9)}$ and $\scriptstyle{(10)}$ all have smooth crepant resolutions, and correspondingly $\FTfour$ is a Lagrangian theory. From the magnetic quivers, we can show that they each have 5d global symmetry   $\mathfrak{g}_F=\mathfrak{e}_8 \oplus \mathfrak{su}(2)$.

\subsection{Case $(b_\alpha)= (N-1,N-1, N-1)$}
Lastly, consider the condition
\be
\label{Diophantine-eq4}
{N-1\ov p_1} +{N-1\ov p_2}+{N-1\ov p_3}  =N~.
\ee
For any $N$, we have the solution $(p_\alpha)=(2,2, N-1)$, but it turns out this is equivalent to the second series in \eqref{series su ii}, namely
\be
\text{XIV}_2\, (N, 2,2,N-1) \; \cong \; \text{VIII}_2\,(N, N, 2, 2) \; \cong \; \text{IV}\,(N, 2, N, 2)~.  
\ee
We also find the sporadic solutions
\be
\begin{tabular}{ c|c |c c  c} % centered columns (4 columns)
\scriptsize{(\#)}& $N$ & $p_1$ & $p_2$ & $p_3$ \\%heading
 \hline % inserts single horizontal line
 \scriptsize{(11)}&7 & 2  & 3   & 3    \\
 \scriptsize{(12)}&13  & 2  & 3   & 4    \\
 \scriptsize{(13)}&  31  & 2  & 3   & 5    
\end{tabular}
\ee
These three models have smooth resolutions and $\FTfour$ is Lagrangian. The 5d SCFT for model  $\scriptstyle{(11)}$ has $r=3$ and $\mathfrak{g}_F=\mathfrak{e}_6 \oplus \mathfrak{su}(2)$. The model  $\scriptstyle{(12)}$ has rank $r=6$ and $\mathfrak{g}_F=\mathfrak{e}_7 \oplus \mathfrak{su}(2)$. Finally, the model $\scriptstyle{(13)}$ has $r=15$ and $\mathfrak{g}_F \supset \mathfrak{e}_8$. 
 In appendix~\ref{app:Diophantine}, we prove that there are no more solutions to the equations \eqref{Diophantine-eq2}, \eqref{Diophantine-eq3} and \eqref{Diophantine-eq4}. This completes our classification.

%%%%%%%%%%%%%%%%%%%%%%%%%%%%%%%%
%%%%%%%%%%%%%%%%%%%%%%%%%%%%%%%%

\section{$D_p^b(G)$-Trinions SCFTs for $G = D, E$}
\label{sec7:DandE}
In this section, we generalize the four-dimensional trinion SCFTs to the cases with a central gauge group of type $D$ or $E$. We focus on the $D_p^b(G)$ trinions that can be engineered by threefold singularities in Type IIB. This leads us to discover a number of new S-dualities between SCFTs with different gauge groups, which we discuss in section~\ref{sec:dualities}. 

\subsection{Trinion Geometries for Generic $G$}

The analysis of section \ref{secquiverinesSU} can be generalized to the case of gaugings of three $D_p^b(G)$ theories with $G$ of special orthogonal or exceptional type. In  the general case, however,  we just have a Landau-Ginzburg description instead of a threefold singularity \cite{Cecotti:2010fi, DelZotto:2015rca}. The LG superpotential reads 
\be\label{superpot}
 \mathcal{W}= w^2+f(t,x)+\sum_{\alpha=1}^3 z_\alpha^{p_\alpha} f_{b_\alpha}(x,t)~,
\ee 
where $ W_{G}(w,x,t)\equiv w^2+f(t,x)=0$ is the ADE singularity of type $G$ and $f_{b_\alpha}(x,t)z_\alpha^{p_\alpha}$ denotes the $z$-dependent part of the threefold singularity describing $D_{p_\alpha}^{b_\alpha}(G)$,  shown in table~\ref{sing}.  
From a LG model point of view,  $w$ in \eqref{superpot} is a massive field and can be integrated out, leaving five fields. The conformality condition reads 
\be\label{beta0} 
\frac{b_1}{p_1}+\frac{b_2}{p_2}+\frac{b_3}{p_3}=h^{\vee}(G)~.
\ee 
Importantly, if the singularity $\CW=0$ is non-isolated, one needs to add marginal terms to \eqref{superpot} in order to have an isolated singularity at the origin, so that the LG model is well-defined.

We will consider the special case when the LG model is equivalent to a model with at most four fields. This happens whenever at least one of the fields $z_\alpha$ is massive. For $p=2$ or $p=1$ and an appropriate choice of $b$, we have:
\be\label{mass terms G}
\CW \; \supset z^{p} f_{b}(x,t) =   z^2 \qquad \text{or}\qquad   \CW \; \supset z^{p} f_{b}(x,t) = t z\, ,\;  x z \,.
\ee
Such massive fields can be integrated out.  If we choose $p=2$ for one of the legs, we are therefore considering a theory $\FTfour$ of the form
\be
 \begin{tikzpicture}[x=.6cm,y=.6cm]
\node at (-3,0) {$D_{p_2}^{b_2} (G)$};
\node at (0,0) {$G$};
\node at (3,0) {$D_{p_3}^{b_3}(G)$};
\node at (0,2) {$D_{2}^{h^\vee} (G)$};
\draw[ligne, black](-1.7,0)--(-.5,0);
\draw[ligne, black](.5,0)--(1.7,0);
\draw[ligne, black](0,0.5)--(0,1.3);
\end{tikzpicture}
\ee
The constraint (\ref{beta0}) now implies that the other two $D_p^b(G)$ sectors should satisfy the relation 
\be\label{D2G cond}
\frac{b_2}{p_2}+\frac{b_3}{p_3}=\frac{h^{\vee}(G)}{2} \,.
\ee 
Since we are now left with only four fields, the system can be understood as a threefold compactification in Type IIB. 
The second option is $p=1$, which gives a mass term \eqref{mass terms G} for $b\neq h^{\vee}(G)$, since in that case the $f_b(x,t)$ function appearing in table \ref{sing} is always linear in either $x$ or $t$, we can simply choose the $D_{p_1}^{b_1}(G)$ theory to be $D_1^{b}(G)$ with $b\neq h^{\vee}(G)$ (we remind the reader that $D_1^b(G)$ is trivial for $b= h^{\vee}(G)$) and, due to \eqref{beta0}, we take the other two  $D_{p}^{b}(G)$ models to satisy the constraint 
\be\label{D1G cond}
\frac{b_2}{p_2}+\frac{b_3}{p_3}=h^{\vee}(G)-b.
\ee 
Under these conditions, we can integrate out both $z_1$ and either $x$ or $t$ (the variable appearing in $f_{b_1}(x,t)$). We can now simply introduce a new massive field $y$ which enters quadratically in the superpotential. The threefold singularity is now given by a hypersurface in the four variables $z_2$, $ z_3$, $y$, and $x$ or $t$ (the variable we have not integrated out). 
We can therefore also consider the family of trinions
\be
 \begin{tikzpicture}[x=.6cm,y=.6cm]
\node at (-3,0) {$D_{p_2}^{b_2} (G)$};
\node at (0,0) {$G$};
\node at (3,0) {$D_{p_3}^{b_3}(G)$};
\node at (0,2) {$D_{1}^{b} (G)$};
\draw[ligne, black](-1.7,0)--(-.5,0);
\draw[ligne, black](.5,0)--(1.7,0);
\draw[ligne, black](0,0.5)--(0,1.3);
\end{tikzpicture}
\ee
satisfying the condition \eqref{D1G cond}.
 In the following, we list all such models by solving equations \eqref{D2G cond} and \eqref{D1G cond} and integrating out the massive LG fields.   We collect the resulting theefold singularities and their basic properties in various tables in appendix~\ref{App:Tables}.

%%%%%%%%%%%%%%%%%%%%%%%%%
\subsection{$D_p^b(G)$-Trinions with $b_1= h^\vee$ and $p_1=2$}
\label{sec:DpG-1}

\subsubsection{$b_2=b_3= h^\vee$}
\label{subsec:b2b3h}
If we choose $b_1=b_2= h^\vee$, we have the universal solutions:
\be
(p_2, p_3) = (4,4), (3,6)~,
\ee
for any $G$.  They are the only solutions, in this case. For $G=SU(N)$, these are the $E_7^{(N-4)}$ and $E_8^{(N-6)}$ series. For $SO(2N)$, the IHS equation reads
\be
x^{N-1}+x t^2+z_2^{p_2}+z_3^{p_3}=0\,.
\ee
The basic properties of these models, for $G=SO(2N)$ and $G=E_n$, are listed in table \ref{tab:b2b3h}.

\subsubsection{Other Sporadic Models}

For $G=SO(2N)$ (so that $h^\vee = 2N-2$), we also have the sporadic models in table \ref{tab:SporadicSO}.
For $G=E_6$, we have $h^\vee=12$ and the allowed values of $b$ are $12,9,8$. The sporadic models are in table \ref{tab:sporadicE6}.
For $G=E_7$, we have $h^\vee=18$ and the allowed values of $b$ are $18$ and $14$. We then find the sporadic models in table \ref{tab:sporadicE7}.
For $G=E_8$, we have $h^\vee=30$ and the allowed values of $b$ are $30,24,20$, with the sporadic models in table \ref{tab:sporadicE8}.

\subsection{$D_p^b(G)$-Trinions with $b_1\neq h^\vee$ and $p_1=1$}
\label{sec:DpG-2}

Let us also consider models with $b_1\neq h^\vee$ and $p_1=1$. 
For $G=SO(2N)$, we have to assign $b_1=N$, and we have the  models when $b_2=b_3=2N-2$ shown in table \ref{tab:Bounty}. 
If $b_2=2N-2$, $b_3=N$, then one needs to take 
\be
\mc{W}=x^{N-1}+x t^2+z_1 t+z_2^{p_2}+z_3^{p_3}t\,,
\ee
add marginal terms to make it isolated, and then integrate out $z_1$ and $t$ (a similar procedure should be applied to the $b_2=b_3=N$ cases as well). We list the resulting singular equations and the data of the corresponding theories in table  \ref{tab:Bounty}. Note that many of these models are equivalent to $E_k^{(n)}$ theories or to other models introduced above.

For $G=E_6$, we can choose either $b_1=9$ or 8, and the models are shown in table \ref{tab:Twix}. 
Likewise, for $G=E_7$ we need to set $b_1=14$, and the corresponding models can be found in table \ref{tab:Snickers}. Finally, 
for $G=E_8$, we can choose either $b_1=24$ or 20, as shown in table \ref{tab:Ferrero}.

\subsection{Matching the CB Spectrum to the Spectrum of the Singularity}
We can provide detailed checks of the above results by comparing the 4d CB spectrum obtained from the threefold singularity in Type IIB \cite{Xie:2017pfl, Closset:2020scj} to the CB spectrum of the individual $D_p^b(G)$ legs. 

Let us illustrate this by presenting one example in more detail. Let us choose the first model in table \ref{tab:Twix}. The trinion takes the form
\bea\label{E6 trinion expl}
\begin{tikzpicture}[x=.6cm,y=.6cm]
\node at (-2.3,0) {$D_5$};
\node at (0,0) {$E_6$};
\node at (2.3,0) {$D_{20}$};
\node at (0,2) {MN$_6$};
\draw[ligne, black](-1.6,0)--(-0.6,0);
\draw[ligne, black](0.6,0)--(1.6,0);
\draw[ligne, black](0,0.5)--(0,1.3);
\end{tikzpicture}\,.
\eea
Here we used the fact that $D_1^9(E_6)$ is the $E_6$ MN theory. The three legs have the CB spectrum
\bea
&D_1^9(E_6)\, : &&  \Delta= \{3\}~,\\
&D_5^{12}(E_6) \,:  &&\Delta= 
 \left\{\frac{6}{5},\frac{9}{5},\frac{12}{5},\frac{13}{5},\frac{16}{5},  \frac{18}{5},\frac{21}{5},\frac{24}{5},\frac{28}{5},\frac{33}{5},\frac{36}{5},\frac{48}{5} \right\}~,  \\
   &D_{20}^{12}(E_6)\,: &&\Delta= \left\{ \frac{6}{5},\frac{6}{5},\frac{6}{5},\frac{7}{5},\frac{7}{5},\frac{7}{5},\frac{9}{5},\frac{9}{5},\frac{9}{5},2,2,\frac{12}{5},\frac{12}{5},\frac{12}{5},\frac{13}{5},\frac{13}{5},3,3,3, \frac{16}{5},\frac{16}{5}, \right.\\
         &&&\qquad  \frac{18}{5},\frac{18}{5}, \frac{18}{5}, \frac{19}{5},\frac{19}{5},\frac{21}{5},\frac{21}{5},\frac{21}{5},\frac{22}{5},\frac{22}{5},\frac{24}{5},\frac{24}{5},\frac{24}{5},5, \frac{27}{5},\frac{27}{5},\frac{27}{5},\frac{28}{5},\\
         &&&\qquad  \left. 6,6, \frac{31}{5},\frac{33}{5},\frac{33}{5},\frac{34}{5},\frac{36}{5},\frac{36}{5},\frac{37}{5},\frac{39}{5},\frac{39}{5},\frac{42}{5},\frac{42}{5},9,\frac{48}{5},\frac{51}{5},\frac{54}{5},\frac{57}{5}\right\}~,
\eea
the union of these spectra, plus the  $E_6$ Casimirs, which give $\Delta= \{ 2, 5, 6, 8, 9, 12\}$, result in the spectrum of the $D_p^b(E_6)$ trinion \eqref{E6 trinion expl}, which is a theory of rank $\h r= 76$. On the other hand, the LG description of the trinion is
\be
\CW= x^3 + t^4 + t z_1 +  z_2^5 +z_3^{20}~.
\ee 
Integrating out the massive fields $t$ and $z_1$, we obtain the threefold singularity
\be
y^2+ x^3 +  z_2^5 +z_3^{20}=0~.
\ee
Its singularity spectrum precisely reproduce the CB spectrum spelled out above.

%%%%%%%%%%%%%%%%%%%%%%%%%%%
%%%%%%%%%%%%%%%%%%%%%%%%%%%

\section{S-Dualities between 4d $D_p^b(G)$-Trinions}
\label{sec:dualities}

\subsection{S-Dualities from Geometric Equivalence}

By comparing the results of sections \ref{sec:SUN} and \ref{sec7:DandE}, we observe that, in a number of examples, the same  singularity simultaneously describes more than one gauging of three $D_p^b(G)$ theories, sometimes involving different gauge groups $G$. The reason for this is mundane from the 2d LG perspective. By the 2d/4d correspondence \cite{Cecotti:2010fi}, the LG superpotential  $\CW$ \eqref{superpot} corresponds to a $G$-gauging of three $D_p^b(G)$ theories. It can then happen that the $\CW$ for two distinct $D_p^b(G)$ trinions are equal, up to massive fields that can be decoupled. For simplicity, we have focussed on LG models with at most 4 non-trivial fields, that can be interpreted as canonical singularities $\MG$ in Type IIB. 

The consequence of this simple observation for the 4d $\CN=2$ SCFTs $\FTfour$ is rather interesting. Since the 4d theory is determined unambigiously by the singularity (or by $\CW$), the distinct-looking $D_p^b(G)$ trinions associated to a single $\MG$ are actually exactly equivalent. We have thus discovered a new type of S-duality among such 4d SCFTs.  The equivalent $D_p^b(G)$ trinions describe two S-duality frames of the same theory $\FTfour$, and the gauge groups manifest in each description become weakly coupled at different cusps of the conformal manifold. From the field-theoretic point of view, these dualities appear rather miraculous. By inspection of the list of theories in section \ref{sec:SUN} and section \ref{sec7:DandE}, we then identify a large number of these S-dualities. We list them below and briefly comment on some of their properties. We hope to return to an in-depth analysis of these theories in the future.

\medskip
\noindent
The list of S-dualities can be simplified by observing that, for a fixed gauge group $G$, we have the following equivalence relation between $D_p^b(G)$ theories
\be
\label{leg-equiv}
D_{p_1}^{b_1}(G)\equiv D_{p_2}^{b_2}(G)\qquad \text{if}\qquad \frac{b_1}{p_1}=\frac{b_2}{p_2}\,.
\ee 
This follows from the geometric engineering of section~\ref{subsec:DpSummary}, since the scaling dimensions of $D_p^{b}(G)$ only depend on ${b\ov p}$.

In order to describe the structure of the conformal manifold of $\FTfour$, we will need a class of  isolated ({\it i.e.} with trivial conformal manifold) 4d SCFTs dubbed $R_{2,N}$ (for $N$ odd) in \cite{Chacaltana:2010ks}. This is a class $\CS$ trinion with punctures $[k-1,k-1,1]$, $[k-1,k-1,1]$ and $[1^N]$, with $N= 2k-1$.
This theory has CB operators of dimension $3,5,\dots,N-2,N$ and global symmetry $SO(2N+4)_{2N}\times U(1)$, where the subscript $2N$ denotes the flavor central charge. We also notice that the 3d mirrors (or magnetic quivers) of $R_{2,N}$ theories appear in (\ref{R2N5d}), which tells us that $R_{2,N}$ is the dimensional reduction of the 5d SCFT which UV completes $SU(k)_{\pm\frac{3}{2}}+(2k+1)\bm{F}$. This family of theories is relevant because it arises in the description of the conformal manifold of $D_{2n}(SO(2n+2))$ theories. More precisely, we find that $D_{2n}(SO(2n+2))$ is equivalent to
\be\label{lagrSO} \boxed{1}-SU(2)-SU(3)-\boxed{4}\leftarrow SU(4)\rightarrow R_{2,5}\leftarrow SO(8)\rightarrow\dots \leftarrow SO(2n)\rightarrow R_{2,2n-1},\ee 
where the notation $A\leftarrow G \rightarrow B$ means that we glue together the SCFTs $A$ and $B$ by gauging a $G$ subgroup of their global symmetry.

\subsection{A List of New S-dualities}

We will now discuss in detail several examples of S-dualities we can obtain with this method.We find it convenient to organise the list according to $r$, the rank of the 5d SCFT $\FT$ associated with the same singularity.

\paragraph{Rank-1 $E_8$ model.} As our first example, the rank-1 $E_8$ theory in 5d is described by the singularity 
\be 
x^6+z_1^2+z_2^3+z_3^6=0~.
\ee 
Modulo the trivial equivalences \eqref{leg-equiv}, there are actually two $D_p^b(G)$ trinions that describe the same theory $\FTfour$
\bea
\begin{tikzpicture}[x=.7cm,y=.7cm]
\node at (-1.8,0.5) {$\FTXfour{E_8}=$};
\draw[ligne, black](1,0)--(8,0);
\draw[ligne, black](3,0)--(3,1);
\node[SUd] at (1,0) [label=above:{{\scriptsize$2$}}] {};
\node[SUd] at (2,0) [label=above:{{\scriptsize$4$}}] {};
\node[SUd] at (3,0) [label=below:{{\scriptsize$6$}}] {};
\node[SUd] at (4,0) [label=above:{{\scriptsize$5$}}] {};
\node[SUd] at (5,0) [label=above:{{\scriptsize$4$}}] {};
\node[SUd] at (6,0) [label=above:{{\scriptsize$3$}}] {};
\node[SUd] at (7,0) [label=above:{{\scriptsize$2$}}] {};
\node[flavor] at (8,0) [label=above:{{\scriptsize$1$}}] {};
\node[SUd] at (3,1) [label=above:{{\scriptsize$3$}}] {};
\end{tikzpicture}
\;\;
 \begin{tikzpicture}[x=.6cm,y=.6cm]
 \node at (-4, 0){$\equiv$};
\node at (-3,0) {$D_6$};
\node at (0,0) {$SO(8)$};
\node at (3,0) {$D_6$};
\node at (0,2) {$D_{1}^{4}$};
\draw[ligne, black](-2.3,0)--(-1.3,0);
\draw[ligne, black](1.3,0)--(2.3,0);
\draw[ligne, black](0,0.5)--(0,1.3);
\end{tikzpicture}
\eea
In the first S-duality frame, which is fully Lagrangian,   we clearly see that the conformal manifold has dimension 8. By using the equivalence \eqref{lagrSO} $D_6(SO(8))\equiv\boxed{1}-SU(2)-SU(3)-SU(4)\rightarrow R_{2,5}$ and the fact that $D_1^4(SO(8))$ is equivalent to $SU(2)$ $N_f=4$ SQCD, we easily see that the $SO(8)$ trinions as well have a eight-dimensional conformal manifold, in which the $SU(5)$ and $SU(6)$ groups of the unitary trinion are replaced by $SU(2)$ and $SO(8)$. In the $SO(8)$ trinion description the theory is not fully Lagrangian, as it includes two copies of the strongly-coupled $R_{2,5}$ theory. Indeed the gauge couplings of the two trinions tend to zero at different cusps of the conformal manifold.

\paragraph{Rank-2 $E_6$ model.}  The singularity that realizes the rank-2 $E_6$ 5d theory with $(r,f,d_H)=(2,6,23)$ leads in 4d to the duality
\be
\ba
& \begin{tikzpicture}[x=.7cm,y=.7cm]
\node at (-1.8,0.5) {$\FTXfour{E_6^{(3)}}=$};
\draw[ligne, black](1,0)--(5,0);
\draw[ligne, black](3,0)--(3,2);
\node[SUd] at (1,0) [label=below:{{\scriptsize$2$}}] {};
\node[SUd] at (2,0) [label=below:{{\scriptsize$4$}}] {};
\node[SUd] at (3,0) [label=below:{{\scriptsize$6$}}] {};
\node[SUd] at (4,0) [label=below:{{\scriptsize$4$}}] {};
\node[SUd] at (5,0) [label=below:{{\scriptsize$2$}}] {};
\node[SUd] at (3,1) [label=left:{{\scriptsize$4$}}] {};
\node[SUd] at (3,2) [label=left:{{\scriptsize$2$}}] {};
\node at (6,.6) {$\equiv$};
\end{tikzpicture}
\begin{tikzpicture}[x=.6cm,y=.6cm]
\node at (-3,0) {$D_3$};
\node at (0,0) {$SO(8)$};
\node at (3,0) {$D_6$};
\node at (0,2) {$D_2$};
\draw[ligne, black](-2.3,0)--(-1.3,0);
\draw[ligne, black](1.3,0)--(2.3,0);
\draw[ligne, black](0,0.5)--(0,1.3);
\end{tikzpicture}\,.
\ea
\ee 
In both S-duality frames, the conformal manifold has dimension seven. In the special unitary trinion this is obvious, and in the $SO(8)$ trinion this follows from the fact that $D_2(SO(8))$ is isolated (it is the rank-1 $E_6$ theory), $D_6(SO(8))$ has three marginal couplings, as we have already explained, and $D_3(SO(8))$ is the Lagrangian theory $SO(4)-Sp(2)-\boxed{4}$.

\paragraph{Rank-2 $E_7$ model.}  A similar result holds for the singularity that realizes the rank-2 $E_7$ theory with $(r,f,d_H)=(2,7,35)$:
\be
\ba
& \begin{tikzpicture}[x=.7cm,y=.7cm]
\node at (-1.5,0.5) {$\FTXfour{E_7^{(4)}}=$};
\draw[ligne, black](1,0)--(7,0);
\draw[ligne, black](4,0)--(4,1);
\node[SUd] at (1,0) [label=below:{{\scriptsize$2$}}] {};
\node[SUd] at (2,0) [label=below:{{\scriptsize$4$}}] {};
\node[SUd] at (3,0) [label=below:{{\scriptsize$6$}}] {};
\node[SUd] at (4,0) [label=below:{{\scriptsize$8$}}] {};
\node[SUd] at (5,0) [label=below:{{\scriptsize$6$}}] {};
\node[SUd] at (6,0) [label=below:{{\scriptsize$4$}}] {};
\node[SUd] at (7,0) [label=below:{{\scriptsize$2$}}] {};
\node[SUd] at (4,1) [label=left:{{\scriptsize$4$}}] {};
\node at (8,.3) {$\equiv$};
\end{tikzpicture}
\begin{tikzpicture}[x=.6cm,y=.6cm]
\node at (-3,0) {$D_4$};
\node at (0,0) {$SO(10)$};
\node at (3,0) {$D_8$};
\node at (0,2) {$D_1^5$};
\draw[ligne, black](-2.3,0)--(-1.3,0);
\draw[ligne, black](1.3,0)--(2.3,0);
\draw[ligne, black](0,0.5)--(0,1.3);
\end{tikzpicture}
\ea
\ee
Once again we can match the dimension of the conformal manifold on the two sides. The special unitary side is easy to understand since the theory is Lagrangian, whereas the $SO(10)$ trinion is more subtle: $D_1^5(SO(10))$ is the rank-1 $E_6$ theory, $D_8(SO(10))\equiv R_{2,7}\leftarrow SO(8)\rightarrow D_6(SO(8))$ from (\ref{lagrSO}) and $D_4(SO(10))=\boxed{1}-SU(2)-SO(6)-Sp(3)-\boxed{5}$.

\paragraph{Rank-2 $E_8$ model and triality.}  In the case of the $r=2$ $E_8$ theory, {\it i.e.} the hypersurface $x^{12}+z_1^2+z_2^3+z_3^6=0$,  we interestingly find a triality
\bea
& \begin{tikzpicture}[x=.7cm,y=.7cm]
\node at (-0.5,0.5) {$\FTXfour{E_8^{(6)}}=$};
\draw[ligne, black](1,0)--(8,0);
\draw[ligne, black](3,0)--(3,1);
\node[SUd] at (1,0) [label=above:{{\scriptsize$4$}}] {};
\node[SUd] at (2,0) [label=above:{{\scriptsize$8$}}] {};
\node[SUd] at (3,0) [label=below:{{\scriptsize$12$}}] {};
\node[SUd] at (4,0) [label=above:{{\scriptsize$10$}}] {};
\node[SUd] at (5,0) [label=above:{{\scriptsize$8$}}] {};
\node[SUd] at (6,0) [label=above:{{\scriptsize$6$}}] {};
\node[SUd] at (7,0) [label=above:{{\scriptsize$4$}}] {};
\node[SUd] at (8,0) [label=above:{{\scriptsize$2$}}] {};
\node[SUd] at (3,1) [label=above:{{\scriptsize$6$}}] {};
\end{tikzpicture}\cr 
&\qquad \qquad 
 \begin{tikzpicture}[x=.6cm,y=.6cm]
\node at (-5,.3) {$\equiv$};
\node at (-3,0) {$D_3$};
\node at (0,0) {$SO(14)$};
\node at (3,0) {$D_{12}$};
\node at (0,2) {$D_{1}^{7}$};
\draw[ligne, black](-2.3,0)--(-1.3,0);
\draw[ligne, black](1.3,0)--(2.3,0);
\draw[ligne, black](0,0.5)--(0,1.3);
\end{tikzpicture}
 \begin{tikzpicture}[x=.6cm,y=.6cm]
\node at (-3.2,0) {$\equiv$};
\node at (-2.3,0) {$D_6$};
\node at (0,0) {$E_6$};
\node at (2.3,0) {$D_{12}$};
\node at (0,2) {$D_{1}^{9}$};
\draw[ligne, black](-1.6,0)--(-0.6,0);
\draw[ligne, black](0.6,0)--(1.6,0);
\draw[ligne, black](0,0.5)--(0,1.3);
\end{tikzpicture}
\eea
In the $SO(14)$ trinion, the nine-dimensional conformal manifold arises as follows: $D_1^7(SO(14))$ is isolated, being equivalent to $R_{2,5}$. $D_3(SO(14))$ is equivalent to the Lagrangian theory $SO(6)-Sp(4)-\boxed{7}$ and finally we have $D_{12}(SO(14))$, which fits in the family (\ref{lagrSO}) and has a six-dimensional conformal manifold.

\paragraph{Sporadic model $\scriptstyle{(2)}$ with $r=2$.}  The singularity associated with the 5d theory with $(r,f,d_H)=(2,8,46)$ in table~\ref{tab:trinion series 1} leads to the following S-duality
\be
\ba
& \begin{tikzpicture}[x=.7cm,y=.7cm]
\node at (-5.8,0.5) {$\FTfour=$};
\draw[ligne, black](-4,0)--(3,0);
\draw[ligne, black](0,0)--(0,1);
%central node:
\node[SUd] at (0,0) [label=below:{{\scriptsize$10$}}] {};
%left:
\node[SUd] at (-1,0) [label=below:{{\scriptsize$8$}}] {};
\node[SUd] at (-2,0) [label=below:{{\scriptsize$6$}}] {};
\node[SUd] at (-3,0) [label=below:{{\scriptsize$4$}}] {};
\node[SUd] at (-4,0) [label=below:{{\scriptsize$2$}}] {};
%up:
\node[SUd] at (0,1) [label=left:{{\scriptsize$5$}}] {};
%right:
\node[SUd] at (1,0) [label=below:{{\scriptsize$7$}}] {};
\node[SUd] at (2,0) [label=below:{{\scriptsize$4$}}] {};
\node[flavor] at (3,0) [label=below:{{\scriptsize$1$}}] {};
\node at (4,0.3) {$\equiv$};
\end{tikzpicture}
\begin{tikzpicture}[x=.6cm,y=.6cm]
\node at (-3,0) {$D_{10}$};
\node at (0,0) {$SO(12)$};
\node at (3,0) {$D_2^6$};
\node at (0,2) {$D_1^6$};
\draw[ligne, black](-2.3,0)--(-1.3,0);
\draw[ligne, black](1.3,0)--(2.3,0);
\draw[ligne, black](0,0.5)--(0,1.3);
\end{tikzpicture}
\ea\,.
\ee

\paragraph{Rank-3 $E_7$ model.}  We also find an S-duality based on the $r=3$ $E_7$ theory %with $(r,f,d_H)=(3,7,53)$:
\be
\ba
& \begin{tikzpicture}[x=.7cm,y=.7cm]
\node at (-1.8,0.5) {$\FTXfour{E_7^{(8)}}=$};
\draw[ligne, black](1,0)--(7,0);
\draw[ligne, black](4,0)--(4,1);
\node[SUd] at (1,0) [label=below:{{\scriptsize$3$}}] {};
\node[SUd] at (2,0) [label=below:{{\scriptsize$6$}}] {};
\node[SUd] at (3,0) [label=below:{{\scriptsize$9$}}] {};
\node[SUd] at (4,0) [label=below:{{\scriptsize$12$}}] {};
\node[SUd] at (5,0) [label=below:{{\scriptsize$9$}}] {};
\node[SUd] at (6,0) [label=below:{{\scriptsize$6$}}] {};
\node[SUd] at (7,0) [label=below:{{\scriptsize$3$}}] {};
\node[SUd] at (4,1) [label=left:{{\scriptsize$6$}}] {};
\node at (8,0.3) {$\equiv$};
\end{tikzpicture}
\begin{tikzpicture}[x=.6cm,y=.6cm]
\node at (-2.3,0) {$D_4$};
\node at (0,0) {$E_6$};
\node at (2.3,0) {$D_{12}$};
\node at (0,2) {$D_1^8$};
\draw[ligne, black](-1.6,0)--(-0.6,0);
\draw[ligne, black](0.6,0)--(1.6,0);
\draw[ligne, black](0,0.5)--(0,1.3);
\end{tikzpicture}
\ea\,.
\ee

\paragraph{Rank-3 $E_8$ model.}
Likewise, the rank-3 $E_8$ theory leads to the S-duality
\be
 \begin{tikzpicture}[x=.7cm,y=.7cm]
\node at (-1.8,0.5) {$\FTXfour{E_8^{(12)}}=$};
\draw[ligne, black](1,0)--(8,0);
\draw[ligne, black](3,0)--(3,1);
\node[SUd] at (1,0) [label=above:{{\scriptsize$6$}}] {};
\node[SUd] at (2,0) [label=above:{{\scriptsize$12$}}] {};
\node[SUd] at (3,0) [label=below:{{\scriptsize$18$}}] {};
\node[SUd] at (4,0) [label=above:{{\scriptsize$15$}}] {};
\node[SUd] at (5,0) [label=above:{{\scriptsize$12$}}] {};
\node[SUd] at (6,0) [label=above:{{\scriptsize$9$}}] {};
\node[SUd] at (7,0) [label=above:{{\scriptsize$6$}}] {};
\node[SUd] at (8,0) [label=above:{{\scriptsize$3$}}] {};
\node[SUd] at (3,1) [label=above:{{\scriptsize$9$}}] {};
\node at (9,0.3) {$\equiv$};
\end{tikzpicture}
\begin{tikzpicture}[x=.6cm,y=.6cm]
\node at (-2.3,0) {$D_6$};
\node at (0,0) {$E_7$};
\node at (2.3,0) {$D_{18}$};
\node at (0,2) {$D_1^{14}$};
\draw[ligne, black](-1.6,0)--(-0.6,0);
\draw[ligne, black](0.6,0)--(1.6,0);
\draw[ligne, black](0,0.5)--(0,1.6);
\end{tikzpicture}\,.
\ee

\paragraph{Rank-3 model.} The singularity $x^9 + y^2 + z_2^3 + z_3^9=0$ associated with the rank-3  theories appearing in table~\ref{tab:Bounty} and \ref{tab:Snickers}, respectively, with $(r,f, d_H)= (3,0,64)$,  implies the duality
\be
\ba
& \begin{tikzpicture}[x=.6cm,y=.6cm]
\node at (-5.8,0.5) {$\FTfour=$};
\node at (-3,0) {$D_3$};
\node at (0,0) {$SO(20)$};
\node at (3,0) {$D_9$};
\node at (0,2) {$D_1^{10}$};
\draw[ligne, black](-2.3,0)--(-1.3,0);
\draw[ligne, black](1.3,0)--(2.3,0);
\draw[ligne, black](0,0.5)--(0,1.3);
\node at (4,0.3) {$\equiv$};
\end{tikzpicture}
\begin{tikzpicture}[x=.6cm,y=.6cm]
\node at (-2.3,0) {$D_9$};
\node at (0,0) {$E_7$};
\node at (2.3,0) {$D_9$};
\node at (0,2) {$D_1^{14}$};
\draw[ligne, black](-1.6,0)--(-0.6,0);
\draw[ligne, black](0.6,0)--(1.6,0);
\draw[ligne, black](0,0.5)--(0,1.6);
\end{tikzpicture}
\ea\,.
\ee
Notice that the $SO(20)$ trinion is a fully Lagrangian SCFT, given by the orthosymplectic quiver
\bea
  \begin{tikzpicture}[x=0.8cm,y=0.8cm]
%\node at (-3,0.5) {$\MQfive  = $};
\draw[ligne, black](-2,0)--(8,0);
\draw[ligne, black](0,0)--(0,1);
\node[bd] at (-2,0) [label=below:{{\scriptsize$\Spin(8)$}}] {};
\node[bd] at (-1,0) [label=above:{{\scriptsize$Sp(6)$}}] {};
\node[bd] at (0,0) [label=below:{{\scriptsize$\Spin(20)$}}] {};
\node[bd] at (0,1) [label=above:{{\qquad\;\scriptsize$Sp(4)$}}] {};
\node[bd] at (1,0) [label=above:{{\scriptsize$Sp(8)$}}] {};
\node[bd] at (2,0) [label=below:{{\scriptsize$\Spin(16)$}}] {};
\node[bd] at (3,0) [label=above:{{\scriptsize$Sp(6)$}}] {};
\node[bd] at (4,0) [label=below:{{\scriptsize$\Spin(12)$}}] {};
\node[bd] at (5,0) [label=above:{{\scriptsize$Sp(4)$}}] {};
\node[bd] at (6,0) [label=below:{{\scriptsize$\Spin(8)$}}] {};
\node[bd] at (7,0) [label=above:{{\scriptsize$Sp(2)$}}] {};
\node[bd] at (8,0) [label=below:{{\scriptsize$\Spin(4)$}}] {};
\end{tikzpicture}
\eea
with a conformal manifold of dimension 13. This quiver has an electric one-form symmetry $\frak{f}= \Z_2^7$, in agreement with the geometry.  This is an interesting example of an S-duality that does not involve an $SU(N)$ trinion.

\medskip\noindent
We also find three curious examples associated with 5d SCFTs of rank larger than three:

\paragraph{Rank-4 model.} The rank-4 theory with $(r,f,d_H)=(4,5,40)$ appears in table~\ref{tab:Bounty} and in table~\ref{tab:Twix}, corresponding to the singularity $t^4 +y^2 +z_2^6 +z_3^6=0$ which results in the S-duality
\be
\ba
& \begin{tikzpicture}[x=.6cm,y=.6cm]
\node at (-5.8,0.5) {$\FTfour=$};
\node at (-3,0) {$D_4$};
\node at (0,0) {$SO(14)$};
\node at (3,0) {$D_6$};
\node at (0,2) {$D_1^7$};
\draw[ligne, black](-2.3,0)--(-1.3,0);
\draw[ligne, black](1.3,0)--(2.3,0);
\draw[ligne, black](0,0.5)--(0,1.3);
\node at (4,0.3) {$\equiv$};
\end{tikzpicture}
\begin{tikzpicture}[x=.6cm,y=.6cm]
\node at (-2.3,0) {$D_6$};
\node at (0,0) {$E_6$};
\node at (2.3,0) {$D_6$};
\node at (0,2) {$D_1^8$};
\draw[ligne, black](-1.6,0)--(-0.6,0);
\draw[ligne, black](0.6,0)--(1.6,0);
\draw[ligne, black](0,0.5)--(0,1.6);
\end{tikzpicture}
\ea\,.
\ee
The conformal manifold is of dimension 7.

\paragraph{Rank-5 $E_8$ model.} 
The rank-5 $E_8$ theory, associated to the singularity $x^3+y^2+z_2^6+z_3^{30}=0$, leads to the duality
\be
 \begin{tikzpicture}[x=.7cm,y=.7cm]
\node at (-1.8,0.5) {$\FTXfour{E_8^{(24)}}=$};
\draw[ligne, black](1,0)--(8,0);
\draw[ligne, black](3,0)--(3,1);
\node[SUd] at (1,0) [label=above:{{\scriptsize$10$}}] {};
\node[SUd] at (2,0) [label=above:{{\scriptsize$20$}}] {};
\node[SUd] at (3,0) [label=below:{{\scriptsize$30$}}] {};
\node[SUd] at (4,0) [label=above:{{\scriptsize$25$}}] {};
\node[SUd] at (5,0) [label=above:{{\scriptsize$20$}}] {};
\node[SUd] at (6,0) [label=above:{{\scriptsize$15$}}] {};
\node[SUd] at (7,0) [label=above:{{\scriptsize$10$}}] {};
\node[SUd] at (8,0) [label=above:{{\scriptsize$5$}}] {};
\node[SUd] at (3,1) [label=above:{{\scriptsize$15$}}] {};
\node at (9,0.3) {$\equiv$};
\end{tikzpicture}
\begin{tikzpicture}[x=.6cm,y=.6cm]
\node at (-2.3,0) {$D_6$};
\node at (0,0) {$E_8$};
\node at (2.3,0) {$D_{30}$};
\node at (0,2) {$D_1^{24}$};
\draw[ligne, black](-1.6,0)--(-0.6,0);
\draw[ligne, black](0.6,0)--(1.6,0);
\draw[ligne, black](0,0.5)--(0,1.6);
\end{tikzpicture}
\ee
Note that $D_1^8(E_8)$ is the rank-1 $E_8$ MN theory. We thus find an S-dual Lagrangian theory of the $E_8$ gauging of the $E_8$ MN theory (with the other two $D_p(E_8)$ tails appended, to total rank being $\h r=141$) as a $E_8$-shaped special unitary quiver.

\paragraph{Rank-6  model.}  Finally, the rank-6 theory with $(r,f,d_H)=(6,0,77)$ appears in table~\ref{tab:Bounty} and table~\ref{tab:Ferrero}, corresponding to the singularity $x^{15} +y^2 +z_2^5 +z_3^3 z_2=0$. This   implies the duality
\be
\ba
& \begin{tikzpicture}[x=.6cm,y=.6cm]
\node at (-5.8,0.5) {$\FTfour=$};
\node at (-3,0) {$D_5$};
\node at (0,0) {$SO(32)$};
\node at (3,0) {$D_2^{16}$};
\node at (0,2) {$D_1^{16}$};
\draw[ligne, black](-2.3,0)--(-1.3,0);
\draw[ligne, black](1.3,0)--(2.3,0);
\draw[ligne, black](0,0.5)--(0,1.3);
\node at (4,0.3) {$\, \equiv$};
\end{tikzpicture}
\begin{tikzpicture}[x=.6cm,y=.6cm]
\node at (-2.3,0) {$D_{15}$};
\node at (0,0) {$E_8$};
\node at (2.3,0) {$D_3^{24}$};
\node at (0,2) {$D_1^{20}$};
\draw[ligne, black](-1.6,0)--(-0.6,0);
\draw[ligne, black](0.6,0)--(1.6,0);
\draw[ligne, black](0,0.5)--(0,1.6);
\end{tikzpicture}
\ea
\ee
We note that the $SO(32)$ trinion is actually a Lagrangian theory
\bea
  \begin{tikzpicture}[x=0.8cm,y=0.8cm]
%\node at (-3,0.5) {$\MQfive  = $};
\draw[ligne, black](-3,0)--(4,0);
\draw[ligne, black](0,0)--(0,1);
\node[bd] at (-3,0) [label=above:{{\scriptsize$Sp(3)$}}] {};
\node[bd] at (-2,0) [label=below:{{\scriptsize$\Spin(16)$}}] {};
\node[bd] at (-1,0) [label=above:{{\scriptsize$Sp(11)$}}] {};
\node[bd] at (0,0) [label=below:{{\scriptsize$\Spin(32)$}}] {};
\node[bd] at (0,1) [label=above:{{\qquad\;\scriptsize$Sp(7)$}}] {};
\node[bd] at (1,0) [label=above:{{\scriptsize$Sp(12)$}}] {};
\node[bd] at (2,0) [label=below:{{\scriptsize$\Spin(20)$}}] {};
\node[bd] at (3,0) [label=above:{{\scriptsize$Sp(6)$}}] {};
\node[bd] at (4,0) [label=below:{{\scriptsize$\Spin(8)$}}] {};
\end{tikzpicture}
\eea
 with a 9-dimensional conformal manifold and an electric one-form symmetry $\frak{f}= \Z_2^5$, in perfect agreement with the geometry.

%%%%%%%%%%%%%%%%%%%%%%%%%%%
%%%%%%%%%%%%%%%%%%%%%%%%%%%
\subsubsection*{Acknowledgements}
We thank  Fabio Apruzzi, Lakshya Bhardwaj, Antoine Bourget, Stefano Cremonesi, Michele Del Zotto, Antonella Grassi, Amihay Hanany, Sheldon Katz, Horia Magureanu, Wolfger Peelaers and Timo Weigand for discussions. 
The work of SG, SSN, Y-NW, is supported by the ERC Consolidator Grant number 682608 ``Higgs bundles: Supersymmetric Gauge Theories and
Geometry (HIGGSBNDL)". SSN acknowledges support also from the Simons Foundation. 
CC is a Royal Society University Research Fellow and a Birmingham Fellow.

%%%%%%%%%%%%%%%%%%%%%%%%%%%%%%%%%%%%
%%%%%%%%%%%%%%%%%%%%%%%%%%%%%%%%%%%%

\appendix

%%%%%%%%%%%%%%%%%%%%%%%%%

\section{Resolutions of Hypersurface Canonical Singularities}
\label{app:resolution}

In this section, we present the detailed resolution sequences and resolved geometry of the isolated hypersurface (IHS) equations. We use the notation and conventions of \cite{Lawrie:2012gg}, see also \cite{Closset:2020scj}. In particular blowups are denoted by 
\be
(x_1^{(i_1)}, x_2^{(i_2)}, x_3^{(i_3)}, x_4^{(i_4)}; \delta) \,,
\ee
which blows up the locus $x_1=x_2= x_3=x_4=0$ with weights $i_j$ and the exceptional divisor is $\delta$. This blowup inserts a weighted projective space given by the weights. If there are not $(i_j)$ labels, the weights are 1. 

\subsection{$E_n^{(k)}$ Theories}

For the $E_n^{(k)}$ theories with rank $r>0$, we review the resolution sequences in \cite{Closset:2020scj}. For $E_6^{(k)}$ theories with the singular equation
\be
x_1^3+x_2^3+x_3^3+x_4^{3+k}=0\,,
\ee
the resolution sequence is
\be
(x_1,x_2,x_3,x_4;\delta_1)\,,\qquad (x_1,x_2,x_3,\delta_i;\delta_{i+1})\ (i=1,\dots,r-1)\,.
\ee
We use $S_i$ to denote the exceptional divisor $\delta_i=0$, and the rank of this theory is $r= \lfloor \frac{k}{3}\rfloor+1$.
The triple intersection numbers among the divisors $S_1,\dots,S_r$ are
\be
\begin{array}{ll}
S_1^3=S_2^3=\dots=S_{r-1}^3=0\ ,\ \quad & S_r^3=3\\
S_i^2\cdot S_{i-1}=-3\ ,\ &S_{i-1}^2\cdot S_i=3\ ,\quad (i=2,\dots,r)\,.
\end{array}
\ee
For any $k\geq 3$, the divisors $S_1,\dots,S_{r-1}$ are ruled surfaces over genus-1 curves, and all the intersection curves $S_i\cdot S_{i-1}$ $(i=2,\dots,r)$ are smooth too. 
For the topology of the divisor $S_r$, we have the following cases:
\begin{enumerate}
\item $k=3r-3$:  the divisor $S_r$ is a smooth $dP_6$.
\item $k=3r-2$: the resolved CY3 is smooth, but the divisor $S_r$ is a singular cubic surface with the following equation:
\be
\label{E6-singular-surface}
x_1^3+x_2^3+x_3^3=0\,,
\ee
which has a point singularity at $x_1=x_2=x_3=0$. 
\item $k=3r-1$: the resolved CY3 has a terminal singularity at the point $x_1=x_2=x_3=\delta_r=0$:
\be
x_1^3+x_2^3+x_3^3+\delta_r^2=0\,.
\ee
The divisor $S_r$ also has the same singular equation in (\ref{E6-singular-surface}).
\end{enumerate}
For $E_7^{(k)}$ theories (rank $r=\lfloor \frac{k}{4}\rfloor+1$) with the singular equation
\be
x_1^2+x_2^4+x_3^4+x_4^{4+k}=0\,,
\ee
the resolution sequence is
\be
(x_1^{(2)},x_2^{(1)},x_3^{(1)},x_4^{(1)};\delta_1)\,,\qquad (x_1^{(2)},x_2^{(1)},x_3^{(1)},\delta_i^{(1)};\delta_{i+1})\,,\qquad  i=1,\dots,r-1\,.
\ee
The triple intersection numbers among the divisors $S_1,\dots,S_r$ are
\be
\begin{array}{ll}
S_1^3=S_2^3=\dots=S_{r-1}^3=0\ ,\ \quad & S_r^3=2\\
S_i^2\cdot S_{i-1}=-2\ ,\ &S_{i-1}^2\cdot S_i=2\ ,\quad (i=2,\dots,r)\,.
\end{array}
\ee
For any $k\geq 4$, the divisors $S_1,\dots,S_{r-1}$ are smooth genus-1 ruled surfaces.
For the topology of the divisor $S_r$, we have the following cases:
\begin{enumerate}
\item{$k=4r-4$: the divisor $S_r$ is a smooth $dP_7$.}
\item{$k=4r-3$: the resolved CY3 is smooth, but the divisor $S_r$ is a singular surface with the following equation:
\be
\label{E7-singular-surface}
x_1^2+x_2^4+x_3^4=0\,,
\ee
which has a point singularity at $x_1=x_2=x_3=0$. }
\item{$k=4r-s$ $(s=1,2)$: the resolved CY3 has a terminal singularity at the point $x_1=x_2=x_3=\delta_r=0$:
\be
x_1^2+x_2^4+x_3^4+\delta_r^{4-s}=0\,.
\ee
The divisor $S_r$ also has the same singular equation in (\ref{E7-singular-surface}).
}
\end{enumerate}
For $E_8^{(k)}$ theories (rank $r=\lfloor \frac{k}{6}\rfloor+1$) with the singular equation
\be
x_1^2+x_2^3+x_3^6+x_4^{6+k}=0\,,
\ee
the resolution sequence is
\be
(x_1^{(3)},x_2^{(2)},x_3^{(1)},x_4^{(1)};\delta_1)\,,\qquad 
(x_1^{(3)},x_2^{(2)},x_3^{(1)},\delta_i^{(1)};\delta_{i+1})\,,\qquad i=1,\dots,r-1\,.
\ee
The triple intersection numbers among the divisors $S_1,\dots,S_r$ are
\be
\begin{array}{ll}
S_1^3=S_2^3=\dots=S_{r-1}^3=0\ ,\ \quad & S_r^3=1\\
S_i^2\cdot S_{i-1}=-1\ ,\ &S_{i-1}^2\cdot S_i=1\ ,\quad (i=2,\dots,r)\,.
\end{array}
\ee
For any $k\geq 6$, the divisors $S_1,\dots,S_{r-1}$ are smooth genus-1 ruled surfaces.
For the topology of the divisor $S_r$, we have the following cases:
\begin{enumerate}
\item{$k=6r-6$: the divisor $S_r$ is a smooth $dP_8$.}
\item{$k=6r-5$: the resolved CY3 is smooth, but the divisor $S_r$ is a singular surface with the following equation:
\be
\label{E8-singular-surface}
x_1^2+x_2^3+x_3^6=0\,,
\ee
which has a point singularity at $x_1=x_2=x_3=0$. }
\item{$k=6r-s$ $(s=1,2,3,4)$: the resolved CY3 has a terminal singularity at the point $x_1=x_2=x_3=\delta_r=0$:
\be
x_1^2+x_2^3+x_3^6+\delta_r^{6-s}=0\,.
\ee
The divisor $S_r$ also has the same singular equation in (\ref{E8-singular-surface}).
}
\end{enumerate}

\subsection{13 Sporadic Trinion Models}
\label{app:SporadicRes}

We now determine the resolution for the 13 sporadic trinion theories from section~\ref{sec:SUN}. 

\paragraph{Sporadic theory $\# (1)$}
Starting from the singular equation
\be
x_1^3 +x_2^4+ x_3^6+x_4^2 x_3=0\,,
\ee
after the resolution
\be
(x_1,x_2,x_3,x_4;\delta_1)\,,
\ee
there is still a residual terminal singularity at $x_1=x_2=\delta_1=x_4=0$:
\be
x_1^3+x_2^4\delta_1+\delta_1^3+x_4^2=0\,,
\ee
which is type II $(2,3,3,4)$, and equivalent to type I $(2,3,3,6)$ ($E_{8}^{(-3)}$).

\paragraph{Sporadic theory $\# (2)$}

Starting from the singular equation
\be
x_1^2+ x_2^5+ x_3^{10}+x_4^3 x_3=0\,,
\ee
the resolution sequence is
\be
(x_1^{(2)},x_2^{(1)},x_3^{(1)},x_4^{(1)};\delta_1)\,,\qquad(x_1^{(3)},x_2^{(1)},x_4^{(2)},\delta_1^{(1)};\delta_2) \,.
\ee
After the resolution, the equation becomes
\be
x_1^2+ x_2^5\delta_1+ x_3^{10}\delta_1^6+x_4^3 x_3=0\,,
\ee
the divisor $\delta_1=0$ is singular at $x_1=x_4=0$:
\be
\delta_1=0:\ x_1^2+x_4^3 x_3=0\,.
\ee
The detailed geometry was already discussed in \cite{Closset:2020scj}, which eventually gives rise to 5d $SU(2)-SU(2)-5\bm{F}$ gauge theory after a flop.

\paragraph{Sporadic theory $\# (3)$}

Starting from the singular equation
\be
x_1^3+ x_2^5+ x_3^{15}+x_4^2 x_3=0\,,
\ee
the resolution sequence is
\be
(x_1,x_2,x_3,x_4;\delta_1)\,,\quad 
(x_1^{(2)},x_2^{(1)},x_4^{(3)},\delta_1^{(1)};\delta_3)\,,\quad 
(x_1^{(2)},x_2^{(1)},x_4^{(3)},\delta_3^{(1)};\delta_4)\,,\quad 
(x_1,x_4,\delta_1;\delta_2)\,.
\ee
After the resolution, the equation becomes
\be
\delta_2 x_1^3+\delta_1^2 \delta_3 x_2^5+\delta_1^{12}\delta_2^{10}\delta_3^6 x_3^{15}+x_3 x_4^2=0\,.
\ee
The divisor $\delta_3=0$ is singular $x_1=x_4=0$
\be
\delta_3=0:\ \delta_2 x_1^3+x_3 x_4^2=0\,.
\ee

\paragraph{Sporadic theory $\# (4)$}

Starting from the singular equation
\be
x_1^3+ x_2^5+x_3^2 x_2+x_4^3 x_2=0\,,
\ee
after the resolution
\be
(x_1,x_2,x_3,x_4;\delta_1)\,,
\ee
there is a residual terminal singularity at $x_1=x_3=x_4=\delta_1=0$:
\be
x_1^3+\delta_1^2+x_3^2+x_4^3\delta_1+x_1 x_3 x_4=0\,,
\ee
which is Type I $(2,2,3,6)$ (corresponding to $E_{8}^{(-4)}$), and another terminal singularity at $x_1=x_2=x_3=\delta_1=0$:
\be
x_1^3+x_2^5\delta_1^2+x_3^2 x_2+x_2\delta_1+x_1 x_3=0\,,
\ee
which is the conifold singularity.

\paragraph{Sporadic theory $\# (5)$}

Starting from the singular equation
\be
x_1^2+x_2^7+x_3^3 x_2+x_4^4 x_2+x_3^2 x_4^2=0\,,
\ee
after the resolution
\be
(x_1^{(2)},x_2^{(1)},x_3^{(1)},x_4^{(1)};\delta_1)\,,
\ee
there is a residual terminal singularity at $x_1=x_2=x_3=\delta_1=0$:
\be
x_1^2+x_2^7\delta_1^3+x_3^3 x_2+x_2\delta_1+x_3^2=0\,,
\ee
which is a conifold singularity. There is also a residual terminal singularity at $x_1=x_3=x_4=\delta_1=0$:
\be
x_1^2+\delta_1^3+x_3^3+x_4^4\delta_1+x_3^2 x_4^2=0\,,
\ee
which is type I $(2,3,3,6)$, the same as $E_{8}^{(-3)}$.

\paragraph{Sporadic theory $\# (6)$}

From the singular equation
\be
x_1^3+x_2^9+x_3^2 x_2+x_4^4 x_2=0\,,
\ee
the resolution sequence is
\be
\ba
(x_1,x_2,x_3,x_4;\delta_1)\,,\quad 
(x_1^{(2)},x_3^{(3)},x_4^{(1)},\delta_1^{(1)};\delta_3)\,,\quad 
(x_1,x_3,\delta_1;\delta_2)\,.
\ea
\ee
The resolved equation and exceptional divisors are smooth.

\paragraph{Sporadic theory $\# (7)$}

From the singular equation
\be
x_1^4+ x_2^{10}+x_3^2 x_2+x_4^3 x_2+x1 x3 x4=0\,,
\ee
after the resolution
\be
\ba
(x_1,x_2,x_3,x_4;\delta_1)\,,\quad
(x_1^{(1)},x_3^{(2)},x_4^{(1)},\delta_1^{(1)};\delta_3)\,,\quad 
(x_3,\delta_1;\delta_2)\,,
\ea
\ee
the equation has a terminal singularity at $x_1=x_3=x_4=\delta_3=0$:
\be
x_1^4\delta_3+\delta_3^3+x_3^2+x_4^3+x_1 x_3 x_4=0\,.
\ee
It is type I $(2,3,3,6)$, which is the same as $E_{8}^{(-3)}$.

\paragraph{Sporadic theory $\# (8)$}

From the singular equation
\be
x_1^2+x_2^{16}+x_3^3 x_2 +x_4^5 x_2+x_3^2
   x_4^2=0\,,
\ee
the resolution sequence is
\be
(x_1^{(2)},x_2^{(1)},x_3^{(1)},x_4^{(1)};\delta_1)\,,\quad 
(x_1^{(3)},x_3^{(2)},x_4^{(1)},\delta_1^{(1)};\delta_3)\,,\quad 
(x_1^{(3)},x_3^{(2)},x_4^{(1)},\delta_3^{(1)};\delta_4)\,,\quad 
(x_3,\delta_1;\delta_2)\,.
\ee
The resolved equation is
\be
x_1^2+\delta_1^{12}\delta_2^{10}\delta_3^6 x_2^{16}+\delta_2 x_2 x_3^3+\delta_1^2 \delta_3 x_2 x_4^5+x_3^2 x_4^2=0\,,
\ee
and the divisor $\delta_3=0$ is singular at $x_1=x_3=0$:
\be
x_1^2+\delta_2 x_2 x_3^3+x_3^2 x_4^2=0\,.
\ee

\paragraph{Sporadic theory $\# (9)$}

From the singular equation
\be
x_1^3+ x_2^{21}+x_3^2 x_2+x_4^5 x_2+x_1 x_3 x_4=0\,,
\ee
the resolution sequence is
\be
\ba
&(x_1,x_2,x_3,x_4;\delta_1)\cr
&(x_1^{(2)},x_3^{(3)},x_4^{(1)},\delta_1^{(1)};\delta_4)\cr
&(x_1^{(2)},x_3^{(3)},x_4^{(1)},\delta_4^{(1)};\delta_6)\cr
&(x_1^{(2)},x_3^{(3)},x_4^{(1)},\delta_6^{(1)};\delta_7)\cr
&(x_1,x_3,\delta_1;\delta_2)\,,\quad 
(x_3,\delta_2;\delta_3)\,,\quad 
(x_1,x_3,\delta_4;\delta_5)\,.
\ea
\ee
The resolved equation is
\be
\delta_2 \delta_5 x_1^3+x_1 x_3 x_4+\delta_1^{18} \delta_2^{16} \delta_3^{15} \delta_4^{12} \delta_5^{10} \delta_6^6 x_2^{21}+\delta_3 x_2 x_3^2+\delta_1^3 \delta_2 \delta_4^2 \delta_6 x_2 x_4^5=0\,.
\ee
The divisor $\delta_6=0$ is singular at $x_1=x_3=0$:
\be
\delta_6=0:\qquad  \delta_2 \delta_5 x_1^3+x_1 x_3 x_4+\delta_3 x_2 x_3^2=0\,.
\ee

\paragraph{Sporadic theory $\# (10)$}

From the singular equation
\be
x_1^5+x_2^{25}+x_3^2 x_2+x_4^3 x_2=0\,,
\ee
the resolution sequence is
\be
\ba
&(x_1,x_2,x_3,x_4;\delta_1)\,,\quad 
(x_1^{(1)},x_3^{(2)},x_4^{(1)},\delta_1^{(1)};\delta_4)\,,\quad 
(x_1^{(1)},x_3^{(3)},x_4^{(2)},\delta_4^{(1)};\delta_7)\,,\cr
&(x_1^{(1)},x_3^{(3)},x_4^{(2)},\delta_7^{(1)};\delta_9)\,,\quad 
(x_1^{(1)},x_3^{(3)},x_4^{(2)},\delta_9^{(1)};\delta_{10})\,,\cr 
&(x_3,\delta_1;\delta_2)\,,\quad 
(x_4,\delta_2;\delta_3)\,,\quad 
(x_3,x_4,\delta_4;\delta_5)\,,\quad 
(x_3,\delta_5;\delta_6)\,,\quad 
(x_3,x_4,\delta_7;\delta_8)\,.
\ea
\ee
The resolved equation is
\be
\delta_1^2 \delta_2 \delta_4^3 \delta_5 \delta_7^2 \delta_9 x_1^5+x_1 x_3 x_4+\delta_1^{22} \delta_2^{21} \delta_3^{20} \delta_4^{18} \delta_5^{16} \delta_6^{15} \delta_7^{12} \delta_8^{10} \delta_9^6 x_2^25+\delta_2 \delta_6 x_2 x_3^2+\delta_1 \delta_3^2 \delta_5 \delta_8 x_2 x_4^3=0\,.
\ee
The divisor $\delta_9=0$ is singular at $x_3=x_4=0$:
\be
\delta_6=0:\ x_1 x_3 x_4+\delta_2 \delta_6 x_2 x_3^2+\delta_1 \delta_3^2 \delta_5 \delta_8 x_2 x_4^3=0\,.
\ee

\paragraph{Sporadic theory $\# (11)$}

From the singular equation
\be
x_1^7+x_2^2 x_1+x_3^3 x_1+x_4^3 x_1+x_2 x_3^2+x_2 x_4^2+x_2 x_3 x_4=0\,,
\ee
the resolution sequence is
\be
\ba
(x_1,x_2,x_3,x_4;\delta_1)\,,\quad 
(x_2^{(2)},x_3^{(1)},x_4^{(1)},\delta_1^{(1)};\delta_3)\,,\quad 
(x_2,\delta_1;\delta_2)\,.
\ea
\ee

The resolved equation and exceptional divisors are smooth.

\paragraph{Sporadic theory $\# (12)$}

From the singular equation
\be
x_1^{13}+x_2^2 x_1+x_3^3 x_1+x_4^4 x_1+x_3 x_4^3+2 x_2 x_3 x_4=0\,,
\ee
the resolution sequence is
\be
\ba
&(x_1,x_2,x_3,x_4;\delta_1)\cr
&(x_2^{(2)},x_3^{(1)},x_4^{(1)},\delta_1^{(1)};\delta_4)\cr
&(x_2^{(3)},x_2^{(2)},x_4^{(1)},\delta_4^{(1)};\delta_6)\cr
&(x_2,\delta_1;\delta_2)\,,\quad (x_3,\delta_2;\delta_3)\,,\quad 
(x_2,x_3,\delta_4;\delta_5)\,.
\ea
\ee
The resolved equation and exceptional divisors are smooth.

\paragraph{Sporadic theory $\# (13)$}

From the singular equation
\be
 x_1^{31}+x_2^2 x_1+x_3^3 x_1+x_4^5 x_1+3 x_2 x_3 x_4=0\,,
\ee
the resolution sequence is
\be
\ba
&(x_1,x_2,x_3,x_4;\delta_1)\,,\quad 
(x_2^{(2)},x_3^{(1)},x_4^{(1)},\delta_1^{(1)};\delta_5)\,,\quad
(x_2^{(3)},x_2^{(2)},x_4^{(1)},\delta_5^{(1)};\delta_9)\cr  
&(x_2^{(3)},x_2^{(2)},x_4^{(1)},\delta_9^{(1)};\delta_{12})\,,\quad 
(x_2^{(3)},x_2^{(2)},x_4^{(1)},\delta_{12}^{(1)};\delta_{14})\,,\quad 
(x_2^{(3)},x_2^{(2)},x_4^{(1)},\delta_{14}^{(1)};\delta_{15})\cr  
&(x_2,\delta_1;\delta_2)\,,\quad 
(x_3,\delta_2;\delta_3)\,,\quad 
(x_2,\delta_3;\delta_4)\,,\cr 
&(x_2,x_3,\delta_5;\delta_6)\,,\quad 
(x_2,\delta_6;\delta_7)\,,\quad 
(x_3,\delta_7;\delta_8)\,,\cr 
&(x_2,x_3,\delta_9;\delta_{10})\,,\quad 
(x_2,\delta_{10};\delta_{11})\,,\quad 
(x_2,x_3,\delta_{12};\delta_{13})\,.
\ea
\ee
The resolved equation is
\be
\ba
&\delta_1^{28} \delta_2^{27} \delta_3^{26} \delta_4^{25} \delta_5^{24} \delta_6^{22} \delta_7^{21} \delta_8^{20} \delta_9^{18} \delta_{10}^{16} \delta_{11}^{15} \delta_{12}^{12} \delta_{13}^{10} \delta_{14}^6 x_1^{31}+\delta_2 \delta_4 \delta_7 \delta_{11} x_1 x_2^2+\cr
&\delta_1 \delta_3^2 \delta_4 \delta_6 \delta_8^2 \delta_{10} \delta_{13} x_1 x_3^3+\delta_1^3 \delta_2^2 \delta_3 \delta_5^4 \delta_6^2 \delta_7 \delta_9^3 \delta_{10} \delta_{12}^2 \delta_{14} x_1 x_4^5+3 x_2 x_3 x_4=0\,.
\ea
\ee
The divisor $\delta_{14}=0$ is singular at $x_2=x_3=0$:
\be
\delta_{14}=0:\ \delta_2 \delta_4 \delta_7 \delta_{11} x_1 x_2^2+\delta_1 \delta_3^2 \delta_4 \delta_6 \delta_8^2 \delta_{10} \delta_{13} x_1 x_3^3+3 x_2 x_3 x_4=0\,.
\ee

\subsection{Type VIII$_2$ $(N,N,2,2)$ series}

Consider the type VIII$_2$ $(N,N,2,2)$ series in table~\ref{tab:trinion series}, which are equivalent to type IV$(N,2,N,2)$. We now resolve these models. 

\paragraph{$N=2k-1$.}

Start with the singular equation
\be
x_1^{2k-1}+x_2^{2k-1}+x_1 x_3^2+x_1 x_4^2+x_2 x_3 x_4=0\,,
\ee
the resolution sequence is
\be
\ba
(x_1,x_2,x_3,x_4;\delta_1)\,,\qquad 
(x_3,x_4,\delta_j;\delta_{j+1})\,,\quad j=1,\dots,k-2 \,.
\ea
\ee
The resolved CY3 and divisors are all smooth.

\paragraph{$N=2k$.}

Start with the singular equation
\be
x_1^{2k}+x_2^{2k}+x_1 x_3^2+x_1 x_4^2+x_2 x_3 x_4=0\,,
\ee
the resolution sequence is
\be
(x_1,x_2,x_3,x_4;\delta_1)\,,\qquad 
(x_3,x_4,\delta_j;\delta_{j+1})\,,\qquad j=1,\dots,k-2\,.
\ee
The resolved CY3 is
\be
(x_1^{2k}+x_2^{2k})\prod_{i=1}^{k-1}\delta_i^{2k-2i-1}+x_1 x_3^2+x_1 x_4^2+x_2 x_3 x_4=0\,.
\ee
There are in total $2k$ conifold singularities at the loci
\be
\delta_{k-1}=x_3=x_4=x_1+x_2 e^{\frac{\pi(2j+1)}{2k}}=0\,, \quad j=1,\dots,2k\,.
\ee

%%%%%%%%%%%%%%%%%%%%%%%%%%%%%%%%%%%%
%%%%%%%%%%%%%%%%%%%%%%%%%%%%%%%%%%%%

\section{Proof of the Completeness of the $SU(N)$ Trinions}
\label{app:Diophantine}

In this section, we prove the completeness of our list of solutions to the conditions \eqref{Diophantine-eq2}, \eqref{Diophantine-eq3} and \eqref{Diophantine-eq4} in section~\ref{sec:SUN}.

\medskip
\noindent 
First of all, it is trivial to see that when $p_1,p_2,p_3\geq 3$, the l.h.s. of \eqref{Diophantine-eq2},\eqref{Diophantine-eq3},\eqref{Diophantine-eq4}  is always strictly smaller than the right hand side. Hence one of $p_\alpha$'s has to be smaller than 3 (equal to 2). 
We start with the equation (\ref{Diophantine-eq2})
\be
\frac{N}{p_1}+\frac{N}{p_2}+\frac{N-1}{p_3}=N\,,
\ee
we assume that $p_1\leq p_2$.

If $p_1=2$, we have
\be
\frac{N}{p_2}+\frac{N-1}{p_3}=\frac{N}{2}\,.
\ee
If $p_2,p_3>3$, then $\frac{N}{p_2}+\frac{N-1}{p_3}<N/2$, and the above equation cannot hold. If $p_2=3$, we have
\be
p_3=\frac{6N-6}{N}<6\,,
\ee
which leads to sporadic solutions $(p_1,p_2,p_3)=(2,3,3),(2,3,4),(2,3,5)$. 

\noindent 
On the other hand, if $p_3=2$, we have the infinite sequence $(p_1,p_2,p_3)=(2,2N,2)$. If $p_3=3$, we have 
\be
p_2=\frac{6N}{N+2}<6\,.
\ee
This leads to sporadic solutions $(p_1,p_2,p_3)=(2,3,3),(2,4,3),(2,5,3)$.

\noindent 
Finally, if $p_3=2$ and $p_1>2$, we have
\be
\frac{1}{p_1}+\frac{1}{p_2}=\frac{N+1}{2N}\,.
\ee
When $p_1,p_2>3$, this equation cannot hold. Hence one has to set $p_1=3$ and
\be
p_2=\frac{6N}{N+3}<6\,.
\ee
It leads to sporadic solutions $(p_1,p_2,p_3)=(3,3,2),(3,4,2),(3,5,2)$. This concludes the enumeration of solutions to (\ref{Diophantine-eq2}).

\noindent 
Next we consider (\ref{Diophantine-eq3})
\be
\frac{N}{p_1}+\frac{N-1}{p_2}+\frac{N-1}{p_3}=N\,,
\ee
we assume that $p_2\leq p_3$.
Then if $p_1=2$, we have
\be
\frac{1}{p_2}+\frac{1}{p_3}=\frac{N}{2(N-1)}> \frac{1}{2}
\ee
If $p_2,p_3>3$, then $\frac{1}{p_2}+\frac{1}{p_3}\leq 1/2$ and the above inequality cannot be satisfied. So we have to set $p_2=2$ or $p_2=3$. For the first case, we get the infinite sequence $(p_1,p_2,p_3)=(2,2,2N-2)$. For the second case, we have
\be
p_3=\frac{6N-6}{N+2}<6\,.
\ee
Thus we have the sporadic solutions $(p_1,p_2,p_3)=(2,3,3),(2,3,4),(2,3,5)$.

\noindent 
If $p_1>2$ and $p_2=2$, then we have 
\be
\frac{N}{p_1}+\frac{N-1}{p_3}=\frac{N+1}{2}\,.
\ee
It is easy to see that if $p_1,p_3>3$, the l.h.s. above is strictly smaller than the r.h.s.. If $p_1=3$, then we have 
\be
p_3=\frac{6N-6}{N+3}<6\,.
\ee
There are the sporadic solutions $(p_1,p_2,p_3)=(3,2,3),(3,2,4),(3,2,5)$. If $p_3=2$, we have the infinite sequence $(p_1,p_2,p_3)=(N,2,2)$. If $p_3=3$, we have
\be
p_1=\frac{6N}{N+5}<6\,.
\ee
This leads to the sporadic solutions $(p_1,p_2,p_3)=(3,2,3),(4,2,3),(5,2,3)$.

\noindent 
For (\ref{Diophantine-eq4})
\be
\frac{N-1}{p_1}+\frac{N-1}{p_2}+\frac{N-1}{p_3}=N\,,
\ee
we assume that $p_1\leq p_2\leq p_3$, and hence we have to set $p_1=2$. Now the equation is simplified to
\be
\frac{1}{p_2}+\frac{1}{p_3}=\frac{N+1}{2N-2}>\frac{1}{2}~.
\ee
If $p_2,p_3>3$, the above inequality cannot be satisfied. So if $p_2=2$, we have the infinite sequence $(p_1,p_2,p_3)=(2,2,N-1)$. If $p_2=3$, we have the equation
\be
p_3=\frac{6N-6}{N+5}<6\,,
\ee
which leads to the sporadic solutions $(p_1,p_2,p_3)=(2,3,3),(2,3,4),(2,3,5)$.

\noindent 
This concludes the enumeration of solutions.

%%%%%%%%%%%%%%%%%%%%%%%%%%%%%%%

\section{Tables of Trinions with $G= D, E$}
\label{App:Tables}

In this appendix we summarize the trinion theories for general $G= D, E$ found in section \ref{sec7:DandE}. 
These are given in the following tables \ref{tab:b2b3h}, \ref{tab:SporadicSO}, \ref{tab:sporadicE6}, \ref{tab:sporadicE7}, \ref{tab:sporadicE8}, \ref{tab:Bounty}, \ref{tab:Twix}, \ref{tab:Snickers}, \ref{tab:Ferrero}.

\begin{table}[hb]
\centering
\begin{tabular}{|c|c||c|c|c||c|c||c|c|}\hline
$SO(2N): \ N$ & $(p_2,p_3)$ & $r$ & $f$  & $d_H$ & $\h r$ & $\h d_H$  & $b_3$ & $\frak{f}$\\
\hline
\hline
$2k$ & $(4,4)$ & $3k-3$ & 6 & $9k+3$ & $9k-3$ & $3k+3$ & 0 & 0 \\
$4k+1$ & $(4,4)$ & $6k-3$ & 3 & $18k+6$ & $18k+3$ & $6k$ & 6 & $\mb{Z}_2^3$\\
$4k+3$ & $(4,4)$ & $6k+1$ & 7 & $18k+17$ & $18k+10$ & $6k+8$ & 0 & 0 \\
$6k$ & $(3,6)$ & $6k-2$ & $4$ & $30k+2$ & $30k-2$ & $6k+2$ & 0 & 0 \\
$6k+1$ & $(3,6)$ & $6k-2$ & 2 & $30k+6$ & $30k+4$ & $6k$ & 8 & $\mb{Z}_2^4$ \\
$6k+2$ & $(3,6)$ & $6k$ & 4 & $30k+12$ & $30k+8$ & $6k+4$ & 0 & 0\\
$6k+3$ & $(3,6)$ & $6k$ & 2 & $30k+16$ & $30k+14$ & $6k+2$ & 0  & $\mb{Z}_2$\\
$6k+4$ & $(3,6)$ & $6k+2$ & 6 & $30k+23$ & $30k+17$ & $6k+8$  & 2 & $\mb{Z}_2$\\
$6k+5$ & $(3,6)$ & $6k+2$ & 2 & $30k+26$ & $30k+24$ & $6k+4$ &  0 & $\mb{Z}_2$ \\
\hline
\end{tabular}

\bigskip
\begin{tabular}{|c|c|c||c|c|c||c|c||c|c|}\hline
$G$ & $(p_2,p_3)$ & $F$ & $r$ & $f$  & $d_H$ & $\h r$ & $\h d_H$  & $b_3$ & $\frak{f}$\\
\hline
\hline
$E_6$ & $(4,4)$ & $x^3+t^4+z_2^4+z_3^4$ & 3 & 0 & 27 & 27 & 3 & 12 & $\mb{Z}_3^3$\\
$E_6$ & $(3,6)$ & $x^3+t^4+z_2^3+z_3^6$ & 5 & 4 & 32 & 28 & 9 & 2 & $\mb{Z}_2$\\
$E_7$ & $(4,4)$ & $x^3+x t^3+z_2^4+z_3^4$ & 9 & 3 & 33 & 30 & 12 & 0 & 0 \\
$E_7$ & $(3,6)$ & $x^3+x t^3+z_2^3+z_3^6$ & 8 & 2 & 36 & 34 & 10 & 4 & $\mb{Z}_3$\\
$E_8$ & $(4,4)$ & $x^3+t^5+z_2^4+z_3^4$ & 12 & 0 & 36 & 36 & 12 & 0 & 0\\
$E_8$ & $(3,6)$ & $x^3+t^5+z_2^3+z_3^6$ & 12 & 0 & 40 & 40 & 12 & 8 & $\mb{Z}_5$\\
\hline
\end{tabular}
\caption{Trinion models of the type described in section \ref{subsec:b2b3h}. with $b_2= b_3 = h^\vee$. \label{tab:b2b3h}}
\end{table}

\begin{table}[hb]
 \begin{tabular}{| c||c |c c|c  |ccc|cc|cc|} \hline
$(b_2, b_3)$& $G$ & $p_2$ & $p_3$ & $F$ & $r$ & $f$ & $d_H$ & $\h r$ & $\h d_H$  & $b_3$ & $\frak{f}$\\%heading
 \hline\hline % inserts single horizontal line
$(2N{-}2, N)$&$SO(8)$& 3 & 4 & $x^3+x t^2+z_2^3+z_3^4 t$ & 2 & 6 & 23 & 17 & 8 & 2 & $\mb{Z}_2$\\
&$SO(8)$& 6 & 2 & $x^3+x t^2+z_2^6+z_3^2 t$ & 2 & 6 & 23 & 17 & 8 & 2 & $\mb{Z}_2$\\
&$SO(12)$& 5 & 2 & $x^5+x t^2+z_2^5+z_3^2 t$ & 4 & 4 & 30 & 26 & 8 & 4 & $\mb{Z}_2^2$\\
\hline\hline
$(N, N)$&$SO(8)$ & 2 & 4 & $x^3+x t^2+z_2^2 t+z_3^4 t+z_3^6$ & 2 & 6 & 23 & 17 & 8 & 2 & $\mb{Z}_2$\\
&$SO(12)$& 2 & 3 & $x^5+x t^2+z_2^2 t+z_3^3 t+z_3^5$ & 4 & 4 & 30 & 26 & 8 & 4 & $\mb{Z}_2^2$\\
\hline
\end{tabular} 
\caption{Sporadic models for $G=SO(N)$. Note that there are only two inequivalent theories in the table above, and the one with $r=2$ is the same as the rank-2 $E_6$ theory.
\label{tab:SporadicSO}}
\end{table}

\begin{table}
\centering
 \begin{tabular}{| c||c |c c|c  |ccc|cc|cc|} \hline
$(b_2, b_3)$& $G$ & $p_2$ & $p_3$ & $F$ & $r$ & $f$ & $d_H$ & $\h r$ & $\h d_H$  & $b_3$ & $\frak{f}$\\%heading
 \hline\hline % inserts single horizontal line
$(12,9)$&$E_6$ & 4 & 3 & $x^3+t^4+z_2^4+z_3^3 t$ & 3 & 0 & 27 & 27 & 3 & 12 & $\mb{Z}_3^3$ \\
&$E_6$ & 8 & 2 & $x^3+t^4+z_2^8+z_3^2 t$ & 3 & 0 & 35 & 35 & 3 & 4 & $\mb{Z}_3^2$ \\
\hline\hline
$(12,8)$ &$E_6$ & 3 & 4 & $x^3+t^4+z_2^3+z_3^4 x$ & 5 & 4 & 32 & 28 & 9 & 2 & $\mb{Z}_2$\\
                &$E_6$ & 6 & 2 & $x^3+t^4+z_2^6+z_3^2 x$ & 5 & 4 & 32 & 28 & 9 & 2 & $\mb{Z}_2$\\
\hline\hline
$(8,8)$ &$E_6$ & 2 & 4 & $x^3+t^4+z_2^2 x+z_3^4 x+z_3^6$ & 5 & 4 & 32 & 28 & 9 & 0 & $\mb{Z}_2$\\
\hline\hline
$(9,9)$ &$E_6$ & 2 & 6 & $x^3+t^4+z_2^2 t+z_3^6 t+z_3^8$ & 3 & 0 & 35 & 35 & 3 & 2 & $\mb{Z}_3^2$\\
                &$E_6$ & 3 & 3 & $x^3+t^4+z_2^3 t+z_3^3 t+z_3^4$ & 3 & 0 & 27 & 27 & 3 & 4 & $\mb{Z}_3^3$\\
\hline
\end{tabular}
\caption{Sporadic models for $G=E_6$. Note that there are no solutions with $(b_2, b_3)=(8,9)$, and the two solutions for $(b_2,b_3)=(12,8)$ are the same.
 \label{tab:sporadicE6}}
\end{table}

\begin{table}[!htbp]
\centering
 \begin{tabular}{| c||c |c c|c  |ccc|cc|cc|} \hline
$(b_2, b_3)$& $G$ & $p_2$ & $p_3$ & $F$ & $r$ & $f$ & $d_H$ & $\h r$ & $\h d_H$  & $b_3$ & $\frak{f}$\\%heading
 \hline\hline % inserts single horizontal line
$(18,14)$&$E_7$ & 9 & 2 & $x^3+x t^3+z_2^9+z_3^2 t$ & 6 & 2 & 45 & 43 & 8 & 6 & $\mb{Z}_2^3$\\
\hline\hline
$(14,14)$ &$E_7$ & 2 & 7 & $x^3+x t^3+z_2^2 t+z_3^7 t+z_3^9$ & 6 & 2 & 45 & 43 & 8 & 6 & $\mb{Z}_2^3$\\
\hline
\end{tabular}
\caption{Sporadic models for $G=E_7$. The two theories are equivalent. \label{tab:sporadicE7}}
\end{table}

\begin{table}[!htbp]
\centering
 \begin{tabular}{| c||c |c c|c  |ccc|cc|cc|} \hline
$(b_2, b_3)$& $G$ & $p_2$ & $p_3$ & $F$ & $r$ & $f$ & $d_H$ & $\h r$ & $\h d_H$  & $b_3$ & $\frak{f}$\\%heading
 \hline\hline % inserts single horizontal line
$(30,24)$&$E_8$ & 10 & 2 & $x^3+t^5+z_2^{10}+z_3^2 t$ & 10 & 0 & 54 & 54 & 10 & 8 & $\mb{Z}_3^2$ \\
\hline\hline
$(30,20)$&$E_8$ & 3 & 4 & $x^3+t^5+z_2^3+z_3^4 x$ & 12 & 0 & 40 & 40 & 12 & 8 & $\mb{Z}_5$ \\
                   &$E_8$ & 6 & 2 & $x^3+t^5+z_2^6+z_3^2 x$ & 12 & 0 & 40 & 40 & 12 & 8 & $\mb{Z}_5$ \\
\hline\hline
$(24,24)$&$E_8$ & 2 & 8 & $x^3+t^5+z_2^2 t+z_3^8 t+z_3^{10}$ & 10 & 0 & 54 & 54 & 10 & 4 & $\mb{Z}_3^2$\\
\hline\hline
$(20,20)$&$E_8$ & 2 & 4 & $x^3+t^5+z_2^2 x+z_3^4 x+z_3^6$ & 12 & 0 & 40 & 40 & 12 & 0 & $\mb{Z}_5$\\
\hline
\end{tabular}
\caption{Sporadic models with $G= E_8$. 
Note that there are no solutions with $(b_2, b_3)=(24,20)$. The two cases for $(b_2,b_3)=(30,20)$ are equivalent.\label{tab:sporadicE8}}
\end{table}

\begin{table}[!htbp]
\centering
 \begin{tabular}{| c||c |c c|c  |ccc|cc|cc|} \hline
$(b_2, b_3)$& $G$ & $p_2$ & $p_3$ & $F$ & $r$ & $f$ & $d_H$ & $\h r$ & $\h d_H$  & $b_3$ & $\frak{f}$\\%heading
 \hline\hline % inserts single horizontal line
$(2N{-}2, 2N-2)$&$SO(8)$& 4 & 12 & $x^3+y^2+z_2^4+z_3^{12}$ & 0 & 6 & 36 & 30 & 6 & 0 & 0\\
&$SO(8)$& 6 & 6 & $x^3+y^2+z_2^6+z_3^{6}$ & 1 & 8 & 29 & 21 & 9 & 0 & 0\\
&$SO(10)$& 3 & 24 & $x^4+y^2+z_2^3+z_3^{24}$ & 0 & 6 & 72 & 66 & 6 & 0 & 0\\
&$SO(10)$& 4 & 8 & $x^4+y^2+z_2^4+z_3^8$ & 2 & 7 & 35 & 28 & 9 & 2 & $\mb{Z}_2$\\
&$SO(12)$& 3 & 15 & $x^5+y^2+z_2^3+z_3^{15}$ & 0 & 0 & 56 & 56 & 0 & 0 & $\mb{Z}_2^4$\\
&$SO(12)$& 5 & 5 & $x^5+y^2+z_2^5+z_3^{5}$ & 2 & 0 & 32 & 32 & 2 & 12 & $\mb{Z}_2^6$\\
&$SO(14)$& 3 & 12 & $x^6+y^2+z_2^3+z_3^{12}$ & 2 & 8 & 59 & 51 & 10 & 2 & $\mb{Z}_2$\\
&$SO(14)$& 4 & 6 & $x^6+y^2+z_2^4+z_3^{6}$ & 4 & 5 & 40 & 35 & 9 & 4 & $\mb{Z}_2^2$\\
&$SO(20)$& 3 & 9 & $x^9+y^2+z_2^3+z_3^{9}$ & 3 & 0 & 64 & 64 & 3 & 14 & $\mb{Z}_2^7$\\
&$SO(22)$& 4 & 5 & $x^{10}+y^2+z_2^4+z_3^{5}$ & 6 & 4 & 56 & 52 & 10 & 4 & $\mb{Z}_2^2$\\
&$SO(26)$& 3 & 8 & $x^{12}+y^2+z_2^3+z_3^{8}$ & 5 & 6 & 80 & 74 & 11 & 2 & $\mb{Z}_2$\\
&$SO(44)$& 3 & 7 & $x^{21}+y^2+z_2^3+z_3^{7}$ & 6 & 0 & 120 & 120 & 6 & 12 & $\mb{Z}_2^6$\\
\hline
%\end{tabular} 
%\end{table}
%
%\be
% \begin{tabular}{| c||c |c c|c  |ccc|cc|cc|} 
%$(b_2, b_3)$& $G$ & $p_2$ & $p_3$ & $F$ & $r$ & $f$ & $d_H$ & $\h r$ & $\h d_H$  & $b_3$ & $\frak{f}$\\%heading
% \hline\hline % inserts single horizontal line
$(2N-2, N)$&$SO(8)$ & 4 & 8 & $x^3+y^2+z_2^4+z_3^{12}$ & 0 & 6 & 36 & 30 & 6 & 0 & 0\\
&$SO(8)$ & 5 & 5 & $x^3+y^2+z_2^5+z_3^5 x$ & 0 & 0 & 26 & 26 & 0 & 0 & $\mb{Z}_2^2$\\
&$SO(8)$ & 6 & 4 & $x^3+y^2+z_2^6+z_3^6$ & 1 & 8 & 29 & 21 & 9 & 0 & 0\\
&$SO(8)$ & 9 & 3 & $x^3+y^2+z_2^9+z_3^3 x$ & 0 & 0 & 28 & 28 & 0 & 0 & $\mb{Z}_2^3$\\
&$SO(10)$ & 3 & 15 & $x^4+y^2+z_2^3+z_3^{24}$ & 0 & 6 & 72 & 66 & 6 & 0 & 0\\
&$SO(10)$ & 4 & 5 & $x^4+y^2+z_2^4+z_3^8$ & 2 & 7 & 35 & 28 & 9 & 2 & $\mb{Z}_2$\\
&$SO(10)$ & 6 & 3 & $x^4+y^2+z_2^6+z_3^4 z_2$ & 2 & 3 & 30 & 27 & 5 & 0 & 0\\
&$SO(10)$ & 16 & 2 & $x^4+y^2+z_2^{16}+z_3^3 z_2$ & 1 & 7 & 53 & 46 & 8 & 0 & 0\\
&$SO(12)$ & 3 & 9 & $x^5+y^2+z_2^3+z_3^{15}$ & 0 & 0 & 56 & 56 & 0 & 0 & $\mb{Z}_2^4$\\
&$SO(12)$ & 4 & 4 & $x^5+y^2+z_2^4+z_3^5 z_2$ & 2 & 4 & 36 & 32 & 6 & 0 & 0\\
&$SO(12)$ & 5 & 3 & $x^5+y^2+z_2^5+z_3^5$ & 2 & 0 & 32 & 32 & 2 & 12 & $\mb{Z}_2^6$\\
&$SO(12)$ & 10 & 2 & $x^5+y^2+z_2^{10}+z_3^3 z_2$ & 2 & 8 & 46 & 38 & 10 & 0 & 0\\
&$SO(14)$ & 3 & 7 & $x^6+y^2+z_2^3+z_3^{12}$ & 2 & 8 & 59 & 51 & 10 & 2 & $\mb{Z}_2$\\
&$SO(14)$ & 8 & 2  & $x^6+y^2+z_2^8+z_3^{3}z_2$ & 3 & 5 & 45 & 40 & 8 & 0 & 0\\
&$SO(16)$ & 3 & 6 & $x^7+y^2+z_2^3+z_3^9 x$ & 1 & 0 & 57 & 57 & 1 & 0 & $\mb{Z}_2^3$\\
&$SO(16)$ & 7 & 2 & $x^7+y^2+z_2^7+z_3^3 x$ & 3 & 0 & 45 & 45 & 3 & 12 & $\mb{Z}_2^6$\\
&$SO(18)$ & 4 & 3 & $x^8+y^2+z_2^4+z_3^4 z_2$ & 5 & 5 & 48 & 43 & 10 & 2 & $\mb{Z}_2$\\
&$SO(20)$ & 3 & 5 & $x^9+y^2+z_2^3+z_3^9$ & 3 & 0 & 64 & 64 & 3 & 14 & $\mb{Z}_2^7$\\
&$SO(20)$ & 6 & 2 & $x^9+y^2+z_2^6+z_3^3 z_2$ & 5 & 6 & 55 & 49 & 11 & 0 & 0\\
&$SO(32)$ & 3 & 4 & $x^{15}+y^2+z_2^3+z_3^4 z_2$ & 2 & 4 & 72 & 68 & 6 & 4 & $\mb{Z}_5$\\
&$SO(32)$ & 5 & 2 & $x^{15}+y^2+z_2^5+z_3^3 z_2$ & 6 & 0 & 77 & 77 & 6 & 10 & $\mb{Z}_2^5$\\
\hline
$(N, N)$&$SO(8)$ & 3 & 6 & $x^3+y^2+z_2^4 z_3+z_3^{9}$ & 0 & 0 & 28 & 28 & 0 & 0 & $\mb{Z}_2^3$\\
&$SO(8)$ & 4 & 4 & $x^3+y^2+z_2^6+z_3^6$ & 1 & 8 & 29 & 21 & 9 & 0 & 0\\
&$SO(10)$ & 2 & 10 & $x^4+y^2+z_2^3 z_3+z_3^{16}$ & 1 & 7 & 53 & 46 & 8 & 0 & 0\\
&$SO(12)$ & 2 & 6 & $x^5+y^2+z_2^3 z_3+z_3^{10}$ & 2 & 8 & 46 & 38 & 10 & 0 & 0\\
&$SO(12)$ & 3 & 3 & $x^5+y^2+z_2^5+z_3^5$ & 2 & 0 & 32 & 32 & 2 & 12 & $\mb{Z}_2^6$\\
&$SO(16)$ & 2 & 4 & $x^7+y^2+z_2^3 z_3+z_3^7$ & 3 & 0 & 45 & 45 & 3 & 12 & $\mb{Z}_2^6$\\
&$SO(24)$ & 2 & 3 & $x^{11}+y^2+z_2^3 z_3+z_3^4 z_2$ & 5 & 0 & 60 & 60 & 5 & 10 & $\mb{Z}_2^5$\\
\hline
\end{tabular}
\caption{$G=SO(2N)$ models with  $b_1=N$, $p_1=1$. \label{tab:Bounty}}
\end{table}

\begin{table}[!htbp]
\centering
 \begin{tabular}{| c||c c|c  |ccc|cc|cc|} \hline
$(b_1,b_2, b_3)$& $p_2$ & $p_3$ & $F$ & $r$ & $f$ & $d_H$ & $\h r$ & $\h d_H$  & $b_3$ & $\frak{f}$\\%heading
\hline\hline
$(9,12,12)$ & 5 & 20 & $x^3+y^2+z_2^5+z_3^{20}$ & 0 & 0 & 76 & 76  & 0 & 0 & $\mb{Z}_3^2$\\
& 6 & 12 & $x^3+y^2+z_2^6+z_3^{12}$ & 2 & 8 & 59 & 51 & 10 & 2 & $\mb{Z}_2$\\
& 8 & 8 & $x^3+y^2+z_2^8+z_3^8$ & 2 & 0 & 49 & 49 & 2 & 6 & $\mb{Z}_3^3$\\
\hline\hline
$(9,12,9)$ & 5 & 15 & $x^3+y^2+z_2^5+z_3^{20}$ & 0 & 0 & 76 & 76 & 0 & 0 & $\mb{Z}_3^2$ \\
& 6 & 9 & $x^3+y^2+z_2^6+z_3^{12}$ & 2 & 8 & 59 & 51 & 10 & 2 & $\mb{Z}_2$\\
& 7 & 7 & $x^3+y^2+z_2^7+z_3^8 z_2$ & 1 & 0 & 50 & 50 & 1 & 0 & $\mb{Z}_3$ \\
& 8 & 6 & $x^3+y^2+z_2^8+z_3^8$ & 2 & 0 & 49 & 49 & 2 & 6 & $\mb{Z}_3^3$ \\
& 10 & 5 & $x^3+y^2+z_2^{10}+z_3^6 z_2$ & 1 & 0 & 51 & 51 & 1 & 0 & $\mb{Z}_3$ \\
& 16 & 4 & $x^3+y^2+z_2^{16}+z_3^5 z_2$ & 1 & 0 & 65 & 65 & 1 & 0 & $\mb{Z}_3^2$ \\
\hline\hline
$(9,12,8)$ & 6 & 8 & $x^3+y^2+z_2^6+z_3^8 x$ & 2 & 8 & 59 & 51 & 10 & 2 & $\mb{Z}_2$\\
& 12 & 4 & $x^3+y^2+z_2^{12}+z_3^4 x$ & 2 & 8 & 59 & 51 & 10 & 2 & $\mb{Z}_2$\\
& 36 & 3 & $x^3+y^2+z_2^{36}+z_3^3 x$ & 0 & 7 & 126 & 119 & 7 & 0 & 0\\
\hline\hline
$(9,9,9)$ & 4 & 12 & $x^3+y^2+z_2^5 z_3+z_3^{16}$ & 1 & 0 & 65 & 65 & 1 & 0 & $\mb{Z}_3^2$ \\
& 6 & 6 & $x^3+y^2+z_2^8+z_3^8$ & 2 & 0 & 49 & 49 & 2 & 6 & $\mb{Z}_3^3$ \\
\hline\hline
$(9,9,8)$ & 9 & 4 & $x^3+y^2+z_2^{12}+z_3^4 x$ & 2 & 8 & 59 & 51 & 10 & 2 & $\mb{Z}_2$\\
& 27 & 3 & $x^3+y^2+z_2^{36}+z_3^3 x$ & 0 & 7 & 126 & 119 & 7 & 0 & 0\\
\hline\hline
$(9,8,8)$ & 3 & 24 & $x^3+y^2+z_2^3 x+z_3^{24}x+z_3^{36}$ & 0 & 7 & 126 & 119 & 7 & 0 & 0 \\
& 4 & 8 & $x^3+y^2+z_2^4 x+z_3^{8}x+z_3^{12}$ & 2 & 8 & 59 & 51 & 10 & 2 & $\mb{Z}_2$\\
\hline\hline
$(8,12,12)$ & 4 & 12 & $t^4+y^2+z_2^4+z_3^{12}$ & 3 & 7 & 53 & 46 & 10 & 4 & $\mb{Z}_3$\\
& 6 & 6 & $t^4+y^2+z_2^6+z_3^6$ & 4 & 5 & 40 & 35 & 9 & 4 & $\mb{Z}_2^2$\\
\hline\hline
$(8,12,9)$ & 4 & 9 & $t^4+y^2+z_2^4+z_3^9 t$ & 3 & 7 & 53 & 46 & 10 & 4 & $\mb{Z}_3$\\
& 12 & 3 & $t^4+y^2+z_2^{12}+z_3^3 t$ & 3 & 7 & 53 & 46 & 10 & 4 & $\mb{Z}_3$\\
\hline\hline
$(8,12,8)$ & 4 & 8 & $t^4+y^2+z_2^4+z_3^{12}$ & 3 & 7 & 53 & 46 & 10 & 4 & $\mb{Z}_3$\\
& 5 & 5 & $t^4+y^2+z_2^5+z_3^6 z_2$ & 3 & 2 & 40 & 38 & 5 & 0 & 0\\
& 6 & 4 & $t^4+y^2+z_2^6+z_3^6$ & 4 & 5 & 40 & 35 & 9 & 4 & $\mb{Z}_2^2$\\
& 9 & 3 & $t^4+y^2+z_2^9+z_3^4 z_2$ & 4 & 4 & 44 & 40 & 8 & 0 & 0\\
\hline\hline
$(8,9,9)$ & 3 & 9 & $t^4+y^2+z_2^3 t+z_3^9 t+z_3^{12}$ & 3 & 7 & 53 & 46 & 10 & 4 & $\mb{Z}_3$\\
\hline\hline
$(8,9,8)$ & 3 & 8 & $t^4+y^2+z_2^3 t+z_3^{12}$ & 3 & 7 & 53 & 46 & 10 & 4 & $\mb{Z}_3$\\
\hline\hline
$(8,8,8)$ & 3 & 6 & $t^4+y^2+z_2^4 z_3+z_3^9$ & 4 & 4 & 44 & 40 & 8 & 0 & 0\\
& 4 & 4 & $t^4+y^2+z_2^6+z_3^6$ & 4 & 5 & 40 & 35 & 9 & 4 & $\mb{Z}_2^2$\\
\hline
\end{tabular}
\caption{Models for $G=E_6$ with $b_1=8, 9$. \label{tab:Twix}}
\end{table}

\begin{table}[!htbp]
\centering
 \begin{tabular}{| c||c c|c  |ccc|cc|cc|} \hline
$(b_2, b_3)$& $p_2$ & $p_3$ & $F$ & $r$ & $f$ & $d_H$ & $\h r$ & $\h d_H$  & $b_3$ & $\frak{f}$\\%heading
\hline\hline
$(18,18)$ & 5 & 45 & $x^3+y^2+z_2^5+z_3^{45}$ & 0 & 0 & 176 & 176 & 0 & 0 & $\mb{Z}_2^4$\\
& 6 & 18 & $x^3+y^2+z_2^6+z_3^{18}$ & 3 & 8 & 89 & 81 & 11 & 4 & $\mb{Z}_3$\\
& 9 & 9 & $x^3+y^2+z_2^9+z_3^9$ & 3 & 0 & 64 & 64 & 3 & 14 & $\mb{Z}_2^7$\\
\hline\hline
$(18,14)$ & 5 & 35 & $x^3+y^2+z_2^5+z_3^{45}$ & 0 & 0 & 176 & 176 & 0 & 0 & $\mb{Z}_2^4$\\
& 6 & 14 & $x^3+y^2+z_2^6+z_3^{18}$ & 3 & 8 & 89 & 81 & 11 & 4 & $\mb{Z}_3$\\
& 8 & 8 & $x^3+y^2+z_2^8+z_3^9 z_2$ & 3 & 2 & 66 & 64 & 5 & 0 & 0\\
& 9 & 7 & $x^3+y^2+z_2^9+z_3^9$ & 3 & 0 & 64 & 64 & 3 & 14 & $\mb{Z}_2^7$\\
& 15 & 5 & $x^3+y^2+z_2^{15}+z_3^6 z_2$ & 2 & 0 & 76 & 76 & 2 & 0 & $\mb{Z}_2$\\
& 36 & 4 & $x^3+y^2+z_2^{36}+z_3^5 z_2$ & 1 & 8 & 149 & 141 & 9 & 0 & 0\\
\hline\hline
$(14,14)$ & 4 & 28 & $x^3+y^2+z_2^5 z_3+z_3^{36}$ & 1 & 8 & 149 & 141 & 9 & 0 & 0\\
& 7 & 7 & $x^3+y^2+z_2^9+z_3^9$ & 3 & 0 & 64 & 64 & 3 & 14 & $\mb{Z}_2^7$\\
\hline
\end{tabular}
\caption{$G= E_7$ with $b_1= 14$ models. \label{tab:Snickers}}
\end{table}

\begin{table}[!htbp]
\centering
 \begin{tabular}{| c||c c|c  |ccc|cc|cc|} \hline
$(b_1,b_2, b_3)$& $p_2$ & $p_3$ & $F$ & $r$ & $f$ & $d_H$ & $\h r$ & $\h d_H$  & $b_3$ & $\frak{f}$\\%heading
\hline\hline
$(24,30,30)$ & 6 & 30 & $x^3+y^2+z_2^6+z_3^{30}$ & 5 & 8 & 149 & 141 & 13 & 8 & $\mb{Z}_5$\\
&10 & 10 & $x^3+y^2+z_2^{10}+z_3^{10}$ & 5 & 0 & 81 & 81 & 5 & 16 & $\mb{Z}_3^4$\\
\hline\hline
$(24,30,24)$ & 6 & 24 & $x^3+y^2+z_2^6+z_3^{30}$ & 5 & 8 & 149 & 141 & 13 & 8 & $\mb{Z}_5$\\
& 7 & 14 & $x^3+y^2+z_2^7+z_3^{15} z_2$ & 4 & 0 & 99 & 99 & 4 & 0 & 0\\
& 9 & 9 & $x^3+y^2+z_2^9+z_3^{10} z_2$ & 6 & 0 & 82 & 82 & 6 & 0 & $\mb{Z}_2$\\
& 10 & 8 & $x^3+y^2+z_2^{10}+z_3^{10}$ & 5 & 0 & 81 & 81 & 5 & 16 & $\mb{Z}_3^4$\\
& 15 & 6 & $x^3+y^2+z_2^{15}+z_3^7 z_2$ & 5 & 0 & 91 & 91 & 5 & 12 & $\mb{Z}_2^6$\\
& 25 & 5 & $x^3+y^2+z_2^{25}+z_3^6 z_2$ & 4 & 0 & 126 & 126 & 4 & 0 & 0\\
\hline\hline
$(24,30,20)$ & 6 & 20 & $x^3+y^2+z_2^6+z_3^{20}x$ & 5 & 8 & 149 & 141 & 13 & 8 & $\mb{Z}_{5}$\\
&  15 & 5 & $x^3+y^2+z_2^{15}+z_3^{5}x$ & 5 & 0 & 91 & 91 & 5 & 12 & $\mb{Z}_2^6$\\
& 30 & 4 & $x^3+y^2+z_2^{30}+z_3^{4}x$ & 5 & 8 & 149 & 141 & 13 & 8 & $\mb{Z}_{5}$\\
\hline\hline
$(24,24,24)$ & 5 & 20 & $x^3+y^2+z_2^6 z_3+z_3^{25}$ & 4 & 0 & 126 & 126 & 4 & 0 & 0\\
& 6 & 12 & $x^3+y^2+z_2^7 z_3+z_3^{15}$ & 5 & 0 & 91 & 91 & 5 & 12 & $\mb{Z}_2^6$\\
& 8 & 8 & $x^3+y^2+z_2^{10}+z_3^{10}$ & 5 & 0 & 81 & 81 & 5 & 16 & $\mb{Z}_3^4$\\
\hline\hline
$(24,24,20)$ & 6 & 10 & $x^3+y^2+z_2^7 z_3+z_3^{10}x$ & 5 & 0 & 91 & 91 & 5 & 12 & $\mb{Z}_2^6$\\ 
& 9 & 6 & $x^3+y^2+z_2^{10}z_3+z_3^6 x$ & 6 & 0 & 82 & 82 & 6 & 0 & $\mb{Z}_2$\\
& 12 & 5 & $x^3+y^2+z_2^{15}+z_3^5 x$ & 5 & 0 & 91 & 91 & 5 & 12 & $\mb{Z}_2^6$\\
& 24 & 4 & $x^3+y^2+z_2^{30}+z_3^4 x$ & 5 & 8 & 149 & 141 & 13 & 8 & $\mb{Z}_5$\\
\hline\hline
$(24,20,20)$ & 4 & 20 & $x^3+y^2+z_2^4 x+z_3^{20}x+z_3^{30}$ & 5 & 8 & 149 & 141 & 13 & 8 & $\mb{Z}_{5}$\\
& 5 & 10 & $x^3+y^2+z_2^5 x+z_3^{10}x+z_3^{15}$ & 5 & 0 & 91 & 91 & 5 & 4 & $\mb{Z}_2^6$\\
\hline
%\end{tabular}~
%\ee
%
%%\be
% \begin{tabular}{| c||c c|c  |ccc|cc|cc|} 
%$(b_1,b_2, b_3)$& $p_2$ & $p_3$ & $F$ & $r$ & $f$ & $d_H$ & $\h r$ & $\h d_H$  & $b_3$ & $\frak{f}$\\%heading
%\hline\hline
$(20,30,30)$ & 4 & 12 & $t^5+y^2+z_2^4+z_3^{12}$ & 8 & 0 & 66 & 66 & 8 & 4 & $\mb{Z}_5$\\
& 6 & 6 & $t^5+y^2+z_2^6+z_3^6$ & 6 & 0 & 50 & 50 & 6 & 16 & $\mb{Z}_5^2$\\
\hline\hline
$(20,30,24)$ & 5 & 6 & $t^5+y^2+z_2^5+z_3^6 t$ & 10 & 0 & 52 & 52 & 10 & 4 & $\mb{Z}_2^2$\\
& 15 &3 & $t^5+y^2+z_2^{15}+z_3^3 t$ & 6 & 0 & 77 & 77 & 6 & 10 & $\mb{Z}_2^5$\\
\hline\hline
$(20,30,20)$ & 4 & 8 & $t^5+y^2+z_2^4+z_3^{12}$ & 8 & 0 & 66 & 66 & 8 & 4 & $\mb{Z}_5$\\
& 5 & 5 & $t^5+y^2+z_2^5+z_3^6 z_2$ & 10 & 0 & 52 & 52 & 10 & 4 & $\mb{Z}_2^2$\\
& 6 & 4 & $t^5+y^2+z_2^6+z_3^6$ & 6 & 0 & 50 & 50 & 6 & 16 & $\mb{Z}_5^2$\\
& 9 & 3 & $t^5+y^2+z_2^9+z_3^4 z_2$ & 8 & 0 & 56 & 56 & 8 & 0 & 0 \\
\hline\hline
$(20,24,24)$ & 3 & 12 & $t^5+y^2+z_2^3 t+z_3^{12} t+z_3^{15}$ & 6 & 0 & 77 & 77 & 6 & 10 & $\mb{Z}_2^5$\\
& 4 & 6 & $t^5+y^2+z_2^4 t+z_3^6 t+z_2^5$ & 10 & 0 & 52 & 52 & 10 & 4 & $\mb{Z}_2^2$\\
\hline\hline
$(20,24,20)$ & 3 & 10 & $t^5+y^2+z_2^3 t+z_3^{15}$ & 6 & 0 & 77 & 77 & 6 & 10 & $\mb{Z}_2^5$\\
& 4 & 5 & $t^5+y^2+z_2^4 t+z_3^6 t+z_2^5$ & 10 & 0 & 52 & 52 & 10 & 4 & $\mb{Z}_2^2$\\
\hline\hline
$(20,20,20)$ & 3 & 6 & $t^5+y^2+z_2^4 z_3+z_3^9$ & 8 & 0 & 56 & 56 & 8 & 0 & 0 \\
& 4 & 4 & $t^5+y^2+z_2^6+z_3^6$ & 6 & 0 & 50 & 50 & 6 & 16 & $\mb{Z}_5^2$\\
\hline
\end{tabular}
\caption{$G=E_8$ models with $b_1= 20, 24$. \label{tab:Ferrero}}
\end{table}

\clearpage

%
%\bibliographystyle{utphys}
%%\bibliographystyle{plain}
%\bibliography{bib5d}{}

\bibliography{FM}
\bibliographystyle{JHEP}
\end{document}